\documentclass[twoside,a4paper,10pt]{report} 
\usepackage{latexsym,amsmath,amssymb,graphicx,psfrag,fancyhdr,epsfig} 
\usepackage{cite}

\newcommand{\beq}{\begin{equation}}
\newcommand{\eeq}{\end{equation}}
\newcommand{\bea}{\begin{eqnarray}}
\newcommand{\eea}{\end{eqnarray}}

\begin{document}
\pagestyle{empty}

\pagestyle{empty}
  \vspace{2cm}
  \sffamily\upshape\mdseries
  \noindent
  {\Huge \textbf{Perfect Gauge Actions on \vspace*{0.4cm}\\
    Anisotropic Lattices}}
  
  \vspace{10.5cm}

  \noindent
  Inauguraldissertation \\
  der Philosophisch-naturwissenschaftlichen Fakult\"at \\
  der Universit\"at Bern

  \vspace{1.5cm}

  \noindent
  vorgelegt von

  \vspace{4mm}

  \noindent  {\large \textbf{Philipp R\"ufenacht}}

  \vspace{3mm}

  \noindent
  von Grossh\"ochstetten (BE)

  \vspace{1.5cm}

  \noindent
  Leiter der Arbeit: \parbox[t]{7cm}{Prof.~Dr.~P.~Hasenfratz \\
    Institut f\"ur theoretische Physik \\
    Universit\"at Bern}
  \normalfont
  \cleardoublepage

\pagestyle{fancy}
\pagenumbering{roman}
\tableofcontents
\newpage
\pagestyle{empty}
\cleardoublepage

\pagestyle{fancy}
\pagenumbering{arabic}
\setcounter{page}{1}

\fancyhead[RE]{\nouppercase{\small\it Abstract and Summary}}
\chapter*{Abstract and Summary}
\addcontentsline{toc}{chapter}{Abstract and Summary}

The theory of the strong interaction, Quantum Chromodynamics (QCD), may be studied using perturbation theory, 
the standard tool of quantum field theory, for high energies. Around 1~GeV, the scale of the hadronic world,
however, the coupling constant is increased and perturbative methods fail. In this domain, Lattice
QCD provides a systematic approach to the non-perturbative study of QCD. The theory is formulated on a
discrete Euclidean space-time grid which non-perturbatively regularises QCD and allows for computer
simulations of the theory using Monte Carlo methods. 

The discretisation of the continuum action can be done in many different ways. Concerning the pure gauge
sector of QCD, besides the standard discretisation, the Wilson action, there are several so-called
improved actions which reduce artifacts coming from the discretisation. The most radical scheme of improving
lattice actions, based on Wilson's Renormalisation Group approach, has been suggested by Hasenfratz and
Niedermayer, namely the creation of actions that are classically perfect, i.e.~there are no lattice artifacts
on the solutions of the lattice equations of motion.

In Lattice QCD, the energy of a physical state is measured studying the decay of correlators of creation
and annihilation operators having an overlap with the state under consideration. If the state is heavy,
these correlators decay very fast in time and one has to make sure that the temporal lattice spacing is small
enough such that the signal of the correlator can be accurately traced over a few time slices before it disappears
in the statistical noise. At the same time, one has to pay attention to choose the physical volume of the lattice
large enough such that there are no significant finite-size effects. Both these requirements together lead to
lattices with a large number of lattice sites, which means that it is computationally very expensive to
perform the simulations. The obvious way out of this dilemma is to use a smaller lattice spacing in 
temporal direction compared to the spatial directions, i.e.~using anisotropic lattices.

Anisotropic lattices have been widely used in the last few years. Studies comprising
excited states of nucleons, heavy-quark bound states, heavy meson semi-leptonic decays, long range
properties of the quark-antiquark potential as well as states composed purely of gluons (glueballs) or
gluons and quarks (hybrids) have been performed, mainly using the standard Wilson discretisation of the
anisotropic action or a mean-link and Symanzik improved anisotropic action. A classically perfect
anisotropic action has still been absent.

It is thus the goal of this work to fill this gap, presenting a way of constructing classically perfect
anisotropic gauge actions, building up on a recent parametrisation of the isotropic classically perfect action
(FP action) that includes a rich structure of operators as it bases on plaquettes built from simple gauge links
as well as from smeared links.
The procedure leading to the anisotropic action is examined analytically on scalar fields as well as
in the quadratic approximation for gauge fields. We then construct the $\xi=2$ action valid on coarse configurations
occurring typically in Monte Carlo simulations. Its properties are studied performing measurements of the torelon
dispersion relation (which serves as a means of determining the renormalisation of the bare (input) anisotropy), of the
potential between a static quark and a static antiquark, of the deconfining phase transition and finally determining
the spectrum of low-lying glueballs in pure gauge theory.

The feasibility of iterating the procedure, obtaining a classically perfect action with $\xi=4$, is briefly
checked. Furthermore, we examine properties of the newly created anisotropic actions (as well as of the underlying isotropic
action), such as autocorrelation times of different updates of the gauge configurations in the Monte Carlo
simulation or the computational overhead of the classically perfect action compared to the widely-used mean-link
and Symanzik improved action and to the standard Wilson gauge action.

It turns out that the construction of anisotropic classically perfect gauge actions is feasible. The iteration of
the process yielding higher anisotropies seems to work as well. Measuring the renormalised anisotropy using the
torelon dispersion relation turns out to be stable and unambiguous and shows that the renormalisation of the
anisotropy is moderate and under
good control. The measurements of the static quark-antiquark potential indicate that the violations of rotational
symmetry are small if the (spatial) lattice is not exceptionally coarse. This shows that the parametrisation 
describes accurately the full action, which is known to have good properties concerning rotational symmetry.
The study of the glueball spectrum is facilitated a lot due to the anisotropic nature of the action, even for
(rather small) $\xi=2$. Reliable results, including continuum extrapolations, are obtained for glueball states
having much larger mass than the highest-lying states that could have been resolved with the same amount of
computational work using the isotropic action.
However, the scaling properties of the glueball states, i.e.~the behaviour of the measured energies as the lattice
spacing is increased, seems to be rather unfavourable compared to the mean-link and Symanzik improved anisotropic
action. Especially, this is the case for the lowest-lying scalar glueball which could be caused by the presence
of a critical end point of a line of phase transitions in the fundamental-adjoint coupling plane assumed to 
define the continuum limit of a scalar field theory. Our action includes in its rich structure operators 
transforming according to the adjoint representation. If their coupling (which we do not control applying our method)
lies in a certain region, the effect of the critical end-point on scalar quantities at certain lattice spacings
(sometimes called the ``scalar dip'') may even be
enhanced compared to other (more standard) discretisations with purely fundamental operators.
Concerning the cut-off effects observed in the glueball simulations, we have to add that other effects, such as
effects due to the finite size of the lattices, may also be present and that this issue requires further study.
Furthermore, the computational overhead of the classically perfect anisotropic action compared to the standard Wilson
action and as well compared to the mean-link and Symanzik improved action is considerable and one has to weigh
up the pros and cons carefully, if one is about to choose a gauge action for large-scale simulations.

\fancyhead[RE]{\nouppercase{\small\it Chapter \thechapter.\, Introduction}}
\chapter{Introduction}

This thesis covers work done in collaboration with Urs Wenger and Ferenc Niedermayer.
Results already extensively discussed in Urs Wenger's PhD thesis \cite{Wenger:2000aa}
that are necessary for the work presented here are briefly recapitulated. Parts of
the results have already been published \cite{Niedermayer:2000yx,Niedermayer:2000ts,
Rufenacht:2000px}. 

The aim of this Introduction is to motivate the work that has been done as well as to
present the basic knowledge necessary for the subsequent Chapters.

\section{Motivation}\label{sec:motivation}
The most systematic approach to the non-perturbative study of QCD, the theory of strong interactions,
is Lattice QCD (LQCD). The QCD action is discretised by putting the theory on a space-time lattice.
As QCD, LQCD needs as input the quark masses and an overall scale. Then any Green's function may
be evaluated by taking averages of certain combinations of the lattice fields (measuring operators) on
an ensemble of vacuum samples (lattice configurations). This allows the study of masses (spectroscopy)
as well as the extraction of matrix elements. In contrast to the experiment, parameters
such as quark masses, boundary conditions or sources may be arbitrarily varied in LQCD, which allows
a wide range of studies to be contrasted with experimental results.

In order to measure the energies of states in LQCD, one studies the decay in time of correlators
of respective operators. Thus the distance between lattice sites has to be small enough
such that the signal of the correlators may be traced along several time-slices of the lattice before
it disappears in the statistical noise. 
On the other hand, however, the physical size of the whole lattice has to be large enough such
that there are no significant finite-size effects.
Both these requirements together lead to lattices with a large number of lattice
sites that are computationally expensive. 

The obvious way out of this dilemma is the use of \emph{anisotropic} lattices,
whose spatial lattice spacing $a_s$, the distance between two neighbouring lattice sites in spatial
direction, is larger than $a_t$, the one in temporal direction. The anisotropy or aspect
ratio is conventionally denoted by $\xi=a_s/a_t$.

Anisotropic lattices have been widely used for studying heavy states in gauge theory for
the last few years.
For measurements in quenched QCD see e.g.~\cite{Lee:2000hh} for $N^*$ masses, 
\cite{AliKhan:2000bv,Chen:2000ej,Chen:2000qj} for heavy quarkonia using relativistic
heavy quarks, \cite{Manke:1998qc,AliKhan:1999yz,Manke:1999ru,Drummond:1998fd,Drummond:1999db,Juge:1999aw}
for heavy quarkonia including heavy hybrid states composed of quarks and glue or \cite{Lewis:1999yw,Lewis:2000sv} for heavy-light
mesons employing NRQCD.
Another work \cite{Shigemitsu:2000hj} has studied heavy meson semi-leptonic decays using
NRQCD.
Studies in pure $SU(3)$ gauge theory have mainly examined glueballs
\cite{Morningstar:1999rf,Morningstar:1999dh,Liu:2000ce}.
The study of string-breaking using static quarks in the adjoint representation
\cite{Kallio:2000jc} employed an anisotropic action to be able to have large spatial
separations of the sources on a lattice with a moderate number of lattice sites.
Other studies of static potentials including
higher representations comprise works like \cite{Darmohval:1999ht,Bali:1999hx,Deldar:1999id,Deldar:1999vi}.
Anisotropic lattices have been also used to study (quenched) QCD at finite temperature
\cite{Fujisaki:1997vv,deForcrand:1998fg,deForcrand:1999df,Umeda:2000ym,Namekawa:2001aa} as the fine lattice in temporal direction
allows for a fine variation of $T=1/(a_t N_t)$ without the need of having to increase
the number of spatial lattice sites.
A study of the autocorrelation of spatial and temporal $SU(3)$ gauge operators on
anisotropic lattices has been performed in \cite{Draper:1997mv}.
Perturbative properties of improved anisotropic gauge and fermion actions,
e.g.~the dependence of the couplings $g_s$ and $g_t$ on $\xi$ as well
as the ratio between the renormalised and the bare anisotropy $\eta=\xi_{\text{R}}/
\xi_0$,
have been studied in \cite{Engels:1999tk,Sakai:1998jz,Sakai:1999pr,Sakai:2000jm}. Matching
parameters for the study of matrix elements (mainly for heavy meson semi-leptonic
decays) have been calculated in \cite{Groote:2000jd}.

The discretisation of the continuum theory leads to cut-off effects (or
lattice artifacts) in the results obtained on the lattice. These effects grow
as the lattice spacing $a$ is increased (i.e.~as the lattice gets more and more different
from the continuum). As simulations using anisotropic actions are carried out
on lattices where the spatial lattice is rather coarse --- in order not to 
have too large cut-off effects it is advantageous to use improved discretisations
of the continuum action suppressing such artifacts.
Whereas it has been common to use improved anisotropic gauge actions for some years, above
all in glueball simulations, the most radical concept of improving actions, the construction
of classically perfect actions, has never been applied to this subject.
It is thus the goal of this work to construct anisotropic classically perfect gauge actions\footnote{In the
following, we will often drop the word \emph{classically} for brevity. However, it has to be understood
that the perfect actions in this work are always classically and \emph{not} quantum perfect.}
and to examine their properties.

Different ways of constructing improved gauge actions
are discussed in the remainder of this Chapter. On one hand, there are actions
that are improved employing perturbation theory, on the other hand, the concept of renormalisation group
provides means of improving actions non-perturbatively. One of the most radical ways of improving
actions is the construction of the classically perfect Fixed Point action, which is described in
detail in Sections \ref{sec:fpaction} and \ref{sec:fpcons}.

In Chapter \ref{chap:physobj} 
we present some physical objects that may be studied in the context of pure lattice gauge
theory and explain how we measure them in practice. The dispersion relation of the torelon
is used to determine the renormalised anisotropy of the parametrised action as well as to provide
an estimate for the scale at a given coupling $\beta$. The measurement of the critical temperature
$T_c$ corresponding to the deconfining phase transition of pure gauge theory yields as well 
information about the (temporal) scale of a simulation. To determine the lattice spacing precisely
as well as to obtain information about the rotational invariance of the parametrised FP action, we
employ the static quark-antiquark potential. Finally, we describe bound states of pure glue, the glueballs,
and discuss how they are measured on the lattice.

Chapter \ref{chap:iso} contains a brief
summary about the newly parametrised isotropic FP action \cite{Wenger:2000aa,Niedermayer:2000yx}.
The action is presented as well as simulations studying the deconfining phase transition, the static
quark-antiquark potential and glueballs. The scaling properties of the isotropic action are discussed
as well.

In Chapter \ref{ch:construction} we present the method used in this work to con\-struct
aniso\-tro\-pic classically perfect actions. Instead of repeating the whole construction programme as described at
the end of this Chapter, we use the parametrised isotropic FP action and perform one single,
purely spatial, blocking step to end up with a perfect anisotropic action with anisotropy $\xi=2$.
Furthermore, we describe how the isotropic parametrisation using smeared links may be generalised such
that it may as well account for the anisotropy between spatial and temporal directions.

This approach is then tested on free scalar fields that may be treated analytically as well as
in the quadratic approximation of the FP action, as described in Chapter \ref{chap:pert}. It turns
out that in both cases the ansatz works nicely.

The construction of the $\xi=2$ perfect action in full SU(3) gauge theory is presented in the next
Chapter, including as well results of measurements of torelon energies , the static quark-antiquark potential,
the critical temperature and the glueball spectrum.

Chapter \ref{ch:xi4act} describes how the spatial blocking step may be repeated to obtain actions with
larger anisotropies and presents first results about the renormalised anisotropy of the $\xi=4$
perfect action. It turns out that the iteration of the spatial blocking step seems to be feasible
and poses no additional difficulties.

In Chapter \ref{chap:aiprop} we present results of measurements of autocorrelation times as well as
information about the overhead to be expected in simulations using the perfect anisotropic actions compared
to the standard Wilson action.

Conclusions and prospects are given in Chapter \ref{ch:concl}. 

\section{Perturbatively Improved Gauge Actions}

On very smooth lattice configurations, i.e.~configurations with a very small
lattice spacing $a$ close to the continuum, any discretisation of the continuum
gauge action yields basically the same results, the leading lattice artifacts of
$O(a^2)$ are very small. Increasing the lattice spacing $a$, the form of the discretisation
gets more and more important and simple discretisations of the continuum action (e.g.~the
Wilson action) may lead to large cut-off effects.
It is thus favourable to use more sophisticated --- improved --- actions that suppress the
lattice artifacts.
There exist several ways of constructing improved actions, in the next few sections
we present the most important ones that are used for lattice gauge theories.


One method to construct improved actions is following Symanzik's
proposal \cite{Symanzik:1983dc,Symanzik:1983gh}, killing
the lattice artifacts order by order in $a^2$, adding operators
of corresponding dimensions that cancel the artifacts. This
yields e.g.~the well known L\"uscher-Weisz action \cite{Luscher:1985xn}
which has leading corrections at $O(g^4a^4)$.

A technique very important in the context of this work is mean-field (or tadpole) improvement which
helps to calculate perturbative coefficients, as those appearing in Symanzik improved actions, more
reliably. This is achieved by changing the set up for lattice perturbation theory.
On the lattice, gauge fields are represented in the following way:
\begin{equation}
U_\mu(x)=e^{iagA_{\mu}(x)}=1+iagA_\mu(x)-\frac{a^2g^2}{2}A_\mu^2(x)+\cdots,
\end{equation}
where $U_\mu(x)$ is the gauge link at lattice site $x$. This form obviously gives rise
to local quark-gluon vertices with 1, 2, $\ldots$ gluons. All these vertices except the
lowest one are lattice artifacts. Contracting the two gluons in the term $a^2g^2A_\mu^2(x)/2$
one obtains so-called tadpole diagrams. As it was recognised by Parisi \cite{Parisi:1981aa}
and later by Lepage and Mackenzie \cite{Lepage:1993xa} these artifacts are in fact only
suppressed by powers of $g^2$ as the ultraviolet divergences generated
by the tadpole loops kill the $a^2$ suppressions. These contributions
are made responsible for the poor match between short distance quantities
and their perturbative estimates as well as for large coefficients occurring in
perturbative lattice expansions.

Mean-field (or tadpole) improvement has been proposed to get rid of this kind
of artifacts. It assumes that the lattice fields can be split into an ultraviolet (UV) and
an infrared (IR) part and that the UV part should be integrated out \cite{Parisi:1981aa,Lepage:1993xa}:
\begin{equation}
e^{iagA_\mu(x)}=e^{iag(A_\mu^{\text{IR}}(x)+A_\mu^{\text{UV}}(x))}=u_0e^{iagA_\mu^{\text{IR}}(x)}\equiv u_0 \tilde{U}_\mu(x).
\end{equation}
It just amounts to a simple rescaling of the link variables $U_\mu(x)$ by an overall constant
factor $u_0<1$. Instead of the common expansion parameters, the coupling $g^2$, the hopping parameter
$\kappa$ in the fermion action and the links $U$ one uses
their improved counterparts $\tilde{g}^2=g^2/u_0^4$, $\tilde{\kappa}=\kappa u_0$ and
$\tilde{U}=U/u_0$. There are two common choices for the tadpole factor $u_0$, either the
fourth root of the plaquette expectation value or the expectation value of the link in Landau
gauge. Tadpole improvement may be especially useful for the construction of anisotropic actions because
it leads to a small renormalisation of the anisotropy $\xi$, if this quantity is introduced
perturbatively, as e.g.~in Symanzik improved anisotropic actions. Indeed, the anisotropic action
most widely used \cite{Lee:2000hh,Manke:1998qc,AliKhan:1999yz,Manke:1999ru,Drummond:1998fd,Drummond:1999db,Juge:1999aw,
Lewis:1999yw,Lewis:2000sv,Shigemitsu:2000hj,Morningstar:1999rf,Morningstar:1999dh,Liu:2000ce,Kallio:2000jc,
Deldar:1999id,Deldar:1999vi,Draper:1997mv,Groote:2000jd} in the last few years, initially presented in
\cite{Morningstar:1997ze}, combines mean-link and Symanzik improvement (at tree-level).

Other, non-perturbative methods of improving actions base on the renormalisation group. This
concept, as well as resulting actions are briefly introduced in the next section.

\section{The Renormalisation Group}\label{sec:renorm}

A Field Theory is defined over a large range of scales from small, physical energies up
to the cut-off that is sent to infinity to reach the continuum limit. The fields associated with
very high (unphysical) scales do influence the physical predictions indirectly through a complex cascade
process; however no physical question involves them directly. This makes the connection between
the local form of the interaction and the final predictions obscure. Additionally, the large
number of degrees of freedom present makes the problem technically difficult. That is why one
attempts to integrate out these non-physical high momentum scales in the path integral. The
method accomplishing this, taking into account the effect on the remaining variables exactly, is
called the Renormalisation Group transformation (RGT) \cite{Wilson:1974jj,Wilson:1975mb,Wilson:1975am,Wilson:1993dy,
Ma:1973zu,Ma:1976mt,Kadanoff:1976wy,Niemeyer:1973aa}.

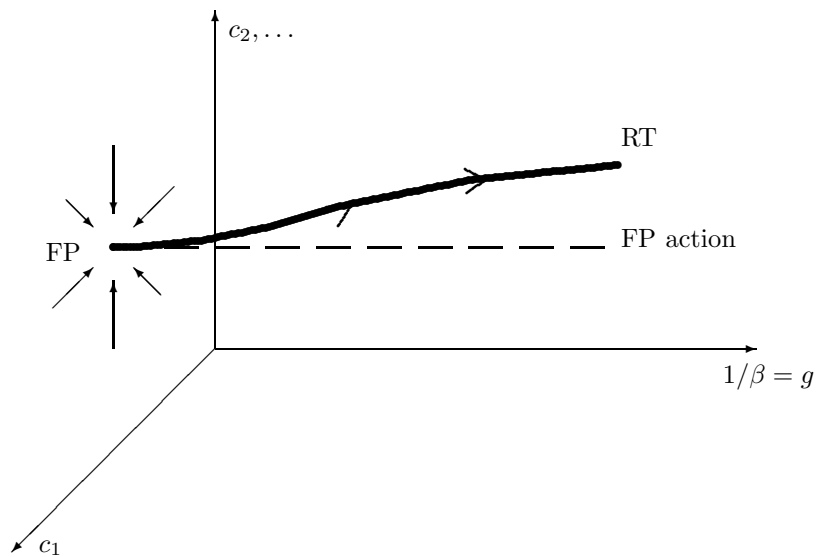
\begin{figure}[htbp]
\setlength{\unitlength}{.9mm}
\begin{center}
\begin{picture}(120,80)
\put(30,30){\vector(1,0){80}}
\put(30,30){\vector(0,1){50}}
\put(30,30){\vector(-1,-1){30}}
\put(4,0){$c_1$}
\put(32,76){$c_2,\ldots$}
\put(105,25){$1/\beta=g$}
\put(5,43){FP}
\put(15,45){\circle*{1}}
\put(15,30){\vector(0,1){10}}
\put(15,60){\vector(0,-1){10}}
\put(6,36){\vector(1,1){6}}
\put(24,54){\vector(-1,-1){6}}
\put(22,38){\vector(-1,1){4}}
\put(8,52){\vector(1,-1){4}}
\multiput(15,45)(.5,0){6}{\circle*{1}}
\multiput(18,45)(.5,.05){20}{\circle*{1}}
\multiput(28,46)(.5,.1){20}{\circle*{1}}
\multiput(38,48)(.5,.15){20}{\circle*{1}}
\multiput(48,51)(.5,.1){20}{\circle*{1}}
\multiput(58,53)(.5,.1){20}{\circle*{1}}
\multiput(68,55)(.5,0.05){44}{\circle*{1}}
\thicklines
\multiput(15,45)(7.5,0){10}{\line(1,0){5}}
\thinlines
\put(90,45){FP action}
\put(90,60){RT}
\multiput(50,51)(-.25,0){15}{\circle*{.5}}
\multiput(50,51)(-.13,-.21){15}{\circle*{.5}}
\multiput(70,55)(-.22,0.11){15}{\circle*{.5}}
\multiput(70,55)(-.20,-.14){15}{\circle*{.5}}
\end{picture}
\end{center}
\caption{Schematic flow diagram for asymptotically free theories.}
\label{fig:flowdiag}
\end{figure}

The sequence of theories defined by the repeated
use of the RGT on an initial theory defines a flow trajectory in the space of couplings. Fixed points (FPs)
of this transformation are theories that reproduce itself under the RGT. Since the correlation length of
the theory scales by the scaling factor of the RGT, its value has to be 0 or $\infty$ at the fixed point.
In Yang-Mills theory there is a non-trivial fixed point (the Gaussian FP) with correlation length $\xi=\infty$ whose
exact location in coupling space $c_1$, $c_2,\ldots$ depends on the RGT used. There is one so-called ``relevant''
coupling whose strength increases as one is starting at this FP and performing RGTs. The flow along
this relevant scaling field whose end-point is the FP is called the renormalised trajectory (RT).
This situation is sketched in Figure \ref{fig:flowdiag}
Simulations performed using an action which is on the exact RT reproduce continuum physics without
any discretisation errors.

There are different approaches to take advantage of this knowledge about field theories in the vicinity of
a fixed point, presented in the next two sections.

\subsection{Renormalisation Group Improved Actions}

Iwasaki \cite{Iwasaki:1983ck} has studied RGTs in perturbation theory and has obtained a
gauge action using two loop-shapes that after a few blockings comes close to the
RT. It shows better rotational invariance and better scaling of $T_c/\sqrt{\sigma}$ than the
standard action. Today, this action is used e.g.~by the CP-PACS collaboration
for MC simulations including domain-wall fermions \cite{AliKhan:2000iv} as it turned out that the Iwasaki action shows
better behaviour concerning chiral symmetry.

Gupta and Patel \cite{Patel:1987ya} have studied the flow of a scale $\sqrt{3}$ RGT and approximate the action
using four parameters and two loop shapes in different representations. Studies of lattice topology have
shown that short distance artifacts are suppressed \cite{Grandy:1995ee,Grandy:1997mm}.

The QCD-TARO collaboration \cite{deForcrand:1997bx,deForcrand:1998jn} has examined the flow of a scale 2 RGT
and evaluated the couplings on blocked lattices using Schwinger-Dyson equations. The blocked action
is reproduced by two couplings corresponding to the simplest Wilson plaquettes $1\times 1$ and $1\times 2$.

\subsection{The Classically Perfect FP Action} \label{sec:fpaction}

Another, more radical, method has been suggested by Hasenfratz and Niedermayer \cite{Hasenfratz:1994sp}:
For asymptotically free theories like SU($N$) Yang-Mills theory, the action at the FP where
the (only) relevant coupling $g=0$ (corresponding to $\beta=\infty$) may be determined using
a saddle-point equation. The FP action is classically perfect, i.e.~there are no lattice artifacts
on the solutions of the lattice equations of motion. This is equivalent to an on-shell Symanzik improvement
at tree-level to all orders in $a$. 

Let us consider SU($N$) pure gauge theory in $d=4$ Euclidean space-time on a periodic lattice.
The partition function reads
\begin{equation}
  Z(\beta) = \int dU e^{-\beta {\cal A}(U)} ,
\end{equation}
where $dU$ is the invariant group measure and ${\cal A}(U)$ a discretisation of the continuum action.
We perform a real space RGT:
\begin{equation}
   e^{-\beta' {\cal A}'(V)} = \int dU \exp\left\{-\beta ({\cal A}(U) + T(U,V))\right\},
\end{equation}
where $V$ is the blocked link variable and $T(U,V)$ is the blocking kernel defining the explicit
form of the transformation. As for asymptotically free theories the FP lies at $\beta=\infty$,
in this limit the path integral can be calculated in the saddle-point 
approximation.
This leads to an equation in classical
field theory defining the FP action which is mapped onto itself by the RGT:
\begin{equation}
  \label{eq:FP_equation}
  {\cal A}^{\text{FP}}(V) = \min_{\{U\}} \left\{{\cal A}^{\text{FP}}(U) + T(U,V)\right\}.
\end{equation}

\section{The Construction of a FP Action for Coarse Configurations} \label{sec:fpcons}

The FP saddle point equation \ref{eq:FP_equation} derived in the last section may be studied analytically up
to quadratic order in the vector potentials \cite{DeGrand:1995ji,Blatter:1996ti} (see Section \ref{sec:quadapp}).
However for solving the FP equation on coarse configurations one has to resort 
to numerical methods. Additionally, although the full FP action is local, it is described
in principle by infinitely many couplings that decrease exponentially with the space-time separation of the
two coupled variables in the action. In order to use
a FP action in a numerical simulation it thus has to be approximated by some parametrisation
with a finite number of couplings.

\subsection{The Method Used}

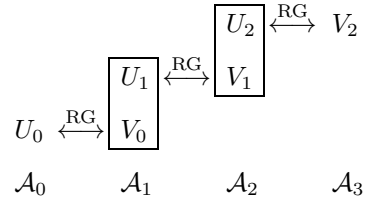
\begin{figure}[h]
  \unitlength2pt
  \begin{center}
    \begin{picture}(100,50)
      \put(20,10){$\mathcal{A}_0$}\put(20,20){$U_0$}
      \put(40,10){$\mathcal{A}_1$}\put(40,30){$U_1$}\put(40,20){$V_0$}
      \put(60,10){$\mathcal{A}_2$}\put(60,40){$U_2$}\put(60,30){$V_1$}
      \put(80,10){$\mathcal{A}_3$}\put(80,40){$V_2$}
      \put(28,20){$\stackrel{\text{RG}}{\longleftrightarrow}$}
      \put(48,30){$\stackrel{\text{RG}}{\longleftrightarrow}$}
      \put(68,40){$\stackrel{\text{RG}}{\longleftrightarrow}$}
      \put(38,18){\framebox(9,17)[lb]}
      \put(58,28){\framebox(9,17)[lb]}
    \end{picture}
    \caption{The cascade procedure used to obtain perfect actions for strongly fluctuating fields $V_2$ starting
with a simple action $\mathcal{A}_0$ suitable for very smooth configurations $U_0$. Coarse configurations
$V_2$ (on the level of MC simulations), $V_1$, $V_0$; fine configurations $U_2$, $U_1$, $U_0$ (of which the latter
may be described with the staring action $\mathcal{A}_0$). The boxes contain configurations on the same level of fluctuations. The actions $\mathcal{A}$ describing the respective configurations are also indicated.
Note, that the number of steps actually used may differ from the number displayed in this Figure.}
    \label{fig:parproc}
  \end{center}
\end{figure}

The construction of the FP action on rough configuration relies on the
observation that the fluctuations of the minimising configurations $U$ in eq.~\ref{eq:FP_equation}
are reduced by a factor of 30 to 40 compared to the initial coarse fields $V$.
That is why after a small number of inverse RGTs (calculating the minimising configurations
$U$ corresponding to coarser configurations $V$, i.e.~using eq.~\ref{eq:FP_equation} as a recursion
relation linking two actions $\mathcal{A}$, $\mathcal{A}'$) the configurations are so smooth
that they can be described accurately by a simple discretisation of the gauge action, e.g.~the
Wilson action. In practise it is not feasible to perform the necessary number of RG steps at once
due to memory and computer time limitations, we therefore resort to a cascade process involving
several single RG steps of first inverse blocking coarse configurations and then parametrising the action
on these configurations using eq.~\ref{eq:FP_equation}. This idea is displayed in Figure \ref{fig:parproc}.
In the following sections we describe
the starting point of the procedure, i.e.~a simple discretisation of the action for smooth fields,
as well as the proceeding to parametrise actions $\mathcal{A}(V)$ at intermediate and final levels,
using eq.~\ref{eq:FP_equation}.

\subsection{The Starting Point}

As a starting point, we use very fine configurations (corresponding to ``$U_0$'' in Figure \ref{fig:parproc})
that are described well by a simple action. As an example, for the parametrisation of the isotropic action
we generate coarse configurations (``$V_0$'') where the plaquette variable $u(V_0(n_B))\leq 0.25$ for all $n_B$
\footnote{Note, that the \emph{average} plaquette variable for configurations occurring in MC simulations amounts
to about $1.5\sim 2$, whereas the maximum values may reach 4.5 which is the upper bound for SU(3) matrices.}
and thus for the corresponding minimised configurations $u(U_0(n))\leq 0.009$ for all $n$. These
fine configurations can be accurately described by the quadratic approximation
of the FP action fulfilling the $O(a^2)$ Symanzik condition, denoted by $\mathcal{A}_0$
\cite{Wenger:2000aa,Niedermayer:2000yx}.

To check how well this action $\mathcal{A}_0$ is working on the minimised configurations, we generate
a set of configurations showing fluctuations of the same magnitude and corresponding configurations which
are identical up to one link which is changed. The change of the action due to this changed link may then be
calculated on one hand using the parametrisation $\mathcal{A}_0$, on the other hand we minimise all the
configurations and use the r.h.s. of eq.~\ref{eq:FP_equation} which rests upon the action on the minimised
configurations whose fluctuations are again smaller by a factor of 30--40, i.e.~this result is expected to
be very close to the exact one. The difference of these two action values is thus a good estimate of the
error of the parametrisation. It turns out that the typical error is about 1--2\% only.

We may now use eq.~\ref{eq:FP_equation} to fix the parameters of an appropriate
parametrisation of the action $\mathcal{A}_1$ valid on configurations with $u(V_0)\leq 0.25$.

This process is now repeated with coarser configurations ($V_1$) whose minimised counterparts ($U_1$)
show fluctuations $u(U_1)\leq 0.25$, describing these minimised configurations on the r.h.s. of
eq.~\ref{eq:FP_equation} by the action $\mathcal{A}_1$. It turns out that step by step
the complexity of the parametrisation (i.e.~the number of free parameters) has to be increased
in order to be able to describe the action well on the coarser and coarser fields.

\subsection{Quantities to be Fitted}\label{sec:fitquant}

Performing the numerical minimisation of the configurations $V$ is an expensive task,
even for small configurations. It is thus very useful to be able to fit not only one
quantity per configuration (the total action $\mathcal{A}(V)$) but to include as well the
derivatives of the gauge action with respect to the gauge links in a given colour direction
$a$:
\begin{equation}
\frac{\delta\mathcal{A}(V)}{\delta V_\mu^a(n)},\quad \mu=1,\ldots,4,\quad a=1,\ldots, N_c^2-1,
\end{equation}
where $N_c$ denotes the number of colours. These quantities are easily accessible and
deliver $V\cdot (N_c^2-1)\cdot 4$ additional residues for the fit per configuration. It is
understood that these quantities are not completely independent, but still the increase
of useful data per configuration is dramatic.

The intermediate actions $\mathcal{A}_1$, $\mathcal{A}_2$ etc. still fulfill the $O(a^2)$
Symanzik conditions in order to be valid as well for configurations with smaller fluctuations ---
however, in the last step yielding the final parametrisation of the FP action for coarse fields, this
requirement is dropped to get an effective FP action describing well the typical MC configurations
using a compact set of parameters. Additionally, we include scale-invariant instanton solutions
\cite{DeGrand:1996ih,DeGrand:1996zb,DeGrand:1997gu,Farchioni_Papa:98} to make the action
performing well concerning topology.

Adjusting the weights of the different quantities entering into the fit, we may tune the
importance of the respective quantity in the determination of the action, see also
Section \ref{sec:linfit} 

In the last step we include into the fit several sets of 10--40 configurations each at different
values of the coupling $\beta$ to obtain a parametrisation which is valid on a certain range
of couplings suitable for MC simulations.

In the following Sections we describe the steps performed to parametrise a perfect action
at a certain level of the cascade procedure.

\subsection{The Non-Linear Fit}\label{sec:nlfit}

Before we may describe the actual fitting procedure used for the parametrisation of the isotropic FP
action as well as for the perfect anisotropic actions, let us briefly explain some of the main features
of our parametrisation for perfect actions. For a more detailed account on this issue, see Section \ref{sec:par}.

The action is a mixed polynomial of traces of simple loops (plaquettes) built from simple gauge links
as well as from (APE-like) smeared links. The parameters in the smearing enter the fitting procedure
\emph{non-linearly}, the
parameters in the polynomial enter the fit \emph{linearly}. This distinction is crucial for the following description
of the actual fitting process. The main advantage of this parametrisation ansatz is the rich structure
including a large number of different loops at moderate computational cost; however, the couplings 
of the loops in the action are not independent from each other but complicated combinations of the non-linear
and linear parameters appearing in our parametrisation.

First we perform a full non-linear $\chi^2$ fit varying the non-linear parameters describing the
smearing as well as the linear parameters accounting for the composition of the action in
terms of simple and smeared plaquettes. The linear part of the fit is performed exactly, its
computational cost is negligible. The non-linear part is done using a Simplex algorithm
which is rather slow, however the danger of missing the true minimum is small. A rough
estimate of the number $N$ of non-linear fitting steps (where every step includes the recalculation
of all the residues) as a function of the number $n_{\text{nl}}$ of non-linear parameters is 
\begin{equation}
N=50(20)e^{0.23(4)n_{\text{nl}}}.
\end{equation}
Due to this behaviour ($N$ increases roughly by a factor of 10 for 10 additional non-linear parameters),
it is not easily possible to increase the number of free non-linear parameters above $15\sim 20$ in
the full non-linear fit.

The non-linear fit is performed for different sized sets of non-zero non-linear parameters
with a (fixed) large number of free linear parameters. The only quantities fitted at this stage are
the derivatives of the action with respect to the coarse couplings
to obtain a ``basic'' value of $\chi^2$ which may be used to estimate the quality of the
fits in the next step where additional quantities are taken into account or the number of
linear parameters is decreased. Restrictions such as the norm of the action or Symanzik conditions
are applied already to the non-linear fit, however.

The number and the values of non-linear parameters of the parametrised action are fixed 
looking at the $\chi^2$ of the fit --- which in principle is rather delicate, as it is unknown
whether e.g.~a 10~\% decrease of $\chi^2$ when introducing an additional free parameter is a sign
of importance of this extension or whether this set is already large enough to account for the
FP nature of the action. These questions can be answered only at the end of the whole procedure
when the final parametrisation is checked on physical quantities. However, whether the
parametrisation is stable, i.e.~whether the number of data points is already large enough to
account for a parametrisation that does not depend on the actual configurations in the fit,
may be --- and has to be --- checked after the non-linear fit, measuring $\chi^2$ on independent
sets of configurations.

\begin{table}
\renewcommand{\arraystretch}{1.3}
  \begin{center}
    \begin{tabular*}{10.5cm}[c]{r@{\extracolsep{\fill}}rrrrrr}
\hline\vspace{-0.05cm}
\# $\eta$\phantom{ } & \# $c_i$ & $i$ & $\max(k+l)_{\text{sp}}$ & $\max(k+l)_{\text{tm}}$ & $\chi_{\text{d}}^2$ & $\xi_{\text{R}}$\\
\hline
4 & 1 & 4 & 4 & 4 & 0.0250 & 1.63(2)\\
2 & 3 & 3 & 4 & 4 & 0.0238 & \\
4 & 3 & 3 & 4 & 4 & 0.0144 & 1.912(9)\\
\hline
\end{tabular*}
    \caption{Example for the behaviour of $\chi^2$ in the full non-linear fit (for the $\xi=2$ perfect
action), depending on the number (and kind) of non-linear parameters. Values of the (physical) renormalised
anisotropy $\xi_{\text{R}}$ measured later are also given.}
\label{tab:fitqual2}
  \end{center}
\end{table}

To give an example, we copy Table \ref{tab:fitqual2} from Section \ref{sec:compext}.
In this case, the 40\% decrease of $\chi^2$ between the \mbox{\# $c_i$ = 1} and the \mbox{\# $c_i$ = 3} non-linear
parameter set (see Section \ref{sec:par} for an explanation)
seems to be vital for an appropriate parametrisation of the perfect action as the renormalised
anisotropy $\xi_{\text{R}}$ of the first parametrisation deviates considerably from the input
value $\xi=2$ whereas the renormalisation is small for the second, more sophisticated, parametrisation.

\subsection{The Linear Fit}\label{sec:linfit}

\begin{figure}[htbp]
\begin{center}
\includegraphics[width=10cm]{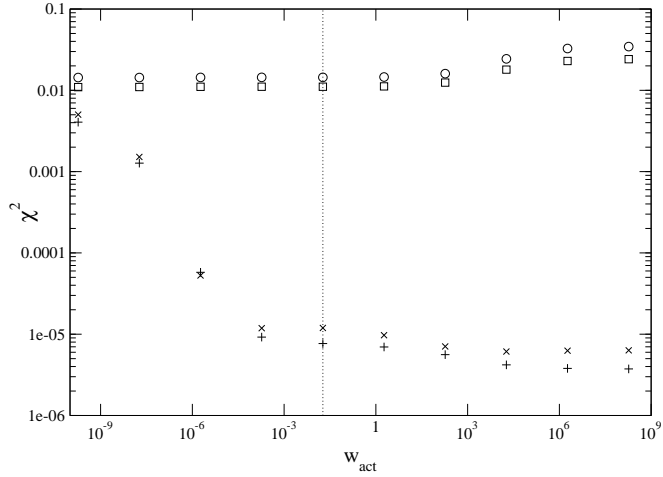}
\end{center}
\caption{The effect of changing $w_{\text{act}}$, the weight of the action compared to the weight of the derivatives, in the linear fit for the $\xi=2$ perfect action.
The values of $\chi^2$ for the derivatives of $\beta=3.5$ (circles) and $\beta=4.0$ (boxes) configurations as well as for
the action values of $\beta=3.5$ (pluses) and $\beta=4.0$ (crosses) configurations are given. The
dotted line marks the final choice of the relative weight $w_{\text{act}}$. The number of parameters remains fixed.}
\label{fig:wact2}
\end{figure}

Having at hand the set of non-linear parameters we may proceed performing fast
linear fitting steps. The action values of the configurations as well as additional data
such as the action of scale-invariant instanton solutions are included at this stage.
The stability of $\chi^2$ is an indication of the ability of the chosen parameter set to
incorporate additional information. At the same time the number of linear parameters
may be decreased until $\chi^2$ starts to rise considerably. Figure \ref{fig:wact2} displays
the change of $\chi^2$ (for the derivatives and the action values of 20 configurations
at $\beta=3.5$ and $\beta=4.0$, respectively) if one changes the weight of the action
value residues compared to the (fixed) weight of the derivatives. The value of $w_{\text{act}}$
finally used is marked by the dotted line. This choice preserves the good value of $\chi^2$ for
the derivatives up to a few per cent while $\chi^2$ of the action values is about twice
the minimum value reachable with a very large weight. Without inclusion of the action values
into the fit, the average relative error of the parametrised action compared to the true one
amounts to about 10\%, with the weight chosen to about 0.5\% and with infinite weight it
might be decreased to about 0.2\% (destroying the good parametrisation of the derivatives completely).

However, in the last step of the cascade process, when one aims at pa\-ra\-me\-trising the final
action to be used in MC simulations, the main difficulty of the choice of the linear set to
be used is the behaviour of the parametrised action in terms of the simple and smeared
plaquettes of the configurations, determined by the linear parameters. It happens that
all the data in the fit is properly described by the parameters, but that varying the
simple or smeared plaquette value slightly above or below the values occurring in the fit
results in fake negative contributions to the action. In a MC simulation, this is a very
dangerous behaviour as this might result in configurations with an artificial structure,
i.e.~the algorithm could try to maximise the number of plaquettes having a negative
contribution to the action etc. One thus has to study the linear behaviour of the action,
e.g.~looking at plots of the action contribution vs. the values of the simple plaquette
$u$ (which is bounded, $0\le u\le 4.5$) and the smeared plaquette $w$ (which in principle
is unbounded as $w$ is not projected to SU(3)). It may be that accidentally none of the
parametrisations behaves well enough for MC simulations. This is especially the case if
conditions on the linear parameters, such as the norm of the action, that are determined
by the values of the non-linear parameters lead to the bad behaviour of the linear part
of the action. In this case we include a number
of constraints $\mathcal{A}(u,w)>0$ (for different pairs of $(u,w)$ at the edge of the
$(u,w)$-region expected to occur in MC simulations) into the \emph{non-linear} fit
in order to get a different set of non-linear parameters allowing for linear parameters
with a good behaviour. (Generally in the space of non-linear parameters several almost
degenerate minima in $\chi^2$ exist for very different sets of parameters, so this does not 
result in a significant rise of $\chi^2$ (of course, this has to be checked). It turns out that
the best way of introducing these constraints into
the non--linear Simplex minimisation is adding the absolute value of negative contributions
to the action at the $(u,w)$ pairs multiplied with a large factor (we use $10^6$). ``Hard''
constraints with a step function at $\mathcal{A}(u,w)<0$ may lead to failure of 
the simplex algorithm as in this case it is not directed towards 
the region of the parameter space, where the action is behaving as desired.

\begin{figure}[htbp]
\psfrag{action}{\hspace{-0.5cm}$\mathcal{A}(u,w)$}
\psfrag{u}{$u$}
\psfrag{w}{$w$}
\begin{center}
\includegraphics[width=8cm]{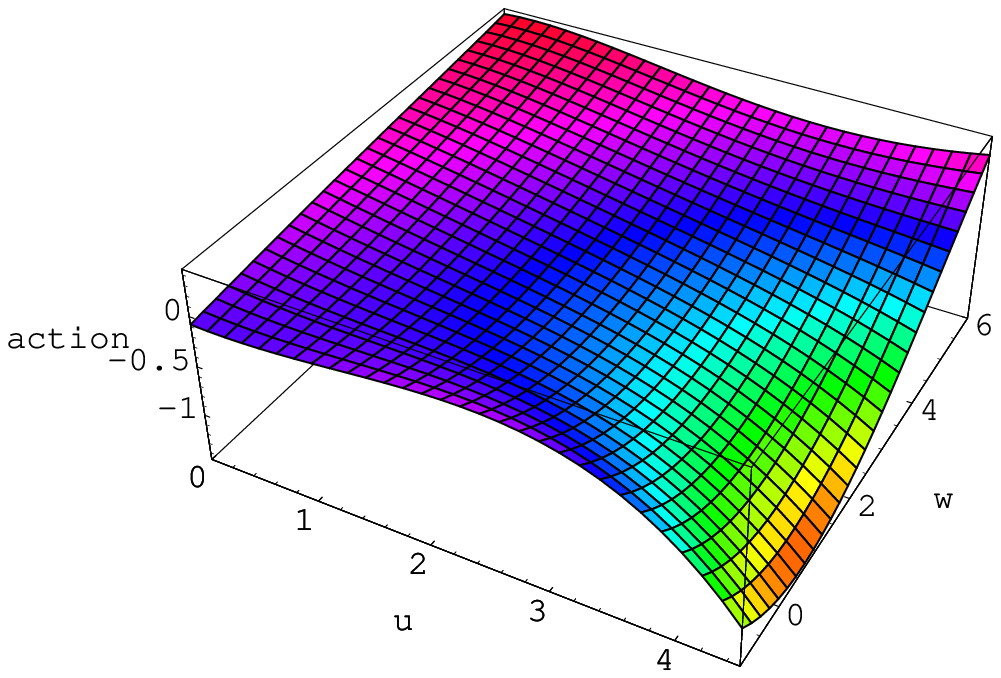}
\includegraphics[width=8cm]{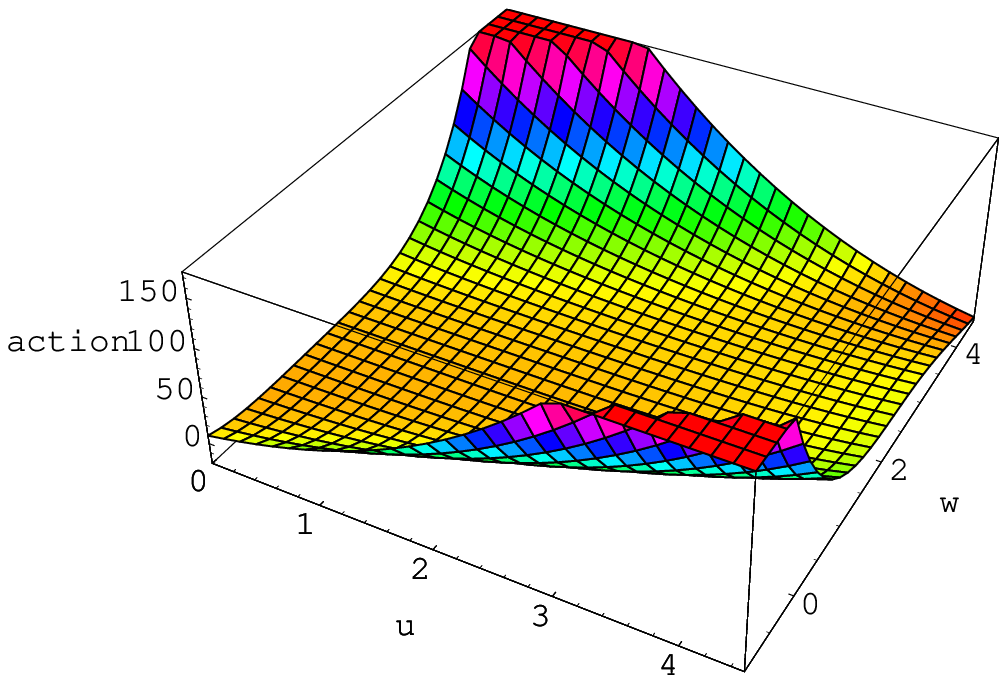}
\end{center}
\caption{A linear set of parameters with bad behaviour (danger of generating artificial
configurations with a lot of negatively contributing plaquettes in a MC run). \emph{Upper plot}: Spatial parameters,
\emph{lower plot}: temporal parameters.}
\label{fig:linbeh_bad}
\end{figure}

\begin{figure}[htbp]
\begin{center}
\psfrag{action}{\hspace{-0.5cm}$\mathcal{A}(u,w)$}
\psfrag{u}{$u$}
\psfrag{w}{$w$}
\includegraphics[width=8cm]{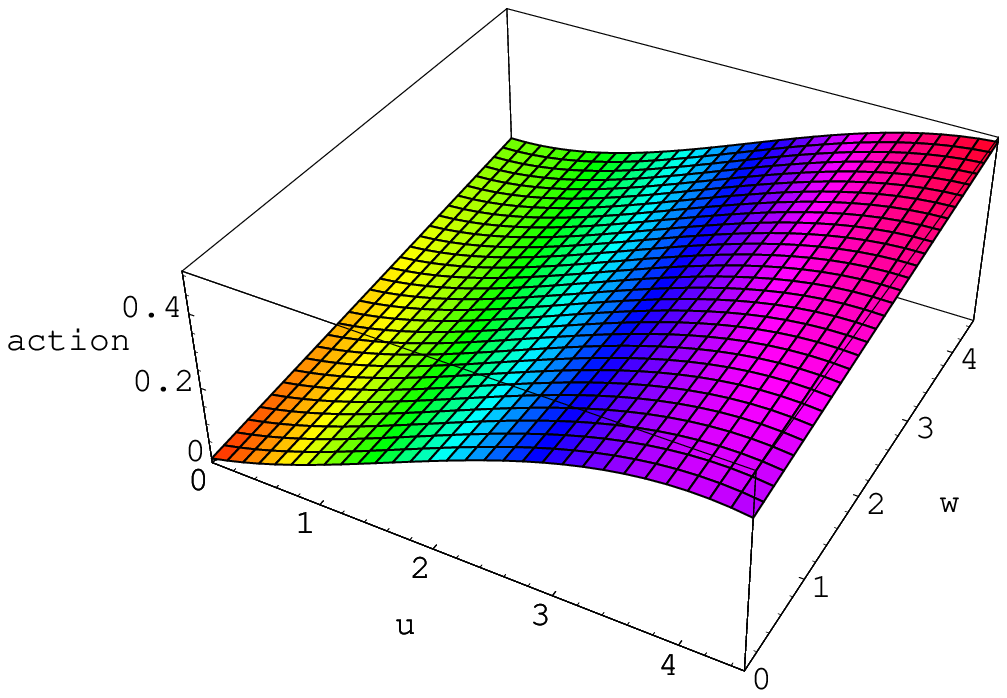}
\includegraphics[width=8cm]{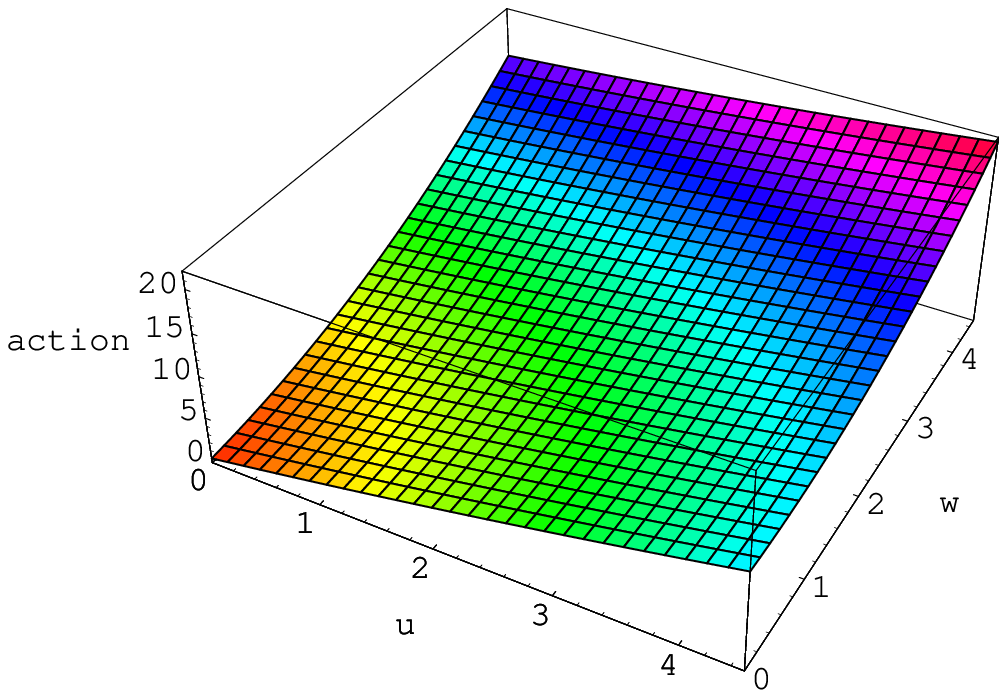}
\end{center}
\caption{A linear set of parameters with good behaviour. \emph{Upper plot}: Spatial parameters,
\emph{lower plot}: temporal parameters.}
\label{fig:linbeh_good}
\end{figure}

Figures \ref{fig:linbeh_bad} and \ref{fig:linbeh_good} show linear sets behaving badly and
well. These examples are taken from the fit of the $\xi=4$ perfect action (see Chapter \ref{ch:xi4act})
where it is necessary to force several points $(u,w)$ in the non-linear fit to result in a
positive contribution to the action. The first figure (standard non-linear fit) shows that in this
case the spatial linear parameters lead to negative contributions of spatial plaquettes in a large
region of $(u,w)$ values whereas the temporal part of the action is o.k. The second
figure shows the behaviour of the parameters finally chosen, where conditions on the action contributions
are included into the non-linear fit.
There are no dangerous regions with negative contributions to the action. Pilot MC runs prove that indeed
no abnormal behaviour of the action is observed.

\fancyhead[RE]{\nouppercase{\small\it Chapter \thechapter.\, Physical Objects in Pure Lattice Gauge Theory}}
\chapter{Physical Objects in Pure Lattice Gauge Theory} \label{chap:physobj}

There are several interesting objects existing in pure gauge theory that
can be studied non-perturbatively on the lattice. Some of them, the ones we
study by Monte Carlo simulation, are presented in this chapter, together
with experimental results (if present). They comprise the torelon (including
non-zero momenta), the deconfining phase transition, the static quark-antiquark
potential and glueballs. Furthermore, we explain how these
states are measured on the lattice and what techniques we use to determine
the energies of the states or the critical temperature of the phase transition respectively.
For the results of the simulations we refer to Chapter \ref{chap:iso} for
a recapitulation of results obtained using the isotropic FP action, to Chapter
\ref{ch:xi2act} for $\xi=a_s/a_t=2$ perfect action results and to Chapter \ref{ch:xi4act}
for first results obtained with the $\xi=4$ perfect action.

\section{The Torelon}\label{sec:torelon}

\subsection{Torelons in the Continuum}

\begin{figure}[htbp]
\begin{center}
\includegraphics[width=3cm,angle=-90]{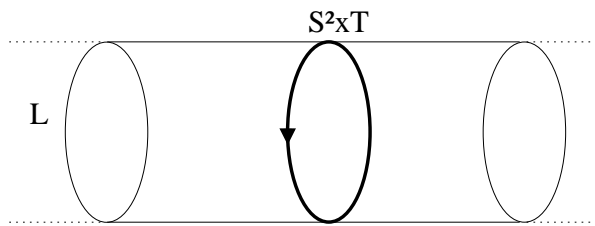}
\end{center}
\caption{A creation operator for torelons.}
\label{fig:tln}
\end{figure}

The torelon is a closed gluon flux tube encircling the periodic spatial boundary of size $L$ of a torus. It
is created and annihilated by closed gauge operators winding around this boundary, see Figure \ref{fig:tln}.

String models which are a good approximation for long flux tubes predict its energy to behave as
\begin{equation} \label{eq:stringtorelon}
E(L)=\sigma L + \frac{\pi}{3L},
\end{equation}
where $\sigma$ is the string tension and the second term describes the string fluctuation for
a bosonic string \cite{Luscher:1981ac,Isgur:1985bm,Perantonis:1988wj,Perantonis:1989uz}.
In contrast to the static quark-antiquark potential (see Section \ref{sec:qbarq}) where the string models the gluon
flux between static colour sources, there is no Coulomb part corresponding to gluon exchange.\footnote{Note, that
despite the absence of the Coulomb part, extracting the string tension from torelon measurements is difficult.
This is due to the large torelon energies for long strings (large $L$) where the relevance of the string model
term $\pi/3L$ decreases.}
The string models assume no restriction on the transverse modes of the string, the transverse extension
$S$ of the volume thus has to be large enough in order to be able to apply their predictions.

To learn more about QCD in a finite volume, see \cite{vanBaal:2000zc}.

\subsection{Torelons on the Lattice}\label{sec:smearing}

On the periodic lattice, a torelon associated
to a point in the $(x,y)$-plane is created by a Polyakov loop winding around the lattice in $z$ direction
starting at that point.
Because of the relatively small lattice sizes necessary and the rather
simple operators that can be used, torelon measurements on the lattice have quite a long history 
\cite{Michael:1987cj,Berg:1988eg,Michael:1988be,Berg:1989cp,Vohwinkel:1989vn,Michael:1989vh,Hasenfratz:1990tp,Michael:1990fx,Michael:1992nc,Michael:1994ej}. 
In order to be able to compare the results to string models as well as not to have too large transverse momenta
we choose to work
on an $S^2\times L\times T$ lattice where $S$ and $T$ are large, whereas the length of the $z$-direction
$L$ is chosen to be rather short, because long torelons are more difficult to measure due to their higher energies
(see eq.~\ref{eq:stringtorelon}).

As the colour flux is wandering as it loops around the lattice, it
is useful to employ (iteratively) APE smeared links \cite{Albanese:1987ds}
to model this, which improves the overlap of the operators with the ground state.
One step of APE smearing acts on a gauge link $U_j(x)$ as
\begin{eqnarray}
\label{eq:APE}
  {\cal S}_1 U_j(x) & \equiv & {\cal P}_{\text{SU(3)}} 
   \Big\{ U_j(x) + \lambda_s \sum_{k \neq j}
  (U_k(x) U_j(x+\hat k) U_k^\dagger(x+\hat j)  \\
 & &  \hspace{2cm}  + U_k^\dagger(x-\hat k)
  U_j(x-\hat k) U_k(x-\hat k+\hat j))  \Big\}, \nonumber 
\end{eqnarray}
the original link variable is replaced by itself plus a sum of the four neighbouring spatial staples and
then projected back to SU(3). The smeared and projected links $S_1U_j(x)$ have the same symmetry properties
under gauge transformations, charge conjugations or reflections and permutations of the coordinate axes. The
smearing is used iteratively allowing to measure operators on configurations of different smearing levels, ${\cal S}_n U$.

We thus measure the operators
\begin{equation}
T_n(x,y,t)=\text{Tr} \prod_{z=0}^{L-1}{\cal S}_n U_z(x,y,z,t),
\end{equation}
for all $x$, $y$, $t$, where ${\cal S}_n U_z$ is the $n$ times iteratively smeared link in the longitudinal $z$
direction. By discrete Fourier transform, we project out the state with momentum
$\mathbf{p}=(p_x,p_y)=(n_x,n_y)(2\pi/Sa_s)$:
\begin{equation}
T_n(\mathbf{p},t)=\sum_{x,y}T_n(x,y,t)e^{i(p_xx+p_yy)}
\end{equation}
and build the correlators
\begin{equation}
C_{nn'}(\mathbf{p},t)=\frac{1}{N_t}\sum_{\tau}\langle T_n(\mathbf{p},\tau) T_{n'}(\mathbf{p},\tau+t)\rangle.
\end{equation}
Using these correlators of operators measured on the different smearing levels we obtain the
energy values $a_t E(\mathbf{p})$ employing variational methods (see Appendix \ref{app:vartech}).
The continuum dispersion relation 
\begin{equation}
E^2=\mathbf{p}^2+m^2
\end{equation}
on the lattice becomes (in temporal units)
\begin{equation}
(a_t E)^2 = a_t^2(\mathbf{p}^2+m^2)=\frac{(a_s\mathbf{p})^2}{\xi_{\text{R}}^2}+(ma_t)^2=\frac{1}{\xi_{\text{R}}^2}(n_x^2+n_y^2)\left(\frac{2\pi}{S}\right)^2
+(ma_t)^2,
\end{equation}
where $n_x$, $n_y$ are the components of the (transversal) lattice momentum. On an
anisotropic lattice, this equation allows for the extraction of the renormalised anisotropy
 $\xi_{\text{R}}=a_s/a_t$ (measuring the ``renormalisation of the speed of light'') as well
as the torelon mass $ma_t$, which in turn may be used to get an estimate of
the scale using eq.~\ref{eq:stringtorelon} and known values of the string tension $\sigma$.

\section{The Deconfining Phase Transition}\label{sec:deconf}
Pure lattice gauge theory is invariant under a global unitary transformation which transforms all temporal links
$U_4(\vec{x},t)$ of a given timeslice (fixed $t$) as
\begin{equation}
U_4(\vec{x},t)\longrightarrow zU_4(\vec{x},t),
\end{equation}
where $z$ is an element of the center $\mathcal{Z}(N_c)$ of the gauge group SU($N_c$). For SU(3),
$z=\exp(2\pi il/3)$, $l=$0, 1, 2.

To study this symmetry, we look at the Polyakov loop (or Wilson line)

\begin{equation} \label{eq:polloop}
P(\vec{x})\equiv \text{Tr} \prod_{t=0}^{N_t-1} U_4(\vec{x},t),
\end{equation}
which is a loop wrapping around the lattice in temporal direction, representing a single static quark at finite temperature, as the temperature $T$ is proportional to
the inverse of the temporal extension $N_t$, $T\propto 1/N_t$.

The correlation of the Polyakov loop $P(\vec{x})$ and its adjoint $P^\dagger(0)$ is
related to the free energy $F_{q\bar{q}}$ of
a static quark separated from a static antiquark by the distance $\vec{x}$ according to
\begin{equation}
\langle P(\vec{x})P^\dagger(0)\rangle \mbox{ }\propto \mbox{ }\exp(-F_{q\bar{q}}/T),
\end{equation}
where $T$ denotes the temperature.
For large separation, cluster decomposition requires
\begin{equation}
\langle P(\vec{x})P^\dagger(0)\rangle\longrightarrow |\langle P(0)\rangle |^2.
\end{equation}

For low temperatures below the critical temperature, $T<T_c$, center symmetry holds and thus
$\langle P(0)\rangle=0$ which allows for increasing
$F_{q\bar{q}}(\vec{x})$ with growing distance $|\vec{x}|$, the static quarks are confined.
For temperatures above $T_c$, center symmetry is broken, $\langle P(0)\rangle\ne 0$ and thus the free energy
of the static quarks has to vanish, $F_{q\bar{q}}(\vec{x})\equiv 0$, the quarks are free and form, together
with the gauge degrees of freedom, the quark-gluon plasma.

\begin{figure}[htbp]
\begin{center}
\begin{tabular}{cc}
\includegraphics[width=6cm]{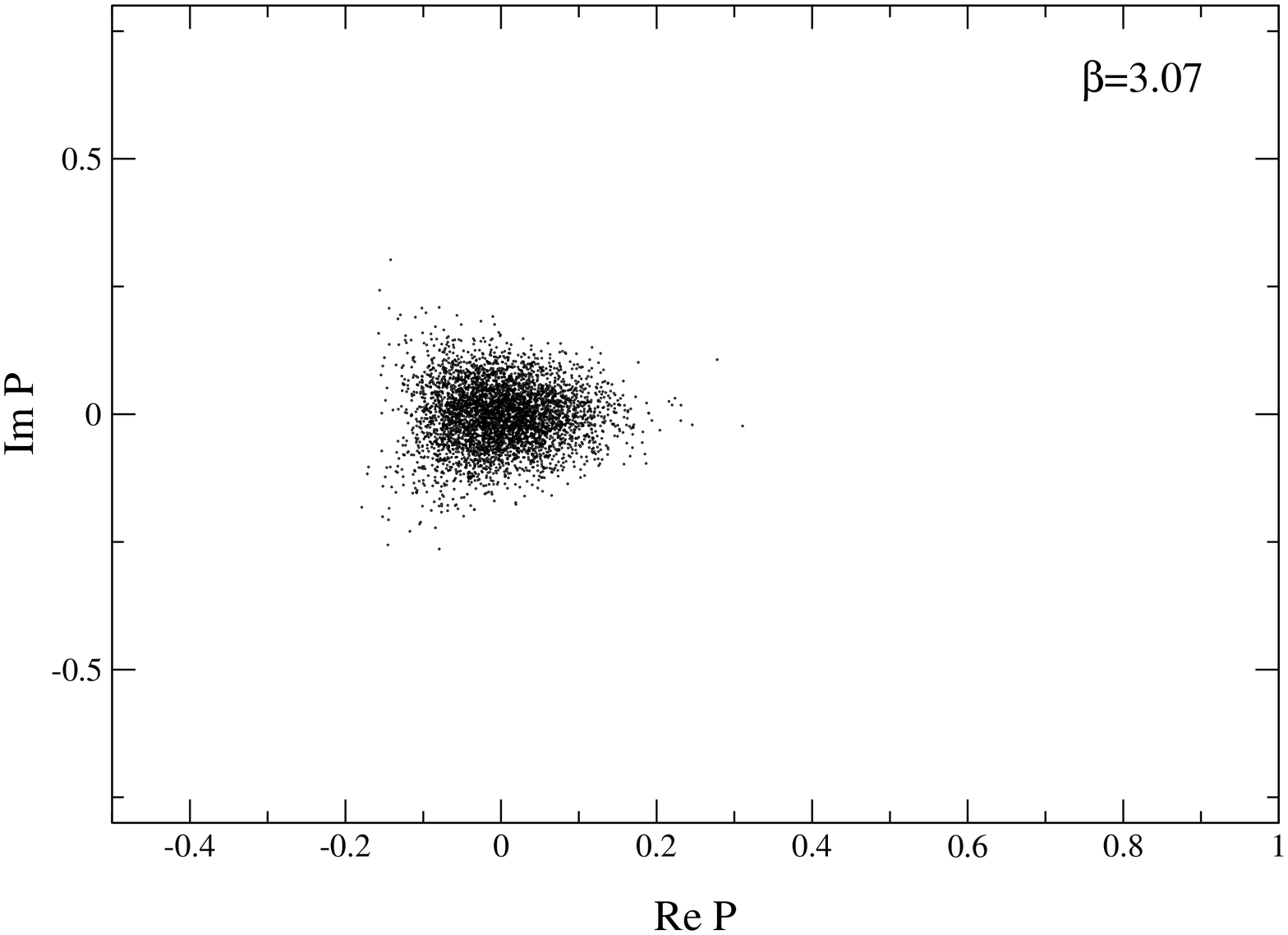} & \includegraphics[width=6cm]{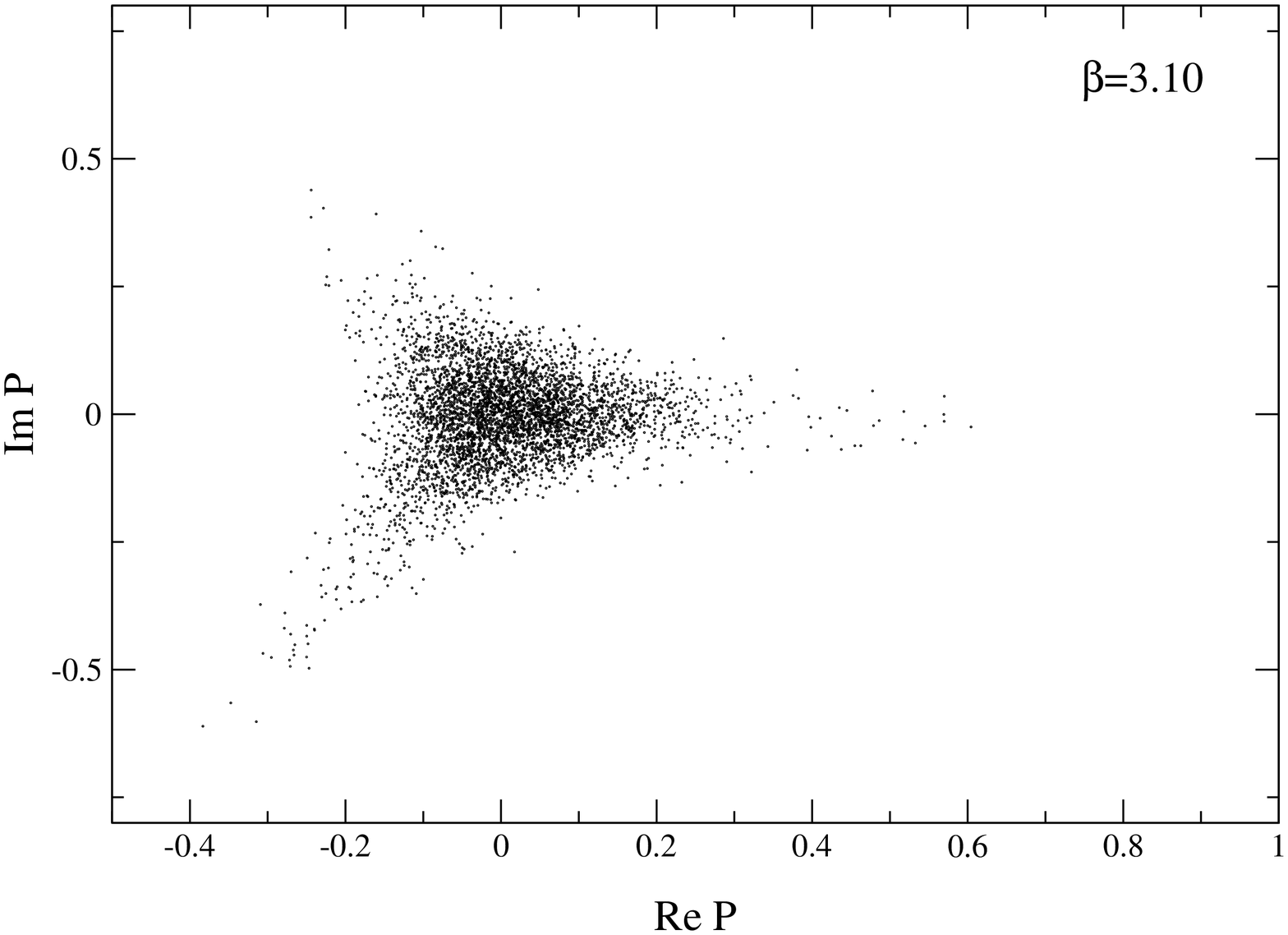}\\
\includegraphics[width=6cm]{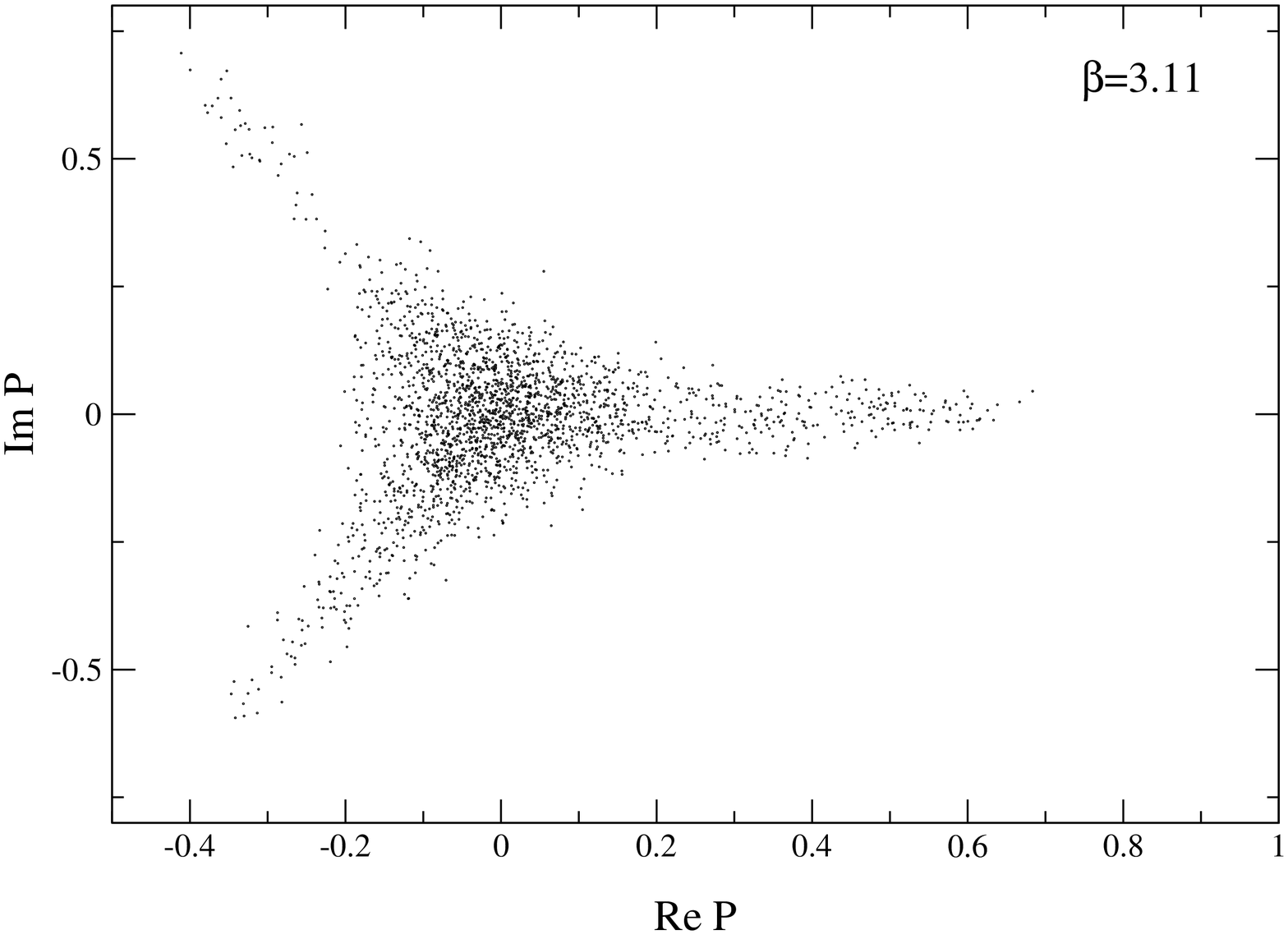} & \includegraphics[width=6cm]{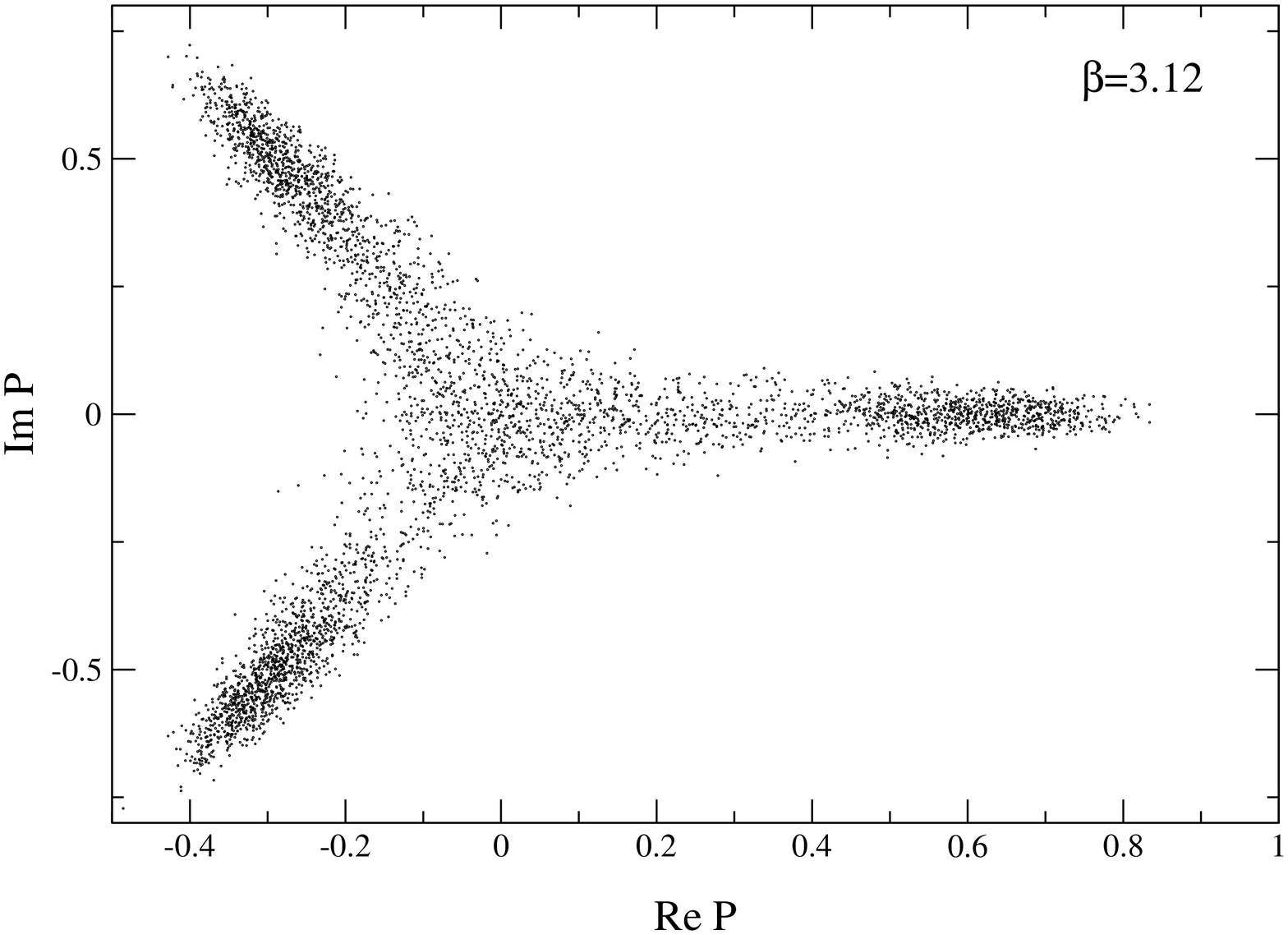}\\
\includegraphics[width=6cm]{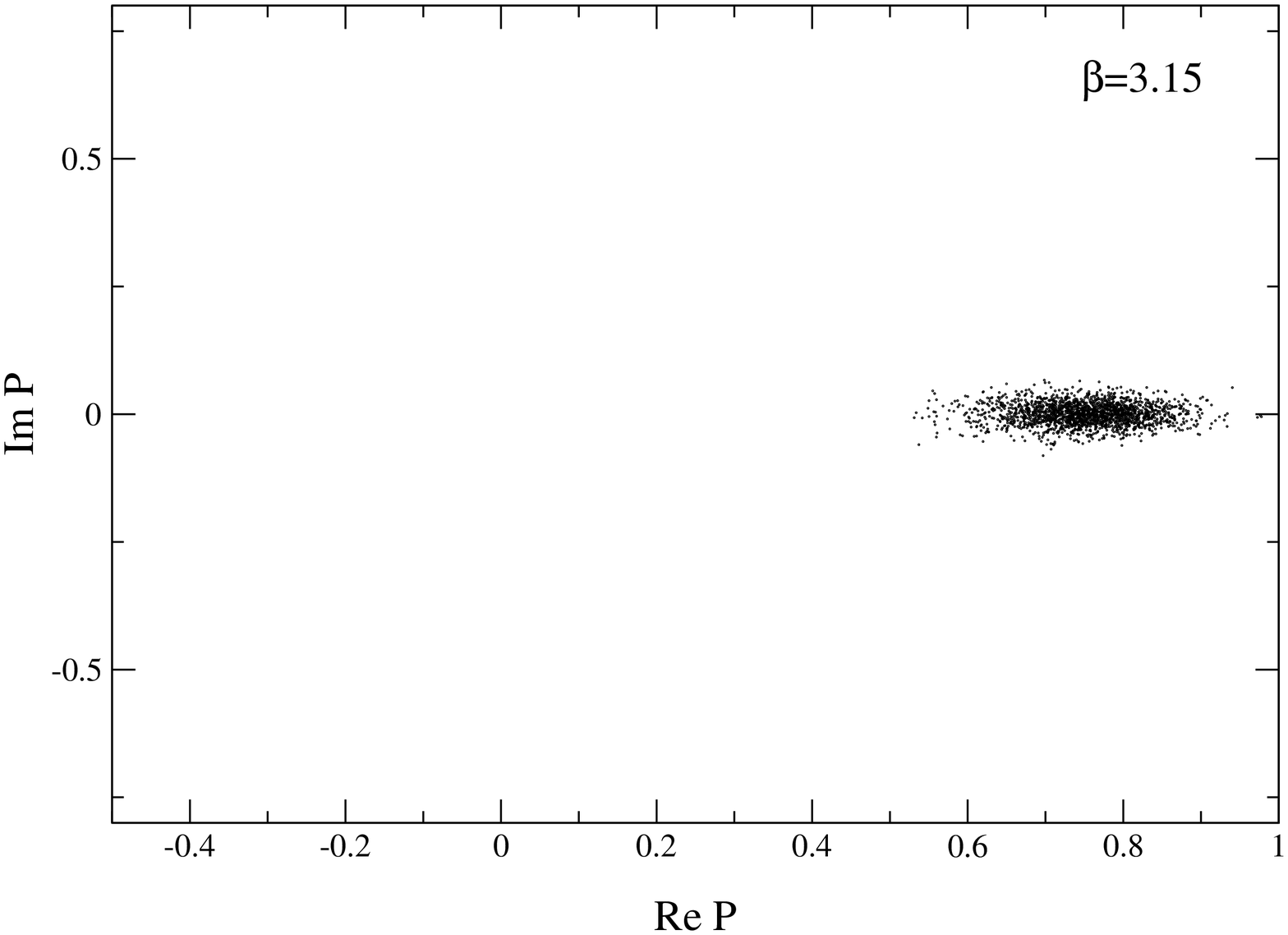} & \\
\end{tabular}
\end{center}
\caption{The distribution of the Polyakov loop at different couplings $\beta$ for the $\xi=2$
perfect action, $N_t=5$, $N_s=9$. The critical value of the coupling has been determined to $\beta_c=3.118(1)$.
The system is heated up such that the center symmetry is broken as $\beta>\beta_c$. For clearness, only
(randomly chosen) 10\% of the data are displayed.}
\label{fig:polscat}
\end{figure}

Figure \ref{fig:polscat} displays the distribution of the Polyakov loop for different couplings $\beta$ (for
the $\xi=2$ perfect action, see Chapter \ref{ch:xi2act}). The critical coupling on the lattice used has been
determined to $\beta_c=3.118(1)$. It is clearly visible that below the critical coupling the Polyakov loop
values are close to zero whereas above $\beta_c$ they choose one of the three degenerate vacua. The coexistence
of the two phases may be observed in the plot for $\beta=3.12$ which is very close to $\beta_c$. 

The critical temperature $T_c$ may be determined by the location of the peak in the susceptibility
\begin{equation}
\label{eq:susc}
\chi\equiv N_s^3 ( \langle |P|^2\rangle - \langle |P|\rangle^2 ).
\end{equation}

There is strong evidence for this transition to be first order for SU(3)
\cite{Kajantie:1981wh,Montvay:1982jj,Kogut:1983rt,Kogut:1983mn,Ukawa:1990tn},
however in finite volume (on lattices
used in MC calculations) the $\delta$-peak of the susceptibility $\chi$ is washed out and limits the
precision with which one may determine the critical temperature $T_c$ or, equivalently, the critical coupling 
$\beta_c$ corresponding to a fixed temporal extension $L_t=aN_t$ of the lattice.

\subsection{Determining the Peak of the Susceptibility} \label{sec:suscpeak}

As mentioned in the last section, the critical coupling $\beta_c$ is determined by
locating the peak of the susceptibility $\chi$ (eq.~\ref{eq:susc}). To accomplish this,
simulations at different values of $\beta$ near the estimated critical coupling $\beta_c$ have to be
performed. For all these simulations a certain statistics has to be reached and it may thus
be very time consuming to approach $\beta_c$ closer and closer performing simulations at
more and more values of $\beta$. As the Monte Carlo simulation may exhibit critical slowing down, i.e.~a
dramatic rise of the autocorrelation times (see Section \ref{sec:autocor}) near the phase transition, the procedure
gets even more expensive. A nice solution to this problem is the use of reweighting
techniques that allow extrapolating or interpolating results obtained at a given coupling
or temperature to other (rather close) values of these parameters. To determine the peak
of the Polyakov loop susceptibility we employ Ferrenberg-Swendsen reweighting
\cite{Ferrenberg:1988yz,Ferrenberg:1989ui} and thus present the main idea of this procedure
very briefly
(for a more extensive presentation, see Appendix C of \cite{Wenger:2000aa}).

All the information about a statistical system at temperature $T$ is contained in the
partition function
\begin{equation}
Z(\beta)=\sum_{\{ \sigma\} }e^{-\beta S(\sigma)},
\end{equation}
where $\{\sigma\}$ denotes the set of all possible configurations of the system, $S(\sigma)$
is the energy for a given configuration $\sigma$. The same partition function may be written
as
\begin{equation}
Z(\beta)=\sum_S W(S)e^{-\beta S},
\end{equation}
where $W(S)$ is the density of states at the energy $S$, also called spectral density
function. This function is universal, it is the same function for every temperature and
thus contains in principle all the information about the system at \emph{any} temperature $T$.
In practice, our numerical results at a given value of $\beta$ will allow us to estimate $W(S)$
only for a finite range of energies $S$ occurring often enough in simulations at the given coupling.
However, if we are studying a phase transition, due
to the broad probability distribution of the examined states in simulations near criticality,
$\beta$ being close enough to $\beta_c$, it is possible to estimate the spectral density
function in a large range of energies.

Let us now consider $n$ measurements of an observable obtained in a numerical simulation at
coupling $\beta$. We plot a histogram of the action values corresponding to the measurements
and denote the number of measurements corresponding to one point of the histogram by $N(S)$.
Trivially, $\sum_SN(S)=n$, where $S$ now denotes a certain range of energies binned together
to build one point of the histogram. We can estimate the probability of a configuration to
have energy $S$ at coupling $\beta$:
\begin{equation}
P_{\beta}(S)\equiv\frac{W(S)e^{-\beta S}}{Z(\beta)}\approx \frac{N_{\beta}(S)}{n_{\beta}}.
\end{equation}
Using this we have an estimate of the spectral density function $W(S)$ of the system:
\begin{equation}
W(S)\approx \frac{N_{\beta}(S)}{n_{\beta}}Z(\beta)e^{\beta S}.
\end{equation}

Suppose now, we have performed MC runs at $K$ different values of $\beta$ ($\beta_1$, $\beta_2,
\ldots$, $\beta_K$) where the frequencies $N_k(S)$ have been measured. For every run we
estimate the spectral density function
\begin{equation}
W_k(S)\approx \frac{N_k(S)}{n_k}Z(\beta_k)e^{\beta_k S}.
\end{equation}
We know however that $W(S)$ is universal, there should thus be a unique function
$\bar{W}(S)$. Using the measured estimates of the functions $W_k(S)$, this unique
function can now be constructed in an optimal way following the
procedure by Ferrenberg and Swendsen \cite{Ferrenberg:1988yz,Ferrenberg:1989ui}, which
we do not describe here. Once we know this unique function we may calculate the partition
function at an arbitrary value of $\beta$:
\begin{equation}
\bar{Z}(\beta)=\sum_S \bar{W}(S)e^{-\beta S}.
\end{equation}
We are now ready to estimate the expectation value of any observable at an arbitrary coupling
$\beta$ using
\begin{equation}
\langle\mathcal{O}\rangle_{\beta}=\frac{1}{\bar{Z}(\beta)}\sum_S\bar{\mathcal{O}}(S)\bar{W}(S)e^{-\beta S}.
\end{equation}

\begin{figure}[htbp]
\begin{center}
\includegraphics[width=10cm]{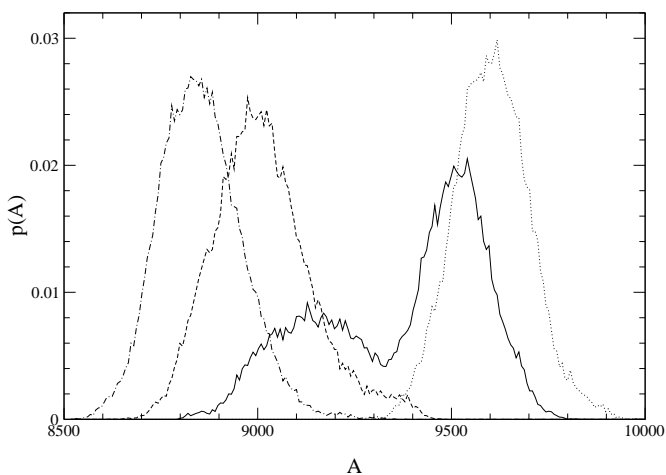}
\end{center}
\caption{The action (energy) distribution at 4 different values of $\beta_c$ on the $4\times 8^3$ lattice for the
$\xi=2$ perfect action. \emph{Dotted}: $\beta=3.02$, \emph{solid}: $\beta=3.03$, \emph{dashed}: $\beta=3.04$,
\emph{dot-dashed}: $\beta=3.05$.}
\label{fig:actdist}
\end{figure}

\begin{figure}[htbp]
\begin{center}
\includegraphics[width=10cm]{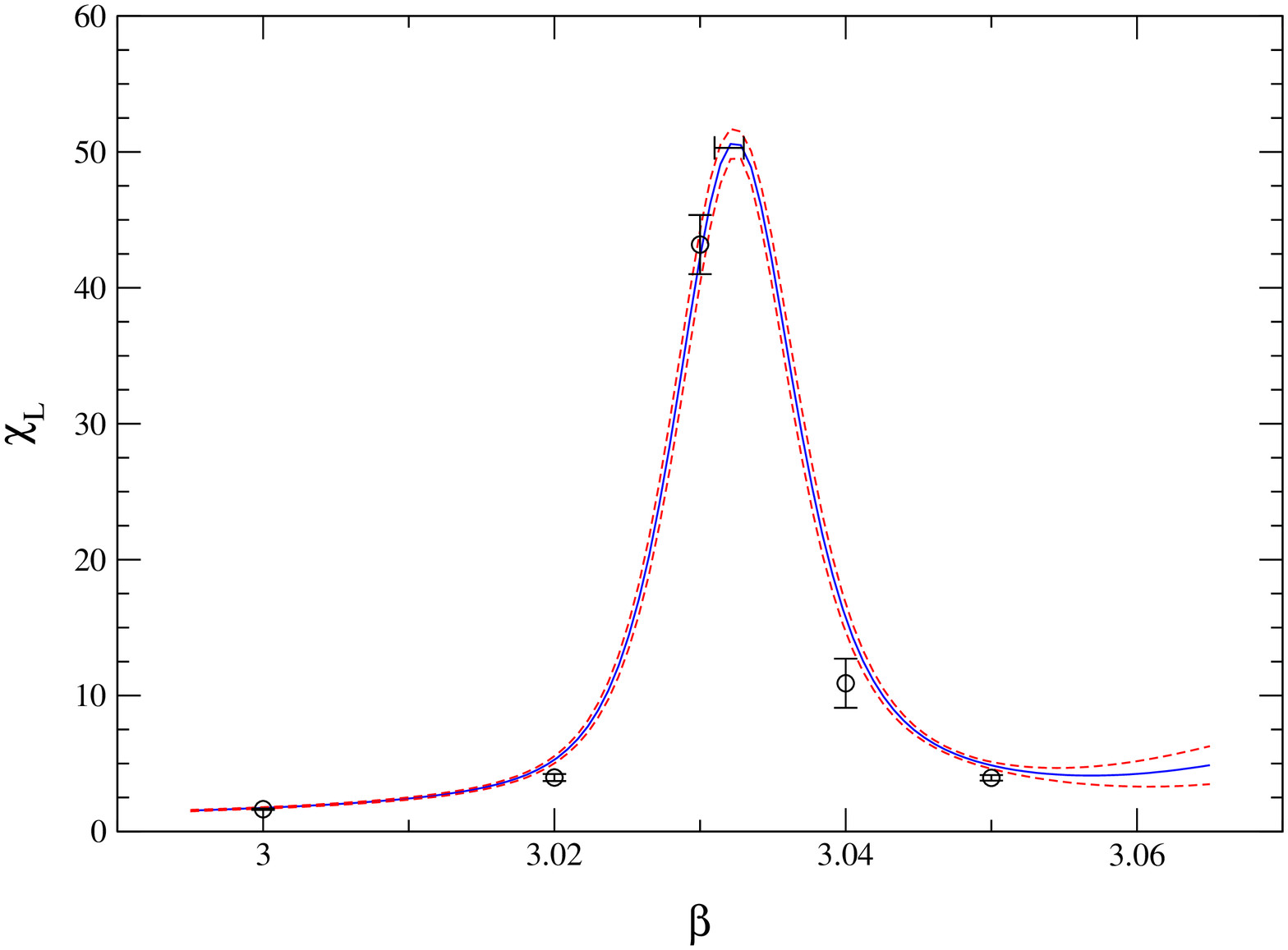}
\end{center}
\caption{The Polyakov loop susceptibility on the $4\times 8^3$ lattice for the
$\xi=2$ perfect action. The solid curves are the interpolations using Ferrenberg-Swendsen
reweighting, the dashed lines indicate the bootstrap error.}
\label{fig:suscrew}
\end{figure}

Figure \ref{fig:actdist} shows the action (energy) histograms for four values of the coupling
$\beta$ in the vicinity of the phase transition. At each coupling, 45000-55600 measurements have been
performed. Note the double peak structure which is
prominent for the $\beta=3.03$ distribution (solid line) closest to $\beta_c=3.032(1)$ and which starts
to evolve for the $\beta=3.04$ distribution (dashed line).
Figure \ref{fig:suscrew} displays the corresponding result of a reweighting procedure for the Polyakov loop
susceptibility. Notice,
that the error of the peak position (which corresponds to the critical coupling $\beta_c$) is small,
in this example the error of $\beta_c$ is as small as $\Delta\beta_c=0.001$.

\begin{figure}[htbp]
\begin{center}
\includegraphics[width=12cm]{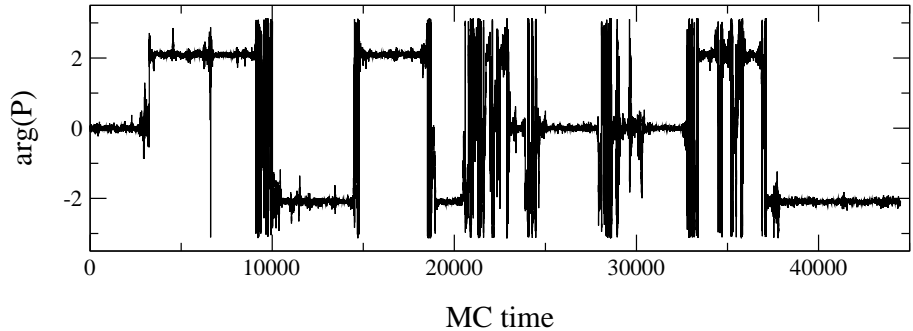}
\end{center}
\caption{MC history of the Polyakov loop measurements on a $5\times 9^3$ lattice at $\beta=3.12$
($\beta_c$=3.118). The argument of the Polyakov loop $P$ is given, $\text{arg}(P)=0, \pm 2\pi/3$ for a
configuration being in the deconfined phase.}
\label{fig:mchist}
\end{figure}

Figure \ref{fig:mchist} displays the MC history of the angle of the Polyakov loop $\text{arg}(P)$ for
$\beta$ close to $\beta_c$. It becomes apparent that most of the information about the phase (and
thus about the Polyakov loop susceptibility $\chi_P$) is contained in the number of flips between the
phases as well as in the time the configuration is staying in one of the phases. An important quantity in this
respect is the persistence time $\tau_p$ which is the MC time (number of sweeps) divided by the number
of flip-flops (change to another phase and another change back to the original phase). Because of critical
slowing down, the autocorrelation time of the operators has to be considered as well. Near the phase transition
knowledge about these quantities is crucial to be able to estimate the quality of the statistics given the number
of sweeps performed.

\section{The Static Quark-Antiquark Potential}\label{sec:qbarq}

The potential between a quark and an antiquark, both carrying colour charge, may be studied even in pure
gauge theory, creating and annihilating a static quark and a static antiquark in the gauge background.
As this potential is rather easy to measure and as it may be compared to experimental data obtained by
studying bound states of a quark and an antiquark, the heavy mesons, it provides an excellent mean of setting
the scale of lattice simulations in the absence of dynamical quarks (i.e.~fermionic theories in the quenched approximation
or pure Yang-Mills theory). The scale is needed if lattice results, such as masses, should be converted
into physical units. In this Section, we will thus present sources of experimental data coming from light
mesons in Section \ref{sec:lightmeson} and from heavy mesons in Section \ref{sec:heavymeson} as well as how the potential
may be examined in pure lattice gauge theory (Sections \ref{sec:latpot}, \ref{sec:measqbarq}) and how it is used to set
the scale in Section \ref{sec:setscale}.

\subsection{Light Mesons}\label{sec:lightmeson}

The elementary constituents of hadronic matter, the quarks, carry colour charge which enables them to build
bound states, held together by the strong force. The simplest such state consists of a quark and a corresponding
antiquark. Since the
early sixties it has been noticed that these states, the mesons, as well as bound states of three
quarks, the baryons, having mass $m$ and spin $J$ group themselves into almost linear, so-called Regge
trajectories \cite{Regge:1959aa,Chew:1961aa,Brink:1973aa}, i.e.~if one plots $m^2$ vs. $J$ the
connection is linear. This can be observed up to spins as high as $J=11/2$. The behaviour is thus mainly
\begin{equation}\label{eq:regge1}
J(m)=\alpha(0)+\alpha'm^2,
\end{equation}
where $\alpha(0)$ is known as the Regge intersect and
\begin{equation}\label{eq:regge2}
\alpha'=\frac{1}{2\pi\sigma}
\end{equation}
is the Regge slope with $\sigma$, the ``string tension''. The explanation for this name stems from
a simple model that explains the linearity of the Regge trajectories: Imagine a rotating string of length
$2d$ with a constant energy density per unit length $\sigma$. If this string spans between (almost) massless
quarks these quarks will move at (almost) the speed of light $c=1$ with respect to the center of mass. The
velocity at a distance $r$ of the centre is thus given by $v(r)=r/d$. From this we may easily calculate the
energy stored in the string,
\begin{equation}
m=2\int_0^d \frac{dr \sigma}{\sqrt{1-v^2(r)}}=\pi d\sigma,
\end{equation}
and the angular momentum,
\begin{equation}
J=2\int_0^d \frac{dr \sigma r v(r)}{\sqrt{1-v^2(r)}}=\frac{\pi d^2\sigma}{2}=\frac{1}{2\pi\sigma}m^2,
\end{equation}
which is the relation between Regge slope and string tension in eqs.~\ref{eq:regge1}, \ref{eq:regge2}.

The experiment predicts values of 470~MeV $< \sqrt{\sigma} <$ 480~MeV for trajectories starting with
a pseudo-scalar ($\pi$ or $K$) and values around 430~MeV for the other trajectories.

\subsection{Heavy Mesons}\label{sec:heavymeson}

After the discovery of the $J/\psi$ meson consisting of a charm quark and a charm anti-quark in $e^+ e^-$ annihilations
it was suggested to treat such states built out of heavy quarks non-relativistically \cite{Appelquist:1975ya}.
Because of the analogy to positronium in electrodynamics these states have been named quarkonia. They
are composed of charm or bottom quarks, the top quark does not appear as a constituent because its weak
decay rate $t\rightarrow bW^+$ is large (see \cite{Quigg:1997fy}). The best studied states are $\bar{c}c$ and
$\bar{b}b$, less is known about bound states of a $b$ and a $c$ quark. The string tension appearing
in the Regge slope may be determined experimentally, e.g.~through the mass splitting between
different states of the same constituents, e.g.~$1P-1S$ for $\bar{b}b$ which is $\chi_b-\Upsilon$. However,
these measurements effectively probe a range of about 0.2~$\sim$~1~fm where it is still difficult to extract
the string tension $\sigma$ unambiguously.

For sufficiently heavy quarks the characteristic time scale of the relative movement of the
quarks is much larger than the one of the gluonic degrees of freedom. In this case we may
apply the adiabatic (or Born-Oppenheimer) approximation and describe the effect of gluons and
sea quarks by an effective instantaneous interaction potential between the heavy quarks. In this
approximation quarkonia are the positronium of QCD. However, unlike in QED, it is not possible to calculate 
the potential perturbatively and predict the spectrum but we have to solve the inverse problem, namely
guess the form of the potential from the observed spectrum (and decay rates). The Cornell potential
\cite{Eichten:1980ms},
\begin{equation}
V(R)=V_0+\frac{\alpha}{R}+\sigma R
\label{eq:isopot}
\end{equation}
describes the interaction reasonably well in the energy range probed. Furthermore, there is not enough
structure in the measured potential to require a more complex parametrisation. Recent results for the
parameters $\alpha$ and $\sigma$ ($V_0$ is an unphysical constant) are $\alpha\approx -0.51$,
$\sqrt{\sigma}\approx 412$~MeV \cite{Quigg:1979vr} and $\alpha\approx -0.52$, $\sqrt{\sigma}\approx 427$~MeV
\cite{Eichten:1980ms}. However in the range 0.2~fm$<r<$~1~fm which is probed by quarkonia splittings these
parametrisations do not differ significantly even from earlier values $\alpha\approx -0.25$,
$\sqrt{\sigma}\approx 455$~MeV \cite{Eichten:1977jk} because the higher value of the Coulomb coefficient is
compensated for by a smaller slope $\sigma$. Still, it is very interesting to see the
compatibility of these values with the estimate for the string tension $\sigma$ from
Regge trajectories of light mesons.

For an extensive review about QCD potentials and their study on the lattice see ref. \cite{Bali:2000gf}.

\subsection{The Static Quark-Antiquark Potential on the Lattice}\label{sec:latpot}
In lattice gauge theory, the rectangular Wilson loop $\mathcal{W}(R,T)$, having extensions $R$ and $T$ in spatial
and temporal directions respectively, starting at the point $(x,t)$, creates at time $t$ a static pair of a quark $q$ and an antiquark $\bar{q}$ sitting at $x+R$ and $x$ and annihilates it again at time $t+T$.
The corresponding static potential between the colour sources is related to the expectation value of $\mathcal{W}(R,T)$ as
\begin{equation}
\langle\mathcal{W}(R,T)\rangle\propto \exp(-V(R)T)
\end{equation}
for large $T$. The potential can thus be calculated in principle by
\begin{equation}
V(R)=\lim_{T\rightarrow\infty}-\frac{1}{T}\ln\langle\mathcal{W}(R,T)\rangle.
\end{equation}
In pure lattice gauge theory the potential is confining the quarks and from strong coupling
expansion \cite{Balian:1975xw,Munster:1981iv}
it follows that the Wilson loop respects an area law
\begin{equation}
\langle\mathcal{W}(R,T)\rangle\propto \exp(-\sigma RT)\mbox{ for }R\rightarrow\infty,
\end{equation}
where $\sigma$ is again the string tension between the static quark sources.

Due to the asymptotic freedom of QCD, perturbation theory is reliable at short distances. So for small
$R$, it predicts a Coulomb-like interaction $V(R)\propto \alpha/R$ between
the quark and the antiquark. The most simple ansatz to describe
both these properties of the $q\bar{q}$-potential is the Cornell potential,
eq.~\ref{eq:isopot}, mentioned in the last section.
All our measurements of the Wilson loop will be fitted to this ansatz.

\subsection{Setting the Scale Using the Static $\mathbf{q\bar{q}}$ Potential}\label{sec:setscale}

Using the string tension $\sigma$ whose value is in principle known experimentally, one may
set the scale corresponding to the coupling $\beta$ used in the simulation. However this is
plagued by several problems: firstly, it is not very easy to determine the long-range quantity
$\sigma$ on the lattice
due to a poor signal/noise ratio for large values of $R$, the value of $\sigma$ is reached
asymptotically which leads to the demand for large (and thus computationally expensive) lattices, and
finally, the string tension $\sigma$ is not well defined in \emph{full} QCD because if the energy of
the string is large enough it may break and create a light quark-antiquark pair (``string breaking'').
Due to these problems, the hadronic scale $r_0$ -- the so-called Sommer scale -- has been introduced \cite{Sommer:1994ce} through
the force $F(R)$ between the two quarks having an intermediate distance 0.2~fm $\leq r \leq 1.0$~fm.
There are several advantages of the hadronic scale: Firstly, it can be measured much better on the lattice
than the string tension $\sigma$, secondly, in this intermediate distance there is reliable information
about the physical scale from phenomenological potential models \cite{Eichten:1980ms,Richardson:1979bt}. The scale is defined like
\begin{equation} \label{eq:hadscale}
r_c^2 F(r_c) \equiv r_c^2 \left.\frac{dV}{dR}\right|_{r_c} = c,
\end{equation}

where originally $c=1.65$ has been chosen. The corresponding experimental value from bottomonium
phenomenology \cite{Sommer:1994ce,Bali:1997am,Bali:1998pi} is $r_0^{-1}=(394\pm 20)$~MeV and thus $r_0\approx 0.50$~fm.
However, it may be useful to measure the hadronic scale at different distances corresponding
to other values of $c$, that is why we have collected values for $c$ and $r_c$ from high-precision measurements
of the static potential performed with the Wilson action \cite{Bali:1992ab,Edwards:1997xf,Guagnelli:1998ud} listed in 
Table \ref{tab:isocrc}.

\begin{table}[htbp]

\renewcommand{\arraystretch}{1.3}
  \begin{center}
    \begin{tabular}{cc}
      \hline\vspace{-0.05cm}
      $r_c/r_0$ & c \\
      \hline
       0.662(1)     & 0.89 \\
       1.00         & 1.65 \\
       1.65(1)      & 4.00 \\
       2.04(2)      & 6.00 \\
      \hline
    \end{tabular}
    \caption{{}Parameter values for the determination of the hadronic
      scale through eq.~\ref{eq:hadscale}.}
    \label{tab:isocrc}
  \end{center}
\end{table}

\subsection{Measuring the Static Potential on the Lattice}\label{sec:measqbarq}

\begin{figure}[htbp]
\begin{center}
\includegraphics[width=3cm]{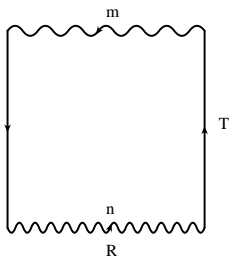}
\end{center}
\caption{The smeared Wilson loop operator $W_{nm}(R,T)$.}
\label{fig:wilson}
\end{figure}

The correlation functions of the strings in MC measurements are always contaminated by high-momentum
fluctuations. The common way to reduce these excited-state contaminations is smoothing the spatial links
employing APE smearing (see Section \ref{sec:smearing}).
The Wilson loop operator $W_{nm}(R,T)$ is constructed as a product of iteratively 
smeared spatial links on smearing level $n$ on time slice $t$ and smearing level $m$ on time slice $t+T$
connecting two spatial points of distance $\vec{R}$ and unsmeared link products of length
$T$ in temporal direction (see Figure \ref{fig:wilson}).
The Wilson loops easiest to measure are the ones parallel to the spatial axes --- in this case the choice
of the shortest paths is unique and the number of paths which could be taken into account and have to be
measured on the lattice is small. However, taking into account separations $\vec{R}$ not parallel to
the spatial axes as well has got the advantages that the discretisation of the separation $|\vec{R}|$ is not as
coarse-grained, i.e.~quantities like $r_0$ that rely on differences on the lattice can be determined more
accurately and more reliably, furthermore the energies of off-axis separated quark-antiquark pairs allow
to estimate effects due to violations of rotational invariance. This is why in some runs we include in the
measurements separations that are multiples of the lattice vectors (1,0,0), (1,1,0), (1,1,1), (2,1,0),
(2,1,1), (2,2,1) (and lattice rotations). The spatial paths are constructed such that all the shortest paths to the
vectors are calculated first, out of these initial six vectors the longer separations are constructed.

Averaging the operators over the three spatial directions as well as over
the whole lattice we get the zero momentum correlation matrix that can be analyzed using the variational method
described in Appendix \ref{app:vartech}.

This method yields projectors to the ground state of the string for each $\vec{R}$ which can be used in
a correlated $\chi^2$ fit to the phenomenological ansatz for the potential, eq.~\ref{eq:isopot}, where $R=|\vec{R}|$.
The hadronic scale $r_c/a$ is then determined by local fits to $V(R)$ including $2\sim 4$ separations and converted
to $r_0/a$ using Table \ref{tab:isocrc} if necessary. Depending on the lattice spacing this can be done for several values of
$c$ yielding information about systematic uncertainties.

\section{Glueballs}\label{sec:glueballs}

\subsection{Introduction}
The particles mediating the strong interaction of QCD, the gluons, carry colour charge and thus interact
with each other, unlike, e.g., their counterpart in electromagnetism, the photons, which have zero electric
charge.
This fact has also been established experimentally, studying 4-jet events in $e^+ e^-$ annihilation.
The spectrum of QCD may thus contain bound states of (mainly) gluons, called glueballs.
These states are described by the quantum numbers $J$ denoting the (integral) spin,
$P$ denoting the eigenvalue $\pm 1$ of the state under parity and $C$, denoting the eigenvalue $\pm 1$ under
charge conjugation. Thus the eigenstates of the Hamiltonian corresponding to glueball states are labeled $|J^{PC}\rangle$.

Due to their non-perturbative
nature, glueballs can be theoretically studied most reliably doing numerical simulations of QCD on
a space-time lattice. As (pure) glueballs are composed entirely of glue, the study using the quenched
approximation of QCD (where fermions, if present, are infinitely heavy) makes sense.\footnote{The question, whether \emph{pure} Yang-Mills Theory has a mass gap is itself a very important
problem. To prove this rigorously is considered one of the seven Millennium Problems, formulated
by the Clay Mathematics Institute (www.claymath.org); the mathematical proof is awarded \$1 million.}
 
Experimentally, there is evidence for exotic glueballs or hybrid particles consisting of quarks and gluonic
excitations. These states (often called ``oddballs'') have exotic (or ``odd'') quantum numbers (e.g., $0^{+-}$ or $1^{-+}$) and cannot be explained by simple bound states of purely quarks.
Due to this, they cannot mix with mesons and are thus particularly interesting to study in
pure gauge theory. However, these states are found to be high lying 
\cite{Michael:1989jr,Bali:1993fb,Morningstar:1997ff,Morningstar:1999rf} (above twice the mass of
the lightest glueball observed). This may lead to ambiguities in the analysis of the experimental
results because these states may mix with bound states of lighter particles.

The lighter glueballs with conventional quantum numbers are difficult to distinguish from
the dense background of conventional meson states observed in experiments.
They are expected to be created in ``gluon-rich'' processes, such as radiative $J/\psi$ decays, central
production (two hadrons passing each other ``nearly untouched'' without valence quark exchange) or
$\bar{p}p$-annihilation.
Currently, there is an ongoing debate whether light glueballs (above all the scalar $0^{++}$ which is the
lightest state in pure lattice gauge theory with a mass of about 1.6~GeV) have been observed
experimentally at about the mass that is predicted by quenched simulations on the lattice, whether the lightest glueball
is much lighter (below 1~GeV) and very broad\cite{Minkowski:1998mf}, or whether glueballs have not been
observed at all in experiments\cite{Klempt:2000ud}. There are mainly two reasons for this uncertainty.
On one hand,
the experimental data seem not yet to be accurate and complete enough, despite large efforts in the
last years, driven by the lattice results; on the other hand, lattice simulations with high
statistics, measuring glueball states, have been performed only in the quenched approximation,
where the quarks are infinitely heavy and thus static. Decreasing the sea (dynamical) quark
mass (finally down to the physical value) will allow to track the glueball states as sea
quark effects are increased. It may turn out, that indeed the glueball mass is lighter
than the one measured in pure gauge theory (for partially quenched results possibly indicating this
see \cite{Michael:1999rs,McNeile:2000xx,Bernard:2001av}). It may even happen that by ``switching
on'' the sea quarks the scalar glueball acquires a very large width and thus decays (almost)
instantaneously to $q\bar{q}$ states, i.e.~it ceases to exist physically.
However, the (partially) unquenched results are rather indecisive yet.

Following the explanation that is dominant at this time, the observed low lying scalar mesons
$f_0(1370)$, $f_0(1500)$ and $f_0(1710)$
are all mixtures of glue and mesonic components as $d\bar{d}$, $u\bar{u}$ and $s\bar{s}$. The
proposed decompositions of the wave functions into these contributions \cite{Amsler:1995tu,Amsler:1996td,Lee:1999kv,Li:2000yn,Close:2000yk,Close:2001ga,Celenza:2000uk,Strohmeier-Presicek:1999yv} are rather different,
however they share some robust common features.
Studies in pure gauge theory may help separating the pure gauge part from the mesonic part
of these states.

\subsection{Glueballs on the Lattice}\label{sec:latgb}

\subsubsection{States}
Glueballs in the continuum are rotationally invariant and have a certain (integral) spin $J$. On the lattice, the
rotational $O(3)$ symmetry is broken, only its discrete cubic subgroup $O_h$ survives the discretisation. Therefore,
the eigenstates of the transfer matrix are classified according to the five irreducible representations of $O_h$:
$A_1$, $A_2$, $E$, $T_1$, $T_2$ with dimensions 1, 1, 2, 3, 3 respectively. Their transformation properties may be
described by polynomials in the components $x$, $y$, $z$ of an $O(3)$ vector as follows: $A_1\sim\{1\}$,
$A_2\sim\{xyz\}$, $E\sim\{x^2-z^2,y^2-z^2\}$, $T_1\sim\{x,y,z\}$, $T_2\sim\{xy,xz,yz\}$.
Generally, an $O(3)$ representation with spin $J$ splits into several representations of the cubic group. Since $O_h$
is a subgroup of $O(3)$, any representation $D_J$ with spin $J$ in the continuum induces a so-called subduced
representation $D_J \downarrow O_h$ on the lattice. This subduced representation no longer has to be irreducible but is a direct
sum of irreducible representations $\Gamma^p$ of $O_h$:
\begin{equation}
D_J \downarrow O_h = \Gamma^1 \oplus \Gamma^2 \oplus \cdots.
\end{equation}
Table \ref{tab:subdrep} lists the subduced representations of $D_J$ for $J=1,\ldots,6$. The spin $J=2$ state for example
splits up into the 2-dimensional representation $E$ and the 3-dimensional representation $T_2$. Approaching the continuum,
rotational symmetry is expected to be restored and thus, as a consequence the mass splitting of these two states will
disappear and the two representations form together the 5 states of a spin $J=2$ object.

\begin{table}
\begin{center}
\renewcommand{\arraystretch}{1.5}
\begin{tabular*}{\textwidth}{c@{\extracolsep{\fill}}ccccccc}
\hline
$\Gamma^p$ & $D_0$ & $D_1$ & $D_2$ & $D_3$ & $D_4$ & $D_5$ & $D_6$\\\hline
$A_1$ & 1 & 0 & 0 & 0 & 1 & 0 & 1\\
$A_2$ & 0 & 0 & 0 & 1 & 0 & 0 & 1\\
$E$   & 0 & 0 & 1 & 0 & 1 & 1 & 1\\
$T_1$ & 0 & 1 & 0 & 1 & 1 & 2 & 1\\
$T_2$ & 0 & 0 & 1 & 1 & 1 & 1 & 2\\
\hline
\end{tabular*}
\end{center}
\caption{The composition of the subduced representation $D_J \downarrow O_h$ in terms of the irreducible representations of the
cubic group $O_h$.}
\label{tab:subdrep}
\end{table}

\subsubsection{Operators}
Pure glue physical states on the lattice are created and annihilated applying gauge invariant operators to the pure
gauge vacuum. In our simulations, we use space-like Wilson loops in the fundamental representation of SU(3). Since we
do not aim at measuring non-zero momentum glueballs we consider only translationally invariant operators, i.e.~operators
averaged in space. 

\begin{figure}[p]
\begin{center}
  \epsfig{file=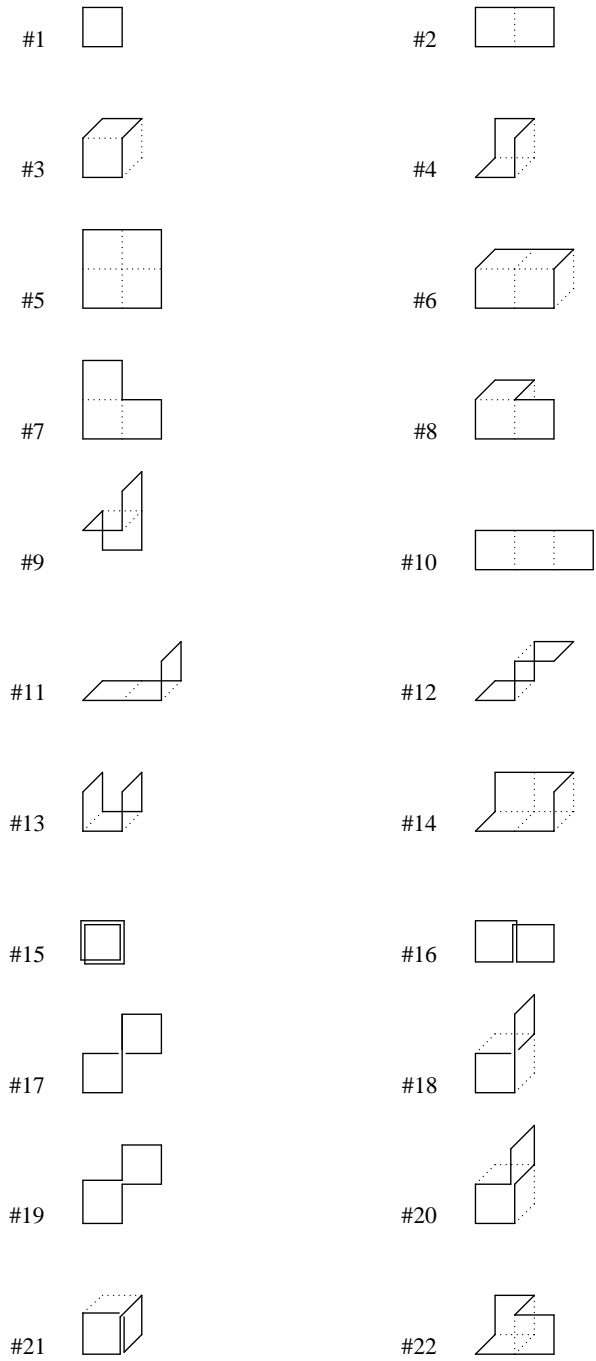,width=8cm}
\end{center}
\caption{All space-like Wilson loop shapes up to length 8.}
\label{fig:loopshapes}
\end{figure}

It is computationally feasible to measure Wilson loop operators up to length 8. The composition of the irreducible representations
$\Gamma^{PC}$ of the cubic group in terms of these 22 loop shapes has been done already in Ref. \cite{Berg:1983kp}. 
Figure \ref{fig:loopshapes} displays all Wilson loop shapes up to length 8 together with the numbering which will be used in the
forthcoming sections. Table \ref{tab:shaporient} lists the number of orientations of the different loop shapes. 
Tables \ref{tab:irrcont1}, \ref{tab:irrcont2} list the irreducible contents of of the $C=+$ and $C=-$ representations of the cubic group,
respectively. Certain operators contribute to the two- or three-dimensional representations with two or three different
polarisations, in analogy to different magnetic quantum numbers $m$ for a given angular momentum $l$ in the O(3) group.
Measuring all these polarisations may suppress statistical noise more than just increasing statistics since the different
polarisations of a loop shape are expected to be anti-correlated. Examining the measured correlators of different
polarisations of the same representation, it turns out that in most cases one of the polarisations is measured very well,
whereas the others exhibit a bad signal/noise ratio.

\begin{table}[h]
\centering

\begin{tabular*}{\textwidth}{l@{\extracolsep{\fill}}|rrrrrrrrrrrr}
loop shape \# \phantom{} & 1 & 2  & 3  & 4 & 5 & 6  & 7  & 8  & 9  & 10 & 11 \\
\hline
dimension $d$   \phantom{} & 6 & 12 & 24 & 8 & 6 & 24 & 24 & 96 & 48 & 12 & 48 \\
\\
loop shape \# \phantom{} & 12 & 13 & 14 & 15 & 16 & 17 & 18 & 19 & 20 & 21 & 22 \\
\hline
 dimension $d$   \phantom{}  & 24 & 12 & 24 & 6  & 12 & 12 & 48 & 12 & 48 & 24 & 96 \\
\end{tabular*}

\caption{Dimension $d$ of the representation of $O_h \times C$ on the loop shapes,
  i.e. number of different orientations.}
\label{tab:shaporient}
\end{table}

\begin{table}[htbp]
  \begin{center}
    \renewcommand{\arraystretch}{1.3}
    \begin{tabular*}{\textwidth}{l@{\extracolsep{\fill}}cccccccccc}
\hline
loop \phantom{ } & $A_1^{++}$ & $A_2^{++}$ & $E^{++}$ & $T_1^{++}$ &
$T_2^{++}$ & $A_1^{-+}$ & $A_2^{-+}$ & $E^{-+}$ & $T_1^{-+}$ & $T_2^{-+}$ \\
\hline
\#1  & 1 & 0 & 1 & 0 & 0 & 0 & 0 & 0 & 0 & 0 \\
\#2  & 1 & 1 & 2 & 0 & 0 & 0 & 0 & 0 & 0 & 0 \\
\#3  & 1 & 0 & 1 & 0 & 1 & 0 & 0 & 0 & 1 & 1 \\
\#4  & 1 & 0 & 0 & 0 & 1 & 0 & 0 & 0 & 0 & 0 \\
\#5  & 1 & 0 & 1 & 0 & 0 & 0 & 0 & 0 & 0 & 0 \\
\#6  & 1 & 0 & 1 & 0 & 1 & 0 & 0 & 0 & 1 & 1 \\
\#7  & 1 & 0 & 1 & 0 & 1 & 0 & 0 & 0 & 1 & 1 \\
\#8  & 1 & 1 & 2 & 3 & 3 & 1 & 1 & 2 & 3 & 3 \\
\#9  & 1 & 0 & 1 & 1 & 2 & 1 & 0 & 1 & 1 & 2 \\
\#10 & 1 & 1 & 2 & 0 & 0 & 0 & 0 & 0 & 0 & 0 \\
\#11 & 1 & 1 & 2 & 1 & 1 & 0 & 0 & 0 & 2 & 2 \\
\#12 & 1 & 1 & 2 & 1 & 1 & 0 & 0 & 0 & 0 & 0 \\
\#13 & 1 & 0 & 1 & 0 & 0 & 0 & 0 & 0 & 0 & 1 \\
\#14 & 1 & 0 & 1 & 1 & 2 & 0 & 0 & 0 & 0 & 0 \\
\#15 & 1 & 0 & 1 & 0 & 0 & 0 & 0 & 0 & 0 & 0 \\
\#16 & 1 & 1 & 2 & 0 & 0 & 0 & 0 & 0 & 0 & 0 \\
\#17 & 1 & 0 & 1 & 0 & 1 & 0 & 0 & 0 & 0 & 0 \\
\#18 & 1 & 0 & 1 & 1 & 2 & 1 & 0 & 1 & 1 & 2 \\
\#19 & 1 & 0 & 1 & 0 & 1 & 0 & 0 & 0 & 0 & 0 \\
\#20 & 1 & 0 & 1 & 1 & 2 & 1 & 0 & 1 & 1 & 2 \\
\#21 & 1 & 0 & 1 & 0 & 1 & 0 & 0 & 0 & 1 & 1 \\
\#22 & 1 & 1 & 2 & 3 & 3 & 1 & 1 & 2 & 3 & 3 \\
\hline
    \end{tabular*}
    \caption{Irreducible contents of the $C$-parity plus representations of the symmetry group of the cube on Wilson loops up to length eight.}
    \label{tab:irrcont1}
  \end{center}
\end{table}

\begin{table}[htbp]
  \renewcommand{\arraystretch}{1.3}
  \begin{center}
    \begin{tabular*}{\textwidth}{l@{\extracolsep{\fill}}cccccccccc}
\hline
loop \phantom{ } & $A_1^{+-}$ & $A_2^{+-}$ & $E^{+-}$ & $T_1^{+-}$ &
$T_2^{+-}$ & $A_1^{--}$ & $A_2^{--}$ & $E^{--}$ & $T_1^{--}$ & $T_2^{--}$ \\
\hline
\#1  & 0 & 0 & 0 & 1 & 0 & 0 & 0 & 0 & 0 & 0 \\
\#2  & 0 & 0 & 0 & 1 & 1 & 0 & 0 & 0 & 0 & 0 \\
\#3  & 0 & 0 & 0 & 1 & 1 & 1 & 0 & 1 & 0 & 1 \\
\#4  & 0 & 1 & 0 & 1 & 0 & 0 & 0 & 0 & 0 & 0 \\
\#5  & 0 & 0 & 0 & 1 & 0 & 0 & 0 & 0 & 0 & 0 \\
\#6  & 0 & 0 & 0 & 1 & 1 & 1 & 0 & 1 & 0 & 1 \\
\#7  & 0 & 1 & 1 & 1 & 0 & 0 & 0 & 0 & 1 & 1 \\
\#8  & 1 & 1 & 2 & 3 & 3 & 1 & 1 & 2 & 3 & 3 \\
\#9  & 0 & 1 & 1 & 2 & 1 & 0 & 1 & 1 & 2 & 1 \\
\#10 & 0 & 0 & 0 & 1 & 1 & 0 & 0 & 0 & 0 & 0 \\
\#11 & 0 & 0 & 0 & 2 & 2 & 1 & 1 & 2 & 1 & 1 \\
\#12 & 0 & 0 & 0 & 2 & 2 & 0 & 0 & 0 & 0 & 0 \\
\#13 & 0 & 0 & 0 & 1 & 0 & 0 & 1 & 1 & 0 & 0 \\
\#14 & 0 & 1 & 1 & 2 & 1 & 0 & 0 & 0 & 0 & 0 \\
\#15 & 0 & 0 & 0 & 1 & 0 & 0 & 0 & 0 & 0 & 0 \\
\#16 & 0 & 0 & 0 & 0 & 0 & 0 & 0 & 0 & 1 & 1 \\
\#17 & 0 & 0 & 0 & 0 & 0 & 0 & 0 & 0 & 1 & 1 \\
\#18 & 0 & 1 & 1 & 2 & 1 & 0 & 1 & 1 & 2 & 1 \\
\#19 & 0 & 1 & 1 & 1 & 0 & 0 & 0 & 0 & 0 & 0 \\
\#20 & 1 & 0 & 1 & 1 & 2 & 1 & 0 & 1 & 1 & 2 \\
\#21 & 0 & 0 & 0 & 1 & 1 & 0 & 1 & 1 & 1 & 0 \\
\#22 & 1 & 1 & 2 & 3 & 3 & 1 & 1 & 2 & 3 & 3 \\
\hline
    \end{tabular*}
    \caption{Irreducible contents of the $C$-parity minus representations of the symmetry group of the cube on Wilson loops up to length eight.}
    \label{tab:irrcont2}
  \end{center}
\end{table}

The accuracy of the measurements of different operators may differ considerably; firstly, because
shapes with larger multiplicity show smaller statistical noise, secondly, because operators
on different smearing schemes may exhibit less or more fluctuations. To give an impression of the
differences between various operators, Tables \ref{impshapes1}, \ref{impshapes2} display correlator
``lifetimes'' obtained from the glueball measurements at coupling $\beta=3.3$, with a temporal lattice
spacing $a_t\approx 0.08$~fm~$\approx$~(2.4~GeV)$^{-1}$. The correlator lifetime $\tau_{\text{corr}}$ is (conventionally) defined such that at 
time $t=\tau_{\text{corr}}$ the relative (bootstrap) error of the correlator is 25\%. The relative
error is linearly interpolated between $t<\tau_{\text{corr}}$ and $t+1 > \tau_{\text{corr}}$.
Table \ref{impshapes1} compares the shapes contributing to different glueball representations, measured
on smearing level 1 (3 times smeared) at $\beta=3.3$, Table \ref{impshapes2} compares the same quantities
measured on smearing level 3 (9 times smeared). For shapes which contribute with more than one projection
to a given representation, the best measured projection is displayed --- usually the other polarisations
are measured much worse due to the anti-correlation.

The operators measured on different smearing levels are again used in the variational method described
in Appendix \ref{app:vartech}. The scalar representation $A_1^{++}$ picks up a vacuum expectation value due to
having the same quantum numbers as the vacuum. The standard procedure is to measure these vacuum expectation values
and subtract them from the correlation matrix elements as
\begin{equation}
C_{ij}(t)=\frac{1}{N_t}\sum_{\tau}\langle O_i(\tau) O_j(\tau+t)\rangle-\langle O_i\rangle \langle O_j\rangle,
\end{equation}
however during the analysis of our measurements, it turns out to be better to treat the vacuum on the same
footing as the other states in the
vacuum channel, i.e.~as a state having mass 0. We thus just cut out the vacuum state obtained from solving
the initial generalised eigenvalue problem (see Appendix \ref{app:vartech}) and perform the fit on the remaining
states.

\begin{table}
\renewcommand{\arraystretch}{1.3}
\begin{center}
\setlength{\tabcolsep}{1pt}
\begin{tabular*}{\textwidth}{c@{\extracolsep{\fill}}ccccccccccc}
\hline
loop \phantom{ } & $A_1^{++}$ & $A_2^{++}$ & $E^{++}$ & $T_1^{++}$ & $T_2^{++}$ & $A_1^{-+}$ & $E^{-+}$ & $T_2^{-+}$ & $A_2^{+-}$ & $T_1^{+-}$ & $T_2^{+-}$\\
\hline
 \#1 & 4.0 &     & 2.2 &     &     &     &     &     &     & 1.2 &   \\
 \#2 & 4.2 & 2.0 & 2.7 &     &     &     &     &     &     & 1.7 & 1.1 \\
 \#3 & 4.0 &     & 2.3 &     & 4.0 &     &     & 2.1 &     & 1.6 & 1.5 \\
 \#4 & 5.0 &     &     &     & 4.0 &     &     &     & 1.4 & 1.6 &   \\
 \#5 & 5.0 &     & 3.4 &     &     &     &     &     &     & 2.1 &   \\
 \#6 & 4.4 &     & 3.0 &     & 4.1 &     &     & 2.1 &     & 2.1 & 2.1 \\
 \#7 & 4.6 &     & 3.1 &     & 2.1 &     &     & 1.0 & 1.3 & 2.0 &   \\
 \#8 & 4.2 & 2.0 & 3.0 & 2.0 & 4.1 & 3.0 & 2.2 & 2.2 & 0.6 & 2.0 & 2.0 \\
 \#9 & 4.0 &     & 1.4 & 1.8 & 4.1 & 2.9 & 2.1 & 2.2 & 1.9 & 2.0 & 1.2 \\
\#10 & 4.5 & 2.1 & 3.1 &     &     &     &     &     &     & 1.8 & 1.4 \\
\#11 & 4.2 & 1.9 & 2.8 & 2.0 & 4.2 &     &     & 2.2 &     & 1.8 & 1.4 \\
\#12 & 4.2 & 1.9 & 3.0 & 2.0 & 4.1 &     &     &     &     & 2.0 & 1.6 \\
\#13 & 3.9 &     & 2.4 &     &     &     &     & 2.0 &     & 1.5 &   \\
\#14 & 4.5 &     & 2.8 & 2.0 & 4.2 &     &     &     & 1.5 & 2.1 & 2.1 \\
\#15 & 4.0 &     & 2.3 &     &     &     &     &     &     & 1.3 &   \\
\#16 & 3.0 & 2.0 & 1.9 &     &     &     &     &     &     &     &   \\
\#17 & 3.0 &     & 2.0 &     & 3.0 &     &     &     &     &     &   \\
\#18 & 3.9 &     & 2.2 & 1.5 & 3.6 & 2.9 & 2.1 & 2.4 & 1.0 & 1.7 & 1.9 \\
\#19 & 4.2 &     & 2.8 &     & 2.0 &     &     &     & 1.1 & 1.9 &   \\
\#20 & 4.0 &     & 2.3 & 1.5 & 3.6 & 2.9 & 2.2 & 2.4 &     & 2.0 & 1.5 \\
\#21 & 3.8 &     & 2.1 &     & 4.0 &     &     & 2.1 &     & 1.6 & 1.6 \\
\#22 & 4.2 & 2.0 & 3.1 & 2.0 & 4.2 & 3.0 & 2.2 & 2.6 & 1.6 & 2.0 & 2.0 \\
\hline
\end{tabular*}
\caption{``Lifetimes'' of glueball correlators at $\beta=3.5$ (first smearing scheme, 3 times smeared); see text.}
\label{impshapes1}
\end{center}
\end{table}

\begin{table}
\renewcommand{\arraystretch}{1.3}
\begin{center}
\setlength{\tabcolsep}{1pt}
\begin{tabular*}{\textwidth}{c@{\extracolsep{\fill}}ccccccccccc}
\hline
loop \phantom{ } & $A_1^{++}$ & $A_2^{++}$ & $E^{++}$ & $T_1^{++}$ & $T_2^{++}$ & $A_1^{-+}$ & $E^{-+}$ & $T_2^{-+}$ & $A_2^{+-}$ & $T_1^{+-}$ & $T_2^{+-}$\\
\hline
 \#1 & 6.0 &     & 4.8 &     &     &     &     &     &     & 0   &   \\
 \#2 & 6.1 & 1.4 & 5.1 &     &     &     &     &     &     & 2.4 & 0   \\
 \#3 & 6.0 &     & 4.9 &     & 5.9 &     &     & 3.9 &     & 1.8 & 0.4 \\
 \#4 & 6.0 &     &     &     & 5.9 &     &     &     & 0.2 & 1.8 &   \\
 \#5 & 6.3 &     & 5.2 &     &     &     &     &     &     & 3.0 &   \\
 \#6 & 6.2 &     & 5.2 &     & 6.1 &     &     & 4.1 &     & 3.0 & 3.1 \\
 \#7 & 6.2 &     & 5.1 &     & 3.1 &     &     & 0   & 0.2 & 3.0 &   \\
 \#8 & 6.1 & 3.0 & 5.1 & 2.4 & 6.1 & 4.2 & 3.4 & 4.0 & 0   & 3.0 & 2.6 \\
 \#9 & 6.0 &     & 2.3 & 2.4 & 6.1 & 4.2 & 3.4 & 4.0 & 1.1 & 3.0 & 0.2 \\
\#10 & 6.2 & 3.1 & 5.2 &     &     &     &     &     &     & 3.1 & 1.3 \\
\#11 & 6.1 & 2.7 & 5.1 & 2.6 & 6.0 &     &     & 4.0 &     & 3.0 & 2.1 \\
\#12 & 6.1 & 2.0 & 5.1 & 2.3 & 6.1 &     &     &     &     & 3.0 & 1.6 \\
\#13 & 5.9 &     & 4.9 &     &     &     &     & 3.9 &     & 0.6 &   \\
\#14 & 6.2 &     & 5.2 & 2.6 & 6.1 &     &     &     & 2.2 & 3.0 & 2.7 \\
\#15 & 6.0 &     & 4.8 &     &     &     &     &     &     & 2.2 &   \\
\#16 & 5.1 & 1.4 & 2.2 &     &     &     &     &     &     &     &   \\
\#17 & 5.1 &     & 2.6 &     & 3.1 &     &     &     &     &     &   \\
\#18 & 5.9 &     & 4.7 & 2.5 & 6.1 & 4.2 & 3.4 & 4.0 & 0   & 2.1 & 0.7 \\
\#19 & 6.2 &     & 5.1 &     & 3.1 &     &     &     & 0   & 2.3 &   \\
\#20 & 6.0 &     & 5.0 & 2.5 & 6.1 & 4.2 & 3.4 & 4.0 &     & 2.2 & 0.7 \\
\#21 & 5.9 &     & 4.7 &     & 5.9 &     &     & 3.9 &     & 1.6 & 0.4 \\
\#22 & 6.1 & 3.0 & 5.1 & 2.6 & 6.1 & 4.2 & 3.5 & 4.0 & 2.3 & 3.0 & 2.2 \\
\hline
\end{tabular*}
\caption{``Lifetimes'' of glueball correlators at $\beta=3.5$ (third smearing scheme, 9 times smeared); see text.}
\label{impshapes2}
\end{center}
\end{table}

\subsubsection{Why Anisotropic Lattices?}

Due to the rather high mass of glueball states and large vacuum fluctuations of the operators
the signal to noise ratio decreases very rapidly as the temporal separation of the source and
the sink is increased.
This demands a small lattice spacing such that the signal can be traced over several slices
increasing the computational cost drastically because the physical volume still has to be kept large
enough ($La\gtrsim 1.2$~fm) in order to avoid severe finite-volume effects. 

The common way to ensure good resolution of the signal on a lattice of moderate size
is using \emph{anisotropic} lattices.

The history of glueball measurements on the lattice employing anisotropic gauge actions is not
very long, despite this the most reliable results from pure gluodynamics stem from anisotropic simulations.
The first study, using a tree-level and tadpole improved action has been performed by Morningstar and
Peardon in 1996 \cite{Morningstar:1997ix}. This was followed by additional studies of the same group
\cite{Morningstar:1997ff,Morningstar:1998du,Morningstar:1999rf,Morningstar:1999dh}.
In 1997, the Kentucky Glueball Collaboration studied glueball matrix elements on anisotropic lattices 
\cite{Dong:1998vi}. A recent study of the glueball spectrum, using the same action as Morningstar and
Peardon, has been performed by Liu \cite{Liu:2000ce,Liu:2000ee}.

\fancyhead[RE]{\nouppercase{\small\it Chapter \thechapter.\, Properties of the Isotropic Action}}
\chapter{Recapitulation: Properties of the Isotropic Action} \label{chap:iso}

As the anisotropic perfect action is based on the isotropic FP action pa\-ra\-met\-ri\-sed using ``fat'' links
it is useful to recapitulate briefly the properties and the results of measurements of physical quantities with
the isotropic action \cite{Wenger:2000aa,Niedermayer:2000yx,Niedermayer:2000ts}.

The measurements that have been performed using the isotropic action comprise the critical temperature
$T_c$ of the deconfining phase transition, the static quark-antiquark potential as well as the lower
lying part of the glueball spectrum. These quantities and techniques used to measure them are presented
in detail in Chapter \ref{chap:physobj}.

This Chapter is organised as follows: In Section \ref{sec:isoact} the action is presented, Section \ref{sec:isotc}
contains results about the deconfining phase transition, in Section \ref{sec:isopot} the results about
the static $q\bar{q}$ potential are summarised, in Section \ref{sec:isoscaling} we report about
scaling tests of the FP action, and finally, in Section \ref{sec:isogb} we present the
results about glueball spectroscopy. For additional information, such as run parameters and
more detailed results, consult \cite{Wenger:2000aa,Niedermayer:2000yx}.

\section{The Isotropic Action} \label{sec:isoact}

The isotropic FP action has been constructed using the block transformation introduced in \cite{Blatter:1996ti}
(see Section \ref{sec:spablock}) which has been optimised for a short interaction range of the FP action as well as for improved rotational
invariance compared to the standard Swendsen-blocking which uses long staples.

The action has been parametrised using the parametrisation presented in Section \ref{sec:isopar}, where the
non-linear parameters $\eta$, $c_i$ have been chosen to be constant (no $x_\mu(n)$-dependence), $i=1$, 2, 3 and
the linear parameters $p_{kl}$ in the mixed polynomial of the standard and smeared staples $u$, $w$ are
non-zero for $0<k+l\leq 4$. The values of the parameters are given in Appendix \ref{app:isoact}.

\section{The Deconfining Phase Transition} \label{sec:isotc}

The properties of the deconfining phase transition in pure Yang-Mills theory are described
in Section \ref{sec:deconf}. On the isotropic lattice we have measured the critical
couplings $\beta_c$ corresponding to temporal extensions $N_t=2,3,4$ of the lattice.

\subsection{Details of the Simulations Performed}

To determine the susceptibility $\chi$, we measure, in the equilibrated system, the Polyakov loops averaged over the lattice 
\begin{equation}
P\equiv \frac{1}{N_s^3}\sum_{\vec{x}}P(\vec{x}),
\end{equation}
where $P(\vec{x})$ is defined in eq.~\ref{eq:polloop},
as well as the action values of the corresponding configurations.
The measurement is performed after each sweep of updates as it is
computationally inexpensive. Both values are stored for later use
in the spectral-density reweighting procedure, described in Section \ref{sec:suscpeak}

We have performed a large number of simulations on lattices with temporal extensions $N_t=$ 2, 3, 4 at 3 to 6
different $\beta$ values near the estimated critical couplings $\beta_c(N_t)$. Additionally, various spatial extensions
$N_s/N_t=2.5\ldots 5$ have been explored in order to be able to examine the finite size scaling of $\beta_c$. The configurations
have been generated by alternating Metropolis and pseudo-over-relaxation steps (see Sections \ref{sec:Metropolis},
\ref{sec:POR}). 

The critical coupling, i.e.~the location of the peak of the susceptibility, is determined employing
the spectral density reweighting method which allows to calculate observables away from the actual
$\beta$ values at which the simulations have been performed \cite{McDonald:1967,Falcioni:1982cz,Ferrenberg:1989ui,
Ferrenberg:1988yz} (see Section \ref{sec:suscpeak}).

To get the value of $\beta_c$ for infinite spatial volume we use the finite-size scaling law
for first order phase transitions:
\begin{equation}
\beta_c(N_t,N_s)=\beta_c(N_t,\infty)-h\left(\frac{N_t}{N_s}\right)^3,
\label{eq:fofss}
\end{equation}
where $h$ denotes a universal quantity in principle independent of $N_t$.

\subsection{Results}

In Table \ref{tab:isobetac} we display the values of $\beta_c(V_s)$ together with the infinite volume
extrapolations according to Eq.~\ref{eq:fofss}. Studying the deconfining
phase transition provides accurate information about the scale at couplings $\beta_c$. This has been
used in scaling tests (see Section \ref{sec:isoscaling}) where the critical
temperature $T_c\propto 1/a$ is compared to the hadronic scale
$r_0\propto a$ whose determination for the isotropic action is described in the following Section.

\begin{table}[htbp]
\renewcommand{\arraystretch}{1.3}
  \begin{center}
    \begin{tabular*}{\textwidth}[c]{c@{\extracolsep{\fill}}ccc}
      \hline\vspace{-0.05cm} 
        $N_s$ & $\beta_c(N_t=2)$ & $\beta_c(N_t=3)$ &
        $\beta_c(N_t=4)$ \\
      \hline
        6        & 2.3552(24) &            &            \\
        8        & 2.3585(12) & 2.6826(23) &            \\
        10       & 2.3593(7)  & 2.6816(12) & 2.9119(31) \\
        12       &            & 2.6803(10) & 2.9173(20) \\
        14       &            &            & 2.9222(20) \\
      \hline
        $\infty$ & 2.3606(13) & 2.6796(18) & 2.9273(35) \\
         $h$     &  0.14(9)   & -0.05(7)   & 0.25(9)   \\
      \hline   
    \end{tabular*}
    \caption{Results of the critical couplings $\beta_c$ from 
      the peak location of the Polyakov loop susceptibility and 
      the corresponding infinite volume limit obtained according 
      to eq.~\ref{eq:fofss}. 
      The finite size scaling constant $h$ is also given.}
    \label{tab:isobetac}
  \end{center}
\end{table}

\section{The Static Quark-Antiquark Potential} \label{sec:isopot}

The static quark-antiquark potential as described in Section \ref{sec:qbarq} is used
to set the scale of our simulations (determining the lattice spacing $a$) as well as
to examine the scaling properties of the FP action, i.e.~measuring lattice artifacts.

\subsection{Details of the Simulation}

Simulations with the isotropic FP action including measurements of the static quark-antiquark potential
have been performed at six different values of $\beta$, three of them corresponding to the critical couplings
indicated in Section \ref{sec:isotc}. The updates have been performed using alternating Metropolis and
pseudo-over-relaxation sweeps. Based on earlier observations in \cite{Bali:1992ab,Edwards:1997xf,Bali:1993ru}
the spatial extent of the lattice has always been chosen to be at least (1.5~fm)$^3$.

The measurements have been performed as described in Section \ref{sec:measqbarq},
for measuring the potential we have chosen $\lambda_s=0.2$ in the APE smearing of the operators
(see Section \ref{sec:smearing}) and used smearing levels $n=$~0, 1, 2, 3, 4.

\subsection{Results}\label{sec:r0res}

It turns out that on coarse lattices with $a_s\gtrsim 0.15$~fm systematic
ambiguities occur determining $r_0$ on different fitting ranges or using different (reasonably chosen)
values of $c$, see Section \ref{sec:setscale}. The main reason for this is the difficulty emerging to define the derivative $dV/dR$ having
at hand discrete values of $R$. Estimates for these systematic ambiguities are displayed together with the
results for $r_0/a$ in Table \ref{tab:isor0}. A possible solution to this problem is including off-axis
separations of the quark and the anti-quark which makes the measurements rather expensive. However, violations
of rotational symmetry show up in this case and allow the estimation of systematic errors in $r_0$ if the result
is changed beyond statistical uncertainty.

\begin{table}[htbp]
  \begin{center}
    \renewcommand{\arraystretch}{1.3}
    \begin{tabular}{ccl}
\hline\vspace{-0.05cm}
      $\beta$ & $N_\tau$ &  \phantom{2cm}$r_0/a$     \\
\hline
        3.400 &          &  $4.833(39)(\genfrac{}{}{0pt}{}{+18}{-22})$ \\ 
        3.150 &          &  $3.717(23)(\genfrac{}{}{0pt}{}{+16}{-17})$ \\
        2.927 &    4     &  $2.969(14)(\genfrac{}{}{0pt}{}{+5}{-14})$  \\
        2.860 &          &  $2.740(10)(\genfrac{}{}{0pt}{}{+17}{-31})$ \\
        2.680 &    3     &  $2.237(7)(\genfrac{}{}{0pt}{}{+11}{-33})$  \\
        2.361 &    2     &  $1.500(5)(\genfrac{}{}{0pt}{}{+29}{-14})$  \\
\hline
      \end{tabular}
    \caption{{}The hadronic scale $r_0/a$ determined from local fits
    to the potential. The first error denotes the statistical error
    and the second is the estimate of systematic ambiguities (see text).}
    \label{tab:isor0}
  \end{center}
\end{table}

\begin{figure}[htbp]
\begin{center}
\includegraphics[width=8cm]{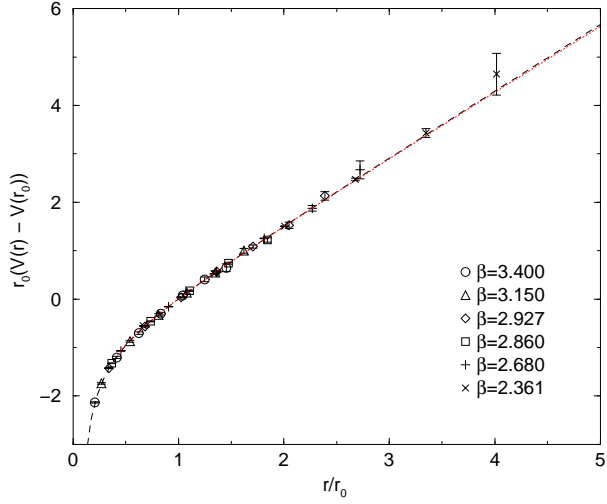}
\end{center}
\caption{The static quark-antiquark potential $V(R)$ expressed in terms of the hadronic scale $r_0$. To match
the curves at $r=r_0$ the unphysical constant $r_0 V(r_0)$ is subtracted for each value of $\beta$. The dashed
line is a fit of all the data to the Cornell potential, eq.~\ref{eq:isopot}, the dotted line stems from ref. \cite{Juge:1997nc}.}
\label{fig:isopot}
\end{figure}

The full static $q\bar{q}$ potentials $a V(R)$ measured on lattices with different $\beta$, i.e.~different lattice
spacings $a$, may be compared directly if $V(R)$ is measured in terms of $r_0$ and the unphysical constant
$r_0 V_0$ is eliminated by subtracting e.g.~$r_0 V(r_0)$ for each lattice spacing. This yields Figure \ref{fig:isopot}
and it turns out that all the measurements agree excellently. Also, the result of \cite{Juge:1997nc}, represented by
the dotted line, is hardly distinguishable from our measurements.

\subsection{The Interpolating Formula}

The information about the scale obtained in the potential measurements may be nicely
collected in an (ad-hoc) interpolating formula that connects the bare coupling $\beta$ to the lattice spacing
(in units of the hadronic scale $r_0$) $a/r_0$. Our fit (analogous to the ones in 
\cite{Edwards:1997xf,Guagnelli:1998ud}) yields
\begin{equation}
\ln(a/r_0)=-1.1622(24)-1.0848(95)(\beta-3)+0.156(17)(\beta-3)^2
\end{equation}
which describes very well the data points in the range $2.361\leq\beta\leq 3.4$.

\section{Scaling Properties} \label{sec:isoscaling}

The parametrised FP action is assumed to reduce lattice artifacts considerably. In order to check
this assumption one may study the scaling properties of renormalisation group invariant quantities,
measuring them on lattices of different lattice spacings. Dimensionless ratios or products of physical
quantities which are RG invariant are constant if there are no scaling violations (due to lattice
artifacts), it is therefore very convenient to study such observables.

Having at hand the physical quantities $r_0$, the hadronic scale, and $\sigma$, the string tension, from
the potential measurements as well as the critical temperature $T_c$ from the study of the deconfining
phase transition, we may build up the dimensionless combinations $T_c/\sqrt{\sigma}$, $r_0T_c$ and $r_0\sqrt{\sigma}$. 

In this section we present the results for the FP action and compare them to results obtained for the Wilson
action and different other improved actions as far as they are available.

\subsection{$\mathbf{T_c/\sqrt{\sigma}}$}

The results for $T_c/\sqrt{\sigma}$, the critical temperature of the deconfining phase transition in terms of
the string tension, measured on lattices of different $\beta$ are compared to the continuum extrapolations for
other gauge actions in Table \ref{tab:tcsigma}. The Wilson action data stem from \cite{Beinlich:1997ia} where
results from \cite{Boyd:1996bx} have been used. From the same reference we state the value for the $1\times2$
tree-level improved action. ``DBW2'' stands for ``doubly blocked from Wilson in two coupling space'', the
results of this action stem from \cite{deForcrand:1999bi}. Iwasaki \cite{Iwasaki:1997sn} has performed measurements
with the RG improved (Iwasaki) action \cite{Iwasaki:1985we}. Finally, the results
obtained using a tree level and tadpole improved action are from \cite{Bliss:1996wy}. The continuum extrapolations for all
these different actions stem from a systematic reanalysis \cite{Teper:1998kw} of the data.

\begin{table}[htbp]
  \begin{center}
    \renewcommand{\arraystretch}{1.3}
    \begin{tabular}{lcl}
\hline\vspace{-0.05cm}
 action   & $\beta$  & $T_c/\sqrt{\sigma}$ \\
\hline
FP action                         & 2.927    & 0.624(7)  \\
                                  & 2.680    & 0.622(8)  \\
                                  & 2.361    & 0.628(11) \\    
\hline
Wilson   \cite{Beinlich:1997ia}   & $\infty$ & 0.630(5)  \\
$1\times2$ tree-level improved \cite{Beinlich:1997ia} & $\infty$ & 0.634(8)  \\
DBW2  \cite{deForcrand:1999bi}    & $\infty$ & 0.627(12) \\
RG improved \cite{Iwasaki:1997sn}     & $\infty$ & 0.651(12) \\
tree-level/tadpole improved \cite{Bliss:1996wy}         & $\infty$ & 0.659(8) \\
\hline
    \end{tabular}
    \caption{{}Results of the deconfining temperature in units of the
    string tension obtained with the FP action and continuum values
    for different other actions. }\label{tab:tcsigma}
  \end{center}
\end{table}

\begin{figure}[htbp]
\begin{center}
\includegraphics[width=7cm,angle=-90]{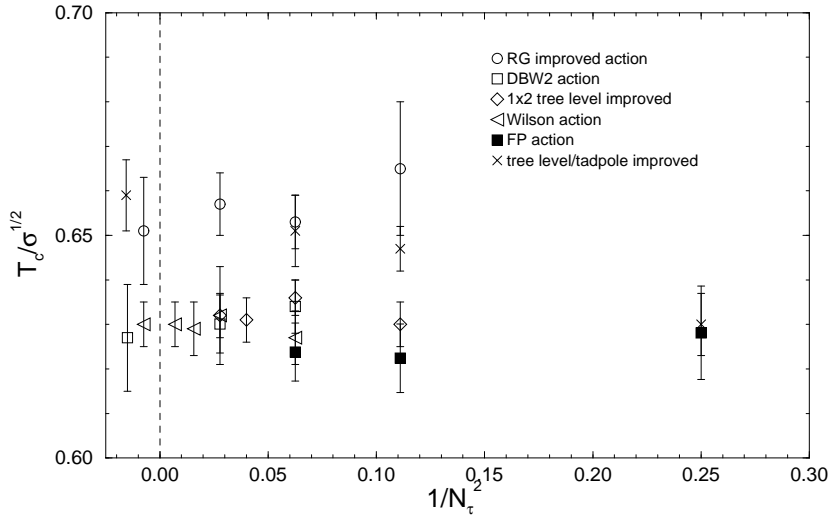}
\end{center}
\caption{The dimensionless quantity $T_c/\sqrt{\sigma}$ vs. $1/N_t^2$ for different actions.}
\label{fig:tcsigma}
\end{figure}

These values, together with those for finite lattice spacings, are displayed in Figure \ref{fig:tcsigma}.
Our measurements are all compatible with the continuum extrapolation of the
Wilson data and we observe no scaling violations of the FP action within statistical errors even
on very coarse lattices corresponding to $N_t$=2. The continuum results obtained using the RG improved action
as well as the tree-level and tadpole improved action lying considerably higher (about $3\sigma$
for the latter) may be explained by ambiguities occurring with the measurement of the string tension $\sigma$.
Depending on the separations $r$ of the static quarks taken into account performing the fit to the Cornell
potential, eq.~\ref{eq:isopot}, the obtained value of the string tension may vary considerably (together with
variations of the Coulomb parameter $\alpha$), such that determinations by different groups may differ
systematically if the methods do not coincide. Especially, Bliss et al. \cite{Bliss:1996wy}, having
studied the treelevel and tadpole improved action, determined the string tension $\sigma$ using results coming
from torelon measurements and employing the string formula, eq.~\ref{eq:stringtorelon}.

\subsection{$\mathbf{r_0T_c}$}

\begin{figure}[htbp]
\begin{center}
\includegraphics[width=8cm,angle=-90]{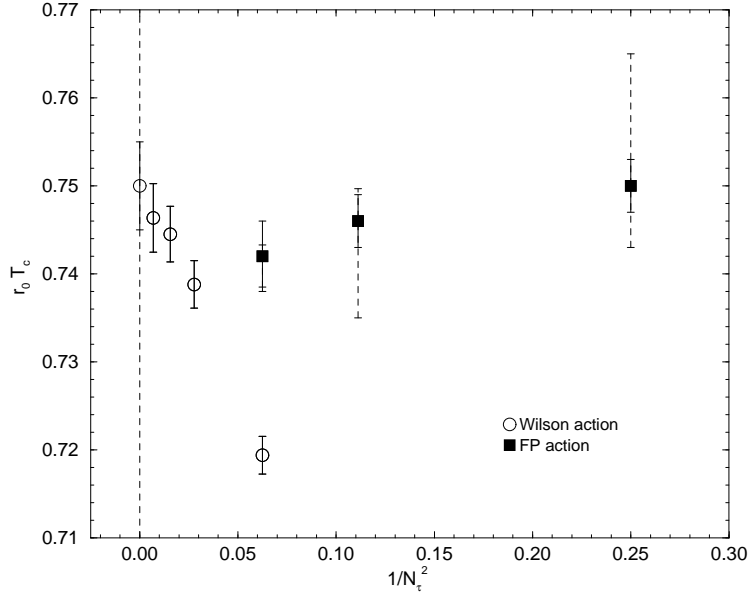}
\end{center}
\caption{The dimensionless quantity $r_0T_c$ vs. $1/N_t^2$ for the FP action compared to the
Wilson action \cite{Beinlich:1997ia,Guagnelli:1998ud}, the dashed error bars indicate the
systematic error from ambiguities in calculating the force $F(R)=V'(R)$ on coarse lattices.}
\label{fig:r0tc}
\end{figure}

Whereas it was easy to find data for the quantity $T_c/\sqrt{\sigma}$ presented in the last section,
precise determinations of $r_0/a$ are missing in the literature except for the Wilson action
 \cite{Edwards:1997xf,Guagnelli:1998ud} and thus we are not able to compare our results for $r_0 T_c$
to a large number of other actions. Our results compared to those obtained with the Wilson action are
displayed in Table \ref{tab:r0tc}. The Wilson results for the critical couplings $\beta_c(N_t)$, $N_t=$ 4,
6, 8, 12  have been taken from \cite{Beinlich:1997ia} and the corresponding values of $r_0/a$ have been calculated
using the interpolating formula from \cite{Guagnelli:1998ud}. The continuum value has been extrapolated
using a linear fit in the leading correction term $1/N_t^2$ on $N_t=$ 4, 6, 8. Figure \ref{fig:r0tc} displays the
same quantities; including the systematic ambiguities on coarse lattices (dashed lines), see Section \ref{sec:r0res}.

\begin{table}[htbp]
  \begin{center}
    \renewcommand{\arraystretch}{1.3}
    \begin{tabular}{ccc}
      \hline\vspace{-0.05cm}
       $N_\tau$ & Wilson action & FP action \\
      \hline 
       2    &          &  0.750(3) \\
       3    &          &  0.746(3) \\
       4    & 0.719(2) &  0.742(4) \\
       6    & 0.739(3) &           \\
       8    & 0.745(3) &           \\
       12   & 0.746(4) &           \\
       $\infty$& 0.750(5) &         \\
      \hline
    \end{tabular}
    \caption{{}Results for the critical temperature in terms of the
    hadronic scale, $r_0 T_c$, from measurements with the Wilson
    action \cite{Beinlich:1997ia,Guagnelli:1998ud}, and the FP action.} 
    \label{tab:r0tc}
  \end{center}
\end{table}

Compared to the Wilson action, which shows scaling violation of about 4~\% at $N_t=4$ the FP
action performs much better in this scaling test exhibiting scaling within 1~\% even on very
coarse lattices corresponding to $N_t=3$ and 2. However this statement is softened due to the
large systematic uncertainties at larger lattice spacings.

\subsection{$\mathbf{r_0\sqrt{\sigma}}$}\label{sec:r0sigma}

\begin{figure}[htbp]
  \begin{center}
    \includegraphics[width=9cm,angle=-90]{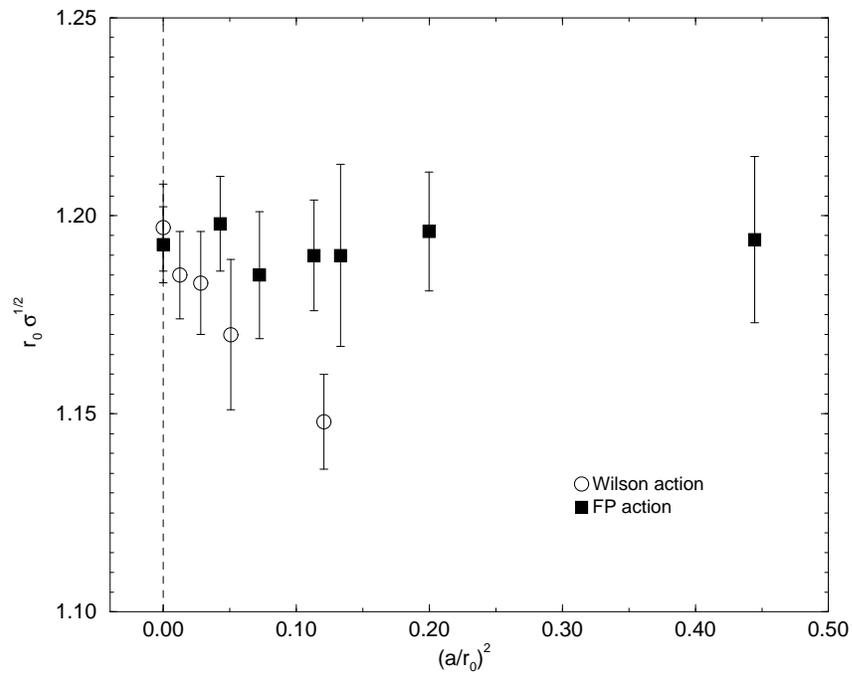}
    \caption{{}Scaling behaviour of $r_0 \sqrt{\sigma}$ for the Wilson
    action \cite{Guagnelli:1998ud,Edwards:1997xf,Teper:1998kw} (empty circles) and the FP action (filled squares).}
    \label{fig:r0sigma}
  \end{center}
\end{figure}

The dimensionless product $r_0\sqrt{\sigma}$ contains two (physical) quantities obtained in the fit of
the static quark-antiquark potential. However its scaling properties are not trivial as $r_0$ probes
the intermediate range and $\sigma$ the long range properties of the potential. We are thus using the
values of $r_0/a$ obtained from local fits to the potential (Table \ref{tab:isor0}) , whereas the values of $\sigma a^2$ are
determined from global fits including large values of $R$. Table \ref{tab:r0sigma} collects the resulting
values of $r_0\sqrt{\sigma}$ together with results from the Wilson action calculated using the interpolating
formula for $r_0/a$ from \cite{Guagnelli:1998ud} and the parametrisation for the string tension from \cite{Edwards:1997xf}.
The continuum extrapolation of the Wilson data has been performed in \cite{Teper:1998kw}.
The same data is displayed in Figure \ref{fig:r0sigma}. The error bars are purely statistical and dominated by the errors
of the string tension --- the systematic ambiguities in $r_0$ may thus be neglected in this case.

\begin{table}[htbp]
  \begin{center}
    \renewcommand{\arraystretch}{1.3}
    \begin{tabular*}{8cm}[c]{l@{\extracolsep{\fill}}lll}
      \hline
\multicolumn{2}{c}{Wilson action} & \multicolumn{2}{c}{FP action} \\\cline{1-2}\cline{3-4}
$\beta$ &  $r_0 \sqrt{\sigma}$ & $\beta$ &  $r_0 \sqrt{\sigma}$ \\
\hline
5.6925  &  1.148(12) & 2.361   &  1.194(21) \\
5.8941  &  1.170(19) & 2.680   &  1.196(15) \\
6.0624  &  1.183(13) & 2.860   &  1.190(23) \\
6.3380  &  1.185(11) & 2.927   &  1.191(12) \\
        &            & 3.150   &  1.185(16) \\
        &            & 3.400   &  1.198(12) \\
$\infty$&  1.197(11) & $\infty$&  1.193(10) \\
\hline
\end{tabular*}
    \caption{{}$r_0 \sqrt{\sigma}$ for the Wilson \cite{Guagnelli:1998ud,Edwards:1997xf,Teper:1998kw} and the FP action.}
    \label{tab:r0sigma}
  \end{center}
\end{table}

Again, the Wilson action shows scaling violations of about 4~\% for $N_t=4$ whereas the FP action exhibits no
artifacts even on lattices as coarse as $\beta=2.361$ (corresponding to criticality for $N_t=2$).

\section{Glueball Spectroscopy} \label{sec:isogb}

The physical nature of glueballs and how they can be studied on the lattice is described in Section \ref{sec:glueballs}.
Numerical simulations using the methods described there have been performed using the isotropic
FP action and will be described in this Section.

\subsection{Details of the Simulation}

The simulations have been performed on three different lattice spacings in the range 0.10~fm~$\leq a \leq$~0.18~fm
and volumes between (1.4~fm)$^3$ and (1.8~fm)$^3$. The updates have been performed using alternating Metropolis
and pseudo-over-relaxation sweeps (see Sections \ref{sec:Metropolis}, \ref{sec:POR}). For the listing of all
the simulation parameters see Table \ref{tab:isogbsimpar}.

\begin{table}[htbp]
\renewcommand{\arraystretch}{1.3}
  \begin{center}
    \begin{tabular*}{\textwidth}[c]{c@{\extracolsep{\fill}}cccc}
      \hline\vspace{-0.05cm} 
      $\beta$ & lattice & $r_0/a$ & $a$ [fm] & \# sweeps / measurements\\
      \hline
      2.86 & $10^4$ & 2.740(10) & 0.18 & 32000 / 8000\\
      3.15 & $12^4$ & 3.717(23) & 0.13 & 101000 / 10100\\
      3.40 & $14^4$ & 4.833(39) & 0.10 & 86520 / 14420\\
      \hline
    \end{tabular*}
    \caption{Run parameters for the glueball measurements using the isotropic FP
      action.}
    \label{tab:isogbsimpar}
  \end{center}
\end{table}

In the first simulation, performed on a coarse lattice with $\beta=2.86$, only the
mass of the lowest-lying $A_1^{++}$ channel has been determined measuring the
loop shapes 2, 4, 7, 10, 18 (see Figure \ref{fig:loopshapes}) which had been shown
to be measured well and having a large overlap with the state in a short pilot run.
For the other two simulations all 22 Wilson loop shapes up to length 8 are measured
and projected to all 20 irreducible glueball representations. To kill ultraviolet
fluctuations and to enhance the overlap with the glueball states the measurements
are performed on five smearing levels ${\cal S}_n U$, $n=2,4,\ldots,10$ with
smearing parameter $\lambda_s=0.2$ (see Section \ref{sec:smearing}). 

It turns out that there are operators which are measured notoriously bad, i.e.~that
have a small signal/noise ratio. As there are representations with contributions
of as many as 145 operators (the representation $T_1^{+-}$ on five smearing levels)
this may destabilise the fitting procedure described in Appendix \ref{app:vartech}.
That is why we remove badly measured operators right from the start of the analysis.

For the scalar glueball, the vacuum expectation values are not subtracted from the
operators, but the vacuum state is treated like the other states in the vacuum channel,
see Section \ref{sec:latgb}.

\subsection{Results}

The full (and rather lengthy) lists of results are contained in \cite{Niedermayer:2000yx},
we only display the masses of the lowest-lying and best measured states in units of
the hadronic scale $r_0$ compared to the results of other groups in Tables \ref{tab:isogbres1},
\ref{tab:isogbres2}. The results denoted by UKQCD, Teper and GF11 have been obtained using the
Wilson action whereas the results of Morningstar and Peardon (M\&P) and Liu have been
measured using a tree-level/tadpole improved anisotropic action.

\begin{table}[htbp]
  \begin{center}
    \renewcommand{\arraystretch}{1.3}
    \begin{tabular*}{\textwidth}[c]{l@{\extracolsep{\fill}}lll}
      \hline\vspace{-0.05cm}
      Collab. & $r_0 m_{0^{++}}$ & $r_0 m_{2^{++}}$  & year    \\
      \hline
      UKQCD \cite{Bali:1993fb}        & 4.05(16)    & 5.84(18)   & 1993 \\
      Teper \cite{Teper:1998kw}       & 4.35(11)    & 6.18(21)   & 1998 \\
      GF11 \cite{Vaccarino:1999ku}    & 4.33(10)    & 6.04(18)   & 1999 \\
      M\&P \cite{Morningstar:1999rf}  & 4.21(15)    & 5.85(8)    & 1999 \\
      Liu  \cite{Liu:2000ce}          & 4.23(22)    & 5.85(23)   & 2000 \\
      \hline
      FP action                       & 4.12(21)    &[5.96(24)]  & 2000 \\
      \hline
    \end{tabular*}
    \caption{{}Comparison of the two lowest glueball masses in units
    of $r_0$. Our $2^{++}$ value is not extrapolated to the continuum
    but is the mass obtained at a lattice spacing $a=0.10$ fm.}
    \label{tab:isogbres1}
  \end{center}
\end{table}

\begin{table}[htbp]
  \begin{center}
    \renewcommand{\arraystretch}{1.3}
    \begin{tabular*}{\textwidth}[c]{l@{\extracolsep{\fill}}llll}
      \hline\vspace{-0.05cm}
    Collab. & $r_0 m_{0^{-+}}$ & $r_0 m_{2^{-+}}$ & $r_0 m_{1^{+-}}$ & year \\
      \hline
    Teper \cite{Teper:1998kw}     & 5.94(68) & 8.42(78) & 7.84(62) & 1998 \\
    M\&P \cite{Morningstar:1999rf}& 6.33(13) & 7.55(11) & 7.18(11) & 1999 \\
    \hline
    FP action                     &[6.74(42)]&[8.00(35)]&[7.93(78)]& 2000 \\
    \hline
    \end{tabular*}
    \caption{{}Comparison of glueball masses in units of $r_0$. 
      Values in brackets denote masses obtained at a lattice spacing
      $a=0.10$ fm and are not extrapolated to the continuum.}
    \label{tab:isogbres2}
  \end{center}
\end{table}

\begin{figure}[htbp]
  \begin{center}
    \includegraphics[width=9cm,angle=-90]{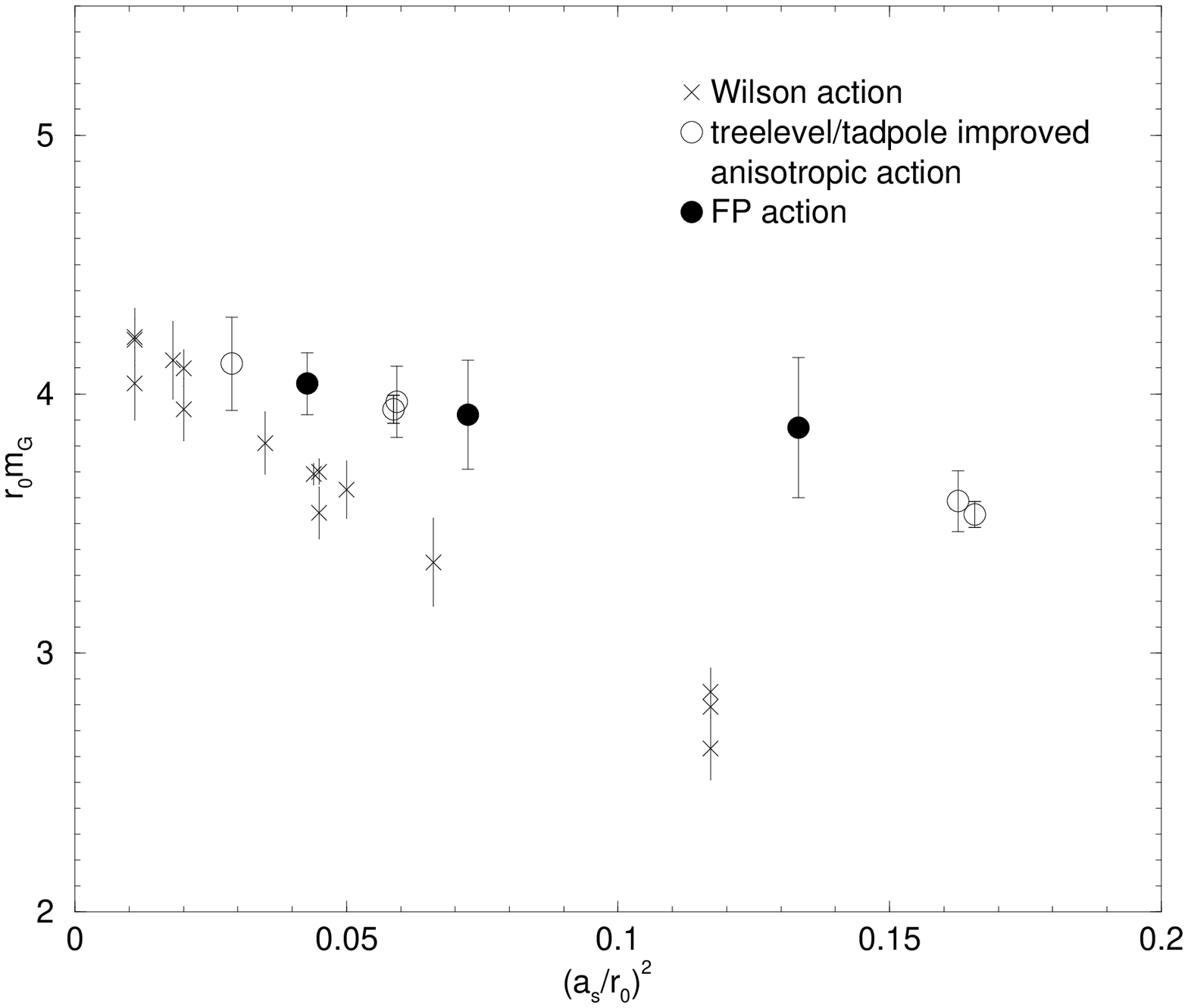}
    \caption{{}Mass estimates for the $0^{++}$ glueball. 
      Results from simulations of the Wilson action (crosses) and a 
      tree level/tadpole improved anisotropic action (empty circles) 
      are shown together with the results obtained with the FP action 
      (filled circles).}
    \label{fig:iso0pp}
  \end{center}
\end{figure}

\begin{figure}[htbp]
  \begin{center}
    \includegraphics[width=9cm,angle=-90]{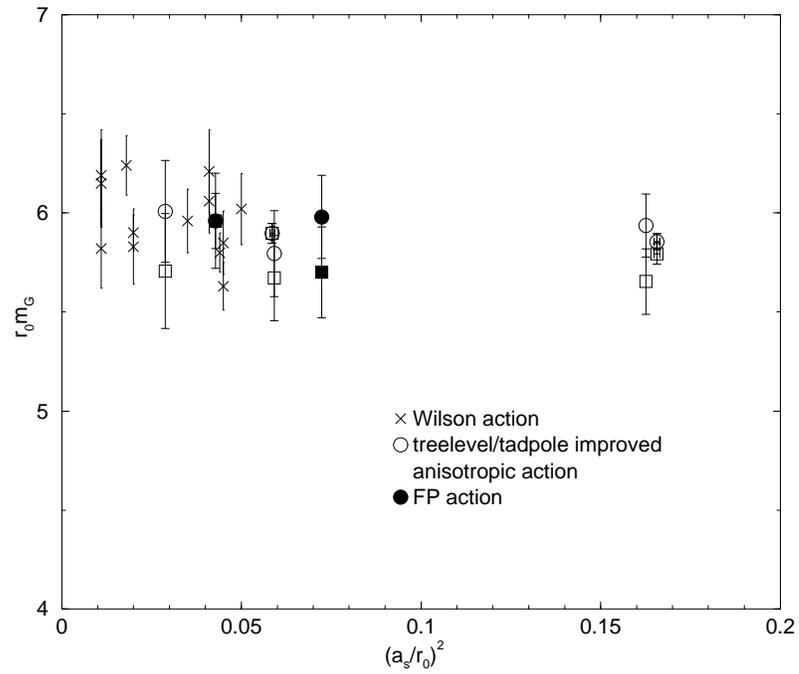}
    \caption{{}Mass estimates for the $2^{++}$ glueball.
    Results from simulations of the Wilson action (crosses)
    and a tree level/tadpole improved anisotropic action (empty
    symbols) are shown together with the results obtained with the FP
    action (filled symbols). Squares and circles denote the $E^{++}$
    and $T_2^{++}$ mass estimates, respectively.}
    \label{fig:iso2pp}
  \end{center}
\end{figure}

The results for the scalar ($0^{++}$) glueball are displayed together with the ones of other groups
in Figure \ref{fig:iso0pp}, the results for the $2^{++}$ glueball which is composed of the two
representations $E^{++}$ and $T_2^{++}$ in Figure \ref{fig:iso2pp}. 

\fancyhead[RE]{\nouppercase{\small\it Chapter \thechapter.\, Construction of Perfect Anisotropic Actions}}
\chapter{The Construction of Perfect Anisotropic Actions} \label{ch:construction}

In this Chapter we describe our concept of generating anisotropic perfect actions without having to
repeat the whole lengthy procedure described in Section \ref{sec:fpcons}, making use of the parametrisation
of the isotropic FP action \cite{Wenger:2000aa,Niedermayer:2000yx,Niedermayer:2000ts}. In Section \ref{sec:consmeth}
we describe two methods one might think of, in the following we concentrate on the one of the two ways which 
is presented more deeply in Section \ref{sec:spablock}. The parametrisation ansatz, for the isotropic as well as for the
anisotropic case is finally presented in Section \ref{sec:par}.

\section{Methods}\label{sec:consmeth}
\begin{figure}
\begin{center}
\includegraphics[width=3cm]{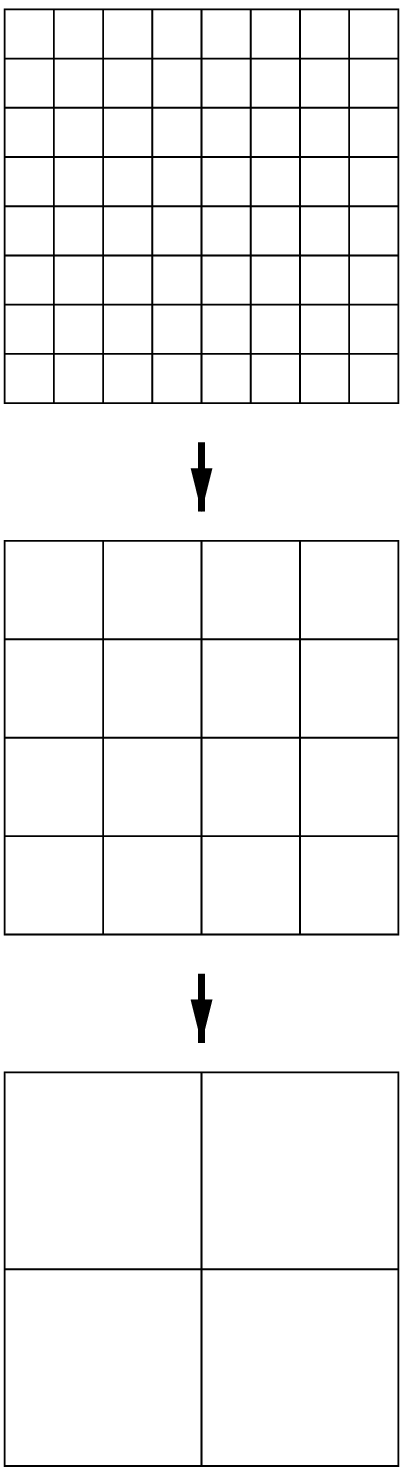}
\includegraphics[width=3cm]{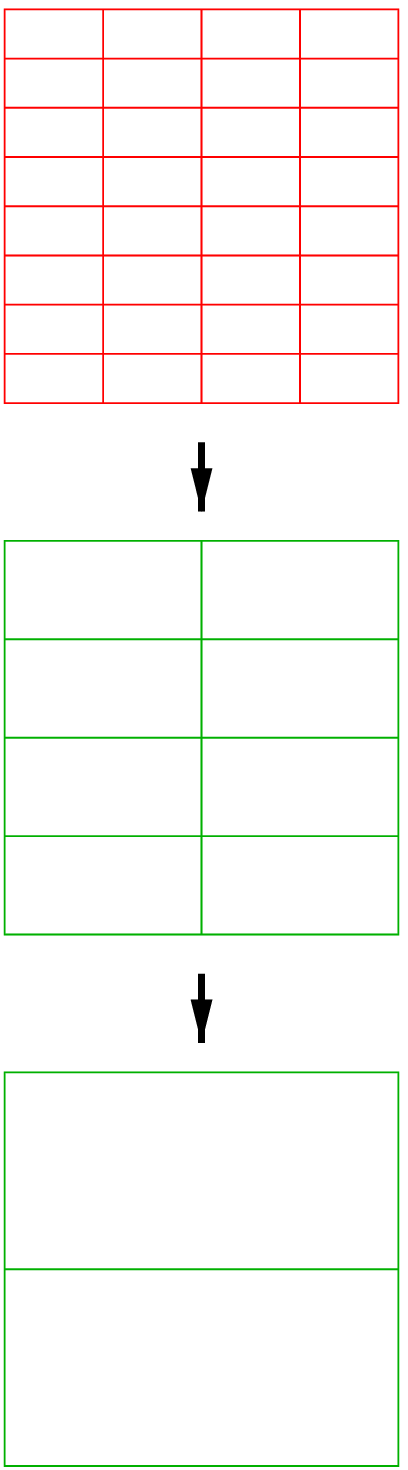}
\includegraphics[width=3cm]{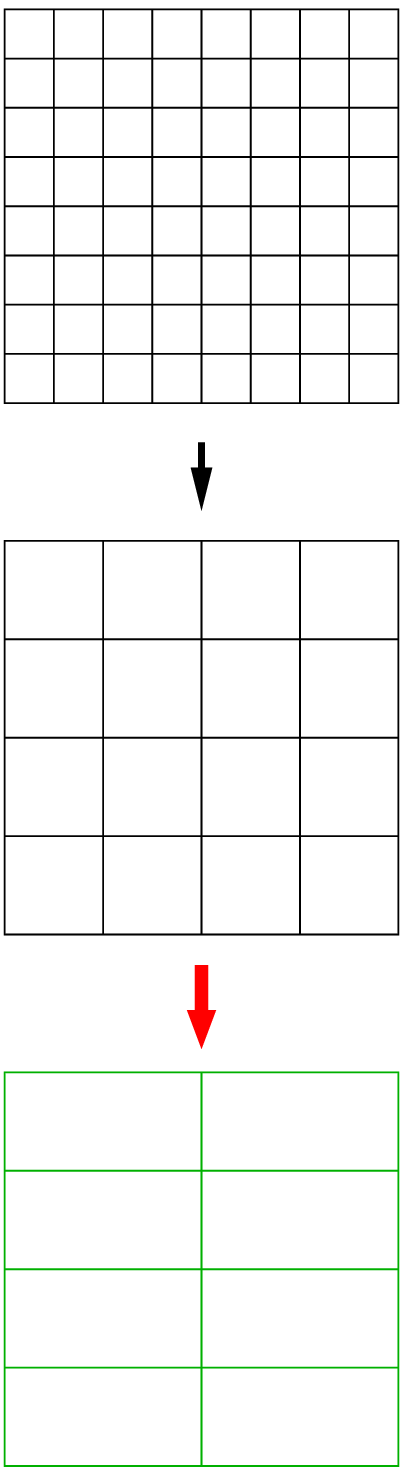}
\end{center}
\caption{Different methods of creating a FP action. \emph{left}: isotropic action; \emph{center}: anisotropic action,
using an anisotropic starting action; \emph{right}: anisotropic action, performing purely spatial blocking steps in the
end.}
\label{fig:blmethods}
\end{figure}

The method for constructing FP actions (see Section \ref{sec:fpaction}) may be straightforwardly generalised
to anisotropic lattices in two different ways, illustrated, together with the
isotropic method, in Figure \ref{fig:blmethods}: 
Firstly, one might repeat the procedure used to generate
the isotropic FP action (see Section \ref{sec:fpcons}), starting with an \emph{anisotropic} action 
$\mathcal{A}_0^{\text{ai}}$ which
behaves well on very small fluctuations and doing the same cascade process
involving blocking steps and parametrisations as before. Of course, the
parametrisation has to be adapted such that it may account for the space/time
asymmetry.

Secondly, one may start with an \emph{isotropic}
parametrised FP action on rather large fluctuations (e.g. with the action
$\mathcal{A}_5^{444}$, see Appendix \ref{app:actions}) and perform a small number of purely \emph{spatial} blocking
steps. If a scale 2 block transformation is used, the resulting actions will
have anisotropies $\xi=$~2, 4, 8, $\ldots$.
This second approach may be performed much
faster than the first one because the number of minimisation/fitting steps is
smaller and one does not have to construct and check a new (anisotropic)
starting action $\mathcal{A}_0^{\text{ai}}$.
It is because of this we concentrate in this work on the spatial blocking approach
described in the next section.

\section{Spatial Blocking} \label{sec:spablock}
To perform the blocking only spatially, doubling the lattice spacing in spatial direction while
leaving the temporal lattice spacing unchanged,
we modify the blocking kernel $T(U,V)$  of the symmetric RGT 3 transformation \cite{Blatter:1996ti},
appearing in eq.~\ref{eq:FP_equation}
\begin{equation}\label{eq:blockkern}
T(U,V)=\sum_{n_B,\mu}\left( \mathcal{N}_\mu^\infty(n_B)-\frac{\kappa}{N}\text{Re Tr} [V_\mu(n_B)Q_\mu^\dag(n_B)]\right),
\end{equation}
with the normalisation
\begin{equation}\label{eq:rgtnorm}
\mathcal{N}_\mu^\infty(n_B)=\max_{W\in \text{SU(N)}}\left\{\frac{\kappa}{N}\text{Re Tr}[WQ_\mu^\dag(n_B)]\right\}
\end{equation}
and the blocked link
\begin{equation}
Q_\mu(n_B)=W_\mu(2n_B)W_\mu(2n_B+\hat{\mu}),
\end{equation}
where $W_\mu(n)$ denotes a smeared (fuzzy) fine link, which is constructed
as follows. In addition to the simple staples we build as well
diagonal staples by first going in the planar or spatial diagonal directions
orthogonal to the direction $\mu$ of the final smeared link, then doing
a step in direction $\mu$ and finally returning along the corresponding diagonal
to $n+\hat{\mu}$. To be specific, let us first build matrices $W^{(m)}(n,n')$
connecting the sites $n$ and $n'$ where $n'$ is a neighbouring site orthogonal
to the smeared link $W_\mu(n)$ to be built:

\begin{subequations}
\begin{eqnarray} \label{eq:type3ww} 
&& W^{(0)}(n,n) = 1 \, , 
                                           \label{eq:type3wa}    \\
&& W^{(1)}(n,n+\hat{\nu}) = U_{\nu}(n) \, , 
                                           \label{eq:type3wb}    \\
&& W^{(2)}(n,n+\hat{\nu}+\hat{\rho}) = \frac{1}{2}
\left( U_{\nu}(n)U_{\rho}(n+\hat{\nu}) + 
       U_{\rho}(n)U_{\nu}(n+\hat{\rho}) \right)\, ,
                                           \label{eq:type3wc}    \\
&& W^{(3)}(n,n+\hat{\nu}+\hat{\rho}+\hat{\lambda}) = 
\frac{1}{6} \left(
U_{\nu}(n)U_{\rho}(n+\hat{\nu}) U_{\lambda}(n+\hat{\nu}+\hat{\rho})
+ \mbox{perms.} \right) \, .\nonumber\\
&&                                           \label{eq:type3wd}
\end{eqnarray}
\end{subequations} 
Here $\nu$, $\rho$ and $\lambda$ go over all (positive and negative)
directions different from $\mu$ and from each other.
In eqs.~\ref{eq:type3wc}, \ref{eq:type3wd}
the sum is taken over all shortest paths leading to the 
endpoint $n'$ of the corresponding diagonal.
The fuzzy link operator $W_\mu(n)$ is then constructed by a modified Swendsen 
smearing \cite{Swendsen:1984vu,Hasenfratz:1984hx,Bowler:1985hv,Bowler:1986rx,Akemi:1993ue}:
\begin{equation} \label{eq:type3w} 
W_\mu(n) = \sum_{m=0}^{3} \sum_{n'} c_{m} W^{(m)}(n,n')
U_{\mu}(n') W^{(m)}(n'+\hat{\mu},n+\hat{\mu}).
\end{equation} 
The coefficients $c_m$ are free parameters subject to the constraint: 
\begin{equation} \label{eq:sumc}
c_0 + 6 \, c_1 + 12 \, c_2 + 8 \, c_3 = 1.
\end{equation}
This normalisation condition ensures that for a trivial field
configuration, $W_\mu(n)$ is equal to the unit matrix.
These parameters $c_m$ have been optimised for a short ranged
quadratic FP action \cite{Blatter:1996ti} and take on the values
$c_1$=0.07, $c_2$=0.016, $c_3$=0.008 and thus $c_0$=0.324.

To modify the isotropic definition, in order to end up with a purely spatial blocking, we set
\begin{equation}
 Q_4(n_B)=W_4(2n_B)
\end{equation}
thus doing only a smearing and
no blocking in temporal direction. (In the case of spatial blocking the expression ``$2n_B$'' stands for
$(2n_B^1,2n_B^2,2n_B^3,n_B^4)$.)
Additionally, we have the freedom to choose different values of $\kappa$ for spatial ($\kappa_s$) and
temporal ($\kappa_t$) links (see also Appendix \ref{app:scal}).

\section{The Parametrisation} \label{sec:par}

As already stated, the FP-type actions, described in principle by an infinite number of
couplings, have to be parametrised in order to be useful for numerical simulations.
We have
shown that a parametrisation of the isotropic FP gauge action including APE-like smearing behaves much
better and is much more flexible compared to common parametrisations using traces of closed loops with
comparable computational cost \cite{Wenger:2000aa,Niedermayer:2000yx,Niedermayer:2000ts}. Thus the starting point for the parametrisation of the anisotropic perfect
gauge action is our isotropic parametrisation, presented in Section \ref{sec:isopar}. We extend this ansatz 
to suit anisotropic actions in Section \ref{sec:anisopar}. The different extensions will be compared 
in Section \ref{sec:compext}.

\subsection{The Isotropic Parametrisation} \label{sec:isopar}

The isotropic action has been parametrised using mixed polynomials of traces of simple loops (plaquettes) built
from single gauge links as well as from (APE-like) smeared links. In this Section, the parametrisation will be
described in detail.

We introduce the notation $S_\mu^{(\nu)}(n)$ for the sum of two staples of gauge links connecting two lattice sites
in direction $\mu$ lying in the $\mu\nu$-plane:
\begin{equation}
S_\mu^{(\nu)}(n)=U_\nu(n)U_\mu(n+\hat\nu)U_\nu^\dagger(n+\hat\mu)+U_\nu^\dagger(n-\hat\nu)U_\mu(n-\hat\nu)U_\nu(n-\hat\nu+\hat\mu).
\end{equation}

Our parametrisation may depend on local fluctuations measured by $x_\mu(n)$ which is defined like
\begin{equation}
x_\mu(n)=\text{Re Tr}(Q_\mu^{\text{s}}(n)U_\mu^\dagger(n)),
\end{equation}
with the symmetrically smeared link
\begin{equation}
Q_\mu^{\text{s}}(n)=\frac{1}{6}\sum_{\lambda\ne\mu}S_\mu^{(\lambda)}(n)-U_\mu(n).
\end{equation}
This parameter is negative, $-4.5\leq x_\mu(n) \leq 0$, it vanishes for trivial gauge configurations: 
$x_\mu(n)\equiv 0$.

To build a plaquette in a plane $\mu\nu$ from smeared links it is convenient to introduce
\emph{asymmetrically} smeared links. First define\footnote{The argument $n$ is suppressed in the following.}
\begin{equation}
\label{eq:asym1}
Q_\mu^{(\nu)}=\frac{1}{4}\left(\sum_{\lambda\ne \mu,\nu} S_\mu^{(\lambda)}+\eta(x_\mu)S_\mu^{(\nu)}\right)
- \left(1+\frac{1}{2}\eta(x_\mu)\right)U_\mu.
\end{equation}
Out of these sums of matrices connecting two neighboring points $n$, $n+\hat\mu$, we build the asymmetrically smeared links 
\begin{equation}
\label{eq:smlconstr}
W_\mu^{(\nu)}=U_\mu+c_1(x_\mu)Q_\mu^{(\nu)}+c_2(x_\mu)Q_\mu^{(\nu)}U_\mu^\dagger Q_\mu^{(\nu)}+\cdots,
\end{equation}
where $\eta(x)$, $c_i(x)$ are polynomials with free coefficients (determined later by the fit to the FP action):
\begin{eqnarray}
\eta(x) & = & \eta^{(0)}+\eta^{(1)}x+\eta^{(2)}x^2+\cdots,\\
c_i(x) & = & c_i^{(0)}+c_i^{(1)}x+c_i^{(2)}x^2+\cdots.
\end{eqnarray}

Of course, these asymmetrically smeared links, built out of a large number of paths connecting the neighbouring lattice
sites are no longer elements of the SU(3) gauge group. They might be projected back to SU(3), however this task increases
the computational cost
in actual numerical simulations, and additionally our studies have shown that projection
reduces the degrees of freedom in defining the action, we are thus using the smeared links $W_\mu^{(\nu)}$ as they
are.

We are now able to build a smeared plaquette like
\begin{equation}
w_{\mu\nu}=\text{Re Tr}(1-W_{\mu\nu}^{\text{pl}}),
\end{equation}
as well as the ordinary plaquette
\begin{equation}
u_{\mu\nu}=\text{Re Tr}(1-U_{\mu\nu}^{\text{pl}}),
\end{equation}
where
\begin{equation}
W_{\mu\nu}^{\text{pl}}(n)=W_\mu^{(\nu)}(n)W_\nu^{(\mu)}(n+\hat\mu)W_\mu^{(\nu)\dagger}(n+\hat\nu)W_\nu^{(\mu)\dagger}(n)
\end{equation}
and
\begin{equation}
U_{\mu\nu}^{\text{pl}}(n)=U_\mu(n)U_\nu(n+\hat\mu)U_\mu^\dagger(n+\hat\nu)U_\nu^\dagger(n).
\end{equation}

Finally, the action is built out of these plaquette variables as
\begin{equation}
\label{eq:isoacpol}
\mathcal{A}[U]=\frac{1}{N_c}\sum_n\sum_{\mu<\nu}\sum_{k,l}p_{kl}u_{\mu\nu}(n)^k w_{\mu\nu}(n)^l,
\end{equation}
where the coefficients $p_{kl}$ are again free parameters defined by the fit to the FP action.

For the actual fitting procedure (see Sections \ref{sec:nlfit}, \ref{sec:linfit}) it is very important
to note that the parameters $p_{kl}$ enter linearly, whereas the parameters $c_i$ as well as $\eta$
enter non-linearly.

\subsection{The Anisotropic Extensions} \label{sec:anisopar}

The isotropic parametrisation presented in the last section does not distinguish between spatial and
temporal directions on the lattice. It is thus not able to describe anisotropic actions at all. However
there are some straightforward ways of extending it to parametrise anisotropic perfect
actions.

Firstly, the coefficients $p_{kl}$ in eq.~\ref{eq:isoacpol} may be chosen differently depending on the
orientation of the plaquette $\mu\nu$, i.e., $p_{kl}^{\text{sp}}$ for $\mu\nu \in \{12, 13, 23\}$
(spatial plaquettes) and
$p_{kl}^{\text{tm}}$ for $\mu\nu \in \{14, 24, 34\}$ (temporal plaquettes).

\begin{figure}
\begin{center}
\begin{tabular*}{\textwidth}[c]{c@{\extracolsep{\fill}}ccc}
\includegraphics[width=2.25cm]{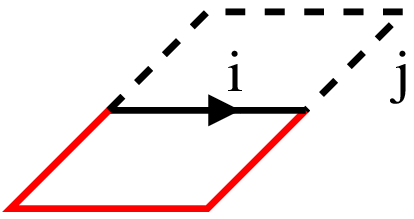} &
\includegraphics[width=2.25cm]{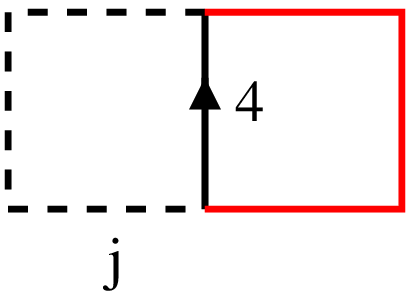} &
\includegraphics[width=1.6875cm]{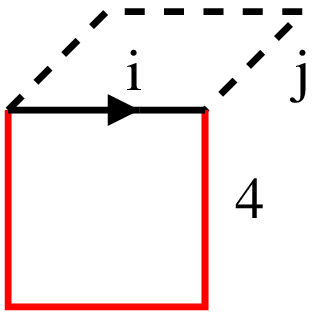} &
\includegraphics[width=1.35cm]{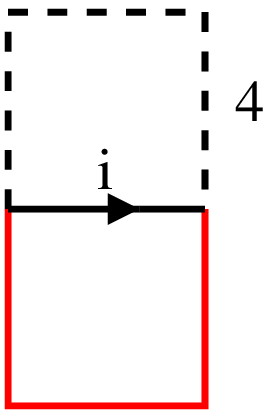}\\
$\eta_1$ & $\eta_2$ & $\eta_3$ & $\eta_4$\\
\end{tabular*}
\end{center}

\caption{Asymmetry in the construction of the smeared links matrices $Q_\mu^{(\nu)}$. \emph{From left to right}:
spatial staple contributing to a spatial smeared link in a spatial plaquette ($\eta_1$),
temporal staple contributing to a temporal smeared link ($\eta_2$),
temporal staple contributing to a spatial smeared link in a spatial plaquette ($\eta_3$),
temporal staple contributing to a spatial smeared link ($\eta_4$).}
\label{fig:etas}
\end{figure}

Secondly, the parameter $\eta$ entering in eq.~\ref{eq:asym1} describing the asymmetry between different
staples contributing to a smeared link $W_\mu^{(\nu)}$ depending on their orientation may be generalised
to distinguish between spatial and temporal links contributing to spatial and temporal smeared plaquettes 
$W_{\mu\nu}^{\text{pl}}$:
\begin{eqnarray}
Q_i^{(j)} & = & \frac{1}{4}(\sum_{k\ne i,j} S_i^{(k)}+\eta_1 S_i^{(j)}+\eta_3 S_i^{(4)})-(\frac{1}{2}
+\frac{1}{2}\eta_1+\frac{1}{2}\eta_3)U_\mu,\\
Q_i^{(4)} & = & \frac{1}{4}(\sum_{\lambda\ne i,4} S_i^{(\lambda)}+\eta_4 S_i^{(4)})-(1
+\frac{1}{2}\eta_4)U_\mu,\\
Q_4^{(j)} & = & \frac{1}{4}(\sum_{\lambda\ne 4,j} S_4^{(\lambda)}+\eta_2 S_4^{(j)})-(1
+\frac{1}{2}\eta_2)U_\mu,\\
\end{eqnarray}
where the anisotropic parameters $\eta_1,\ldots,\eta_4$ may be again polynomials in the
local fluctuation $x_\mu$. These situations are depicted in Figure \ref{fig:etas}.

\begin{figure}
\begin{center}
\begin{tabular*}{\textwidth}[c]{c@{\extracolsep{\fill}}cc}
\includegraphics[width=1.6875cm]{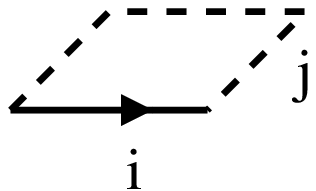}\hspace{1cm} &
\includegraphics[width=1.35cm]{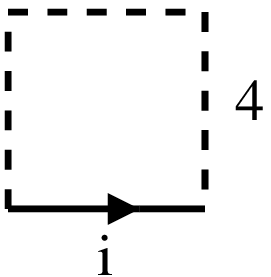}\hspace{1cm} &
\includegraphics[width=1.35cm]{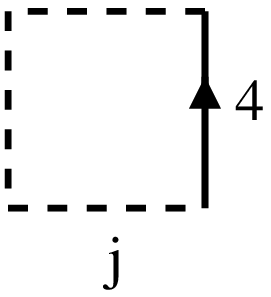}\\
$c_{i1}$ & $c_{i2}$ & $c_{i3}$\\
\end{tabular*}
\end{center}
\caption{Asymmetry in the construction of the smeared links $W_\mu^{(\nu)}$ out of the matrices
$Q_\mu^{(\nu)}$. \emph{From left to right}:
spatial $Q_\mu$ finally contributing to a spatial smeared plaquette ($c_{i1}$),
spatial $Q_\mu$ finally contributing to a temporal smeared plaquette ($c_{i2}$),
temporal $Q_4$ ($c_{i3}$).}
\label{fig:cis}
\end{figure}

Finally, also the construction of the smeared links $W_\mu^{(\nu)}$ out of the matrices
$Q_\mu^{(\nu)}$, described by the parameters $c_i$ in eq.~\ref{eq:smlconstr} is generalised such that these parameters
are chosen differently for temporal links (always contributing to temporal plaquettes),
spatial links contributing to spatial plaquettes and spatial links contributing to temporal
plaquettes, introducing $c_{i1}$, $c_{i2}$ and $c_{i3}$ respectively, see Figure \ref{fig:cis}.

It is not a priori clear whether all the three extensions are necessary at the same time. Results 
from fits to different extended parametrisations are given in Section \ref{sec:compext}.

\subsection{The Normalisation Condition}

Our anisotropic action shall have the correct normalisation and anisotropy in the continuum limit. This can be
accomplished demanding the following two normalisation conditions to be exactly fulfilled:

\begin{eqnarray}
p^{\text st}_{01}+p^{\text st}_{10}+2p^{\text st}_{01}c_{13}+p^{\text st}_{01}c_{13}\eta_2+
2p^{\text ss}_{01}c_{11}\eta_3+p^{\text st}_{01}c_{12}\eta_4 & = & \xi,\\
p^{\text ss}_{01}+p^{\text ss}_{10}+2p^{\text ss}_{01}c_{11}+
2p^{\text st}_{01}c_{12}+2p^{\text ss}_{01}c_{11}\eta_1 & = & \frac{1}{\xi},
\end{eqnarray}
where all parameters $\eta$, $c$ denote the constant ($0^{\text{th}}$ order) terms in the polynomials in $x$.
The way how these normalisation conditions are derived is presented in Appendix \ref{app:aisym}.

\subsection{Comparing Different Sets of Parameters} \label{sec:compext}

During the construction of the perfect $\xi=2$ action (see Chapter \ref{ch:xi2act}) we check
the quality of the fits to different anisotropic extensions of the APE parametrisation. As it turns
out, it is indispensable to have the full parameter set including four
different asymmetry parameters $\eta$ as well as distinguishing between different plaquette orientations 
($c_{i1}$, $c_{i2}$, $c_{i3}$) in
building up $W_{\mu\nu}$. Table \ref{tab:fitqual} lists the results of fits to the $\xi=2$ perfect
action as well as measured anisotropies $\xi_R$ using the torelon dispersion relation (see Section \ref{sec:xir}).
It is clear from these results that setting all $c_{ij}$, $j=1,2,3$ equal leads to a large renormalisation of
the anisotropy; as well, setting $\eta_3=\eta_4\equiv 0$ in order to have a positive definite transfer matrix
decreases the quality of the fit considerably. We thus choose to use at the same time anisotropic linear parameters
($p_{kl}^{\text{sp}}$ and $p_{kl}^{\text{tm}}$) as well as non-linear parameters $\eta_1$, $\eta_2$, $\eta_3$,
$\eta_4$ and $c_{i1}$, $c_{i2}$, $c_{i3}$ to parametrise our final anisotropic perfect gauge actions.

The proceeding and the results of the parametrisation of anisotropic perfect actions are presented in
Chapter \ref{ch:xi2act} for the anisotropy $\xi=2$ and Chapter \ref{ch:xi4act} for $\xi=4$.

\begin{table}[htbp]
  \begin{center}
    \renewcommand{\arraystretch}{1.3}
    \begin{tabular*}{10.5cm}[c]{r@{\extracolsep{\fill}}rrrrrr}
\hline
\# $\eta$ & \# $c_i$ & $i$ & $\max(k+l)_{\text{sp}}$ & $\max(k+l)_{\text{tm}}$ & $\chi_{\text{d}}^2$ & $\xi_{\text{R}}$\\
\hline
4 & 1 & 4 & 4 & 4 & 0.0250 & 1.63(2)\\
2 & 3 & 3 & 4 & 4 & 0.0238 & \\
4 & 3 & 3 & 4 & 4 & 0.0144 & 1.912(9)\\
\hline
\end{tabular*}
\caption{Comparison of the accuracy of the fit on derivatives $\chi_{\text{d}}^2$ on 20 configurations
at $\beta=3.5$ and the measured renormalised anisotropy $\xi_R$ (using the torelon dispersion relation)
at $\beta=3.3$ for different choices of the set of non-linear parameters.}
    \label{tab:fitqual}
  \end{center}
\end{table}

\fancyhead[RE]{\nouppercase{\small\it Chapter \thechapter.\, Scalar Fields and Perturbative Tests}}
\chapter{Scalar Fields and Perturbative Tests}\label{chap:pert}

The methods for creating anisotropic actions presented in Chapter \ref{ch:construction}
 are first examined for the case of $d=2$ free scalar field theory. In this theory, the
renormalisation group transformation and thus the calculation of the FP action
may be performed analytically. In Section \ref{sec:quadapp}
we study the quadratic approximation to the FP action in $d=4$ gauge theory. 

It turns out that the ansatz works in both theories and that the results look very similar to what
is obtained for the isotropic FP action.

\section{The Free Scalar Field in d=2}

The FP action for free scalar fields has been studied
earlier \cite{Rufenacht:1998aa}, a recent study also involves anisotropic lattices
and massive fields \cite{Bietenholz:1999kr}.

\subsection{Anisotropic Blocking out of the Continuum}
A simple way to get a FP action for the free scalar field in $d=2$ is to block 
it out of the continuum choosing different lattice spacings $a_s$ and $a_t$ in
temporal and spatial direction, respectively. The FP propagator 
is derived in Appendix \ref{app:scal}. The result is
\begin{eqnarray}\label{eq:fpscal}
D^{\text{FP}}(q) & = & \sum_{l=-\infty}^\infty \frac{\Pi(q)^2}
{\xi^{-1}(q_1+2\pi l_1)^2+\xi(q_2+2\pi l_2)^2} + \frac{1}{\kappa},
\end{eqnarray}
with a $2\pi$-periodic function $\Pi(q)$ defining the blocking. The momentum
$q=(q_1,q_2)$ is expressed in units of the lattice spacing $a$ and runs in the first
Brillouin zone $]-\pi/a,\pi/a]$. The parameter $\kappa$ is due to the generalisation of
the blocking from $\delta$ functions to a Gaussian ($\kappa\rightarrow\infty$ corresponds
to the $\delta$ function), i.e.~it is a sort of a smearing parameter (with extensive smearing
for small values of $\kappa$). For isotropic lattices it can be shown analytically that $\kappa=6$
yields ultra-locality for $d=1$.
The blocking function $\Pi(k)$ is a
dimensionless quantity, while the propagator and $\kappa$ carry the dimension
$m^{-2}$ and $m^2$ respectively. It is purely a matter of convention whether we express (or regulate) these quantities in
terms of $a_s$ or $a_t$, yielding different $\xi$--factors at different
places:
\begin{eqnarray}
D(q)^{\text{FP}} & = & \xi \left( \sum_{l=-\infty}^\infty \frac{\Pi(q)^2}{(q_s+2\pi
  l_s)^2 + \xi^2 (q_t+2\pi l_t)^2} + \frac{1}{\kappa_s} \right),
\end{eqnarray}
where $\kappa_s$ and the propagator are now expressed in terms of $a_s$ and 
\begin{eqnarray}
\label{eq:aiprop}
D(q)^{\text{FP}} & = & \frac{1}{\xi} \left( \sum_{l=-\infty}^\infty \frac{\Pi(q)^2}{\xi^{-2}(q_s+2\pi
  l_s)^2 +  (q_t+2\pi l_t)^2} + \frac{1}{\kappa_t} \right),
\end{eqnarray}
if they are expressed in terms of $a_t$. The values
$\kappa_s a_s^2=\kappa_t a_t^2=6$ are optimal for locality in spatial and temporal
direction, respectively, and we obtain ultra-locality for the $1d$ projection
onto the corresponding axis. As $\kappa\sim m^2$ we get $\kappa_t/\kappa_s=\xi^2$.

The corresponding FP action is defined as:
\begin{equation}
S^{\text{FP}}=\frac{1}{2}\sum_{n,r}\rho(r)\phi_n\phi_{n+r}
\end{equation}
on the scalar field $\phi_r$, where
\begin{equation}
\rho(q)=D^{-1}(q).
\end{equation}

If we require the action to go over to the form of classical field theory for small momenta
$q$, the coefficient of the $\text{O}(q^2)$ term in the action should be 1:
\begin{equation}
\tilde{\rho}(q)=\sum_{n_1=-\infty}^{\infty}\sum_{n_2=-\infty}^{\infty}\rho(n)e^{-iqn}\rightarrow q^2\quad\mbox{for q}\ll 1.
\end{equation}
Expanding $\exp(-iqn)$ and comparing the $\text{O}(q^2)$ terms we obtain the sum rules:
\begin{eqnarray}
\sum_{n_s, n_t} n_s^2 \rho(n_s,n_t) &=& - \frac{2}{\xi}, \\\label{eq:scainorm1}
\sum_{n_s, n_t} n_t^2 \rho(n_s,n_t) &=& - 2 \xi,\label{eq:scainorm2}
\end{eqnarray}
if we express our quantities symmetrically in $a_s a_t$.

Requiring the spectrum to be massless we obtain another (trivial) sum rule:
\begin{eqnarray}
\sum_{n_s, n_t} \rho(n_s,n_t) &=& 0. \\\label{eq:scainorm3}
\end{eqnarray}

\begin{figure}[p]
\begin{center}
\epsfig{file=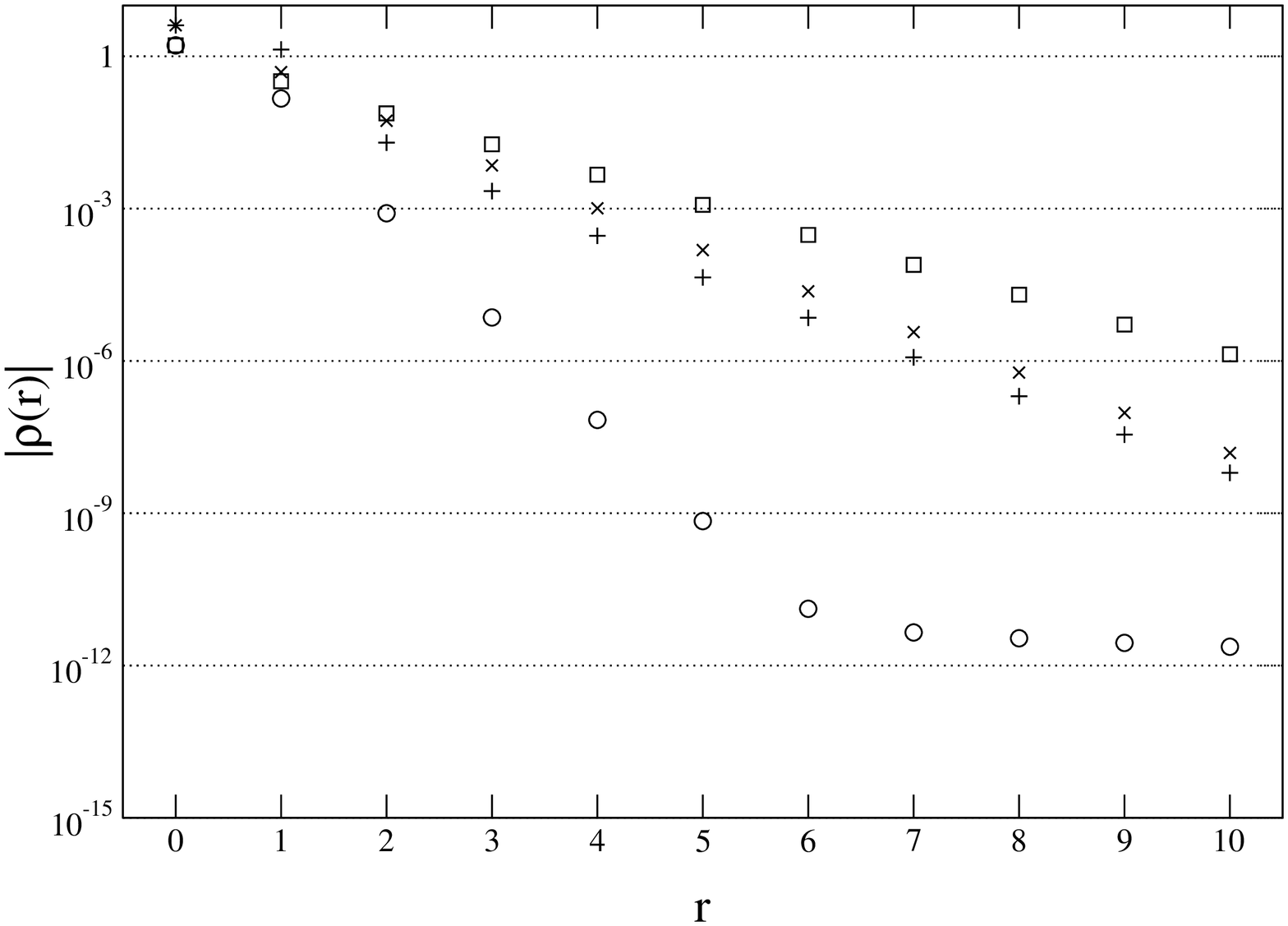,width=10cm}
\end{center}
\caption{Couplings of the $\xi=2$ free scalar field in $d=2$, optimised for
spatial locality ($\kappa=3$, \emph{circles}: on the spatial axis,
\emph{squares}: on the temporal axis) and temporal locality ($\kappa=12$,
\emph{pluses}: on the spatial axis, \emph{crosses}: on the temporal axis).}
\label{fig:sc_coupl_xi2}
\end{figure}

\begin{figure}[p]
\begin{center}
\epsfig{file=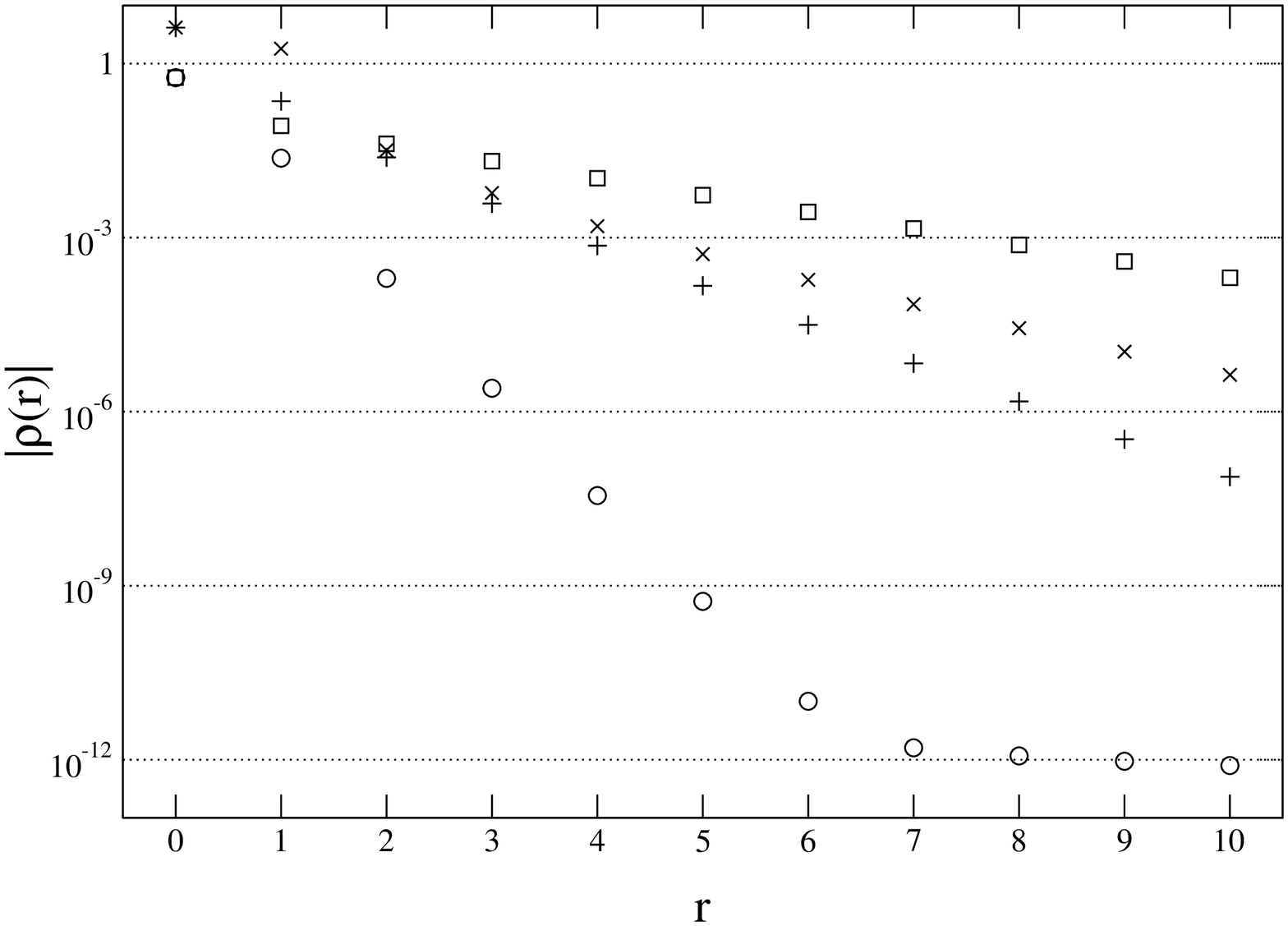,width=10cm}
\end{center}
\caption{Couplings of the $\xi=4$ free scalar field in $d=2$, optimised for
spatial locality ($\kappa=1.5$, \emph{circles}: on the spatial axis,
\emph{squares}: on the temporal axis) and temporal locality ($\kappa=24$,
\emph{pluses}: on the spatial axis, \emph{crosses}: on the temporal axis).}
\label{fig:sc_coupl_xi4}
\end{figure}

\begin{figure}[p]
\begin{center}
\epsfig{file=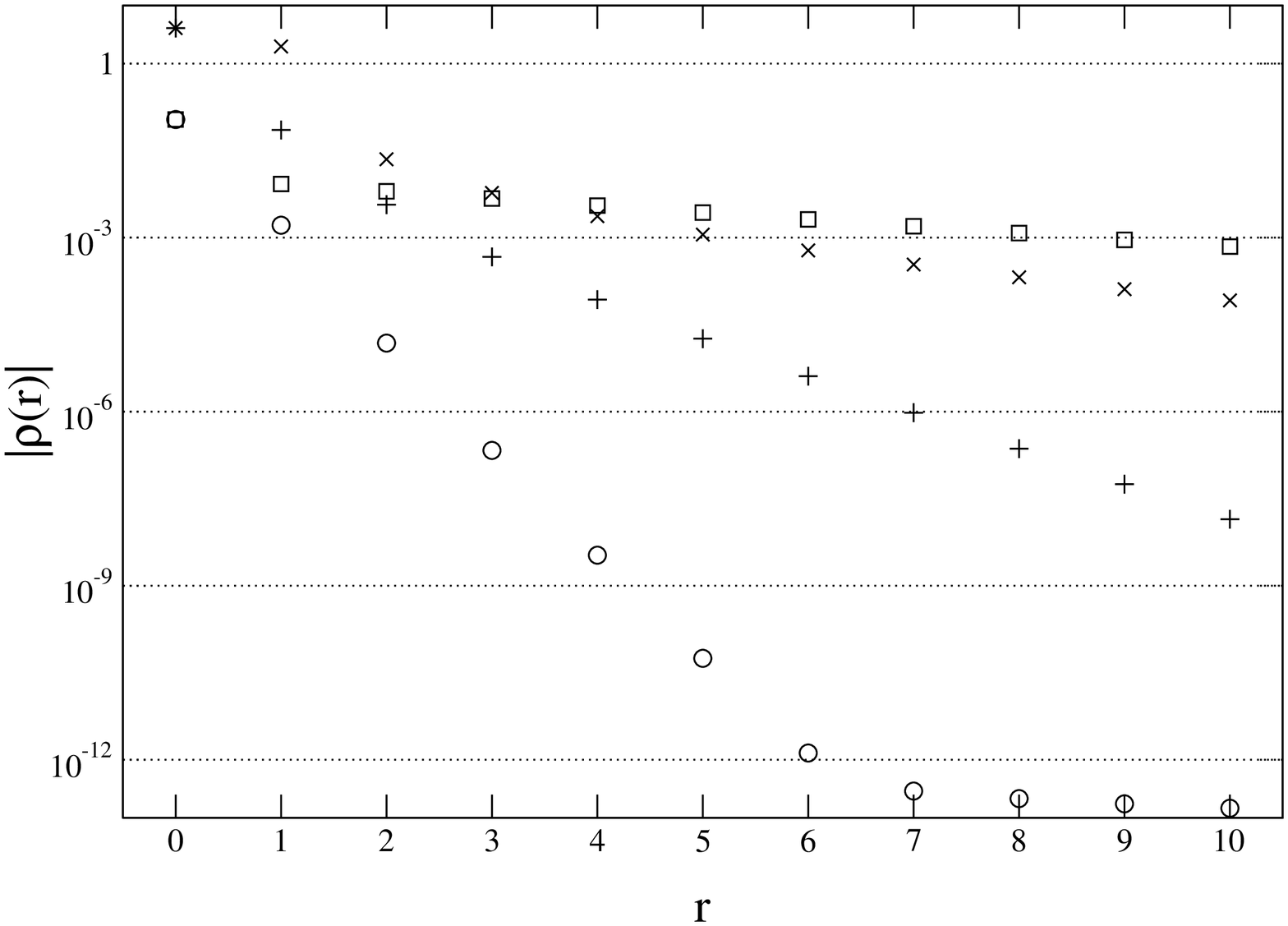,width=10cm}
\end{center}
\caption{Couplings of the $\xi=10$ free scalar field in $d=2$, optimised for
spatial locality ($\kappa=0.6$, \emph{circles}: on the spatial axis,
\emph{squares}: on the temporal axis) and temporal locality ($\kappa=60$,
\emph{pluses}: on the spatial axis, \emph{crosses}: on the temporal axis).}
\label{fig:sc_coupl_xi10}
\end{figure}

Calculating the couplings in position space $\rho(r)$ we can either choose $\kappa=6/\xi$
which yields optimised spatial locality and ultra--locality at the projection to the
temporal axis, or $\kappa=6\xi$ for optimal temporal locality and ultra--locality after
projecting to the spatial axis. The resulting magnitudes of the couplings, $|\rho(r_s,0)|$ and
$|\rho(0,r_t)|$ are displayed in Figures \ref{fig:sc_coupl_xi2}, \ref{fig:sc_coupl_xi4},
\ref{fig:sc_coupl_xi10} for anisotropies $\xi=$~2, 4, 10, where the separations $r_s$ and
$r_t$ are given in lattice units. Note that the physical separation in temporal direction
is a factor $\xi$ smaller compared to the spatial one.

By truncating the action, we introduce $O(a_s^2)$ and $O(a_t^2)$ artifacts.
As for our anisotropic actions $a_s\gg a_t$ we expect the temporal
artifacts to be much smaller than the spatial ones and therefore truncate the action to
nearest--neighbour couplings in temporal direction (``nnt'') which is only possible 
without completely distorting the sum rules, the spectrum etc.
if the action has been optimised for temporal locality choosing $\kappa=6\xi$. This
truncation scheme yields, employing the sum rules eqs.~\ref{eq:scainorm1}--\ref{eq:scainorm3}, 
\begin{eqnarray} 
\rho(n_s,1)=\rho(n_s,-1) & = & \frac{1}{2} \sum_{n_t} n_t^2 \rho^{\text{FP}}(n_s,n_t), \\
\rho(n_s,0) & = & \sum_{n_t} \rho^{\text{FP}}(n_s,n_t) - \sum_{n_t} n_t^2 \rho^{\text{FP}}(n_s,n_t).
\end{eqnarray}

\begin{figure}[p]
\begin{center}
\epsfig{file=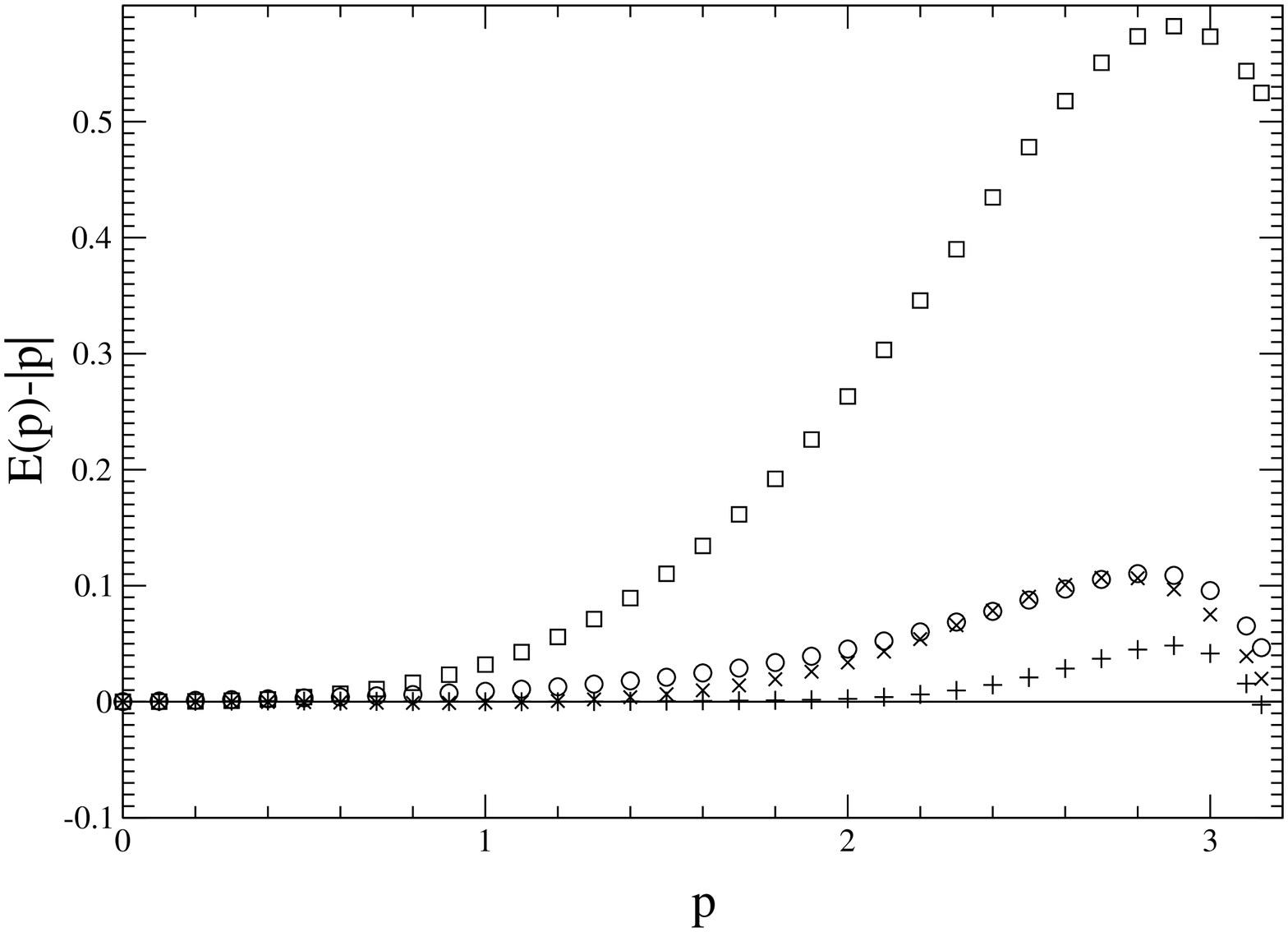,width=10cm}
\end{center}
\caption{Deviations from the exact dispersion relation of the $\xi=2$ perfect action for the $d=2$ free
scalar field. \emph{Circles}: $\kappa=3$, $5\times 5$ couplings, \emph{squares}:
$\kappa=3$, $5 \times $~nnt couplings, \emph{pluses}: $\kappa=12$, $5\times 5$
couplings, \emph{crosses}: $\kappa=12$, $5 \times $~nnt couplings.}
\label{fig:sc_disp_xi2}
\end{figure}

\begin{figure}[p]
\begin{center}
\epsfig{file=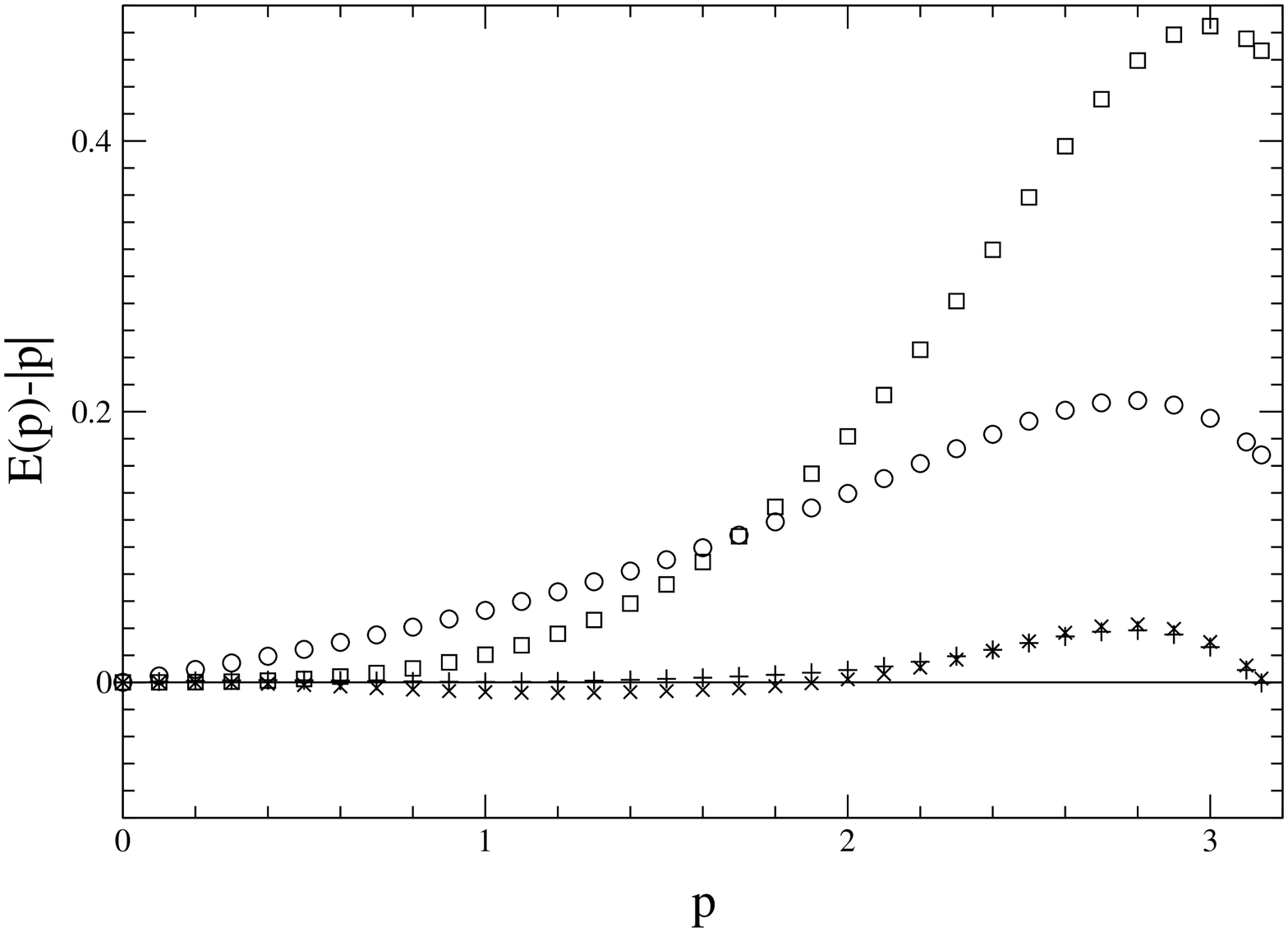,width=10cm}
\end{center}
\caption{Deviations from the exact dispersion relation of the $\xi=4$ perfect action for the $d=2$ free
scalar field. \emph{Circles}: $\kappa=1.5$, $5\times 5$ couplings, \emph{squares}:
$\kappa=1.5$, $5 \times $~nnt couplings, \emph{pluses}: $\kappa=24$, $5\times 5$
couplings, \emph{crosses}: $\kappa=24$, $5 \times $~nnt couplings.}
\label{fig:sc_disp_xi4}
\end{figure}

\begin{figure}[p]
\begin{center}
\epsfig{file=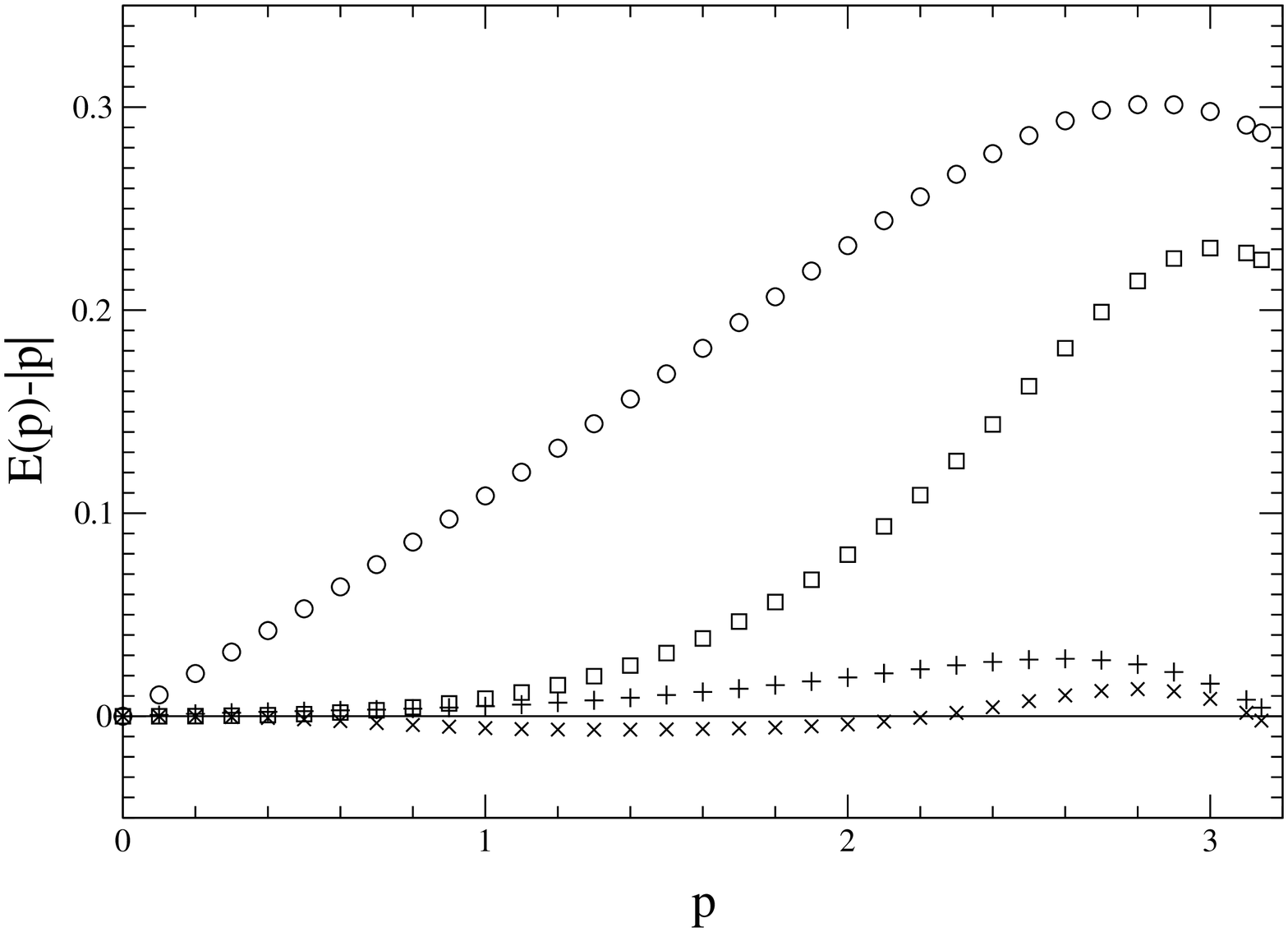,width=10cm}
\end{center}
\caption{Deviations from the exact dispersion relation of the $\xi=10$ perfect action for the $d=2$ free
scalar field. \emph{Circles}: $\kappa=0.6$, $5\times 5$ couplings, \emph{squares}:
$\kappa=0.6$, $5 \times $~nnt couplings, \emph{pluses}: $\kappa=60$, $5\times 5$
couplings, \emph{crosses}: $\kappa=60$, $5 \times $~nnt couplings.}
\label{fig:sc_disp_xi10}
\end{figure}

To estimate the physical quality of the truncated perfect actions we 
display the deviations of the exact spectrum $E(p)=p/\xi$ in the first Brillouin zone
for different truncations and different
values of $\kappa$ in Figures \ref{fig:sc_disp_xi2}, \ref{fig:sc_disp_xi4}, \ref{fig:sc_disp_xi10}.

Comparing the spectra of the symmetrically truncated actions ($5\times 5$) it turns out that indeed the 
temporally optimised couplings ($\kappa=6\xi$) yield much better spectra than the spatially optimised
couplings --- for $\xi=10$ the spectrum is even completely off in the latter case.
As well, truncating these actions to nearest-neighbour in time does only mildly decrease the quality of
the spectra. The larger the anisotropy gets, the more important is good locality of the temporal couplings.

\subsection{Spatial Blocking}
For the ($d=2$) scalar field, we may perform purely spatial blockings (cf. Section \ref{sec:spablock})
analytically and check whether
(in this model) we end up with the perfect anisotropic action, eq.~\ref{eq:fpscal}. We start from the isotropic FP propagator
which we know analytically for a blocking that averages the continuum field isotropically around lattice
sites (without overlapping):
\begin{eqnarray}
D^{\text{FP}}(q) & = & \sum_{l=-\infty}^\infty \frac{1}{(q+2\pi l)^2}\prod_{\mu=1}^2\frac{\sin^2(q_\mu/2)}{(q_\mu/2+\pi
  l_\mu)^2} + \frac{1}{3\kappa},
\label{scfp}
\end{eqnarray}
and perform a single spatial ($\mu=1$) RG step, increasing the spatial lattice spacing by a factor of two:
\begin{eqnarray}
D_{\text{sp}}^{\text{FP}}(q) & = & \frac{1}{2} \sum_{l_1'=0}^1 \cos^2(\frac{q_1}{4}+\pi \frac{l_1'}{2})D^{\text{FP}}(\frac{q_1}{2}+\pi l_1',q_2)+ \frac{1}{6\kappa}
\end{eqnarray}
(note the normalisation of the $\kappa$ factor).
Inserting expression eq.~\ref{scfp} and performing the substitution $q_1/2+\pi l_1'+2\pi l_1\rightarrow q_1/2+\pi l_1$ we get
\begin{eqnarray}
D_{\text{sp}}^{\text{FP}}(q_1,q_2) & = & \frac{1}{2} \sum_{l=-\infty}^\infty \frac{1}{(\frac{q_1}{2}+\pi l_1)^2+(q_2+2\pi l_2)^2}\frac{\sin^2(\frac{q_1}{2})\sin^2(\frac{q_2}{2})}{(\frac{q_1}{2}+\pi l_1)^2(\frac{q_2}{2}+\pi l_2)^2}\nonumber\\
&& + \frac{1}{3\kappa},
\end{eqnarray}
which is the anisotropic $\xi=2$ perfect propagator for this model, compare eq.~\ref{eq:aiprop}.
(Performing another single \emph{temporal} RG
step on this propagator gives back the initial FP propagator $D^{\text{FP}}$, which proves that the normalisation
of both terms has been chosen correctly.) 
Of course this discussion of a single spatial RG step in the $d=2$ scalar theory cannot be generalised
straightforwardly to gauge theory as the RG step cannot be treated analytically there, however these
results indicate that this approach may be feasible as well for gauge theory.

\section{Gauge Fields in the Quadratic Approximation} \label{sec:quadapp}
We will now turn to $\text{SU}(3)$ gauge fields in $d=4$ dimensions.
On small fluctuations we may approximate the FP gauge action quadratically:
\begin{equation}
S^{\text{FP}}=\frac{1}{2N_c}\sum_{n,r}\rho_{\mu\nu}(r)\text{Tr}[A_\mu(n+r)A_\nu(n)]+\text{O}(A^3)
\label{eq:quadapppos}
\end{equation}
neglecting terms of order 3 and higher in the gauge potential.
In momentum space the action reads
\begin{equation}
S^{\text{FP}}=\frac{1}{2N_c}\frac{1}{V}\sum_k\tilde{\rho}_{\mu\nu}(k)\text{Tr}[\tilde{A}_\mu(-k)
\tilde{A}_\nu(k)]+\text{O}(\tilde{A}^3).
\end{equation}

The calculation of the couplings $\rho_{\mu\nu}(r)$ and the propagator $D_{\mu\nu}(r)$ is described
in Appendix \ref{app:quad}.

\subsection{Physical Quantities}

Having at hand the quadratic couplings $\rho_{\mu\nu}$ and the propagator $D_{\mu\nu}$ we may easily calculate
physical quantities, as e.g.~the spectrum of the action or the tree level potential as described in the next
sections

\subsubsection{The Spectrum}
The quadratic FP action, eq.~\ref{eq:quadapppos}, is defined on the lattice. Nevertheless it describes
transverse massless gluons with the exact relativistic spectrum $E(k)=|\vec{k}|$ for $k_i=-\infty..\infty$.
This is a fundamental difference to perturbatively (e.g.~Symanzik) improved actions where the spectrum
shows $O(a^2)$ or $O(a^4)$ lattice artifacts.

The spectrum is determined by looking at the singularities of the propagator $D_{\mu\nu}(k)$: We fix the
momentum $\vec{k}$ and determine the corresponding energy $k_4$. This may be done analytically using exactly
the same arguments as for the isotropic FP action (see \cite{DeGrand:1995ji}) with the result that
the energy behaves as $k_4=-i|\vec{k}|$, where $k=q+2\pi l$, which is the exact continuum relation realised
by a tower of poles in $D_{\mu\nu}$ for any given $\vec{q}$, $q_i\in (0,2\pi)$.

\subsubsection{The Perturbative Potential}\label{sec:pertpot}

Another physical quantity that can be determined using the quadratic couplings is the static quark--anti-quark
potential up to tree level.

\begin{figure}[htbp]
\begin{center}
\includegraphics[width=11cm]{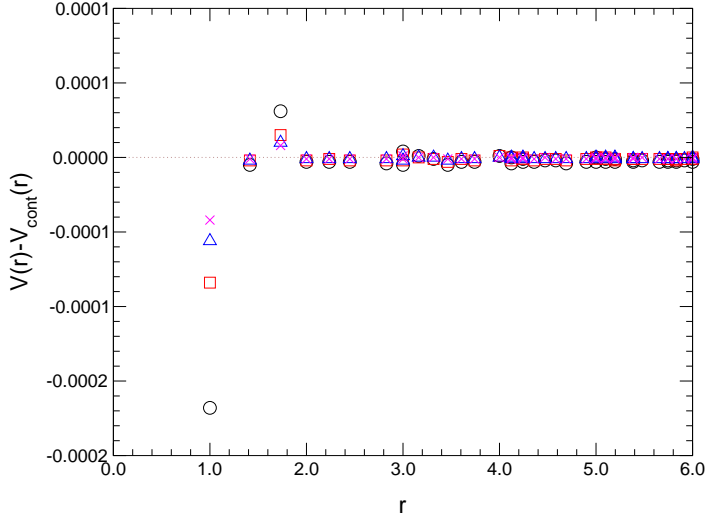}
\end{center}
\caption{The deviations to the exact tree--level quark--anti-quark potential of the quadratic approximation to the perfect
anisotropic action for $\xi=1$ (circles), $\xi=2$ (squares), $\xi=3$ (triangles), $\xi=4$ (crosses).}
\label{fig:ppot}
\end{figure}

\begin{figure}[htbp]
\begin{center}
\includegraphics[width=11cm]{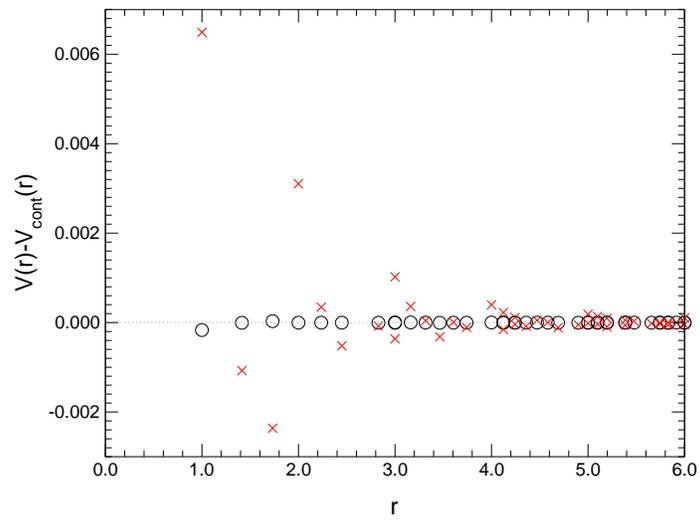}
\end{center}
\caption{The deviations to the exact tree--level quark--anti-quark potential of the Wilson action (crosses) and the FP action (circles).}
\label{fig:ppotwil}
\end{figure}

At the quadratic level the potential in a finite continuum box of size $L_s^3$ is proportional to the
Coulomb potential:
\begin{equation}
V_{\text{cont}}(\vec{x})=\frac{1}{L_s^3}\sum_{\vec{k}\ne 0} e^{i\vec{k}\vec{x}}\frac{1}{(\vec{k})^2},\quad where \quad k_i=\frac{2\pi}{L_s}l_i.
\end{equation}
On the lattice, the corresponding relation reads
\begin{equation}
V_{\text{lat}}(\vec{r})=\frac{1}{N_s^3}\sum_{\vec{k}\ne 0} e^{i\vec{k}\vec{r}}D_{44}(\vec{k},k_4=0),
\end{equation}
where $D_{\mu\nu}(k)$ is the propagator corresponding to the quadratic part of the lattice action.
As this quantity is unchanged from the isotropic case (except a global factor $1/\xi$) the result
is exactly the same as in \cite{Blatter:1996ti}. Figure \ref{fig:ppot} summarises the results for the anisotropic
actions. As $V\sim 1/\xi$ the deviations are accordingly smaller for larger values of the anisotropy $\xi$.
Figure \ref{fig:ppotwil} compares the deviation to the continuum result of the perfect action to the deviation
present with the Wilson action (for the isotropic action; the picture is exactly the same for $\xi\ne 1$) ---
note the difference in the scale of the deviations.
It is self--evident that the perfect action heavily suppresses discretisation errors.
The deviations that are still present are on one hand due to the RGT transformation not averaging exactly
spherical, on the other hand due to remaining quantum fluctuations on the fine lattice.

\fancyhead[RE]{\nouppercase{\small\it Chapter \thechapter.\, The $\mathbf{\xi}=2$ Perfect Action}}
\chapter{The $\mathbf{\xi=}2$ Perfect Action} \label{ch:xi2act}

The main topic of this work is the construction and the application of a classically perfect
gauge action for $\xi=2$ anisotropic lattices. These tasks and their results are presented in
this Chapter. The construction follows the proceeding presented
in Section \ref{sec:fpcons} using the spatial blocking method and the parametrisation described
in Chapter \ref{ch:construction}.

The anisotropy of the resulting action is measured using
the torelon dispersion relation (see Section \ref{sec:torelon}), the (spatial) scale is determined by
the static quark-antiquark potential (see Section \ref{sec:qbarq}), where for some values of the
coupling $\beta$, off-axis separations of the quarks are considered as well, to be able to estimate
violations of rotational symmetry by the parametrised action. Another way of determining the
(temporal) scale is the study of the deconfining phase transition (see Section \ref{sec:deconf}), measuring
the critical couplings $\beta_{\text{crit}}$ for different spatial extensions $N_t$ of the lattice. 

Finally, the glueball spectrum is measured performing simulations at three values of $\beta$ using
the $\xi=2$ action.

On one hand, the study of the classically perfect action with rather small anisotropy
$\xi=2$ is a feasibility test of the ansatz as it employs only one single spatial blocking
step and one single parametrisation of the resulting action. On the other hand, to determine
accurately the continuum limit, one should perform simulations on fine lattices as well ---
using actions with large anisotropies, this may be more expensive than necessary, at least
when the states to be studied are not extraordinarily heavy. The feasibility of additional
spatial blocking steps, followed by parametrisations is briefly (and incompletely) checked in
Chapter \ref{ch:xi4act}.

\section{Construction} \label{sec:xi2constr}
We construct the $\xi=2$ classically perfect anisotropic action starting with the
parametrised isotropic action\cite{Niedermayer:2000yx} performing one spatial blocking step as
described in Chapter \ref{ch:construction}.
The quadratic approximation suggests $\kappa_t=\xi^2 \kappa_s=4\kappa_s$ for keeping
the block transformation as close to the isotropic case as possible.
Furthermore, spatial locality should stay at its
optimum for the ``isotropic'' value of $\kappa_s=8.8$, thus we choose $\kappa_t=35.2$. Indeed, it turns
out that varying $\kappa_t$ away from $\kappa_t=35.2$, keeping $\kappa_s=8.8$ fixed, makes it much more difficult to
parametrise the resulting perfect action such that the renormalised anisotropy $\xi_{\text{R}}$
(see Section \ref{sec:xir}) stays close to the input value $\xi=2$.

Examining this blocking, we observe that coarse configurations obtained in a MC run with the 
ad-hoc anisotropic action (see Appendix \ref{app:adhoc}) are mapped to fine configurations that
are close to isotropic (concerning spatial and temporal plaquette values and expectation values of the Landau gauge-fixed link variables)
for input anisotropy $\xi_{\text{ad-hoc}}\approx$~3.2.

We thus generate 20 configurations each at $\beta_{\text{ad-hoc}}=$~2.5, 3.0, 3.5, 4.0 using the ad-hoc anisotropic
 action\footnote{Note, that these values of $\beta_{\text{ad-hoc}}$, used in this section, may not be directly compared to the values of
$\beta_{\text{perf}}$ corresponding to the parametrised classically perfect action, used in the following sections; but
rather the fluctuations (plaquette values, expectation values of Landau gauge-fixed links etc.) of the gauge configurations
produced in MC runs using the perfect action should lie approximately in the same range as the fluctuations of the initial coarse
configurations produced with the ad-hoc action.} with $\xi_{\text{ad-hoc}}$=3.2. These configurations are spatially blocked using the
isotropic action $\mathcal{A}_5^{444}$ (see Appendix \ref{app:isointact}) to describe the isotropic minimised configuration on the r.h.s. of eq.~\ref{eq:FP_equation}.

To construct the action we follow the procedure described in Sections \ref{sec:fitquant}--\ref{sec:linfit}.
Several non-linear fits are performed for the derivatives of 20 configurations at $\beta=3.5$,
using different sets of parameters. Concerning the non-linear parameters,
we find out that it is necessary to keep all asymmetry parameters $\eta_i$ non-zero. Moreover, it is
crucial to distinguish the $c_i$ parameters for different plaquette orientations, see Table \ref{tab:fitqual}
and Section \ref{sec:compext}. The parameters $c_i$ are chosen to be non-zero for $i=1,2,3$ as adding
additional free $c_4$ parameters does not improve $\chi^2$ significantly. 

Having fixed the non-linear parameters $\eta$ and $c_i$, we include 
the derivatives of 20 configurations at $\beta=4.0$ and the action values of all the configurations
at $\beta=$~3.5, 4.0 in the linear fit, keeping the non-linear parameters fixed. The resulting values of $\chi^2$ and the linear behaviour
of the actions suggest a linear set with $\max(k+l)_{\text{sp}}=4$, $\max(k+l)_{\text{tm}}=3$ where the action
values are included with weights $w_{\text{act}}=0.018$ relative to the derivatives (see also Figure \ref{fig:wact2}).
Using this value, the good parametrisation of the derivatives obtained in the full
nonlinear fit is preserved, at the same
time the mean error of the action value due to the parametrisation is as small as 0.35\%.

It is checked that the
values of $\chi^2$ are still acceptable on sets of configurations down to $\beta=3.0$, besides that they do
not increase significantly for
configurations independent of the ones used in the fitting procedure. The behaviour of the linear parameters
of the action as a function of the simple plaquette values $u$ and smeared plaquette values $w$ is
displayed in Figure \ref{fig:xi2uw}, the parameters of the resulting action are listed in Appendix \ref{app:xi2act}.

\begin{figure}[htbp]
\begin{center}
\psfrag{action}{\hspace{-0.5cm}$\mathcal{A}(u,w)$}
\psfrag{u}{$u$}
\psfrag{w}{$w$}
\includegraphics[width=8cm]{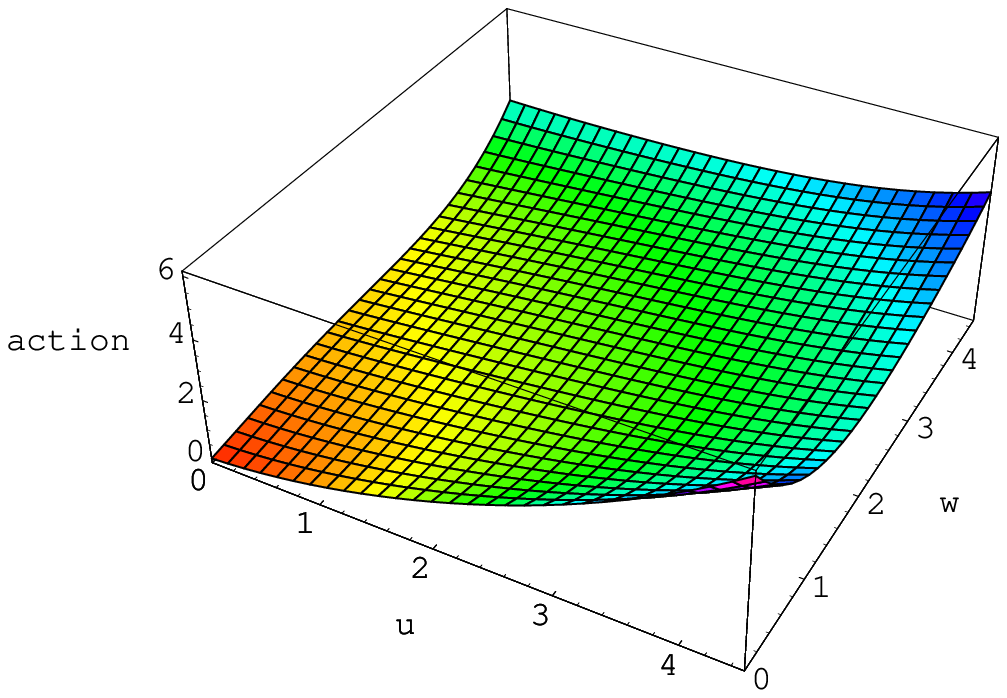}
\includegraphics[width=8cm]{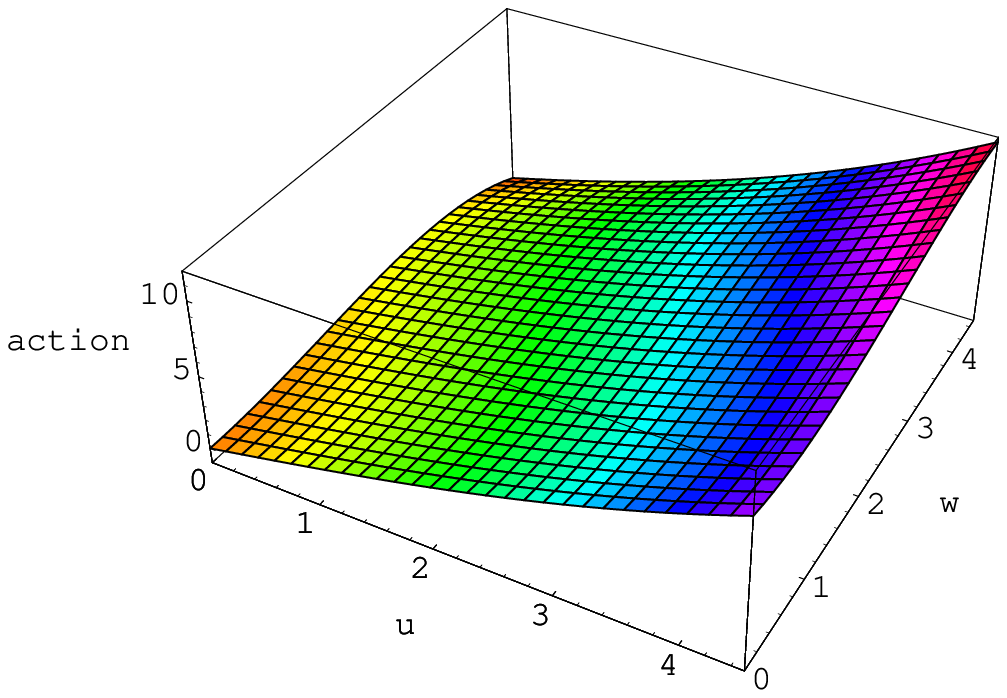}
\end{center}
\caption{The behaviour of the linear parameters $p_{kl}$ as a function of the values of the standard
plaquette ($u$) and the smeared plaquette ($w$) for the parametrisation of the classically perfect 
$\xi=2$ action.}
\label{fig:xi2uw}
\end{figure}

\section{The Renormalised Anisotropy} \label{sec:xir}
The renormalised anisotropy $\xi_R$ of the action presented above is measured using the torelon dispersion relation as
described in Section \ref{sec:torelon}.
The main advantage of this approach is obtaining the renormalised anisotropy $\xi_R$ as
well as an estimate for the scale $a_s$, $a_t$ performing computationally rather inexpensive
measurements.

\subsection{Details of the Simulations and Results}

The parameters of the simulations carried out are collected in Table \ref{tab:xi2tordet}.

\begin{table}[tbhp]
\renewcommand{\arraystretch}{1.3}
  \begin{center}
    \begin{tabular*}{\textwidth}[c]{c@{\extracolsep{\fill}}cc}
      \hline\vspace{-0.05cm} 
      $\beta$ & $S^2\times L\times T$ & \# sweeps / measurements\\
      \hline
      3.00     & $8^2\times 4\times 20$ & 72000 / 14400\\
      3.15$^a$ & $8^2\times 5\times 20$ & 54000 / 10800\\
      3.15$^b$ & $5^2\times 8\times 20$ & 47600 / 9520\\
      3.30     & $12^2\times 6\times 30$ & 35600 / 7120\\
      3.50$^a$ & $12^2\times 8\times 24$ & 39400 / 7880\\
      3.50$^b$ & $14^2\times 6\times 30$ & 36800 / 7360\\
      \hline
    \end{tabular*}
    \caption{Run parameters for the torelon measurements using the $\xi=2$ perfect
      action. The lattice extensions in torelon direction $L$, the extension
      in the two transversal spatial directions $S$ as well as the temporal extension
      $L$ are given.}
    \label{tab:xi2tordet}
  \end{center}
\end{table}

The Polyakov line around the short spatial direction is measured on 
APE-smeared configurations of smearing levels 3, 6, 9, 12, 15 for $\beta\ge 3.3$ and 2, 4, 6, 8, 10 for $\beta\le 3.15$
with smearing parameter $\lambda_s=0.1$ (see Section \ref{sec:smearing}).
The measured energies $E(p^2)$ determined using variational methods (see Appendix \ref{app:vartech}) 
depending on the lattice momentum $p^2$ are
given, together with the number of operators used in the variational method, the fit ranges and
the values of $\chi^2$ per degree of freedom, $\chi^2/N_{\text{DF}}$, in Tables \ref{tab:xi2torcoll}, 
\ref{tab:xi2torcoll2} in Appendix \ref{app:rescoll}. The two simulations denoted by $3.15^b$ and $3.50^b$
are used to check the stability of the method. The first one is carried out on a lattice with small
transversal spatial size (the torelon string measured is even longer), the latter is measuring the energies of a rather
short string. 

\begin{figure}[htbp]
\begin{center}
\includegraphics[width=8cm]{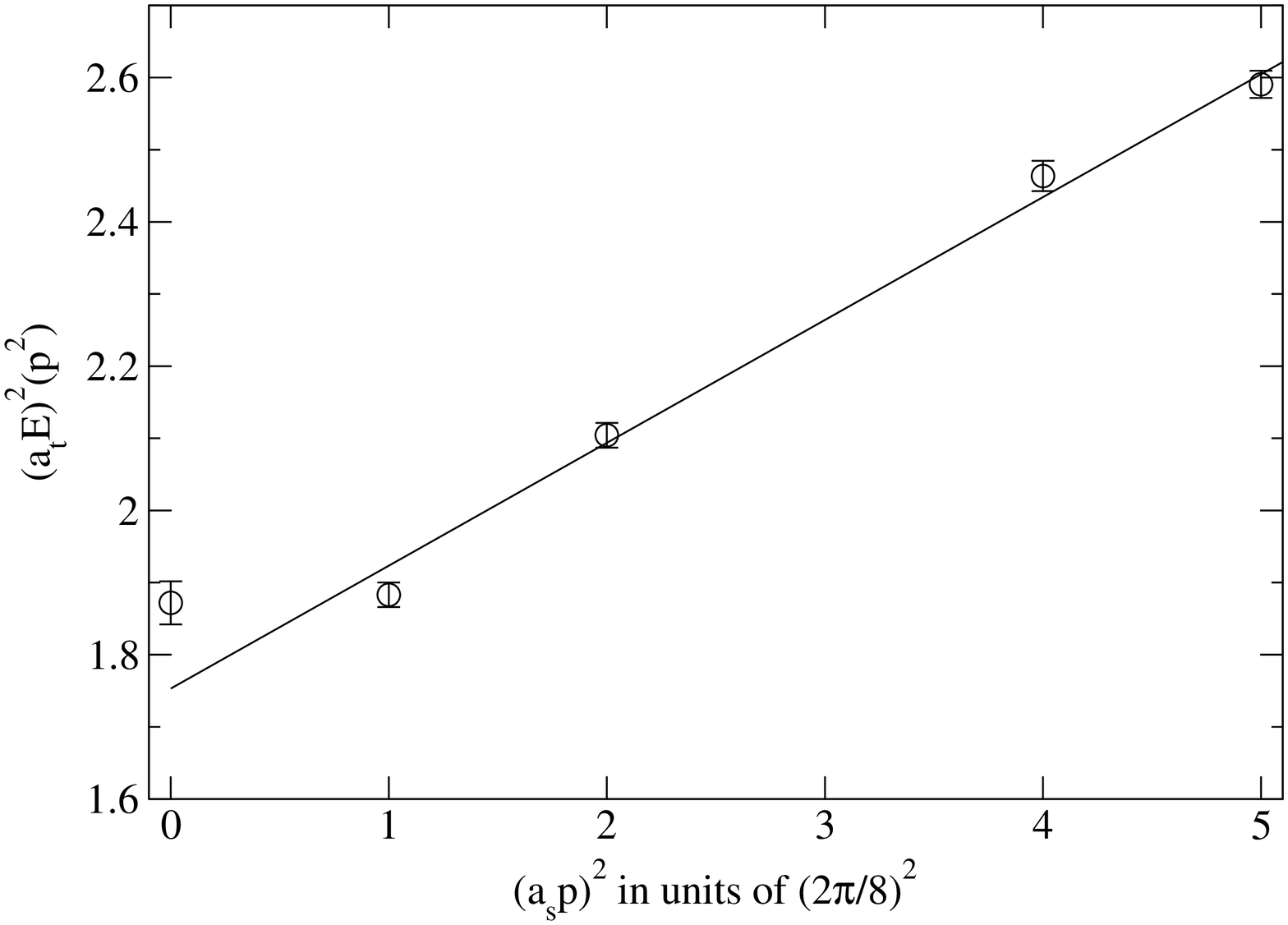}
\end{center}
\caption{Torelon dispersion relation for $\beta=3.0$. The straight line is the correlated fit to $E^2(p)=m_T^2+p^2$ in the range $p^2=0..5$.}
\label{fig:tordisp_b30_05}
\end{figure}

\begin{figure}[htbp]
\begin{center}
\includegraphics[width=8cm]{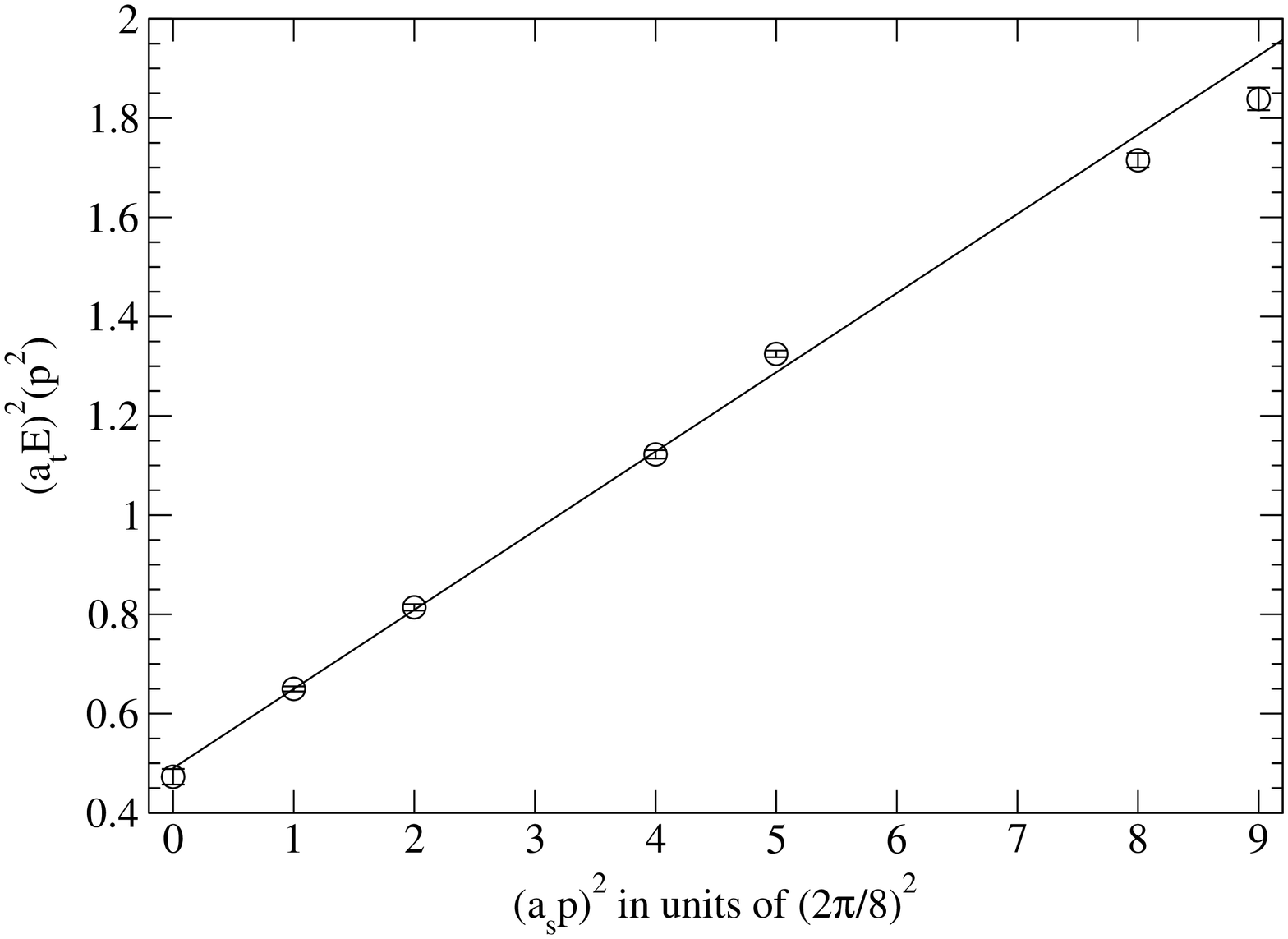}
\end{center}
\caption{Torelon dispersion relation for $\beta=3.15^a$. The straight line is the correlated fit to $E^2(p)=m_T^2+p^2$ in the range $p^2=0..4$.}
\label{fig:tordisp_b315_09}
\end{figure}

\begin{figure}[htbp]
\begin{center}
\includegraphics[width=8cm]{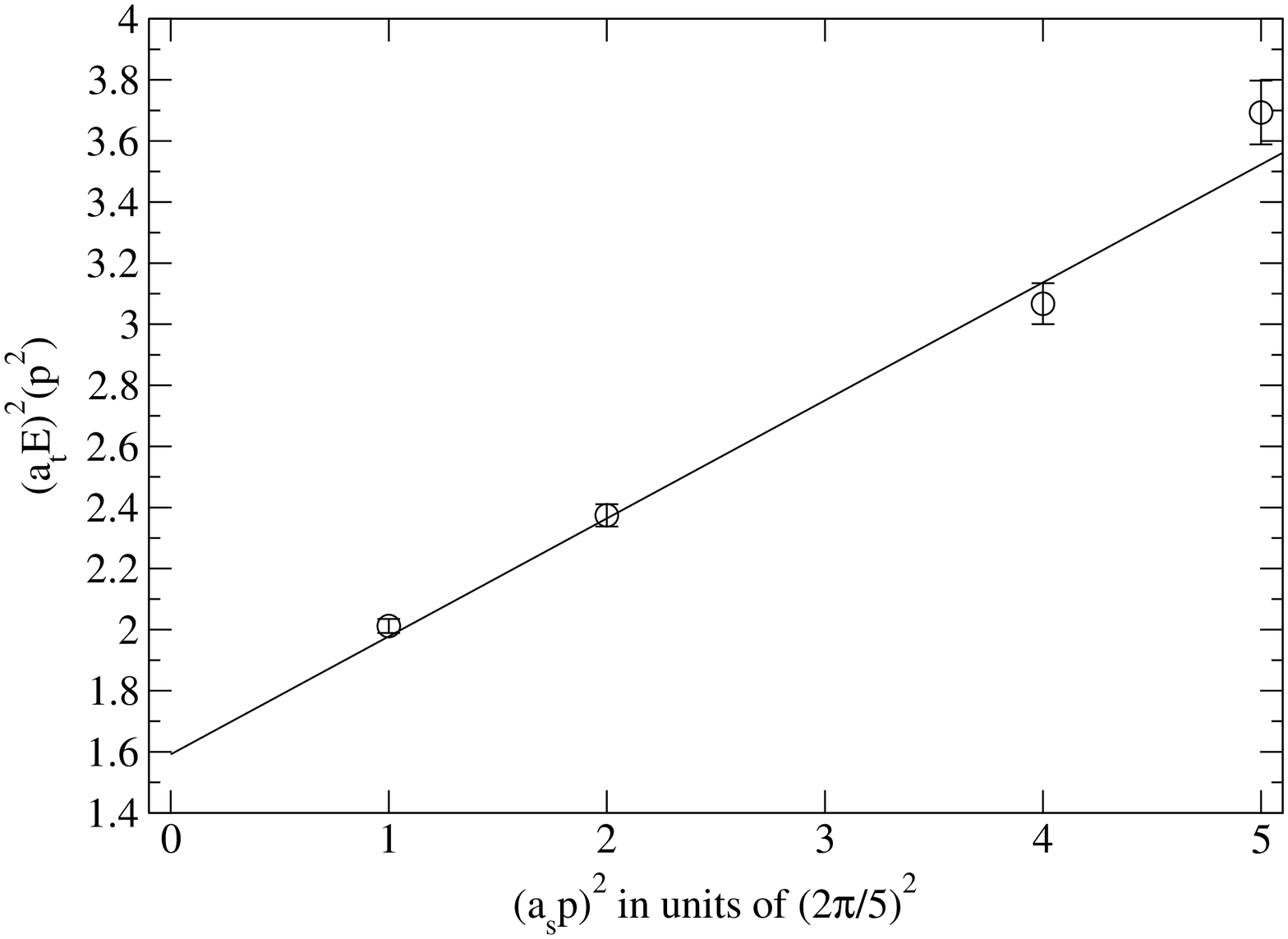}
\end{center}
\caption{Torelon dispersion relation for $\beta=3.15^b$ on a small transverse volume. The straight line is the correlated fit to $E^2(p)=m_T^2+p^2$ in the range $p^2=0..5$.}
\label{fig:tordisp_b315b_05}
\end{figure}

\begin{figure}[htbp]
\begin{center}
\includegraphics[width=8cm]{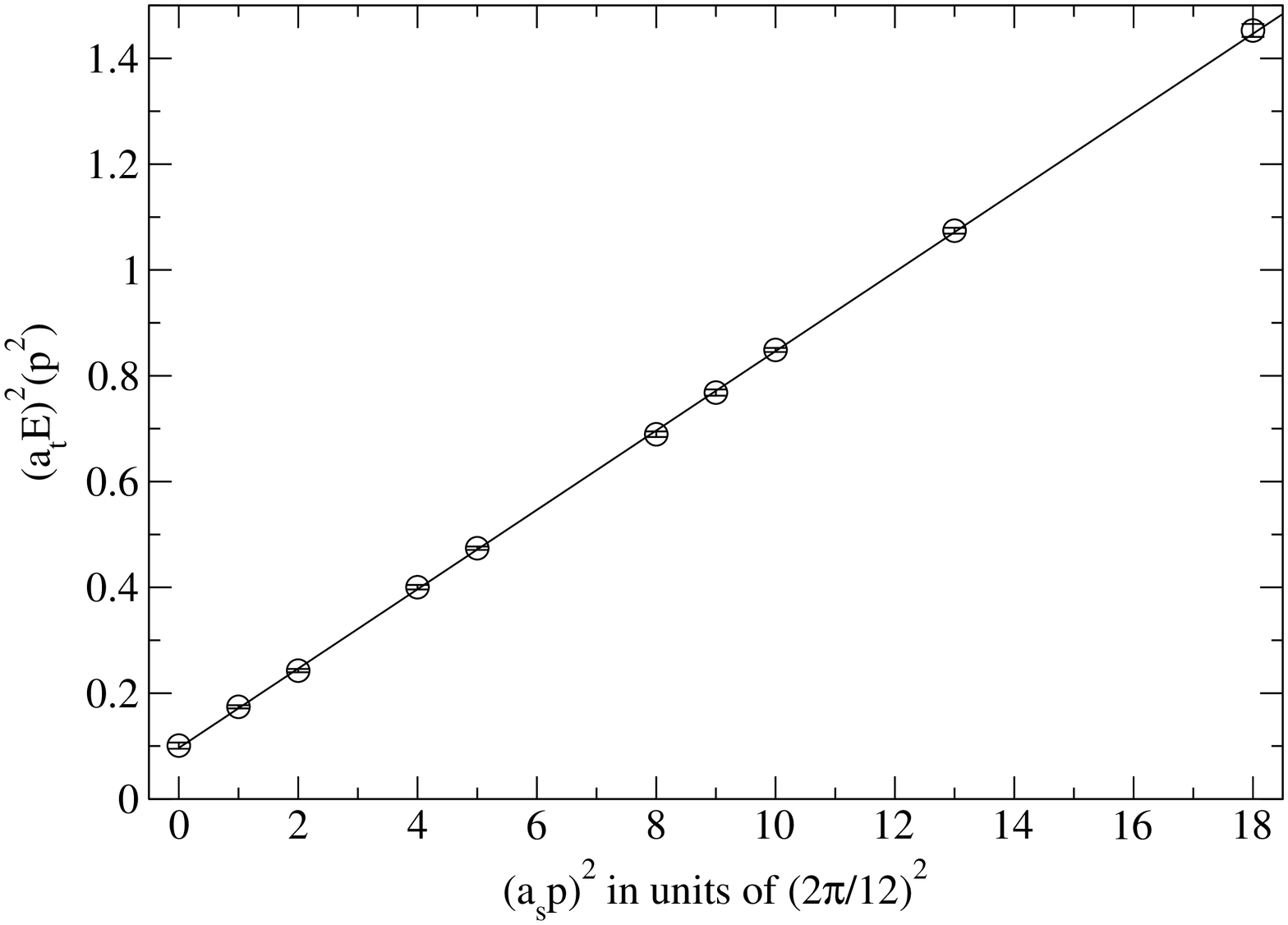}
\end{center}
\caption{Torelon dispersion relation for $\beta=3.3$. The straight line is the correlated fit to $E^2(p)=m_T^2+p^2$ in the range $p^2=1..9$.}
\label{fig:tordisp_b33_018}
\end{figure}

\begin{figure}[htbp]
\begin{center}
\includegraphics[width=8cm]{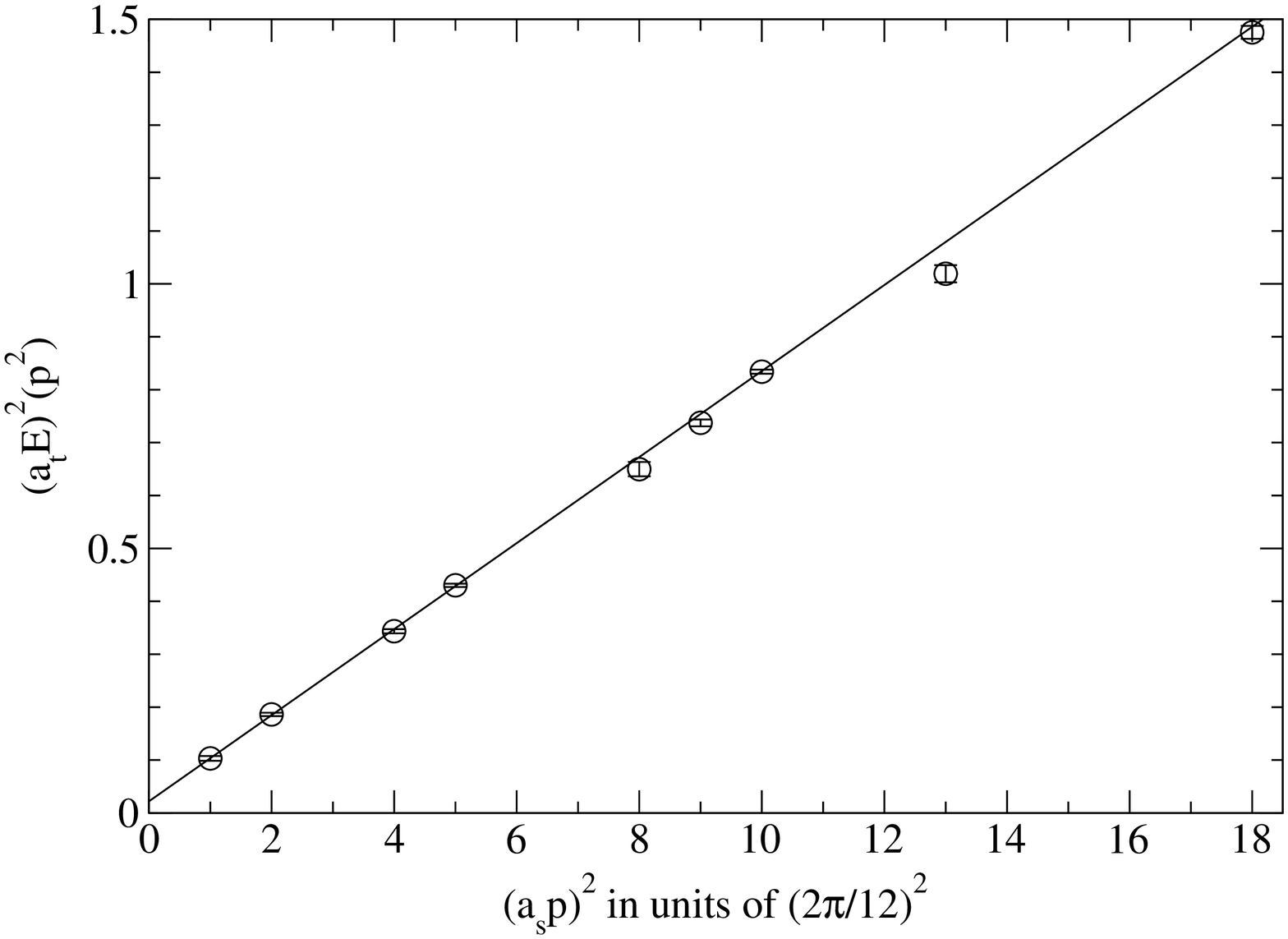}
\end{center}
\caption{Torelon dispersion relation for $\beta=3.5^a$. The straight line is the correlated fit to $E^2(p)=m_T^2+p^2$ in the range $p^2=1..10$.}
\label{fig:tordisp_b35b_018}
\end{figure}

\begin{figure}[htbp]
\begin{center}
\includegraphics[width=8cm]{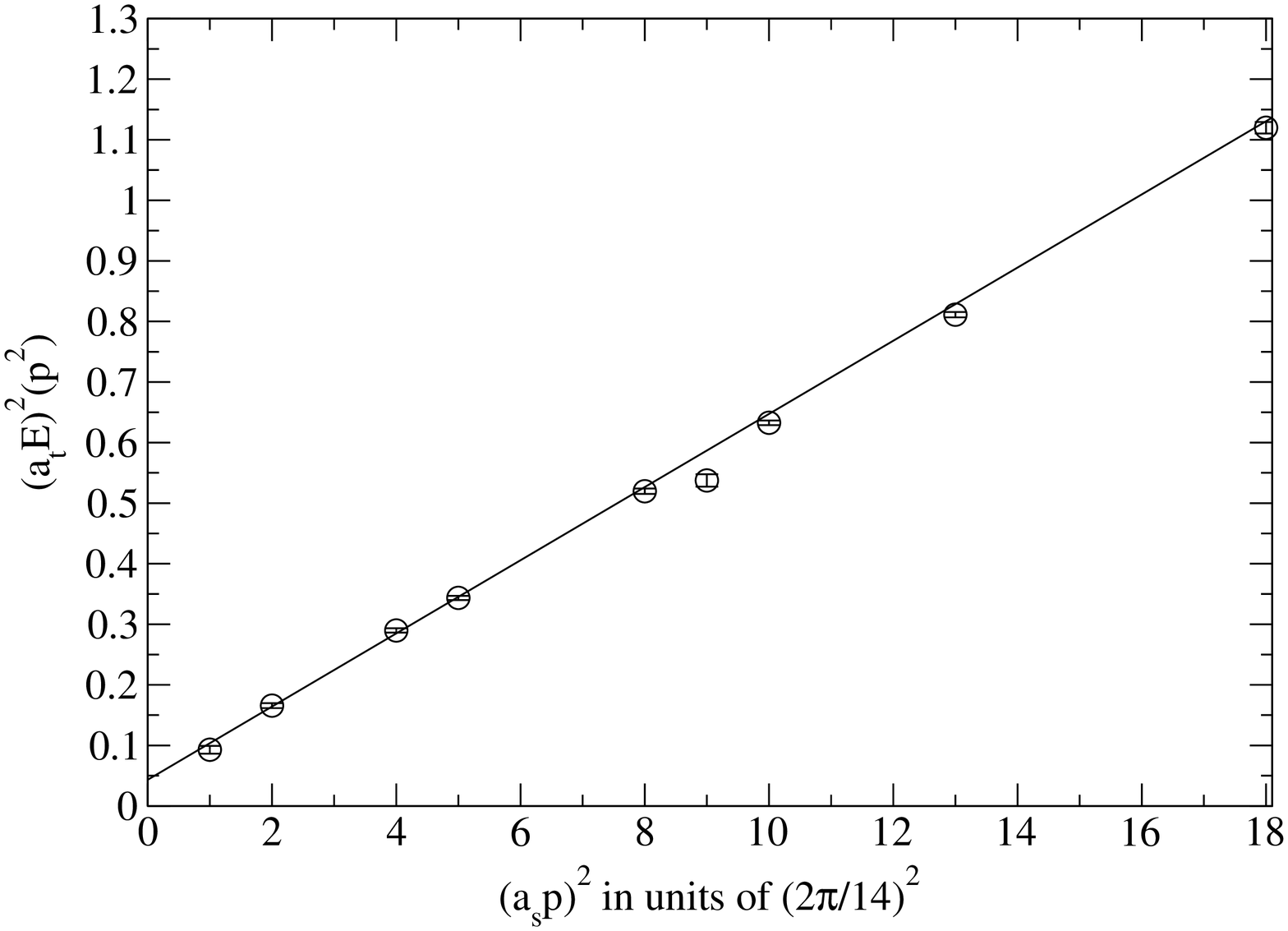}
\end{center}
\caption{Torelon dispersion relation for $\beta=3.5^b$ for a short torelon string. The straight line is the correlated fit to $E^2(p)=m_T^2+p^2$ in the range $p^2=1..8$.}
\label{fig:tordisp_b35_018b}
\end{figure}

Figures \ref{fig:tordisp_b30_05}--\ref{fig:tordisp_b35_018b} display the dispersion relations at the different
values of the coupling $\beta$ including all values of $p$ at which the energies could be determined on the
lattices considered. Note, that the energies of the $p=0$ torelons (the torelon masses) may be hard to determine 
because the effective masses do not reach a plateau within the temporal extent of the lattice; however, using the
non-zero momentum energies, the masses may still be accurately determined.

To determine the renormalised anisotropy $\xi_R$ as well as the torelon mass in units of the
temporal lattice spacing $m a_t$ we employ fully correlated fits over different ranges of $p^2$ and
choose the values depending on $\chi^2/N_{\text{DF}}$ and the precision of the dispersion relation
data in the respective range. The results are given in Table \ref{tab:xi2torres}.

\begin{table}[htbp]
\renewcommand{\arraystretch}{1.3}
  \begin{center}
    \begin{tabular*}{\textwidth}[c]{c@{\extracolsep{\fill}}cccccccc}
      \hline\vspace{-0.05cm}
      $\beta$ & fit range & $\xi_R$ & $m_{\text{T}} a_t$ & $\chi^2/N_{\text{DF}}$\\
      \hline
      3.00     & 0..5 & 1.903(81) & 1.324(98) & 0.39\\
      3.15$^a$ & 0..4 & 1.966(39) & 0.700(35) & 0.83\\
      3.15$^b$ & 0..5 & 2.022(104) & 1.262(84) & 0.62\\
      3.30     & 1..9 & 1.912(9) & 0.311(5) & 0.94\\
      3.50$^a$ & 1..10 & 1.836(9) & 0.149(10) & 0.94\\
      3.50$^b$ & 1..8 & 1.826(16) & 0.208(16) & 1.61\\
      \hline
    \end{tabular*}
    \caption{Results of the torelon simulations using the $\xi=2$ perfect action. The fit
      range in $p^2$ is given in units of $(2\pi/S)^2$.}
    \label{tab:xi2torres}
  \end{center}
\end{table}

Using the finite size relation for the torelon mass corresponding to eq.~\ref{eq:stringtorelon}:
\begin{equation}\label{eq:finite_size_torelon}
M_T(La_s)=(\sigma+\frac{D}{(La_s)^2})La_s
\end{equation}
it is possible to calculate the string tension $\sigma$ and thus the spatial scale $a_s$.
The string model predicts $D=-\frac{\pi}{3}$ for long strings. To estimate the scale using the
results of the torelon simulation, we use this string picture value and stay aware that the estimate
gets worse for short strings. To obtain the hadronic scale $r_0$ and the lattice spacings we employ $r_0\sqrt{\sigma}$=1.193(10) from 
\cite{Niedermayer:2000yx} and use the definition $r_0=0.50$~fm. The results are collected in Table \ref{tab:xi2torscal}.

The scale is measured accurately using the static quark-antiquark potential (see
Section \ref{sec:xi2scale}) such that we can compare our torelon estimates to the true value. This is done
in Table \ref{tab:xi2scalcomp} and it turns out that the deviation of the torelon estimates for the scale
$r_0/a_s$ to the much more reliable potential values does not exceed about 6\% if the length of the string
is not too short (say $La_s\gtrsim 1$~fm). For short strings, the string picture does not apply, we may therefore
not expect good results in this case. But even for very long strings, it is not possible to obtain very accurate
information about the (hadronic) scale. Firstly, long strings are very heavy and thus difficult to measure
(comparable to the very long range region of the static quark-antiquark potential), especially when the anisotropy
$\xi$ is not very large; and secondly, using the string tension $\sigma$ to set the scale is plagued by ambiguities.
Depending on the analysis of the respective potential measurements, $r_0\sqrt{\sigma}$ may vary as much as about 4\%
(see Section \ref{sec:r0sigma}) and additionally, the effective string tension at separations around $r=1$~fm does
not correspond to $\sigma$ occurring with the torelon strings.

There seems to be no problem if the transverse volume is rather small
as in the $\beta=3.15^b$ simulation, except of course the large momenta occurring which make the determination of
the energies more difficult. As well, the estimates of the renormalised anisotropy $\xi_R$ do not show
significant deviations neither for small transverse volume nor for short strings.

\begin{table}[htbp]
\renewcommand{\arraystretch}{1.3}
  \begin{center}
    \begin{tabular*}{\textwidth}[c]{c@{\extracolsep{\fill}}cccc}
      \hline\vspace{-0.05cm}
      $\beta$ & $\sqrt{\sigma}a_s$ & $r_0/a_s$ & $a_s$ [fm] & $a_t$ [fm]\\
      \hline
      3.00     & 0.834(20) & 1.43(5)  & 0.350(12) & 0.184(14)\\
      3.15$^a$ & 0.563(8) & 2.12(5)  & 0.236(6) & 0.120(5)\\
      3.15$^b$ & 0.579(18) & 2.06(8) & 0.243(9) & 0.120(11)\\
      3.30     & 0.3578(21) & 3.33(5) & 0.150(2) & 0.079(1)\\
      3.50$^a$ & 0.225(5) & 5.31(16) & 0.094(3) & 0.051(2)\\
      3.50$^b$ & 0.304(7) & 3.92(12) & 0.128(4) & 0.070(3)\\
      \hline
    \end{tabular*}
    \caption{Estimates of the scale determined from torelon results (see Table \ref{tab:xi2torres}),
     see text.}
    \label{tab:xi2torscal}
  \end{center}
\end{table}

\begin{table}[htbp]
\renewcommand{\arraystretch}{1.3}
  \begin{center}
    \begin{tabular*}{\textwidth}[c]{c@{\extracolsep{\fill}}cccc}
      \hline\vspace{-0.05cm}
      $\beta$ & $La_s$ [fm] & $r_0/a_s$ (Torelon) & $r_0/a_s$ (Potential) & rel. error\\
      \hline
      3.00     & 1.48 & 1.43(5) & 1.353(16) & +5.7\%\\
      3.15$^a$ & 1.23 & 2.12(5) & 2.038(2)  & +4.0\%\\
      3.15$^b$ & 1.96 & 2.06(8) & 2.038(2)  & +1.1\%\\
      3.30     & 0.95 & 3.33(5) & 3.154(8)  & +5.6\%\\
      3.50$^a$ & 0.82 & 5.31(16) & 4.906(21) & +8.2\%\\
      3.50$^b$ & 0.61 & 3.92(12) & 4.906(21) & -20.0\%\\
      \hline
    \end{tabular*}
    \caption{The estimates of the scale obtained from torelon measurements (see. Table \ref{tab:xi2torscal}) compared
to the scale measured using the static quark-antiquark potential (see Table \ref{tab:xi2potres}). The length of the
torelon string $La_s$ in physical units is given as well.}
    \label{tab:xi2scalcomp}
  \end{center}
\end{table}

Finally, we may conclude that determining the renormalised anisotropy using the torelon dispersion relation
is a stable and apparently sensible procedure. There are no manifest problems and it works for coarse
and fine lattices as well. Checks employing different lattice extensions indicate that the determination of
the renormalised anisotropy is very stable. The estimate of the lattice scale, however, is not very accurate,
mainly due to the use of the string picture relation, eq.~\ref{eq:finite_size_torelon}; the (systematic) 
error of the torelon scale is about 5\% for sensible torelon lengths $L\gtrsim 1$~fm with a tendency of
underestimating the lattice spacing $a_s$.

Choosing the lattice size, one has to pay attention on one hand
to the length of the torelon string, i.e.~the extension of the (short) longitudinal direction: if the string
is too short, there is no useful estimate for the scale; if it is much too long, it will be difficult to measure
due to its high energy. On the other hand, the transversal direction in principle may be short in order to make
the simulation cheaper what concerns computational effort, however, if it is too short, the momenta are 
large, and again the measurement becomes difficult. Finally, the temporal extension should be chosen large
enough such that the torelon mass (which may be rather light) may be determined directly.

\section{The Scale}\label{sec:xi2scale}
To set the scale of the action at a given $\beta$-value we measure the
quark-antiquark potential (see Section \ref{sec:qbarq}) and determine the Sommer scale $r_0\approx$~0.50~fm.
For the anisotropic action, we still employ the definition of $r_0$:
\begin{equation}
r^2 V'(r)|_{r=r_0} = 1.65,
\end{equation}
however $V(r)$ is now measured in units of the temporal lattice spacing and thus:
\begin{equation}
-\alpha+\xi\hat{\sigma}(\frac{r_0}{a_s})^2=1.65,
\end{equation}
where $\hat{\sigma}$ denotes the fitted dimensionless value $\hat{\sigma}=\sigma a_s a_t$ of the
string tension. The renormalised anisotropy $\xi$ thus has to be known before the spatial scale $r_0$
may be determined.

The static quark-antiquark potential is measured on lattices of different scales. The measurements
at $\beta=3.00$, 3.30
include the determination of the potential between quarks that are off-axis separated along the lattice
vectors (1,0,0), (1,1,0),
(1,1,1), (2,1,0), (2,1,1), (2,2,1) (and lattice rotations) in order to estimate violations of rotational symmetry.
The rest of the measurements include only on-axis separations due to the large computational cost
(concerning speed and memory) of the off-axis measurement. The parameters of the simulations are given in
Table \ref{tab:xi2potpar}.

\begin{table}[htbp]
\renewcommand{\arraystretch}{1.3}
  \begin{center}
    \begin{tabular*}{\textwidth}[c]{c@{\extracolsep{\fill}}ccc}
      \hline\vspace{-0.05cm} 
      $\beta$ & off-axis sep. & $S^3\times T$ & \# sweeps / measurements\\
      \hline
      3.00 & yes & $8^3\times 16$ & 42000 / 2800\\
      3.15 & no & $10^3\times 20$ & 39800 / 3980\\
      3.30 & yes & $10^3\times 20$ & 27000 / 1800\\
      3.50 & no & $12^3\times 24$ & 36400 / 3640\\
      \hline
    \end{tabular*}
    \caption{Run parameters for the measurements of the static quark-antiquark potential using the $\xi=2$ perfect
      action.}
    \label{tab:xi2potpar}
  \end{center}
\end{table}

The values of the off-axis potential $a_t V(\vec{r})$ at $\beta=3.30$ are collected in Table \ref{tab:xi2offax}
in Appendix \ref{app:rescoll} and displayed in Figure \ref{fig:pot_grnd}, the values of the off-axis potential on
the coarse lattice at $\beta=3.00$ are collected in Table \ref{tab:xi2offax_b300}. The potential values
$a_t V(r)$ of the on-axis simulations are collected in Table \ref{tab:xi2onaxpot}.

\begin{figure}[htbp]
\begin{center}
\includegraphics[width=10cm]{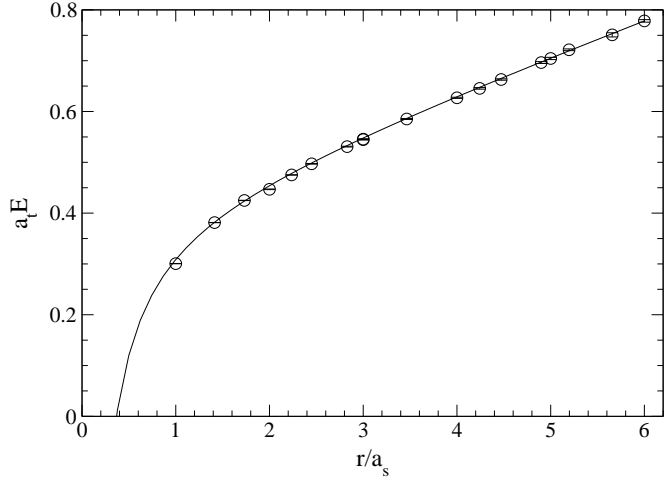}
\end{center}
\caption{Ground state of the static quark-antiquark potential at $\beta=3.3$. Including a global fit
to the Cornell potential in the range $r=\sqrt{2}..5$.}
\label{fig:pot_grnd}
\end{figure}

The parameters $\alpha$ and $\sigma$ (the string tension) in the Cornell potential, eq.~\ref{eq:isopot},
are determined using global fits, the hadronic scale $r_0$ is
determined performing local fits. The fit ranges and results are given in Table \ref{tab:xi2potres}.
In the global fit to the off-axis potential at $\beta=3.30$ we exclude the badly measured separations $2(2,1,0)$ and 
$2(2,1,1)$.

\begin{table}[htbp]
\renewcommand{\arraystretch}{1.3}
  \begin{center}
    \begin{tabular*}{\textwidth}[c]{c@{\extracolsep{\fill}}cccccc}
      \hline\vspace{-0.05cm} 
      $\beta$ & gl. fit & $\alpha$ & $\sigma a_s a_t$ & loc. fit & $r_0/a_s$ & $r_0\sqrt{\sigma}$\\
      \hline
      3.00 & \phantom{ }$1-2\sqrt{6}$ & -0.201(18) & 0.3595(87) & $\sqrt{6}-3$ & 1.353(16) & 1.119(51)\\
      3.15 & $1-5$ & -0.1503(12) & 0.1654(6) & $1-3$ & 2.038(2) & 1.162(15)\\
      3.30 & $\sqrt{2}-5$ & -0.1539(7) & 0.0683(3) & $2\sqrt{2}-2\sqrt{3}$ & 3.154(9) & 1.140(8)\\
      3.50 & $1-6$ & -0.1478(4) & 0.0320(3) & $2-4$ & 4.906(21) & 1.189(14)\\
      \hline
    \end{tabular*}
    \caption{Results of the measurements of the static quark-antiquark potential using the $\xi=2$ perfect
      action. The parameters $\alpha$ and $\sigma$ of the Cornell potential are given together with the
      global fit range chosen to determine them. The hadronic scale $r_0$ is given in spatial lattice units
      together with the local fit range used to determine that quantity. Note, that $c=6.00$ has been chosen for
      $\beta=3.0$ and $c=0.89$ has been chosen
      for $\beta=3.5$ (see Section \ref{sec:setscale}). Additionally, the dimensionless quantity 
      $r_0\sqrt{\sigma}$ is given.}
    \label{tab:xi2potres}
  \end{center}
\end{table}

\begin{figure}[htbp]
\begin{center}
\includegraphics[width=10cm]{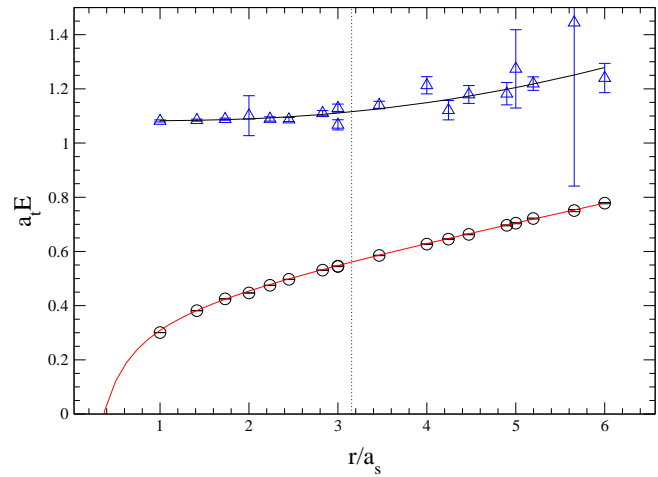}
\end{center}
\caption{Static quark antiquark potential at $\beta=3.3$ including the first excited state $V^*(r)$ at $\beta=3.3$. The ground state values
are again fitted to the ansatz $V(r)=V_0+\alpha/r+\sigma r$, whereas the values of the excited state are fitted
to the (ad-hoc) ansatz $V^*(r)=A/r+B+Cr+Dr^2$. The value of $r_0$ where the determination of the anisotropy is
done, using the gap, is marked by the dotted line.}
\label{pot_exst}
\end{figure}

In principle, it is possible to determine the renormalised anisotropy using the first excited state
of the potential together with the ground state. As can be seen from Figure \ref{pot_exst} however,
for the off-axis potential at $\beta=3.3$, the energy values of the first excited state have large errors.
The result for the separation $a_t(V^*(r_0)-V(r_0)$=0.555(47) thus shows a large error.
Comparing this value to $r_0(V^*(r_0)-V(r_0))\approx 3.25(5)$ of \cite{Morningstar:1998da} one obtains $\frac{r_0}{a_t}=5.86(59)$
and thus $\xi_{\text{R}}=a_s/a_t$=1.86(19) for $\beta=3.3$ which is in agreement with the value determined using the torelon dispersion relation, $\xi_{\text{R}}$=1.912(9). A similar determination for the on-axis potential at $\beta=3.50$ is
even more difficult due to the small number of data. The result is $\xi_{\text{R}}$=1.66(32) which does again agree
with the torelon result, $\xi_{\text{R}}=1.836(9)$. Due to the large errors (caused by the choice of $\beta$ and the operators used) this
method of determining the renormalised anisotropy $\xi_{\text{R}}$ is not suitable in the context of this work.
However, for the determination of the scale as well as of the anisotropy on lattices with finer temporal lattice
spacing (above all with higher anisotropies) this way might be feasible at least if off-axis separations are included in
the measurement.

\begin{figure}[htbp]
\begin{center}
\includegraphics[width=10cm]{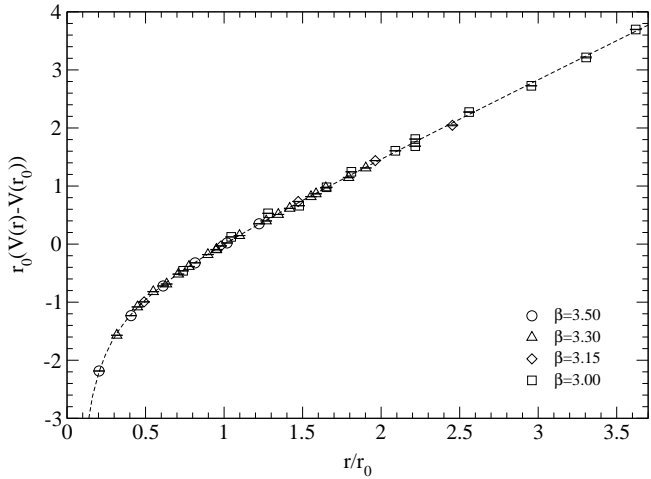}
\end{center}
\caption{The static quark-antiquark potential measurements expressed in RG invariant, dimensionless units $r/r_0$,
$r_0 V$. The unphysical constant $r_0 V(r_0)$ has been subtracted such that all the curves exactly coincide at $r=r_0$.
The dashed line is a global fit to the Cornell potential, eq.~\ref{eq:isopot}.}
\label{fig:allpot_xi2}
\end{figure}

The measurements at $\beta=3.0, 3.3$ including a large number of off-axis separations provide information about
the deviations from rotational invariance. This issue has been addressed in \cite{Blatter:1996ti} for the
renormalisation group used in this work. It turned out that violations of rotational symmetry caused by the
blocking are small (see also Section \ref{sec:pertpot}). That this is still true for the parametrisation employed
is shown by the off-axis potential, see Figure \ref{fig:pot_grnd}. The deviations on the coarse lattice
at $\beta=3.00$ are significantly larger (see Figure \ref{fig:allpot_xi2}) which is no surprise as this coupling
corresponds to a spatial lattice spacing $a_s\approx 0.37$~fm.


The static $q\bar{q}$ potential is an effective test of scaling. Expressing the potential measurements
performed at different couplings $\beta$ in the RG invariant, dimensionless ratios $r/r_0$ and $r_0 V$,
subtracting the unphysical constant $r_0 V(r_0)$ should lead to potentials lying exactly on top of each
other. Deviations indicate either scaling violations or ambiguities in the determination of $r_0$.
Figure \ref{fig:allpot_xi2} which includes all potential measurements shows that the different curves can hardly
be distinguished from each other, except some energies determined on the coarse lattice at $\beta=3.00$
that deviate notably from the curve which is a fit to the Cornell
potential including all the measurements that have been included into the global fits of the single
$\beta$ values, see Table \ref{tab:xi2potres}. The results for the two physical parameters are 
$\alpha=-0.27799(5)$ and $\sigma r_0^2$=1.3690(3).

\section{The Critical Temperature}

Another interesting quantity to study with the $\xi=2$ perfect action is the critical temperature of the
deconfining phase transition, see Section \ref{sec:deconf}. This quantity contains mainly information about the temporal scale of the
lattice at a given coupling $\beta$ as $T_c=1/(N_t a_t)$. Comparing this information to quantities obtained
from the measurements of torelons or the static quark-antiquark potential offers a lot of possibilities of
scaling (and other) tests. Because measurements of these quantities at the determined values of $\beta_{\text{crit}}$
are yet absent and because interpolations in $\beta$ are very difficult, because of the renormalisation of the
anisotropy $\xi$ which is another parameter entering, these tests are a future project.

\begin{figure}[htbp]
\begin{center}
\includegraphics[width=10cm]{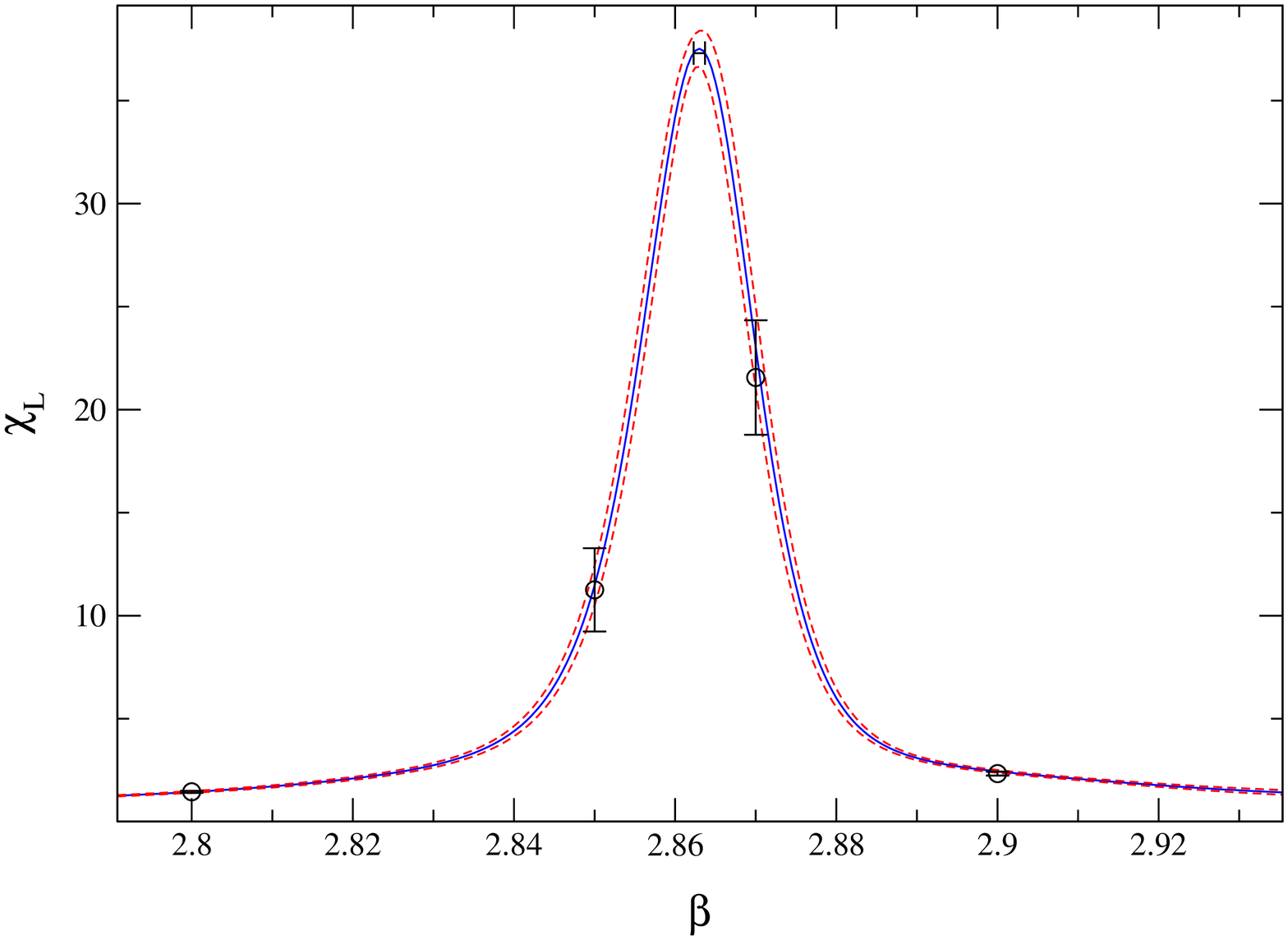}
\end{center}
\caption{The Polyakov loop susceptibility on the $3\times 6^3$ lattice for the
$\xi=2$ perfect action. The solid curves are the interpolations using Ferrenberg-Swendsen
reweighting, the dashed lines indicate the bootstrap error.}
\label{fig:xi2nt3susc}
\end{figure}

\begin{figure}[htbp]
\begin{center}
\includegraphics[width=10cm]{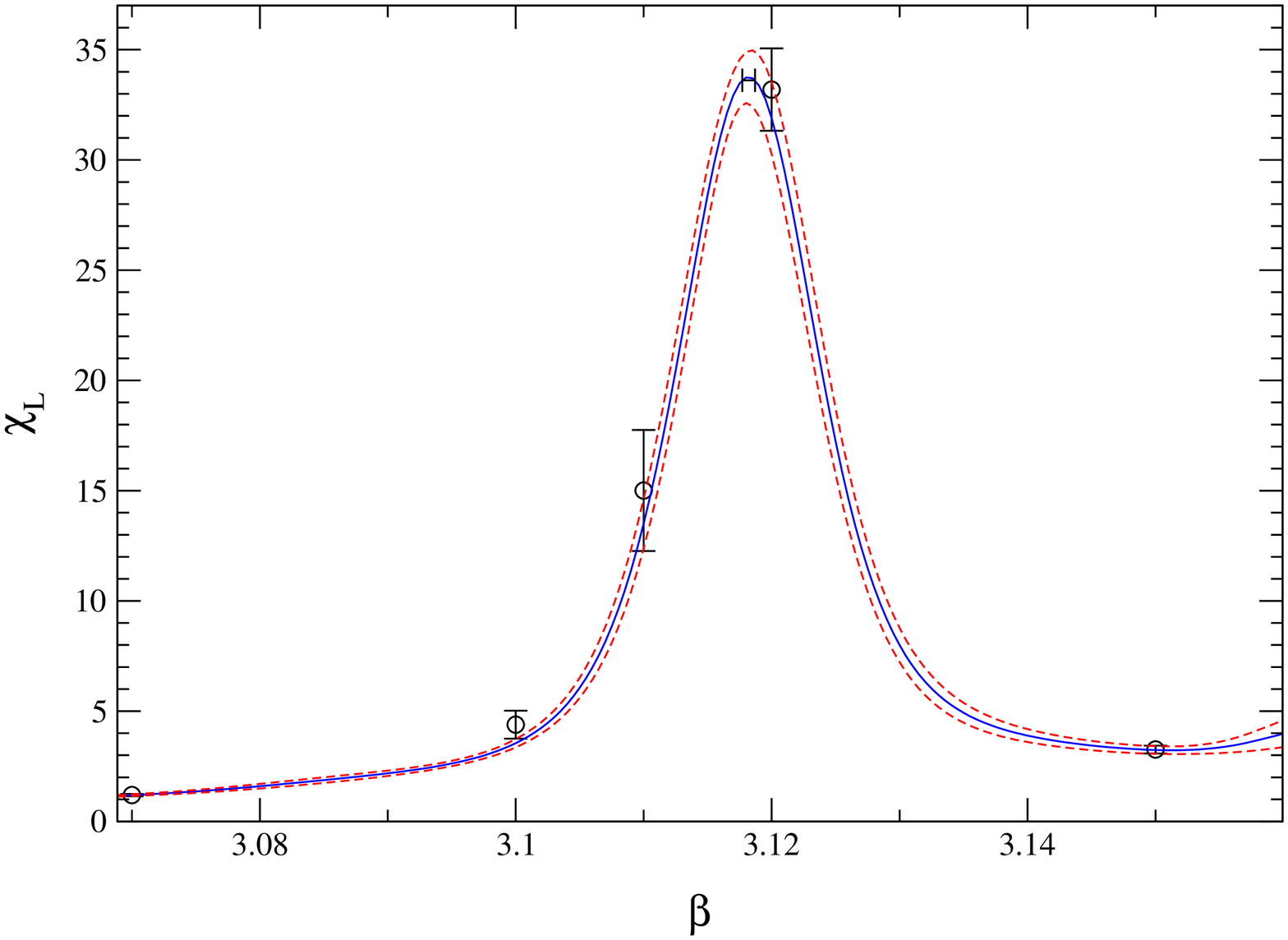}
\end{center}
\caption{The Polyakov loop susceptibility on the $5\times 9^3$ lattice for the
$\xi=2$ perfect action. The solid curves are the interpolations using Ferrenberg-Swendsen
reweighting, the dashed lines indicate the bootstrap error.}
\label{fig:xi2nt5susc}
\end{figure}

\begin{figure}[htbp]
\begin{center}
\includegraphics[width=10cm]{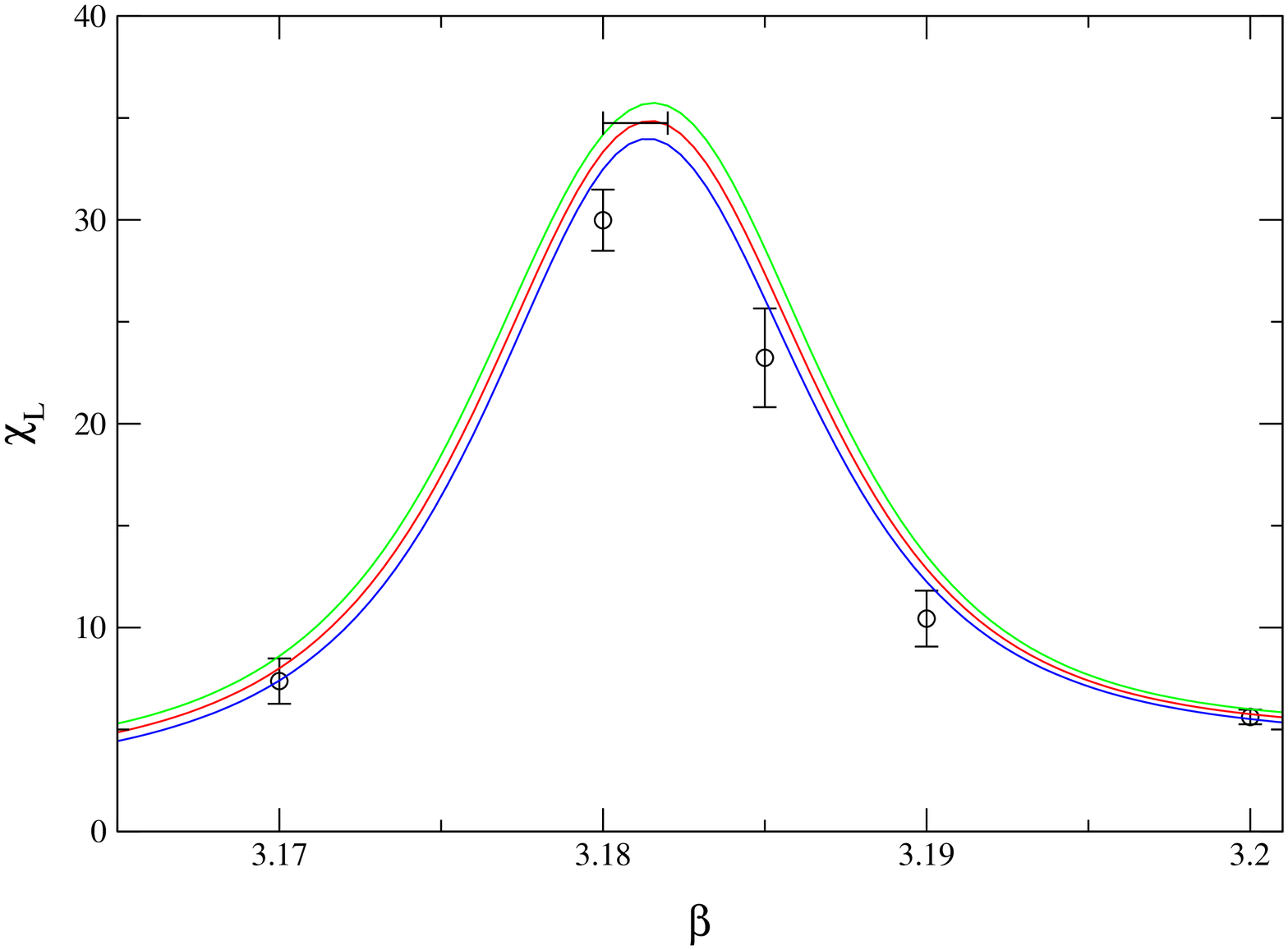}
\end{center}
\caption{The Polyakov loop susceptibility on the $6\times 11^3$ lattice for the
$\xi=2$ perfect action. The solid curves are the interpolations using Ferrenberg-Swendsen
reweighting, the dashed lines indicate the bootstrap error.}
\label{fig:xi2nt6susc}
\end{figure}

\begin{figure}[htbp]
\begin{center}
\includegraphics[width=10cm]{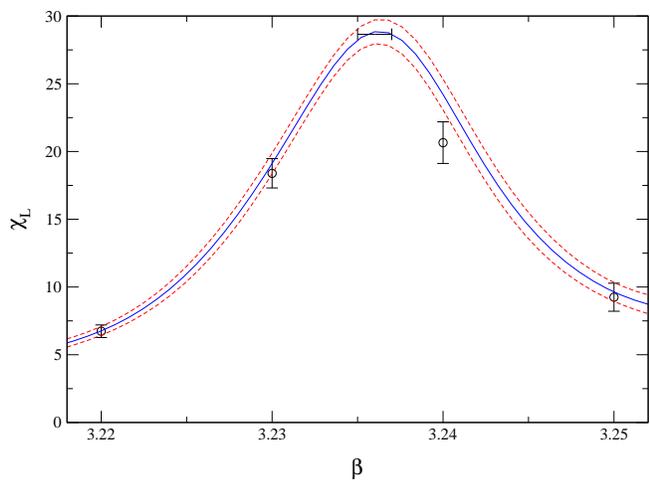}
\end{center}
\caption{The Polyakov loop susceptibility on the $7\times 13^3$ lattice for the
$\xi=2$ perfect action. The solid curves are the interpolations using Ferrenberg-Swendsen
reweighting, the dashed lines indicate the bootstrap error.}
\label{fig:xi2nt7susc}
\end{figure}

In this first study, we decide not to examine the finite-size scaling of $\beta_{\text{crit}}$ but to
choose $L_s/L_t=\xi\cdot N_s/N_t\approx 3.5\sim 4$. Obviously, the anisotropic nature of the action makes the
computational effort of the simulations for exploring the deconfining phase transition smaller.
The effects of the finite volume are (rather conservatively) estimated by using the finite-size scaling relation
for $\beta_{\text{crit}}$, eq.~\ref{eq:fofss}. The parameter $|h|$ is set to the largest value $|h|=0.25$ that
has been observed with the isotropic action, the value of $\xi_{\text{R}}$, used to estimate the spatial
lattice volume, is set to the lowest value appearing in the whole range of $\beta$ values considered.
These finite-volume errors are given together with
the run-time parameters and results of the $\xi=2$ simulations in Table \ref{tab:xi2deconfrtp}.
Figures \ref{fig:xi2nt3susc}--\ref{fig:xi2nt7susc} show that the determination of $\beta_c$ is rather precise,
partly due to the reweighting technique. As well, it shall be understood that in order to make the results
even more precise, a finite-size extrapolation is indispensable.

\begin{table}[htbp]
\renewcommand{\arraystretch}{1.3}
  \begin{center}
    \begin{tabular*}{\textwidth}[c]{c@{\extracolsep{\fill}}cccccc}
      \hline\vspace{-0.05cm}
      lattice size & $\beta$ & sweeps & $\tau_p$ & $\tau_{\text{int}}$ & $\chi_{\text{L}}$ & $\beta_{c}$ \\
      \hline
      $3\times 6^3$ & 2.80 & 45000 & & 70 & 1.45(5) & 2.863(1)(5)\\
                    & 2.85 & 45000 & 410 & 259 & 11.3(20) &\\
                    & 2.87 & 45000 & & 312 & 21.6(28) &\\
                    & 2.90 & 50000 & & 100 & 2.34(10) \\
      \hline
      $4\times 8^3$ & 3.00 & 45000 & & 81 & 1.63(5) & 3.032(1)(5)\\
                    & 3.02 & 45000 & 870 & 102 & 3.97(26) & \\
                    & 3.03 & 45000 & 830 & 372 & 43.2(22) &\\
                    & 3.04 & 45000 & & 188 & 10.9(18) &\\
                    & 3.05 & 55600 & & 144 & 3.94(21) &\\
      \hline
      $5\times 9^3$ & 3.07 & 44000 & & 117 & 0.84(4) & 3.118(1)(6)\\
                    & 3.10 & 45000 & 560 & 154 & 4.38(63) & \\
                    & 3.11 & 20500 & 640 & 198 & 15.0(27) & \\
                    & 3.12 & 44500 & 3700 & 278 & 33.2(19) & \\
                    & 3.15 & 20000 & & 110 & 3.26(18) & \\
      \hline
      $6\times 11^3$ & 3.17 & 45000 & 1440 & 121 & 7.4(11) & 3.181(1)(6)\\
                     & 3.18 & 45000 & 1750 & 201 & 30.0(15) &\\
                     & 3.185 & 45000 & & 232 & 23.2(24) & \\
                     & 3.19 & 45000 & & 123 & 10.4(14) &\\
                     & 3.20 & 45000 & & 100 & 5.62(36) &\\
      \hline
      $7\times 13^3$ & 3.22 & 32400 & 700 & 87 & 6.7(5) & 3.236(1)(6)\\
                     & 3.23 & 44500 & 750 & 129 & 18.4(11) & \\
                     & 3.24 & 29000 & & 149 & 20.7(16) & \\
                     & 3.25 & 22400 & & 83 & 9.2(10) \\
      \hline
    \end{tabular*}
    \caption{Run-time parameters of $\xi=2$ $T_c$-simulations. The number of sweeps, the persistence
time $\tau_p$ (if definable), the integrated autocorrelation time $\tau_{\text{int}}$ and the Polyakov loop
susceptibility $\chi_{\text{L}}$ are given for all values of $\beta$ used in the final reweighting procedure,
as well as the resulting $\beta_c$. The systematic error in the second brackets is a (rather conservative)
estimate of the finite volume effects and has to be considered if the given value should be an estimate of
the infinite-volume critical coupling.} 
    \label{tab:xi2deconfrtp}
  \end{center}
\end{table}

\begin{figure}[htbp]
\begin{center}
\includegraphics[width=10cm]{xi2nt4susc.eps}
\end{center}
\caption{The Polyakov loop susceptibility on the $4\times 8^3$ lattice for the
$\xi=2$ perfect action. The solid curves are the interpolations using Ferrenberg-Swendsen
reweighting, the dashed lines indicate the bootstrap error.}
\label{fig:xi2nt4susc}
\end{figure}

\clearpage
\section{Glueballs}\label{sec:xi2gb}

Finally, our $\xi=2$ classically perfect action is used to determine the glueball spectrum in pure lattice
gauge theory
(cf.~Section \ref{sec:glueballs}). As glueballs are rather heavy, $m_G\gtrsim 1.5$~GeV, to study their masses
the use of anisotropic lattices is particularly useful.

Due to mixing with other states existing in pure gauge theory on a periodic lattice, such as torelons or glueball
pairs, and due to the breaking of the continuum rotational symmetry by the lattice which leads to glueball
states transforming under the cubic group (see Section \ref{sec:glueballs}), the identification and the 
continuum spin assignment of single glueball states are additional vital questions which will be addressed
in Sections \ref{sec:gbident}, \ref{sec:spinident}. Finally, the resulting energies of all the states that can be measured and assigned a value
of continuum spin to, are given in units of the hadronic scale $r_0$ as well as in MeV for convenience.

\subsection{Simulation Parameters}

To allow a continuum extrapolation, at least for the lighter states, we decide to perform simulations at
three different lattice spacings in the range 0.10~fm~$\leq a_s \leq$~0.25~fm in volumes between
(1.2~fm)$^3$ and (2.0~fm)$^3$. The simulation parameters are given in Table \ref{tab:xi2gbpar}.
The gauge fields are updated by performing compound sweeps consisting of 4 pseudo
over-relaxation and 1 Metropolis sweep (see Sections \ref{sec:Metropolis},
\ref{sec:POR}), and after every 2 compound sweeps we measure all 22 loop shapes
up to length 8 (see Figure \ref{fig:loopshapes}). The measurement is performed on APE-smeared configurations
(see Section \ref{sec:smearing}) to spatially enlarge the operators, thus improving the overlap with the glueball
states, and to reduce unphysical high-momentum fluctuations.
On the coarse lattice at $\beta=3.15$ we use smearing levels $2, 4, 6, 8, 10$, on
the other lattices we use levels $3, 6, 9, 12, 15$, always with smearing parameter $\lambda_s=0.1$.
The measured operators are projected into all the irreducible representations of the cubic group. We measure
all possible polarisations to suppress statistical noise (see Section \ref{sec:latgb}).

\begin{table}[htbp]
\renewcommand{\arraystretch}{1.3}
  \begin{center}
    \begin{tabular*}{\textwidth}[c]{c@{\extracolsep{\fill}}cccc}
      \hline\vspace{-0.05cm} 
      $\beta$ & $(a_s/r_0)^2$ & $S^3\times T$ & $V$ & \# sweeps / measurements\\
      \hline
      3.15 & 0.2408(5) & $8^3\times 16$ & (1.96~fm)$^3$ & 241000 / 24100\\
      3.30 & 0.1005(6) & $10^3\times 20$ & (1.59~fm)$^3$ & 99000 / 9900\\
      3.50 & 0.0415(4) & $12^3\times 24$ & (1.22~fm)$^3$ & 115000 / 11500\\
      \hline
    \end{tabular*}
    \caption{Run parameters for the glueball simulations using the $\xi=2$ perfect
      action.}
    \label{tab:xi2gbpar}
  \end{center}
\end{table}

\subsection{Determination of the Energies}

The masses of the lowest states and the first excited states (if possible) in all the representations 
are determined using the variational methods and techniques described in Appendix \ref{app:vartech}.
The results in units of the temporal lattice spacing $a_t$ are listed in 
Tables \ref{tab:b315_fit_results} -- \ref{tab:b350_fit_results} in Appendix \ref{app:rescoll} together
with the number $N$ of operators entering into the fitting procedure, the time slices on which the
initial generalised eigenvalue problem is solved (usually 1/2), the number of operators $M$ 
in the final fit and the corresponding value of $\chi^2/N_{\text{DF}}$. These results are then multiplied
by $\xi_R\cdot r_0/a_s$ to obtain the glueball masses in units of the hadronic scale $r_0$, listed
in Table \ref{tab:gb_r0_results}. 

In order to obtain reliable estimates of the glueball masses from the variational method, one has to pay
attention that among the large number of operators (up to 145 on five smearing levels) 
there are no correlators entering the process that are measured exceptionally bad, i.e.~with
large errors even at small time separation, as this may destabilise the method. To filter out such data,
we calculate the ``correlator lifetimes'' defined in Section \ref{sec:latgb} and drop all correlators whose
lifetime is smaller than a given minimum. Usually, this is repeated with different minimal values, yielding
different sized sets of operators ($2\sim 4$ sets per representation). The $O(100)$ mass determinations using
different parameter sets (see Appendix \ref{app:vartech}) are then carried out on each set of initial operators,
and it turns out that generally the variational method used is stable, provided that the badly measured operators
are absent from the beginning and provided that no important operators are missing. Patently, having such a
window of the number of initial operators may be difficult if there is a rather small number of operators.

For some heavy states, such as the $PC=--$ glueballs, whose masses could only be determined from the $\beta=3.50$
measurements, we have to resort to using the correlators at $t_0=0$ and $t_1=1$ for the solution of the first generalised
eigenvalue problem, eq.~\ref{eq:gen_eigenvalue_problem}, yielding the lowest lying dual vectors $v_n$, $n=1,\ldots,M$,
that are used in the subsequent optimisation. This has the advantage that the correlators from separation $t_0=0$
are positive definite by definition (which is not true for $t_0\ge 1$), thus rendering the second generalised
eigenvalue problem, eq.~\ref{genev_CM}, well defined even if all initial operators are kept, $M=N$. However,
the correlators $C(t=0)$ are under suspicion of containing rather little physical information about the correlation
lengths (which correspond to the \emph{decay} of the correlators), actually they almost only provide information
about the relative normalisation of operators. Obviously, operators with a large signal at $t=0$ might be preferable
because their signal/noise ratio is probably rather better on average, however in general they do not correspond to the
operators having the largest overlap with the lowest-lying states. It turns out that using $t_0=0$, $t_1=1$ one
has to pay much more attention in choosing the fit range $t=t_{\text{min}} \ldots t_{\text{max}}$ because at small
$t$ the contamination of the lowest masses due to higher states is much more significant.

During the analysis, it turns out that the mass determinations for the scalar representation $A_1^{++}$ 
are rather challenging. The ground state receives a larger relative error than comparable states where the operators
are measured about as well; for the first excited state it is very hard to obtain a stable determination, and
the error of its mass may turn out to be huge (e.g. at $\beta=3.50$). Probably, it is the underlying vacuum, having
the same quantum numbers as the glueball state, that is responsible for these troubles. As already specified in
Section \ref{sec:latgb} we treat the vacuum just like another state in this representation, such that the glueball
ground state is effectively a first excited state and the glueball first excited state is effectively a second
excited state of the representation. Attempts of using other ways of getting rid of the vacuum such as the
usual subtraction or the subtraction of large $t$ correlators (that are assumed to contain solely noise, however
correlated to the noise at lower $t$) or even more sophisticated methods (like solving a generalised eigenvalue
problem using large $t$ correlators to dig out the vacuum state) do not at all succeed in improving the situation.

\begin{table}[htbp]
  \begin{center}
    \renewcommand{\arraystretch}{1.5}
    \begin{tabular*}{\textwidth}{c@{\extracolsep{\fill}}cccc}
      \hline
      Channel & $J$ & $\beta=3.15$ & $\beta=3.30$ & $\beta=3.50$\\
      \hline
      $A_1^{++}$   & 0 & 2.58(9) & 3.55(13) & 3.65(15)\\
    ${A_1^{++}}^*$ & 0 & 5.47(53) & 6.83(37) & 6.49(148)\\
      $E^{++}$     & 2 & 5.63(23) & 5.93(15) & 6.08(28)\\
      ${E^{++}}^*$ & 2 & 7.95(73) & 8.76(48) & 10.66(59)\\
      $T_2^{++}$   & 2 & 5.67(25) & 5.82(13) & 6.13(16)\\
    ${T_2^{++}}^*$ & 2 & & 7.97(69) & 10.16(30)\\
      $A_2^{++}$   & 3 & 6.71(73) & 9.11(64) & 10.92(48)\\
    ${A_2^{++}}^*$ &   & & & 14.12(130)\\
      $T_1^{++}$   & 3 & 7.54(68) & 9.00(44) & 11.21(33)\\
    ${T_1^{++}}^*$ &   & 8.69(91) & &\\
\hline
      $A_1^{-+}$   & 0 & 6.36(59) & 6.13(28) & 6.79(27)\\
    ${A_1^{-+}}^*$ &   & & & 10.27(110)\\
      $E^{-+}$     & 2 & 7.41(48) & 8.23(28) & 8.48(29)\\
      ${E^{-+}}^*$ &   & 8.35(125) & & 12.66(84)\\
      $T_2^{-+}$   & 2 & 7.41(49) & 8.18(21) & 8.57(42)\\
    ${T_2^{-+}}^*$ &   & & & 15.08(67)\\
\hline
      $T_1^{+-}$   & 1 & 7.20(39) & 7.69(26) & 9.12(24)\\
      $A_2^{+-}$   & 3 & & & 11.02(68)\\
    ${A_2^{+-}}^*$ &   & & & 15.55(147)\\
      $T_2^{+-}$   &   & 7.75(87) & & 10.29(118)\\
    ${T_2^{+-}}^*$ &   & & & 14.98(139)\\
     $E^{+-}$      &   & & & 12.42(42)\\
\hline
     $T_1^{--}$    &   & & & 10.81(83)\\
     $T_2^{--}$    &   & & & 10.85(83)\\
     $A_2^{--}$    &   & & & 12.18(45)\\
     $A_1^{--}$    &   & & & 15.34(103)\\
    \hline
    \end{tabular*}
    \caption{{}Final glueball mass estimates in terms of the hadronic scale $r_0$, $m_G r_0$ from the measurements
using the perfect $\xi=2$ action. The continuum spin assignment $J$ is given as well.}
    \label{tab:gb_r0_results}
  \end{center}
\end{table}

\subsection{Identification of Glueballs}\label{sec:gbident}

Besides single glueballs, the spectrum of pure gauge theory on a lattice with periodic boundary conditions
also contains states consisting of several glueballs, torelons or mixed states of glueballs and torelons. Although
we expect that the operators used to measure glueball energies couple most strongly to the single glueball states,
other states with similar masses and compatible quantum numbers might mix with them and distort the result,
i.e.~the energy determined by the analysis of
the correlation matrix may be dominated by a multi-glueball or torelon state with smaller energy than the 
single-glueball state to be measured.

In principle, there are several means of determining the nature of a state that has been measured. Firstly, the
simulation may be repeated on lattices of different size, keeping the lattice spacing fixed. As multi-glueball and
torelon states show a finite-volume behaviour very different from single glueballs, such states stand out and may
be dropped from further analysis. Secondly, one may measure additional operators that couple strongly to torelon
or multi-glueball states. Including and excluding these operators in the variational method and studying the coefficients
obtained from the optimisation, the mixing strength is determined and states which do not mix considerably with any of
the additional operators may be safely considered to be single-glueball states. Finally, using the mass estimates for the
low-lying glueballs, one may determine the approximate locations of the lowest-lying multi-glueball states. Also, the
minimal energy of mixing torelons may be estimated using the string formula, eq.~\ref{eq:stringtorelon}.

\subsubsection{Torelons}

The first two of the methods mentioned above require additional work and computer time and go beyond the scope of
this work. The last method, however, can be done rather easily. Let us first calculate the energy of the lowest-lying
torelon states that may interfere with our measurements. Single torelons (see Section \ref{sec:torelon}) transform
non-trivially under $Z_3$ symmetry operations; our operators, closed Wilson loops, however, are invariant under
these transformations. This means that they cannot create single torelon states, however the creation of two torelons
of opposite center charge is possible. We assume that they do not interact considerably and that our lattice extension
is rather large and thus we use the simple formula $E_{2T}\approx 2\sigma L$ to estimate the energy of a torelon pair with
momentum zero, where $\sigma$ is the string tension (measured by the static $q\bar{q}$ potential,
see Section \ref{sec:xi2scale}) and $L$ is the spatial
extent of the lattice. Table \ref{tab:min2tln} lists the minimum energies of torelon pairs to be expected on our lattices.

\begin{table}[htbp]
\begin{center}
  \renewcommand{\arraystretch}{1.5}
  \begin{tabular*}{4cm}{c@{\extracolsep{\fill}}c}
    \hline
    $\beta$ & $r_0 E_{2T}$\\
    \hline
    3.15 & 10.60\\
    3.30 & 8.24\\
    3.50 & 6.92\\
    \hline
  \end{tabular*}
  \caption{{}The minimum energies of momentum zero torelon pairs on the lattices used in the glueball simulations,
given in units of $r_0^{-1}$.}
  \label{tab:min2tln}
\end{center}
\end{table}

Note, that a state composed of two opposite center charge torelons and with total zero momentum is symmetric under
charge conjugation. The operators for $C=-$ states therefore do not create such torelon pairs.

\subsubsection{Multi-Glueball States}

To estimate the lowest energies of multi-glueball states present on our lattices, we follow the method used by Morningstar and Peardon \cite{Morningstar:1999rf}, described in detail in Appendix \ref{app:2gb}.
We assume that the energy of multi-glueball states is approximately given by the sum of the energies of the individual glueballs, i.e.~that there is no substantial interaction. It is therefore clear that the lowest-lying multi-glueball states with zero total
momentum are the two-glueball states, with energy
\begin{equation}
E_{2G}\approx\sqrt{\vec{p}^2+m_1^2}+\sqrt{\vec{p}^2+m_2^2},
\end{equation}
where $m_1$ denotes the rest mass of the first glueball with momentum $\vec{p}$ and $m_2$ denotes the rest mass
of the second glueball with momentum $-\vec{p}$. The masses are taken from our determinations at the respective values
of $\beta$, the (lattice) momenta are chosen such that the energy $E_{2G}$ of the glueball pair contributing to some representation
$\Gamma^{PC}$ is minimised. Note, that glueball pairs contribute to different representations of the cubic group,
not only depending on the representations according to which the single glueballs transform, but also
on the single glueball momentum $\vec{p}$. It happens therefore, that a pair of rather light glueballs with
non-vanishing momenta $\vec{p}$ has a smaller energy than the pairs of glueballs \emph{at rest} contributing to
the same representation $\Gamma^{PC}$.
The details of the analysis are collected in Appendix \ref{app:2gb},
the resulting minimal energies in units of $r_0$ of glueball pairs in all the representations where states can
be measured on our lattices are collected in Tables \ref{tab:gbpair_b315}--\ref{tab:gbpair_b350}.

\begin{figure}[htbp]
\begin{center}
\includegraphics[width=11cm]{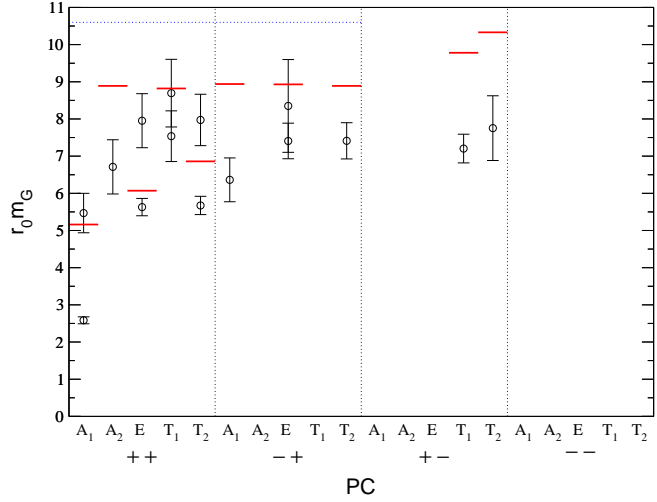}
\end{center}
\caption{The results of the glueball measurements at $\beta=3.15$ converted into units of $r_0$. The lower bounds
of the energies of multi-glueball states (solid lines) or torelon pairs (dotted line) are given as well.}
\label{fig:gbres_b315}
\end{figure}

\begin{figure}[htbp]
\begin{center}
\includegraphics[width=11cm]{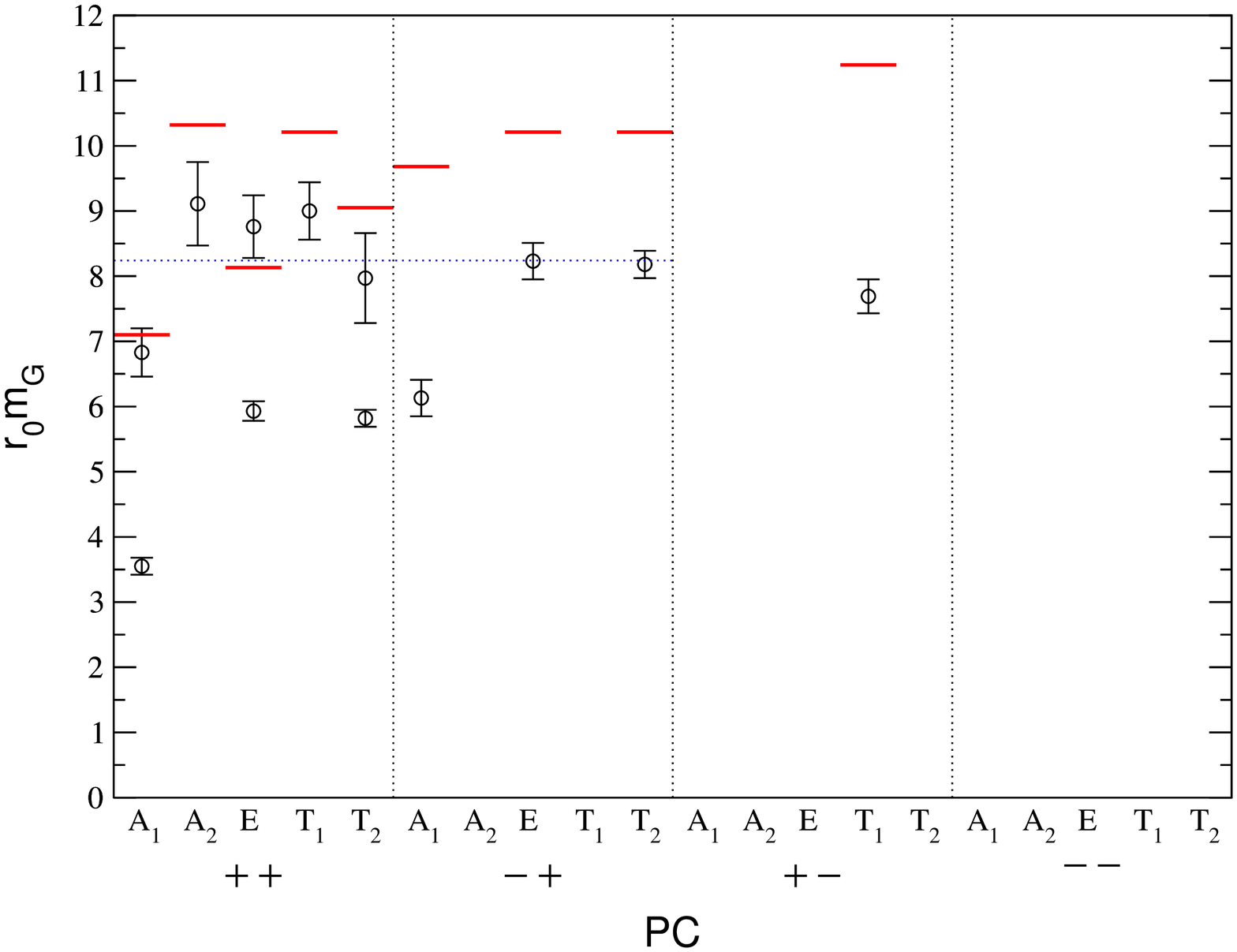}
\end{center}
\caption{The results of the glueball measurements at $\beta=3.30$ converted into units of $r_0$. The lower bounds
of the energies of multi-glueball states (solid lines) or torelon pairs (dotted line) are given as well.}
\label{fig:gbres_b330}
\end{figure}

\begin{figure}[htbp]
\begin{center}
\includegraphics[width=11cm]{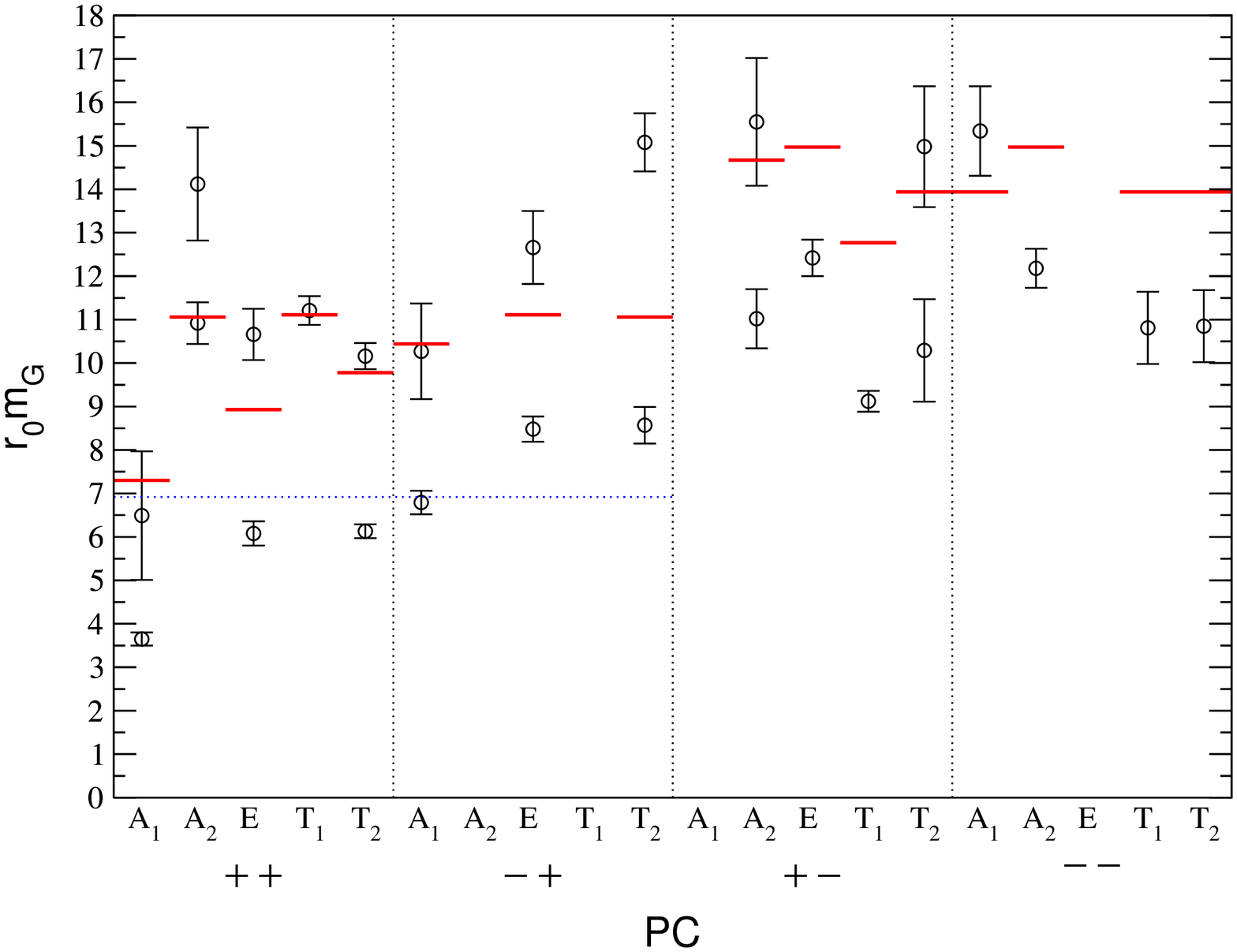}
\end{center}
\caption{The results of the glueball measurements at $\beta=3.50$ converted into units of $r_0$. The lower bounds
of the energies of multi-glueball states (solid lines) or torelon pairs (dotted line) are given as well.}
\label{fig:gbres_b350}
\end{figure}

The lower bounds of the multi-glueball energy region as well as the lower bound for torelon pairs are
indicated in Figures \ref{fig:gbres_b315}--\ref{fig:gbres_b350}. It turns out that all of the excited states measured
and even some of the lowest states in a given representation could be affected by torelon pairs or multi-glueball
states. Of course, it is not at all ruled out that these states are indeed single glueballs, however one has to be
very careful with the interpretation and has to keep in mind that these issues require further study.

\subsection{The Continuum Limit}

In order to remove discretisation errors from glueball masses $r_0 m_G$ that have been obtained on lattices
with finite lattice spacing, we have to perform the continuum limit, i.e.~to extrapolate the results
to $a_s=0$. Having obtained results at three values of the coupling $\beta$, one of them corresponding to rather
coarse lattices, and having no accurate information (e.g.~from perturbation theory) about the behaviour of the
energies depending on the lattice spacing (about the form of the curve used in the extrapolation) this is 
not an easy task and rather ambiguous. 

We perform two different kinds of continuum limits. Firstly, we extrapolate the masses in terms of the hadronic
scale $r_0$, $r_0 m_G = m_G a_t \cdot \xi \cdot r_0/a_s$, secondly, we extrapolate mass ratios $m_{G_1}/m_{G_2}$
of different glueball representations.
Generally, the procedure we resort to, is the following: We include all the
three energies or mass ratios, obtained at $\beta=3.15, 3.30, 3.50$, into the fit and use the form
\begin{equation}\label{eq:gbfitform2}
r_0 m_G\mid_{a_s} = r_0 m_G\mid_{a_s=0} + c_2 \left(\frac{a_s}{r_0}\right)^2.
\end{equation}
Additionally, we perform fits including only the results from the two finer lattices at $\beta= 3.30, 3.50$ using
a constant (if this fits the results reasonably well) or the form stated above. This value of the continuum mass
(or mass ratio) is
then compared to the result of the fit of all the three data points, which is only accepted if the two results
coincide within their errors.

\subsubsection{Masses in Terms of $\mathbf{r_0}$}

For the extrapolation of masses in terms of the hadronic scale $r_0$,
this procedure works very well for the representations $A_2^{++}$, $E^{-+}$, $E^{++}$, ${E^{++}}^*$,
$T_2^{-+}$ and $T_2^{++}$ where we may always fit the results from the two finer lattices to a constant which
agrees with the result of the extrapolation using eq.~\ref{eq:gbfitform2} on all three data points. For the
$A_1^{-+}$ representation, we obtain the best results fitting the masses to a constant, for the fit to two
data points as well as to three data points.

\begin{figure}[htbp]
\begin{center}
\includegraphics[width=12cm]{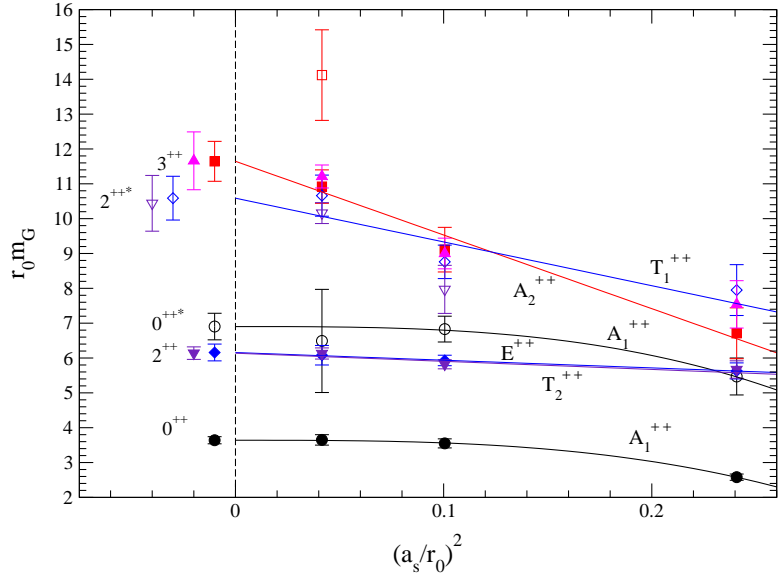}
\end{center}
\caption{Mass estimates of the $PC=++$ glueballs in terms of the hadronic scale $r_0$ against $(a_s/r_0)^2$. The
curves are the continuum limit extrapolations of the form $r_0 m_G+c_2(a_s/r_0)^2$. \emph{Circles}: $A_1^{++}$,
\emph{boxes}: $A_2^{++}$, \emph{diamonds}: $E^{++}$, \emph{upward triangles}: $T_1^{++}$, \emph{downward triangles}:
$T_2^{++}$; \emph{solid symbols}: ground states, \emph{open symbols}: first excited states. Note that the continuum
results of the representations $T_1^{++}$ and ${T_2^{++}}^*$ have been obtained using a fit of glueball mass
ratios.}
\label{fig:gb_ex_pp}
\end{figure}

\begin{figure}[htbp]
\begin{center}
\includegraphics[width=12cm]{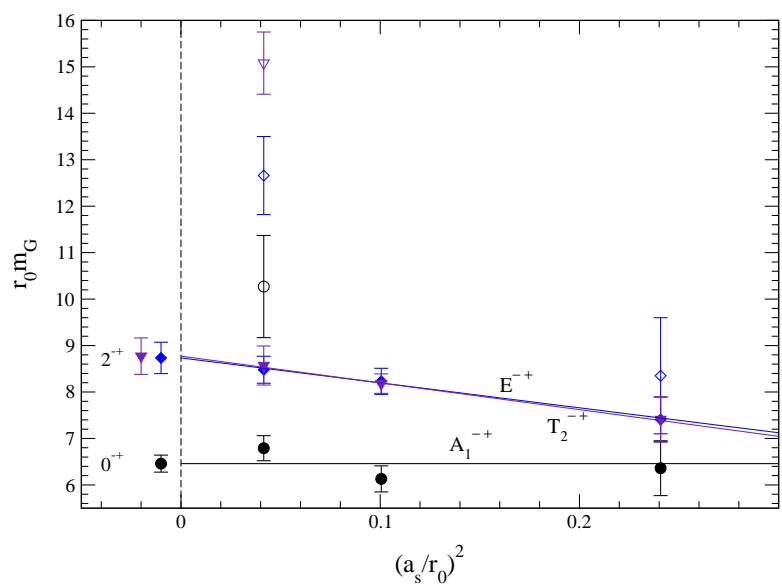}
\end{center}
\caption{Mass estimates of the $PC=-+$ glueballs in terms of the hadronic scale $r_0$ against $(a_s/r_0)^2$. The
curves are the continuum limit extrapolations of the form $r_0 m_G+c_2(a_s/r_0)^2$. \emph{Circles}: $A_1^{-+}$,
\emph{diamonds}: $E^{-+}$, \emph{downward triangles}: $T_2^{-+}$; \emph{solid symbols}: ground states, \emph{open symbols}: first excited states.}
\label{fig:gb_ex_mp}
\end{figure}

Much more difficult is the extrapolation for the scalar glueball and its excited state, $A_1^{++}$ and
${A_1^{++}}^*$. It turns out that the fit has a very bad $\chi^2$ and describes the data very badly if
eq.~\ref{eq:gbfitform2} is used. However, we find good behaviour if we use
\begin{equation}\label{eq:gbfitform6}
r_0 m_G\mid_{a_s} = r_0 m_G\mid_{a_s=0} + c_6 \left(\frac{a_s}{r_0}\right)^6
\end{equation}
for the ground state as well as for the excited state energies and thus decide to use this (ad-hoc) fitting form for
the scalar glueball. The results of the continuum extrapolations for various glueball states are
collected in Table \ref{tab:cont_ex}. In Figures \ref{fig:gb_ex_pp}, \ref{fig:gb_ex_mp} the curves are
shown together with the measured values for the $PC=++$ and the $PC=-+$ sectors where extrapolations
are possible with the present data.

Other representations where energies can be determined at several values of $\beta$, such as
$T_1^{+-}$, $T_1^{++}$ or ${T_2^{++}}^*$ do not yield consistent results using the procedure presented
above and we decide to abstain from performing continuum extrapolations in terms of $r_0 m_G$ for these states.

\subsubsection{Mass Ratios}

The other method of extrapolating to the continuum, namely using mass ratios $m_{G_1}/m_{G_2}$ between
different glueball representations, has got the advantage that uncertainties present in the
determination of $r_0/a_s$ or the renormalised anisotropy $\xi_R$ do not enter into the continuum extrapolation.
Furthermore,
we may assume that several effects such as finite size scaling are more similar for different glueball states
measured on the same lattice
as opposed to completely different quantities such as the static quark-antiquark potential or the torelon
dispersion relation measured on different sized lattices that are used to obtain $r_0/a_s$ and $\xi_R$, respectively.

The common way of extrapolating glueball mass ratios to the continuum is using ratios $m_G/m_{0^{++}}$
(see Table \ref{tab:a1pp_ratio}), 
i.e.~ratios to the scalar glueball mass, which is the lightest mass present. Doing this, we observe that
the extrapolation is difficult, the goodness of the fit is rather bad and the errors are large. 
Furthermore, the resulting continuum mass ratios are not in good agreement to ratios of the continuum
masses coming from the extrapolation of the masses in terms of $r_0$. This is not a surprise, because
the behaviour of the scalar glueball as a function of the lattice spacing is rather difficult, i.e.~there
are large cut-off effects that cannot be described too easily. Additionally, the errors of the mass estimates
of the scalar glueball are quite large. Due to these reasons, we decide to use the well measured masses
of the tensor $T_2^{++}$ representation, which show small errors and seem to scale rather well. These ratios
are listed in Table \ref{tab:t2pp_ratio}.

The procedure of the extrapolation is the same as for the case of $r_0 m_G$ described in the previous
section; the unambiguous results are listed in Table \ref{tab:cont_ex_ratio}.
Again, for the $A_1^{-+}$ states, as well as for the $E^{++}$, the fit to a constant works out best;
the $T_2^{-+}$ glueball can only be reasonably extrapolated using a fitting form analogous to 
eq.~\ref{eq:gbfitform6}. Note, that this way we obtain
continuum results for the representations $T_1^{+-}$, $T_1^{++}$ and ${T_2^{++}}^*$ which is not possible
performing the extrapolation using masses in terms of $r_0$. Furthermore, all the other results listed are in
agreement to the ratios of the continuum masses $r_0 m_G$ within one standard deviation.

In the following, we will use the finite lattice spacing results from the simulation on the finest lattice
at $\beta=3.50$ for the energies of states that cannot be extrapolated to the continuum in one of the
two ways, e.g.~to study
degeneracies or to draw a (rather qualitative) picture of the low-lying glueball spectrum. In tables listing
continuum results, these
values not extrapolated to $a_s=0$ will be stated in square brackets.

\subsection{Continuum Spin Identification}\label{sec:spinident}
Once the extrapolation of the measured glueball energies to the continuum, $a_s=0$, has been performed, what
remains is the assignment of continuum spin $J$ to the different representations of the cubic group $\Gamma$.
From Table \ref{tab:subdrep} we know from which lattice representations $\Gamma^{PC}$ a continuum glueball with quantum
numbers $J^{PC}$ may obtain contributions, where the assignment of parity $P$ and charge conjugation $C$ is
simply $1:1$. Degeneracies (in the continuum limit) between several representations $\Gamma^{PC}$
contributing to the same continuum state $J^{PC}$ are a strong indication for the correctness of the assignment
of all the lattice states involved to the same continuum $J^{PC}$ glueball. Furthermore, we make the assumption that the
mass of the glueballs increases with the spin getting larger.

In the $PC=++$ sector, we observe a single low-lying state, $A_1^{++}$, which is thus assigned $J=0$.
The $E^{++}$ and $T_2^{++}$ states are degenerate to a very high precision as it should happen if these
representations correspond to the five polarisations of a $J=2$ glueball. The excited state ${A_1^{++}}^*$ again
has no degenerate partner and is assigned to an excited state of the continuum $J=0$ glueball. The excited states
${E^{++}}^*$ and ${T_2^{++}}^*$ again turn out to be degenerate and so they are assigned to an excitation of the $J=2$ glueball. The same situation is met with the
representations $A_2^{++}$ and $T_1^{++}$ which are thus assigned $J=3$. The first excitation ${T_2^{++}}^*$
being part of the $J=2$ state, the missing 3 polarisations should
come from the second excitation of $T_2^{++}$ which cannot be measured by our simulations.

In the $PC=-+$ sector, the situation of the $A_1^{-+}$ and the $E^{-+}$ and $T_2^{-+}$ is very similar to 
the one of their partners in the $PC=++$ sector; these states are thus assigned $J=0$ and $J=2$ respectively.
The remaining three, excited states have no apparent degeneracies, there is thus no safe assignment of continuum
spin to these states. The facts that the excited state ${A_1^{-+}}^*$ is rather light and comes with no degenerate
partner, raises the presumption that it is the excited $J=0^{-+}$ glueball.

In the other two sectors, continuum extrapolations of our measured energies are not possible, except for the 
representation $T_1^{+-}$ using mass ratios. However, looking at
the degenerate representations in the $PC=++, -+$ sectors, it is noticeable that the degeneracies are apparent
even for finite lattice spacing at $\beta=3.50, 3.30$. Assuming that this behaviour persists for $PC= +-, --$, at
least for energies not too high, we may assign continuum spin even to some of the remaining states.

In the $PC=+-$ sector, we notice the lowest-lying $T_1^{+-}$ state which has no degenerate partner which suggests
a $J=1$ interpretation. Next, there is the $A_2^{+-}$ which is likely to correspond to $J=3$. 

Finally, in the $PC=--$ sector, there are different possible scenarios. The almost exact degeneracy of $T_1^{--}$
and $T_2^{--}$ suggests them contributing to the $J=3$ continuum state. The rather heavier state $A_2^{--}$ could
still be degenerate with the two latter states (note the high mass and the finite lattice spacing!) and carry
the remaining polarisation. Admittedly, it could as well correspond to an excited $J=3$ state or even to $J=6$.
The final $A_1^{--}$ state is very heavy and indicates that glueballs with even spin and $PC=--$ have large
energies as the representation $A_1$ contributes solely to even numbered spin states up to $J=8$!

\begin{figure}[htbp]
\begin{center}
\includegraphics[width=12cm]{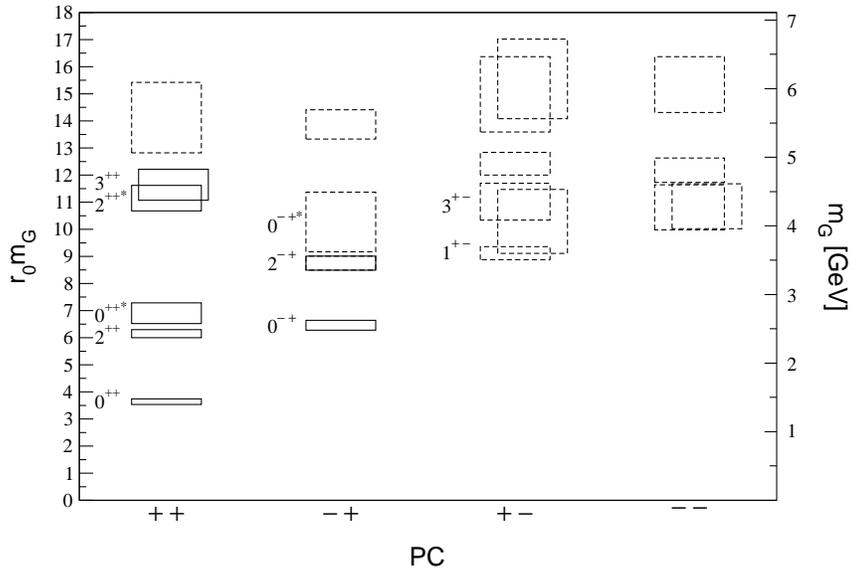}
\end{center}
\caption{{}The mass spectrum of glueballs in SU(3) pure gauge theory as obtained from our simulations using the
$\xi=2$ classically perfect action. The masses are given in units of the hadronic scale $r_0$. The dashed boxes
are used for states that could not be extrapolated to the continuum, instead the mass obtained from the simulation
on the finest lattice at $\beta=3.50$ is given. The most probable assignment of continuum spin is indicated as well.}
\label{fig:all_gbmasses}
\end{figure}

The rather well supported continuum spin assignments presented above are given together with the glueball masses
in Tables \ref{tab:gb_r0_results}, \ref{tab:cont_ex}, \ref{tab:cont_ex_ratio}, the final results on the masses of
these states are collected, together with the masses in MeV, in Table \ref{tab:final_gb}. Figure \ref{fig:all_gbmasses}
provides a rather qualitative picture of the low-lying glueball spectrum as obtained from our $\xi=2$ simulations.

\begin{table}[htbp]
  \begin{center}
    \renewcommand{\arraystretch}{1.5}
    \begin{tabular*}{\textwidth}{l@{\extracolsep{\fill}}lcl}
      \hline
      $J^{PC}$ & $\Gamma^{PC}$ & $m_G r_0$ & $m_G$~[MeV]\\
      \hline
      $0^{++}$     & $A_1^{++}$                   & 3.64(10)     & 1440(40)\\
      ${0^{++}}^*$ & ${A_1^{++}}^*$               & 6.90(38)     & 2730(150)\\
      $2^{++}$     & $E^{++}$, $T_2^{++}$         & 6.15(15)     & 2430(60)\\
      ${2^{++}}^*$ & ${E^{++}}^*$, ${T_2^{++}}^*$ & 10.52(51)    & 4160(200)\\
      ${3^{++}}$   & $A_2^{++}$, $T_1^{++}$       & 11.65(49)    & 4600(190)\\
      ${0^{-+}}$   & $A_1^{-+}$                   & 6.46(18)     & 2550(70)\\
      ${2^{-+}}$   & $E^{-+}$, $T_2^{-+}$         & 8.75(26)     & 3460(110)\\
      ${0^{-+}}^*$ & ${A_1^{-+}}^*$               & [10.27(110)] & [4060(430)]\\
      ${1^{+-}}$   & $T_1^{+-}$                   & 9.45(71)     & 3730(280)\\
      ${3^{+-}}$   & $A_2^{+-}$                   & [11.02(68)]  & [4350(270)]\\
      \hline
      \end{tabular*}
      \caption{{}Final results for the masses of continuum glueballs with quantum numbers $J^{PC}$, obtained
from the lattice representations $\Gamma^{PC}$. For the conversion
to MeV, $r_0^{-1}\approx 0.5$~fm~$\approx 395$~MeV has been used. Quantities in brackets have not been extrapolated to
the continuum but denote results from the simulation on the finest lattice at $\beta=3.50$ (corresponding to
$a_s=0.102$~fm).}
    \label{tab:final_gb}
  \end{center}
\end{table}

\subsection{Discussion}

In order to discuss the results obtained from the glueball measurements, presented in the previous Sections,
let us first list the sources of errors that may be present. Firstly, there are finite size effects; our finest
lattice ($\beta=3.50$) has got a spatial volume of (1.22~fm)$^3$ which is rather small compared to other volumes
used in glueball measurements. The magnitude of finite-size effects depends largely on the quantity studied, more precisely
on the form of the glueball wavefunctions (above all their extension). Due to the three volumes employed having such different
size, finite-size effects do not only systematically shift the lattice results but they can also make reliable continuum
extrapolations much more difficult or even impossible.

Secondly, the mass of the scalar glueball $A_1^{++}$ (above all the ground state) shows very large cut-off effects.
These could be due to the presence
of a critical end point of a line of phase transitions in the fundamental-adjoint coupling plane assumed to 
define the continuum limit of a scalar field theory. Our action includes in its rich structure operators 
transforming in the adjoint representation. If their net coupling (which we do not control during the construction
and the parametrisation of the action)
lies in a certain region, the effect of the critical end-point on scalar quantities at certain lattice spacings
(sometimes called the ``scalar dip'') may even be enhanced compared to other (more standard) discretisations with
purely fundamental operators.

Furthermore, our analysis shows that states consisting of several glueballs or torelon pairs could mix with most of
the higher-lying states occurring in our measurements. This makes the interpretation of the measured states more
difficult, additionally, systematic effects (e.g.~different mass shifts at different values of $\beta$ because
of the different lattice sizes or due to the presence of operators having larger overlap with unwanted states)
may again complicate the continuum extrapolation.

Finally, as always, there is the statistical error. Generally, it does not introduce bias and it can be
estimated quite reliably. However, glueball correlators are notorious for large statistical noise and this noise
may hamper the reliable variational analysis of states that in principle could still be measured as for their energies.

There are several possibilities of improving the measurements. Firstly, and most simple, improving the statistics
may help. This can be seen e.g.~looking at the results of the simulation on the coarsest lattice at $\beta=3.15$.
Although the lattice is coarser, more energies can be determined than on the finer $\beta=3.3$-lattice, mainly
due to the larger statistics. Additionally, a lot of problems stated previously are solved (or at least kept under
control) if for each value of $\beta$, we determine the glueball energies on several lattices of different physical size.
Studying the finite-size scaling creates the possibility of separating single glueball from other (multi-glueball,
torelon pair) states and it provides estimates of finite lattice spacing infinite-volume results to be used in a
safe continuum extrapolation. Another way of improving the measurements is systematically studying the operators 
employed to create and annihilate single glueball states in order to increase their overlap with the single glueballs
to be measured and to decrease their overlap with all the other states. This includes the study of other smearing
techniques, such as e.g.~Teper fuzzing \cite{Teper:1987wt}. Conversely, the introduction of operators
coupling most strongly to the unwanted multi-glueball or torelon-pair states allows for the calculation of mixing
strengths in order to exclude all the unwanted states from further analysis. Another technically rather easy but computationally
expensive way of improving the results is performing simulations on additional values of $\beta$. This allows for a more
detailed study of the continuum limit, especially in the case of non-perturbatively improved actions, where there
is no clearcut information about the
way how the continuum is approached. Naturally, one has to take care not to go too deep into the strong coupling
region as there is not much information about the continuum; performing simulations on two additional values of
$\beta$ corresponding to $(a_s/r_0)^2\approx 0.16, 0.3$ would probably be a good choice in our case. But keep in
mind that for each additional lattice spacing the measurements of the torelon dispersion relation as well as of
the static-quark potential have to be repeated.

There is one error which is not cured by the measures recommended in the last paragraph, namely the effects
on the scalar states, coming from the ``scalar dip''. Morningstar and Peardon improve
the situation adding a two-plaquette adjoint term with a negative coefficient which results in an approach to the
continuum on a trajectory always far away of the dangerous ``dip'' region \cite{Morningstar:1999dh}. In principle,
the classically perfect action 
could be treated the same way: Extract all the operators with adjoint contributions present in the parametrisation,
determine their sign and add another (non-linear) constraint to the fit, namely that the action of all
these operators together corresponds to an adjoint operator with a certain (negative) coupling. Because of the large
freedom in the fit, the inclusion of this single criterion should not impair the quality of the parametrisation
considerably.

\begin{table}[htbp]
  \begin{center}
    \renewcommand{\arraystretch}{1.3}
    \begin{tabular*}{9cm}[c]{l@{\extracolsep{\fill}}ll}
      \hline\vspace{-0.05cm}
    Collab. & $r_0 m_{0^{++}}$ & $r_0 m_{2^{++}}$\\
      \hline
      UKQCD \cite{Bali:1993fb}        & 4.05(16)    & 5.84(18)   \\
      Teper \cite{Teper:1998kw}       & 4.35(11)    & 6.18(21)   \\
      GF11 \cite{Vaccarino:1999ku}    & 4.33(10)    & 6.04(18)   \\
      M\&P \cite{Morningstar:1999rf}  & 4.21(15)    & 5.85(8)    \\
      Liu  \cite{Liu:2000ce}          & 4.23(22)    & 5.85(23)   \\
      \hline
    FP action                         & 4.12(21)    & [5.96(24)] \\
    AICP                              & 3.64(10)    & 6.15(15)   \\
    \hline
    \end{tabular*}
    \caption{{}Comparison of the lowest-lying glueball masses in units of $r_0$. 
      Values in brackets denote masses obtained at a lattice spacing
      $a=0.10$ fm and are not extrapolated to the continuum.}
    \label{tab:cmpgbcont}
  \end{center}
\end{table}

\begin{table}[htbp]
  \begin{center}
    \renewcommand{\arraystretch}{1.3}
    \begin{tabular*}{\textwidth}[c]{l@{\extracolsep{\fill}}llll}
      \hline\vspace{-0.05cm}
    Collab. & $r_0 m_{0^{-+}}$ & $r_0 m_{{0^{++}}^*}$ & $r_0 m_{2^{-+}}$ & $r_0 m_{1^{+-}}$\\
      \hline
      Teper \cite{Teper:1998kw}       & 5.94(68) & 7.86(50) & 8.42(78) & 7.84(62) \\   
      M\&P \cite{Morningstar:1999rf}  & 6.33(13)   & 6.50(51) & 7.55(11)   & 7.18(11)   \\
      \hline
    FP action                         &[6.74(42)]& & [8.00(35)]&[7.93(78)] \\
    AICP                              & 6.46(18) & 6.90(38) & 8.75(26) & 9.45(71) \\
    \hline
    \end{tabular*}
    \caption{{}Comparison of glueball masses in units of $r_0$. 
      Values in brackets denote masses obtained at a lattice spacing
      $a=0.10$ fm and are not extrapolated to the continuum.}
    \label{tab:cmpgbcont2}
  \end{center}
\end{table}

Table \ref{tab:cmpgbcont} compares the newly obtained continuum glueball masses with the result of the isotropic
FP action as well as with results obtained by other collaborations. There is reasonable agreement between the
different determinations of the $2^{++}$, ${0^{++}}^*$, $0^{-+}$, $2^{-+}$ glueball masses. The situation
for the scalar $0^{++}$ and
for the $1^{+-}$ state is different, however. The scalar glueball mass comes out about 5 standard deviations below
the average mass of the other collaborations if only the usual statistical errors are considered. Thinking of all the
possible sources of errors stated above, which are (partly) also present in the analyses of the other groups, we
tend to explain this discrepancy with underestimated (or disregarded) errors. Concerning the $1^{+-}$ glueball,
looking at the masses determined on the lattices ($T_1^{+-}$ in Table \ref{tab:gb_r0_results}), we observe that the masses
determined on the coarser (and larger) lattices at $\beta=3.15, 3.30$ coincide with the (continuum) results
obtained by other groups and do not even show a significant discrepancy, despite the considerable difference between
the lattice spacings. The mass determined on the fine (and small) lattice at $\beta=3.50$ however, is much higher.
This tendency is even amplified by the continuum extrapolation.  We suspect that this state exhibits strong finite-size
effects pushing up its mass in small volumes.

\begin{figure}[htbp]
\begin{center}
\includegraphics[width=12cm]{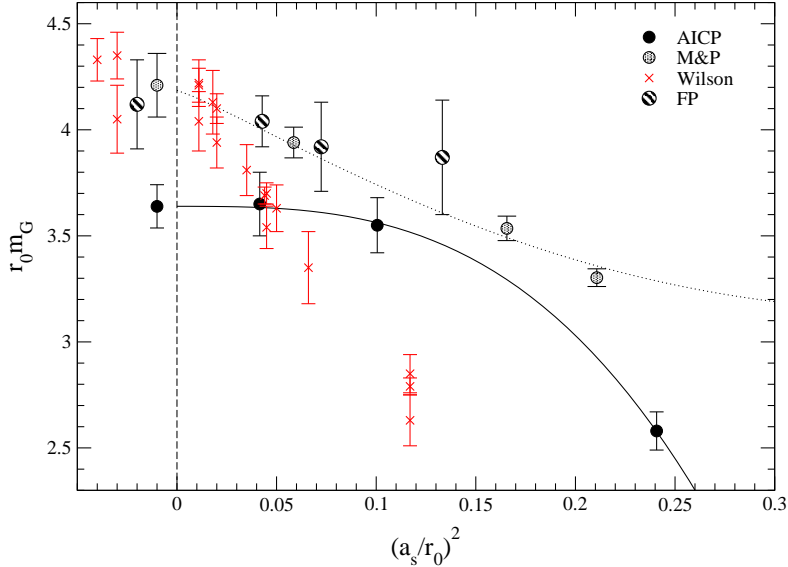}
\end{center}
\caption{{}Lattice results for the scalar ($0^{++}$) glueball mass, together with results of continuum extrapolations obtained
from simulations employing different actions. The different continuum values for the Wilson
action stem from different groups.}
\label{fig:scal_0pp}
\end{figure}

\begin{figure}[htbp]
\begin{center}
\includegraphics[width=12cm]{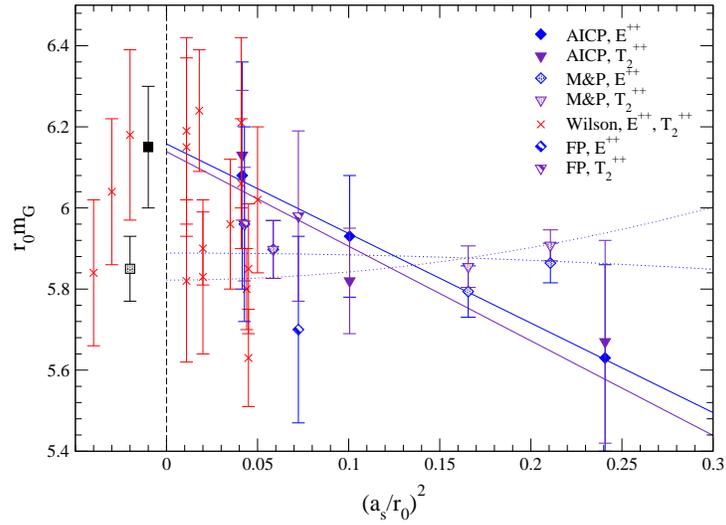}
\end{center}
\caption{{}Lattice results for the tensor ($2^{++}$) glueball mass, together with results of continuum extrapolations obtained
from simulations employing different actions. The continuum extrapolations of the $E^{++}$ and $T_2^{++}$ representations
are averaged in order to get the mass of the single continuum glueball. The different continuum values for the Wilson
action stem from different groups.}
\label{fig:scal_2pp}
\end{figure}

Figures \ref{fig:scal_0pp} and \ref{fig:scal_2pp} compare our measurements of the $A_1^{++}$, $E^{++}$ and $T_2^{++}$
states to measurements obtained by other groups, using different actions, as well as to the isotropic FP action results.
Concerning the scalar glueball, the artifacts at moderate lattice spacings corresponding to $(a_s/r_0)^2\lesssim 0.15$
($a_s \lesssim 0.19$~fm) seem to be rather smaller for the isotropic FP as well as for the anisotropic classically
perfect action compared to the tadpole and treelevel improved action (M\&P) and
certainly much smaller than the artifacts of the Wilson action. The mass obtained on the coarsest lattice at
$\beta=3.15$ (corresponding to $a_s\approx 0.24$~fm) using the classically perfect anisotropic action (AICP),
however, exhibits large cut-off effects of about 30\% (compared to about 20\% for the M\&P action), probably due to
the scalar dip. Concerning the tensor glueball, the picture is not clear, mainly due to the considerable statistical
errors of the measurements, except of the ones obtained by using the tadpole and treelevel improved anisotropic action.

These considerations show again, that in order to obtain fool-proof results measuring glueball masses (and other
quantities) on the lattice, one has to make sure to measure everything that is necessary, taking into account
a certain number of lattices with different couplings $\beta$, lattices with different volumes, measuring the renormalised anisotropy
for all the values of $\beta$ considered and measuring additional operators having large overlaps with states that
could mix.

Being the first application of the FP action technique to anisotropic gauge theory, it goes beyond the scope
of this work to check all the possible sources of errors in the glueball mass determinations. Observing the
difficulties arising with the determination of the glueball spectrum already in pure gauge theory, one can estimate
that reliable determinations in the context of more and more unquenched Lattice QCD (including lighter and lighter
dynamical fermions until, finally, one ends up with quarks having their natural mass) will be a very tough job.
However, this will yield information about the fate of glueballs in the presence of quarks and possibly solve the
big puzzle concerning the presence (or absence) of glueballs in nature. It will thus be really worth the big effort.

\fancyhead[RE]{\nouppercase{\small\it Chapter \thechapter.\, Repeating the Spatial Blocking}}
\chapter{Repeating the Spatial Blocking Step} \label{ch:xi4act}

As noted in Chapter \ref{ch:construction}, the spatial blocking step used to obtain a classically
perfect $\xi=2$ action may be repeated straightforwardly to obtain actions with higher anisotropies.
Using another spatial scale 2 blocking step we create a classically perfect $\xi=4$ gauge action and
parametrise it the same way as the $\xi=2$ action. Furthermore, we measure the renormalised anisotropy
and show that it is indeed feasible to construct such actions to be used in MC simulations for heavy
states such as glueballs. The classically perfect $\xi=4$ action is not yet examined thoroughly,
however the parametrisation is ready (see Appendix \ref{app:xi4act}) and the analyses of the $\xi=2$
action (see Chapter \ref{ch:xi2act}) may be repeated for $\xi=4$.

\section{Perfect Actions for Larger Anisotropies}

Each application of the spatial blocking step (slightly) renormalises the ani\-so\-tro\-py of the
action; the renormalised anisotropies of the final actions thus have to be measured in the end.
However, this is necessary anyhow if
one is interested in comparing results (other than e.g.~mass ratios, where the renormalisation of
the anisotropy cancels out) to other collaborations or to the experiment.

In order to be able to repeat the spatial blocking step getting from \mbox{$\xi=2$} to \mbox{$\xi=4$}
we construct a $\xi=2$ action which is valid on minimised \mbox{$\xi=4$} configurations including
into the fit configurations at different values of $\beta$ covering a large range. The $\xi=2$
action presented in Chapter \ref{ch:xi2act} is not suitable for this task as it is dedicated
to be used solely on largely fluctuating configurations around $\beta=3.0$. The construction and the
parameters of the intermediate $\xi=2$ action are described in Appendix \ref{app:xi2intact}.

Once the fine action to be used on the r.h.s. of the renormalisation group equation, eq.~\ref{eq:FP_equation},
is ready, we
may proceed the same way as for the $\xi=2$ action. We perform the non-linear fit on 2 $\xi_{\text{ad-hoc}}=6$
configurations each at $\beta_{\text{ad-hoc}}=$ 4.0, 3.5, 3.0, 2.5, 2.0. It turns out that in order
to obtain parametrisations that are free of dangerous ``traps'' in the $u$-$w$ plane (see Section \ref{sec:linfit})
we have to include the condition $\mathcal{A}(u,w)>0$ at ten points $(u,w)$ into
the non-linear fit. Studying the values of $\chi^2$ as well as the linear behaviour of the parametrisations
we decide to use a set with $\max(k+l)_{\text{sp}}=3$, $\max(k+l)_{\text{tm}}=2$. The parameters are
given in Appendix \ref{app:xi4act}, the behaviour of the linear parameters is displayed in Figure \ref{fig:linbeh_good}.

\section{The Renormalised Anisotropy} \label{sec:xi4r}

To check whether the construction of the classically perfect $\xi=4$ action worked well and whether the
parametrisation reproduces the input anisotropy, we measure the renormalised anisotropy at one value of
$\beta=3.0$, using the torelon dispersion relation, following the method described in Sections \ref{sec:torelon} and \ref{sec:xir}.
The simulation parameters are collected in Table \ref{tab:xi4tordet}.

\begin{table}[htbp]
\renewcommand{\arraystretch}{1.3}
  \begin{center}
    \begin{tabular*}{\textwidth}[c]{c@{\extracolsep{\fill}}cc}
      \hline\vspace{-0.05cm} 
      $\beta$ & $S^2\times L\times T$ & \# sweeps / measurements\\
      \hline
      3.0 & $8^2\times 4\times 32$ & 54000 / 10800\\
      \hline
    \end{tabular*}
    \caption{Run parameters for the torelon measurements using the $\xi=4$ perfect
      action. The lattice extension in torelon direction $L$, the extension
      in the two transversal spatial directions $S$ as well as the temporal extension
      $L$ are given.}
    \label{tab:xi4tordet}
  \end{center}
\end{table}

Due to the (expected) coarser spatial lattice spacing we do not need as many smearing steps as for
the $\xi=2$ action and thus perform our measurements on configurations of smearing levels 2, 4, 6, 8, 10
keeping $\lambda_s=0.1$ fixed.

\begin{figure}[htbp]
\begin{center}
\includegraphics[width=8cm]{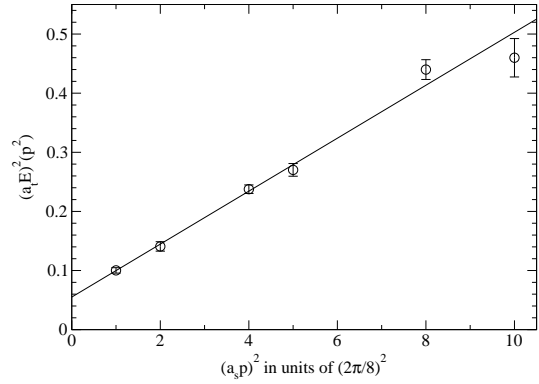}
\end{center}
\caption{Torelon dispersion relation for $\beta=3.0$. The straight line is the correlated fit to $E^2(p)=m_T^2+p^2$ in the range $p^2=1..5$.}
\label{fig:tordisp_xi4_b30}
\end{figure}

Figure \ref{fig:tordisp_xi4_b30} displays the dispersion relation obtained, in Table \ref{tab:xi4tor} 
the measured energies
$E(p^2)$ determined using variational methods (see Appendix \ref{app:vartech}) 
depending on the lattice momentum $p^2$ are
given, together with the number of operators used in the variational method, the fit ranges and
the values of $\chi^2$ per degree of freedom, $\chi^2/N_{\text{DF}}$.
The renormalised anisotropy as well as the torelon mass are determined in exactly the same way
as for the $\xi=2$ action. The results are listed in Table \ref{tab:xi4torres}. Furthermore,
we may again evaluate an estimate of the lattice scale using the string picture relation, 
eq.~\ref{eq:finite_size_torelon}, yielding $\sqrt{\sigma}a_s=0.532(16)$, and thus $r_0/a_s=2.241(86)$
which corresponds to $a_s=0.22$~fm, $a_t=0.060$~fm. Having a torelon of length $La_s\approx 0.89$~fm
the error of the scale estimate is expected to be of about 5-10\% (see Table \ref{tab:xi2scalcomp}).

\begin{table}[htbp]
\renewcommand{\arraystretch}{1.3}
  \begin{center}
    \begin{tabular*}{\textwidth}[c]{c@{\extracolsep{\fill}}cccccccc}
      \hline\vspace{-0.05cm}
      $\beta$ & fit range & $\xi_R$ & $m_{\text{T}} a_t$ & $\chi^2/N_{\text{DF}}$\\
      \hline
      3.0 & 1..5 & 3.71(8) & 0.235(13) & 0.82\\
      \hline
    \end{tabular*}
    \caption{Results of the torelon simulations using the ``$\xi=4$'' perfect action. The fit
      range in $p^2$ is given in units of $(2\pi/S)^2$.}
    \label{tab:xi4torres}
  \end{center}
\end{table}

Concerning the application of the spatial blocking method presented in this work to yield classically
perfect gauge actions for higher anisotropies, we may conclude that the construction including the blocking
as well as the parametrisation works in exactly the same way as the construction of the $\xi=2$ action to
which the main body of this work is dedicated. The resulting action shows a renormalisation of the anisotropy
of about 7\% at the value of $\beta=3.00$ where this has been measured. Calculations of physical observables such
as the static quark-antiquark potential or glueball masses using the $\xi=4$ action are still absent.

\fancyhead[RE]{\nouppercase{\small\it Chapter \thechapter.\, Properties of the Actions}}
\chapter{Properties of the Perfect Anisotropic Actions} \label{chap:aiprop}

In this Chapter, we examine the autocorrelation times and the average change of the gauge links for
different updates in the MC simulation
for anisotropies $\xi=1, 2, 4$. Furthermore, the computational overhead of the parametrisation of
the perfect action is compared to
the standard Wilson action as well as to the widely-used Symanzik and tadpole improved anisotropic
action.

\section{Updating Algorithms}\label{sec:autocor}

There are different methods of updating Monte Carlo (MC) configurations. The two methods we use
for our rather complex parametrised gauge actions are the plain local Metropolis algorithm as well as
pseudo over-relaxation (POR), both described in the next two sections.

In order to define the run parameters of MC simulations one should know about the properties
of the updates, in order to have configurations that have small autocorrelation, to be sure
that the algorithm does not introduce any bias and to estimate the optimal MC time between
measurements of observables.

\subsection{Local Metropolis Update} \label{sec:Metropolis}

A very general method for updating MC ensembles satisfying detailed balance has been introduced
by Metropolis et al. \cite{Metropolis:1953am}. We update link by link, first proposing a new link
$U'\in$ SU(3). The change $U\rightarrow U'$ is accepted with probability 1 if the action of the
proposed configuration is smaller than the initial action, with probability $e^{-\Delta S}$ if
the action increases with the update. In our updating procedure we use proposed links $U'$ that
are products of the initial link $U$ and random SU(3) matrices whose fluctuations are chosen such
that the acceptance rate of the Metropolis sweep stays between 0.3 and 0.7.

There is the possibility of performing multi-hit updates, i.e.~using ``coarser'' trial matrices
that lead to larger changes and compensating for the lower acceptance by trying several
proposed links in a row.

\subsection{Pseudo Over-Relaxation Update} \label{sec:POR}

The method of over-relaxation bases on the fact that for simple actions (as e.g.~the Wilson
gauge action) it is possible to perform large changes in the update of a link without changing
the action $S$, i.e. to perform micro-canonical updates. For SU(3) gauge theory this may be done
executing such updates subsequently on different SU(2) subgroups. 

For our complex parametrisation, exact micro-canonical steps are not easy to implement. That is why
we propose new links that would not change the corresponding part of the anisotropic Wilson action and correct
for the change in the perfect action by adding an accept-/reject step, i.e.~our pseudo over-relaxation (POR)
algorithm is a variant of the Metropolis algorithm presented in the last Section, where the generation
of the proposed links is altered. However, one has to be careful about this procedure because all the
configurations that are reached by subsequent POR updates lie on the same $\mathcal{A}_{\text{Wilson}}=\text{const}.$ 
hyperplane in configuration space. In order to keep ergodicity, POR steps have to be mixed with another
kind of (ergodic) updates, e.g.~with Metropolis steps. Additionally, due to the similarities of the Wilson
action and the FP action, during the thermalisation, when a certain action level shall be reached,
 it may be favourable to use rather less POR steps than
later on, generating configurations at the same coupling $\beta$.

\section{The Autocorrelation}

The autocorrelation is a measure of correlation between an observable with the very same observable
at a later (MC) time. Near phase transitions, the autocorrelations in MC simulations of observables are proportional to
a power $n$ of the lattice correlation length $\xi^{\text{lat}}$:
\begin{equation}\label{eq:autocor}
\tau_{\text{op}}\sim (\xi^{\text{lat}})^n,
\end{equation}
where theoretically $n=2$ for local stochastic (Metropolis) and $n=0$ for cluster or
over-relaxation updates. Eq.~\ref{eq:autocor} shows as well that one has to expect
\emph{critical slowing down}, i.e.~a dramatic increase of the autocorrelation time, if one
simulates near a phase transition with divergent correlation length $\xi$ using local stochastic
updates only.

On the anisotropic lattice there are two different correlation lengths (in lattice
units) $\xi^{\text{lat}}_s$ in spatial and $\xi^{\text{lat}}_t$ in temporal direction, where 
$\xi^{\text{lat}}_s/\xi^{\text{lat}}_t = a_t/a_s = \xi^{-1}$ (where this latter $\xi$ denotes the anisotropy).

Operators that couple strongly to $\xi^{\text{lat}}_t$, i.e.~operators living in the spatial
domain, are thus expected to have autocorrelation times rising with the anisotropy for simulations that
are performed not too far away from phase transitions.

\subsection{The Integrated Autocorrelation Time}

Let us first introduce the (unnormalised) autocorrelation function of an operator $O$, $C_{OO}(t)$:
\begin{equation}
C_{OO}(t)=\langle O(\tau) O(\tau+t) \rangle - \langle O \rangle^2.
\end{equation}
As well, we define the normalised autocorrelation function $\rho(t)$:
\begin{equation}
\rho_{OO}(t)=\frac{C_{OO}(t)}{C_{OO}(0)}.
\end{equation}

Typically, $\rho(t)$ decays exponentially, $\rho(t)\sim \exp(-|t|/\tau)$ for $|t|$ large enough
and we thus define the exponential autocorrelation time $\tau_{\text{exp}}$ 
\begin{equation}
\tau_{\text{exp},O}\equiv \lim_{t\rightarrow\infty}\frac{t}{-\ln|\rho_{OO}(t)|},
\end{equation}
which is the ``lifetime'' of the correlation. This definition is vivid, however what is more
important for the statistics of MC runs is the \emph{integrated} autocorrelation time $\tau_{\text{int}}$
which is describing the loss of statistics due to the subsequent configurations not being completely
independent:
\begin{equation}
\text{var}(\bar{O})\approx \frac{1}{n}(2\tau_{\text{int},O})C(0),
\end{equation}
i.e.~the true variance of the estimated observable $O$ is a factor $2\tau_{\text{int},O}$ larger
than if the configurations were completely independent of each other.

There are different possibilities of estimating $\tau_{\text{int}}$. Firstly, there is an inequation
between the two autocorrelation times \cite{Sokal:1989aa,Sokal:1989ea,Sokal:1992uk}
which holds for algorithms that satisfy detailed balance:
\begin{equation}
\label{eq:tauexpint}
\tau_{\text{int}} \le \frac{1}{2}\left(\frac{1+e^{-1/\tau_{\text{exp}}}}{1-e^{-1/\tau_{\text{exp}}}}\right)
\approx \tau_{\text{exp}}.
\end{equation}

The exponential autocorrelation time $\tau_{\text{exp}}$ can be obtained performing fits to $\rho(r)$.
For a direct estimate of $\tau_{\text{int}}$ on may use the same procedure as used by the QCD-TARO collaboration
in \cite{Akemi:1994aa},
where 
\begin{equation}
\tau_{\text{int}}=\frac{1}{2}\rho(0)+\sum_{t=1}^N \rho(t)\frac{N-t}{N},
\end{equation}
where $N$ is determined such that $\tau_{\text{int}}$ is maximised, at the same time
$N<10$~\% of the total sample. 

Another method has been proposed by Sokal in \cite{Sokal:1989ea,Sokal:1992uk}:
\begin{equation}
\tau_{\text{int}}=\frac{1}{2}\sum_{t=-m}^{m} \frac{R(t)}{R(0)},
\end{equation}
with
\begin{equation}
R(t)=\frac{1}{n-|t|}\sum_{i=1}^{n-|t|}(O_i-\bar{O})(O_{i+|t|}-\bar{O}),
\end{equation}
where $n$ is the number of measurements performed (extent of ``MC time'').
The parameter $m$ defines a ``window'' in which the measurements are taken into account
to calculate the integrated autocorrelation time.
The corresponding variance of the estimate is
\begin{equation}
\text{var}(\tau_{\text{int}})\approx \frac{2(2m+1)}{n}\tau_{\text{int}}^2
\end{equation}
for $\tau\ll m \ll n$.
There are several possibilities choosing
$m$; generally, large $m$ leads to a small bias of the estimated $\tau_{\text{int}}$, however,
the variance is growing with increasing $m$. One possibility is to fix $m=c\tau_{\text{int}}$ with a
factor $c$ depending on the behaviour of $\rho(t)$ ($c\approx 4$ for purely exponential decay,
$c\approx 6$ for asymptotic or pre-asymptotic decays or even $c\approx 10\sim20$ for exceptionally
slow decays). However, analogous to the procedure described before, we choose $m$ such that
the estimate for $\tau_{\text{int}}$ is maximised. It turns out, that there is always a long plateau
of maximal $\tau_{\text{int}}$ and that the bias corresponding to the $m$ chosen is small. That is
why we display the resulting values of the integrated autocorrelation time together with their standard
deviation.

In practice, the effect of autocorrelations of the operators between measurements close in the MC
trajectory is taken into account by binning subsequent measurements into blocks of length 
$T_{\text{MC}}>\tau_{\text{int}}$, where $\tau_{\text{int}}$ should be the largest integrated autocorrelation
time of the operators measured in the simulation. These binned measurements are independent from
each other and yield the correct variance of the observables, if they are used e.g.~in a bootstrap procedure.

\section{The Change in the Update}

A measure for the change of the configuration in a sweep is the following
\begin{equation}\label{eq:duup}
(\Delta U)_{\text{upd}}=\frac{1}{N_{\text{upd}}}\sum_{i=1}^{N_{\text{upd}}}[N_c-\text{Re Tr} (U_i U_i^{'\dagger})],
\end{equation}
which is the average change per gauge link in one update; $N_{\text{upd}}$ is the number of update attempts,
$U$ and $U'$ denote the initial and the
updated link respectively (of course $U'\equiv U$ for rejected steps). It is this change
that is expected to be considerably larger for (pseudo) over-relaxation updates compared to standard
Metropolis steps.

The measure defined above may also be used as a measure for autocorrelations of the most local objects, i.e.~the
gauge links itself, if it is used to compare an initial configuration ($U$) to the configurations at MC
time $t=1,2,3,\ldots$ ($U'$).

\section{Results}

To study the autocorrelation of the two update algorithms used, local Metropolis and POR,
we perform MC runs ($\approx 100000$ sweeps each) on $6^3\times 6\xi$ lattices with $a_s/r_0\approx 0.3$
using different combinations of Metropolis (M) and POR steps: 1 M, 1 M + 1 POR, 1 M + 2 POR, 1 M + 4 POR;
there is no sense in using pure POR updates as they are not ergodic and must not be used alone in actual
simulations. The gauge operators measured are
the simple plaquette, either purely spatial ($p_{\text{ss}}$) or going into temporal direction
($p_{\text{st}}$), as well as the $1\times 2$ rectangle: purely spatial ($r_{\text{ss}}$),
extension 1 in temporal direction ($r_{\text{st1}}$) and extension 2 in temporal direction
($r_{\text{st2}}$).

First, the simulations are performed for the $\xi=2$ perfect action, the results are listed in
Table \ref{tab:xi2ac}. Firstly, it turns out that the statistics necessary is indeed very large
($n\approx 1000 \tau_{\text{int}}$): in the case of pure Metropolis updates where this condition 
is not fulfilled for all the operators, the statistical errors are of about 30\%.
Secondly, one may notice that the inequation between the exponential and the integrated autocorrelation
time, eq.~\ref{eq:tauexpint}, holds --- thus, if one is
interested in an upper bound of $\tau_{\text{int}}$, e.g.~to choose the bin size in a simulation, it
is enough to estimate $\tau_{\text{exp}}$ (for the measured operators) which is rather cheap as this
can be done with much less statistics up to errors of about 10-20\% which normally do not matter.
The combination of standard Metropolis and POR updates which turns out to be optimal for the decorrelation
of small operators, using the $\xi=2$ action, is 1:1.

\begin{table}[htbp]
\renewcommand{\arraystretch}{1.3}
  \begin{center}
    \begin{tabular*}{\textwidth}[c]{lr@{\extracolsep{\fill}}rrrrrrr}
\hline
 & \multicolumn{2}{c}{1 M} & \multicolumn{2}{c}{1 M + 1 POR} & \multicolumn{2}{c}{1 M + 2 POR} & \multicolumn{2}{c}{1 M + 4 POR} \\\cline{2-3}\cline{4-5}\cline{6-7}\cline{8-9}
op. & $\tau_{\text{exp}}$ & $\tau_{\text{int}}$ & $\tau_{\text{exp}}$ & $\tau_{\text{int}}$ & $\tau_{\text{exp}}$ & $\tau_{\text{int}}$ & $\tau_{\text{exp}}$ & $\tau_{\text{int}}$\\
\hline
$p_{ss}$  & 395 & 125(28) & 45 & 31(4) & 59 & 38(5) & 67 & 42(5)\\
$p_{st}$  & 227 & 71(16)  & 36 & 32(4) & 52 & 49(7) & 66 & 61(8)\\
$r_{ss}$  & 392 & 172(39) & 54 & 34(5) & 64 & 38(5) & 72 & 39(4)\\
$r_{st1}$ & 279 & 73(17)  & 39 & 25(3) & 54 & 37(5) & 67 & 45(6)\\
$r_{st2}$ & 280 & 126(29) & 44 & 30(4) & 58 & 40(5) & 70 & 46(5)\\
\hline
\end{tabular*}
\caption{Autocorrelation times for simple gauge operators using different updates with the $\xi=2$ perfect action.
The integrated autocorrelation time is calculated using the method described by Sokal, the parameter $m$ chosen
corresponds to the maximum of the estimated values of $\tau_{\text{int}}$.}
    \label{tab:xi2ac}
  \end{center}
\end{table}

\begin{table}[htbp]
\renewcommand{\arraystretch}{1.3}
  \begin{center}
    \begin{tabular*}{\textwidth}[c]{lr@{\extracolsep{\fill}}rrrrr}
\hline
 & \multicolumn{2}{c}{$\xi=1$} & \multicolumn{2}{c}{$\xi=2$} & \multicolumn{2}{c}{$\xi=4$} \\\cline{2-3}\cline{4-5}\cline{6-7}
op. & $\tau_{\text{exp}}$ & $\tau_{\text{int}}$ & $\tau_{\text{exp}}$ & $\tau_{\text{int}}$ & $\tau_{\text{exp}}$ & $\tau_{\text{int}}$ \\
\hline
$p_{ss}$  & 17 & 15(6) & 45 & 31(4) & 17 & 8(2) \\
$p_{st}$  & 18 & 13(3) & 36 & 32(4) & 17 & 18(4) \\
$r_{ss}$  & 21 & 13(6) & 54 & 34(5) & 19 & 10(4) \\
$r_{st1}$ & 21 & 10(5) & 39 & 25(3) & 18 & 12(2) \\
$r_{st2}$ & 21 & 11(4) & 44 & 30(4) & 18 & 13(3) \\
\hline
\end{tabular*}
\caption{Autocorrelation times for simple gauge operators using alternating Metropolis and POR updates for
actions with different anisotropies $\xi=1,2,4$.
The integrated autocorrelation time is calculated using the method described by Sokal, the parameter $m$ chosen
corresponds to the maximum of the estimated values of $\tau_{\text{int}}$.}
    \label{tab:divxiac}
  \end{center}
\end{table}

The integrated autocorrelation time listed is calculated using the method proposed by Sokal, 
the method of QCD-TARO yields results in excellent agreement, however it seems as if the Sokal method
does a better job in distinguishing between small contributions to the integrated autocorrelation time
and the noise, the method is more stable and the window used to determine $\tau_{\text{int}}$ has a more
reasonable size. Furthermore, because most of the values of the QCD-TARO method lie below the values of the
Sokal method (however, within one standard deviation of the Sokal results), it is more conservative to use
the Sokal results.

\begin{figure}[htbp]
\begin{center}
\epsfig{file=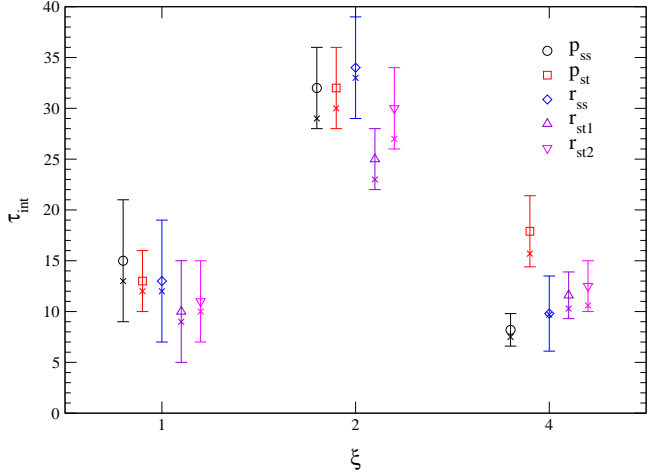,width=10cm}
\end{center}
\caption{Integrated autocorrelation times of different operators using 1 Metropolis + 1 OR updates for the $\xi=1, 2, 4$
actions. The autocorrelation times displayed by the open symbols
as well as the errors have been determined by the method of Sokal, the crosses display the corresponding
results obtained with the method of QCD-TARO.}
\label{fig:divxiac}
\end{figure}

Finally, we compare the autocorrelation times for the different anisotropic actions
looking at the 1 Metropolis + 1 POR update, which turned out to be the most efficient for $\xi=2$,
see Table \ref{tab:divxiac} and Figure \ref{fig:divxiac}.
Rather surprisingly, it turns out that the average autocorrelation times (of these small operators) are
by a factor $\approx 2.5$ larger for the $\xi=2$ action compared to the isotropic and the $\xi=4$ action.
It is not observed, that the spatial operators $p_{ss}$ and $r_{ss}$ have larger autocorrelation times than
the temporal operators, which is probably due to being far from the critical region near the deconfining
phase transition.

\begin{figure}[htbp]
\begin{center}
\epsfig{file=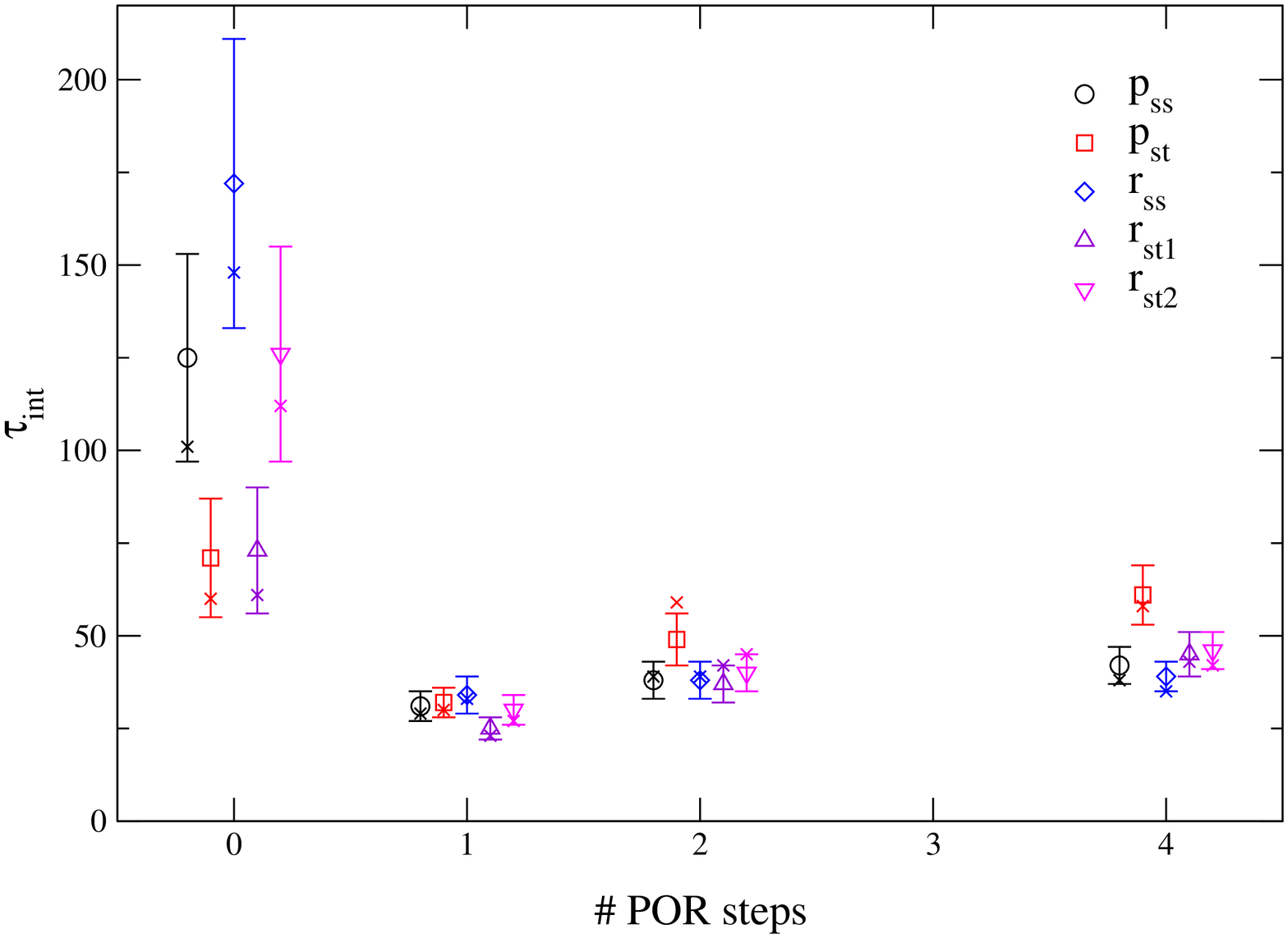,width=10cm}
\end{center}
\caption{Integrated autocorrelation times of different operators using different update algorithms (1
Metropolis step + $n$ POR steps, $n=$ 0, 1, 2, 4) for the $\xi=2$ action. The autocorrelation times displayed by the open symbols
as well as the errors have been determined by the method of Sokal, the crosses display the corresponding
results obtained with the method of QCD-TARO.}
\label{fig:ac_xi2}
\end{figure}

Table \ref{tab:avchange} lists the average change per link for Metropolis and POR updates using
the perfect $\xi=2$ action at different values of the coupling $\beta$. It turns out that the
average change of POR updates is about a factor of 10 larger than in the case of Metropolis updates;
the normalised values are changing slowly with the lattice scale.

\begin{table}[htbp]
\renewcommand{\arraystretch}{1.3}
  \begin{center}
    \begin{tabular*}{\textwidth}[c]{rr@{\extracolsep{\fill}}rrr}
\hline
$\xi$ & $\beta$ & $\langle\Box\rangle$ & $\frac{1}{\langle\Box\rangle}(\Delta U)_{\text{upd}}^{\text{Metropolis}}$ & $\frac{1}{\langle\Box\rangle}(\Delta U)_{\text{upd}}^{\text{POR}}$ \\
\hline
1 & 2.86 & 1.917 & 0.0652 & 0.472\\
  & 3.05 & 1.846 & 0.0527 & 0.486\\
  & 3.40 & 1.735 & 0.0660 & 0.506\\\hline
2 & 3.00 & 1.885 & 0.0546 & 0.613\\
  & 3.15 & 1.702 & 0.0552 & 0.584\\
  & 3.30 & 1.584 & 0.0568 & 0.577\\
  & 3.50 & 1.487 & 0.0572 & 0.574\\\hline
4 & 3.00 & 1.750 & 0.0400 & 0.328\\
\hline
\end{tabular*}
\caption{The average change per link for Metropolis and pseudo over-relaxation updates for the classically perfect
actions with anisotropies $\xi=1, 2, 4$, normalised by the plaquette expectation value $\langle\Box\rangle=\langle N_c-\text{Re Tr } U^{\text{pl}}\rangle$.}
    \label{tab:avchange}
  \end{center}
\end{table}

Comparing configurations separated by $N$ MC updates using the measure eq.~\ref{eq:duup}, we obtain the
following (obvious) behaviour:
\begin{equation}
(\Delta U)_{\text{upd}}(N)=3-\alpha_0 \cdot (1-q)^N,
\end{equation}
where $(\Delta U)_{\text{upd}}=3$ indicates that the configurations are completely independent on the level
of gauge links. The factor $q$ determines the speed of the decorrelation. The results, obtained using only
20 starting configurations for each anisotropy and update and performing 100 sweeps each are displayed in
Table \ref{tab:duupmc}. It turns out that these results are already completely stable. Obviously, these numbers
disclose no secret and it is clear that all the effects apparent in the autocorrelation times, Tables \ref{tab:xi2ac},
\ref{tab:divxiac}, stem from larger range correlations. Note as well, that the measure, eq.~\ref{eq:duup} is not gauge
invariant, i.e.~the gauge transform of a configuration may show to be completely independent,
 $(\Delta U)_{\text{upd}}=3$, of the initial configuration.

\begin{table}[htbp]
\renewcommand{\arraystretch}{1.3}
  \begin{center}
    \begin{tabular*}{\textwidth}[c]{rr@{\extracolsep{\fill}}rrrr}
\hline
$\xi$ & \phantom{ }1 M &  \phantom{ }1 POR & 1 M + 1 POR & 1 M + 2 POR & 1 M + 4 POR\\
 & $q$ & $q$ & $q$ & $q$ & $q$\\
\hline
1 & 0.015 & 0.183 & 0.102 & 0.128 & 0.152\\
2 & 0.012 & 0.174 & 0.094 & 0.121 & 0.143\\
4 & 0.008 & 0.064 & 0.037 & 0.046 & 0.054\\
\hline
\end{tabular*}
\caption{The speed of the link decorrelation $q$ (see text) for different updates at different anisotropies.}
    \label{tab:duupmc}
  \end{center}
\end{table}

\section{The Computational Overhead}
One very important quantity of an improved action is its computational overhead compared to the standard
(i.e. Wilson) action.

To measure the effective gain or loss of an action is a very difficult task that depends very strongly on
the application (spectroscopy, thermodynamics etc.), on the lattice sizes and lattice spacings used and as
well on the architecture of the computers (scalar, vector, parallel and a lot of more subtle details) used.

Here, we will thus simply give raw ratios between the anisotropic perfect action, the widely used Symanzik
and tadpole improved action and the Wilson action. It is clear that our parametrisation which is quite 
sophisticated has a large ``raw'' overhead, mainly due to the large number of operators that are affected
if one link is changed in the MC update. Presently, our parametrisation is probably the most complex one that may
still be considered to be utilised in numerical simulations of pure gauge theory. Although, in principle, the fit to the exact
perfect action could be done much better, e.g.~employing double smearing or including diagonal links into
the smearing, such a procedure would surely make it unsuitable for MC simulations. Of course, if the gauge
action is used together with (quenched or even dynamical) fermions, the computational cost of the gauge part
is rather negligible, even better parametrisations of perfect actions might be imaginable in this case.

Table \ref{tab:overhead} compares the perfect anisotropic action to the Symanzik/tadpole-improved action of
\cite{Morningstar:1997ze} and the Wilson action on lattices of size $8^4$ on an Alpha workstation. The numbers
refer to the pure Metropolis update of each link of the configuration (``1 sweep'').

\begin{table}[htbp]
\renewcommand{\arraystretch}{1.3}
  \begin{center}
    \begin{tabular}[c]{cc@{\extracolsep{\fill}}c}
\hline
action & $t_{\text{sweep}}$ [s] & \phantom{} $t_{\text{act}}/t_{\text{Wilson}}$\\
\hline
Wilson & 0.095 & 1\\
Symanzik & 0.53 & 5.6\\
Perfect & 6.3 & 66\\ 
\hline
\end{tabular}
\caption{The updating time (1 sweep on a $8^4$ lattice) and the computational overhead of the perfect anisotropic action, compared to the Symanzik and tadpole improved
action \cite{Morningstar:1997ze} and the standard Wilson action.}
    \label{tab:overhead}
  \end{center}
\end{table}

\fancyhead[RE]{\nouppercase{\small\it Chapter \thechapter.\, Conclusions and Prospects}}
\chapter{Conclusions and Prospects} \label{ch:concl}

In this work, we have presented the construction and parametrisation of a classically perfect
anisotropic SU(3) gauge action based on the Fixed Point Action technique. The recently 
parametrised isotropic FP action as well as its parametrisation ansatz using mixed polynomials of plaquettes
built from single gauge links as well as from smeared links, have been the starting point. Performing
one and two purely spatial blocking (RG) steps, respectively, starting from the isotropic action, and adapting the parametrisation
to distinguish between spatial and temporal directions, we have obtained parametrised classically perfect
actions with anisotropies $\xi=2$ and 4. That the method works has been checked as well in the context of scalar fields
and in the quadratic approximation for SU(3) gauge theory. The $\xi=2$ action has been tested extensively in measurements
of the torelon dispersion relation, of the static quark-antiquark potential, of the deconfining phase transition
as well as of the low-lying glueball spectrum of pure gauge theory. For the $\xi=4$ action we have checked 
that the renormalisation of the anisotropy remains small and that it exhibits no pathologies in Monte Carlo
simulations. For both actions (and additionally the isotropic FP action) autocorrelation times have been
measured and the computational overhead, compared to the widely-used anisotropic tadpole and tree-level improved
action as well as to the standard Wilson gauge action, has been determined.

The results of the torelon measurements show that the renormalisation of the anisotropy (due to quantum corrections
and parametrisation artifacts) is small (below 10\% for $\xi=2$ and 4). The rotational invariance of the action 
has been examined by measuring the static quark-antiquark potential, including separations corresponding to a large set
of lattice vectors, again with good results. The glueball measurements, including some rather heavy states, confirm
that the use of anisotropic lattices facilitates spectroscopy when heavy states are present. Compared to the isotropic
simulations, with the same amount of computational work it is possible to resolve states with considerably larger
energy, allowing reliable continuum extrapolations from larger ranges of the lattice spacing. The mass of the scalar
glueball with quantum numbers $J^{PC}=0^{++}$ is measured to be considerably lighter (1440(40)~MeV) compared to masses
obtained by other groups (around 1670~MeV); furthermore it shows large cut-off effects when it is measured on a lattice
with spatial lattice spacing $a_s\approx 0.25$~fm, which could be caused by a large sensitivity of the action to a
critical point in the fundamental-adjoint coupling plane, due to adjoint terms in the
action. However, the sources of possible errors in measurements of the glueball spectrum are manifold and there is
no ultimate statement about the goodness of the parametrised classically perfect actions.

During our studies, we have noticed that the examination of scaling properties of lattice gauge actions is a very
delicate problem: Quantities which can be reliably measured and which are not very sensitive to systematic factors
(such as the volume, other
states mixing with the observed ones, the method used for extracting the mass etc.), such as the critical temperature
or the hadronic scale $r_0$ (for moderate lattice spacings)  generally exhibit rather small cut-off
effects and thus demand very large statistics, if the differences between the actions should be explicitly identified.
On the other hand, some glueball states show large cut-off effects which make the glueball spectrum another interesting
object to study scaling violations. As pointed out in this work, however, there are a lot of systematic factors
making the extraction of the pure lattice artifacts difficult.
The task of comparing different actions
is even more difficult if the measurements and analyses are performed independently by different groups which
introduces additional systematic discrepancies and ambiguities.

Let us therefore conclude that despite the rich parametrisation, the correspondence to the classically perfect
FP action and the approximate scale invariance of classical instanton solutions there is no conclusive evidence of the
parametrised classically perfect actions (isotropic as well as anisotropic) behaving significantly better
than other improved actions (such as the tadpole and tree-level improved anisotropic gauge action) --- however,
there is also no evidence for the converse. What is clear and what one has to consider if one is about to plan
a project where the update of the gauge configurations takes a considerable part of the computer time, is the
considerable overhead of the parametrisation compared to other actions. 

For the future, very accurate scaling tests comparing the classically perfect actions to other improved gauge
actions are desirable. On one hand, these can include large statistics measurements of the critical temperature
and the hadronic scale, on the other hand, one might perform very careful simulations determining the glueball
spectrum, including all the limits to be taken and excluding all known sources of errors. These simulations should all be
performed by the same collaboration in order to minimise systematic discrepancies coming from the analyses.

Another promising plan would be to gain control about the adjoint operators present in the parametrisation of
the classically perfect actions in order to circumvent the critical point in the adjoint coupling plane such that
the influence of the scalar dip is minimised.

\fancyhead[RE]{\nouppercase{\small\it Acknowledgements}}
\chapter*{Acknowledgements}
Primarily, I would like to thank my collaborators, Urs Wenger and Ferenc Niedermayer, for the inspiring
cooperation. Furthermore, I am indebted to Peter Hasenfratz for carefully reading the manuscript and to
Peter, Ferenc, Urs and the other members of our group, Simon Hauswirth, Kieran Holland and Thomas J\"org, for many useful 
discussions and a pleasant atmosphere to
work in. Further thanks go to all the other members of the institute for the good social environment. 
Finally, I thank Regula Wyss just for being there.

\begin{appendix}
\fancyhead[RE]{\nouppercase{\small\it Appendix \thechapter.\, The FP Action for Scalar Field Theory}}
\chapter{The FP Action for Scalar Field Theory}\label{app:scal}

A simple way to get the lattice Fixed Point Action for $d=2$ free scalar theory is
performing one single renormalisation group transformation (RGT) with an
infinitely large scale factor starting in the continuum, a procedure called
\emph{blocking out of continuum}. The lattice is laid over the continuum field
which is averaged over certain regions around the future lattice sites 
to obtain the lattice fields. The resulting FP action depends on the form of
the averaging (corresponding to the blocking in finite scale RGTs) but the
procedure presented here will be completely general.

We use the following RGT, accomplishing the task of averaging the continuum
field $\varphi(x)$ to get to the lattice field $\phi_n$, where the form of the
averaging is described by the function $\Pi(x)$
 \begin{equation}
\label{bocblock}
e^{-S^{FP}(\phi)}=\int\!\mathcal{D}\varphi(x)\,e^{-\frac{1}{2}\int\! d^2x\,
  \partial_{\mu}\varphi(x)\partial_{\mu}\varphi(x)-\kappa\sum\limits_n(\phi_n-\int\! d^2x\,
  \Pi(na-x)\varphi(x))^2},
\end{equation}
the parameter $\kappa$ is again allowing the lattice field to fluctuate around the
exact average of the continuum field (largely for small $\kappa$). The averaging
is normalised to $\int d^2x\Pi(x)=1$. To account for an anisotropy, $a_1\ne a_2$, we
perform the calculation using dimensionful quantities, the symbol $a$ is a shorthand
notation for $(a_1,a_2)$.

Our goal is to obtain the FP propagator $D^{\text{FP}}$ on the lattice, we thus
calculate the lattice two-point function $D(n-n')=\langle\phi_n\phi_{n'}\rangle$
in this theory:
\begin{equation}
 D(n-n')=\frac{1}{Z}\int\!\mathcal{D}\phi\,
  e^{-\frac{1}{2}\sum\limits_{n,r}\rho(r)\phi_n\phi_{n+r}}\phi_n\phi_{n'}; 
\end{equation}
inserting eq.~\ref{bocblock} we get
\begin{equation}
 \label{firstboc}\frac{1}{Z}\int\!\mathcal{D}\phi\mathcal{D}\varphi\, e^{-\frac{1}{2}\int\! d^2x\,
  \partial_{\mu}\varphi(x)\partial_{\mu}\varphi(x)-\kappa\sum\limits_n(\phi_n-
\int\! d^2x\, \Pi(na-x)\varphi(x))^2}\phi_n\phi_{n'}. 
\end{equation}
The blocking (averaging) kernel is now Fourier transformed in the continuum:
\begin{equation} K(n)=\int\! d^2x\, \Pi(na-x)\varphi(x): \end{equation}
\begin{equation} \Pi(na-x)=\int\limits_p e^{ip(na-x)}\Pi(p), \end{equation}
\begin{equation} \varphi(x)=\int\limits_k e^{ikx}\varphi(k), \end{equation}
thus
\begin{equation} K(n)=\int d^2x\int\limits_p\int\limits_k e^{ipna}e^{i(k-p)x}\Pi(p)\varphi(k)=\int\limits_k
  e^{ikna}\Pi(k)\varphi(k). \end{equation}
Now we split up the continuum momentum variables $k_{\mu}$ according to:
\begin{equation} k_1=q_1+\frac{2\pi}{a_1}l_1,\quad k_2=q_2+\frac{2\pi}{a_2}l_2\end{equation}
with $q_1$, $q_2$ being lattice momenta in the respective Brillouin zones:
\begin{equation} q_1\in \left(-\frac{\pi}{a_1},\frac{\pi}{a_1}\right),\quad q_2\in
\left(-\frac{\pi}{a_2},\frac{\pi}{a_2}\right)\end{equation}
In the following, we will use the notation $k=q+\frac{2\pi}{a}l$ for convenience,
where $k=(k_1,k_2)$.
Inserting this replacement, we get:
\begin{eqnarray}
\int\limits_{-\infty}^{\infty}\!\frac{d^2k}{(2\pi)^2}\,e^{ikna} & =
&\sum\limits_{l=-\infty}^{\infty}\int\limits_{-\frac{\pi}{a_1}}^{\frac{\pi}{a_1}}\!\frac{dq_1}{2\pi}\,\int\limits_{-\frac{\pi}{a_2}}^{\frac{\pi}{a_2}}\!\frac{dq_2}{2\pi}\,e^{i(q_1+\frac{2\pi}{a_1}l_1)n_1a_1}e^{i(q_2+\frac{2\pi}{a_2}l_2)n_2a_2}\nonumber\\
& = & \int\limits_{q} e^{iqna}\sum\limits_{l=-\infty}^{\infty},\end{eqnarray}
using the shorthand notation
\begin{equation} \int\limits_{q}= \int\limits_{-\frac{\pi}{a_1}}^{\frac{\pi}{a_1}}\!\frac{dq_1}{2\pi}\,\int\limits_{-\frac{\pi}{a_2}}^{\frac{\pi}{a_2}}\!\frac{dq_2}{2\pi}.\end{equation}
Using this, our two-point function reads:
\begin{equation} \label{secboc} \frac{1}{Z}\int\!\mathcal{D}\varphi\int\!\mathcal{D}\phi_n\,
  e^{-\frac{1}{2}\int\! d^2x\,\partial_{\mu}\varphi(x)\partial_{\mu}\varphi(x)-\kappa\sum\limits_n[\phi_n-\int\limits_q e^{iqna}\sum\limits_{l=-\infty}^{\infty}\Pi(q+\frac{2\pi}{a} l)\varphi(q+\frac{2\pi}{a} l)]^2}\phi_n\phi_{n'}. \end{equation}
We replace $\phi_n$ by
\begin{equation} \phi_n^*=\phi_n-\int\limits_q e^{iqna}\stackrel{b(q)}{\overbrace{\sum\limits_{l=-\infty}^{\infty}\Pi(q+\frac{2\pi}{a}
  l)\varphi(q+\frac{2\pi}{a} l)}} \end{equation}
and get:
\begin{equation}
\label{bocint1}
  \frac{1}{Z}\int\!\mathcal{D}\varphi\int\!\mathcal{D}\phi_n^*\,(\phi_n^*+\int\limits_q e^{iqna}b(q))(\phi_{n'}^*+\int\limits_q e^{iqn'a}b(q))e^{-\frac{1}{2}\int\! d^2x\, \partial_{\mu}\varphi(x)\partial_{\mu}\varphi(x)-\kappa\sum\limits_n(\phi_n^*)^2}. \end{equation}
Performing the integration over $\phi_n^*$, the terms linear in $\phi_n^*$
disappear.

If we integrate both sides of eq.~\ref{bocblock} with respect to $\phi_n$ we get
  $Z$ on the l.h.s. and $Z_{\varphi}Z_{\phi}$ on the r.h.s., with
\begin{equation}\label{zvarphi} Z_{\varphi}=\int\!\mathcal{D}\varphi\, e^{-\frac{1}{2}\int
  d^2x\partial_{\mu}\varphi(x)\partial_{\mu}\varphi(x)}\end{equation}
and
\begin{equation} Z_{\phi}=\int\!\mathcal{D}\phi_n\, e^{-\kappa\sum\limits_n(\phi_n-\int\!
  d^2x\,\Pi(na-x)\varphi(x))^2}. \end{equation}
Using this notation we may write eq.~\ref{bocint1} as:
\begin{eqnarray} \frac{Z_{\varphi}}{Z}\int\!\mathcal{D}\phi_n^*\,\phi_n^*\phi_{n'}^* e^{-\kappa\sum\limits_n
  (\phi_n^*)^2} & & \\
+\frac{Z_{\phi}}{Z}\int\!\mathcal{D}\varphi\int\limits_q\! e^{iqna}b(q)\int\limits_q\! e^{iqn'a}b(q)e^{-\frac{1}{2}\int\! d^2x\, \partial_{\mu}\varphi(x)\partial_{\mu}\varphi(x)}. & &
\end{eqnarray}
The first term is just
\begin{equation}
\frac{Z_{\varphi}}{Z}\cdot\frac{1}{\kappa}\delta_{n,n'}\cdot
Z_{\phi}=\frac{1}{\kappa}\delta_{n,n'}=\frac{a_1a_2}{\kappa}\int\limits_q
e^{iq(n-n')a},
\end{equation}
the second term reads
\begin{equation} \frac{Z_{\phi}}{Z}\int\!\mathcal{D}\varphi\left[\int\limits_q
e^{iq(n-n')a}\sum\limits_{l=-\infty}^{\infty}\Pi(k)\varphi(k)\Pi^{\dag}(k)\varphi^{\dag}(k)\right]e^{-\frac{1}{2}\int\! d^2x\, \partial_{\mu}\varphi(x)\partial_{\mu}\varphi(x)},\end{equation}
after inserting the content of $b(q)$. Using eq.~\ref{zvarphi}, the definition
of $Z_{\varphi}$, we get
\begin{equation} \int\!\mathcal{D}\varphi\,\varphi(q+\frac{2\pi}{a} l)\varphi^{\dag}(q+2\pi
l)e^{-\frac{1}{2}\int\!
  d^2x\,\partial_{\mu}\varphi(x)\partial_{\mu}\varphi(x)}=Z_{\varphi}\cdot
d(q+2\pi l),\end{equation}
and thus the second term is just
\begin{equation} \int\limits_q e^{iq(n-n')a}\sum\limits_{l=-\infty}^{\infty}\Pi(q+\frac{2\pi}{a}
l)d(q+\frac{2\pi}{a} l)\Pi^{\dag}(q+\frac{2\pi}{a} l); \end{equation}
together with the first term:
\begin{equation} D(n-n')=\int\limits_q
e^{iq(n-n')a}\left[\frac{1}{\kappa}+\sum\limits_{l=-\infty}^{\infty}\Pi(q+2\pi
  l)d(q+2\pi l)\Pi^{\dag}(q+2\pi l)\right]. \end{equation}
Thus we get $\rho(q)$:
\begin{equation} \label{bocinvrho} \frac{1}{\rho(q)}=\sum\limits_{l=-\infty}^{\infty}\left[\Pi(q+\frac{2\pi}{a} l)d(q+\frac{2\pi}{a}
l)\Pi^{\dag}(q+\frac{2\pi}{a} l)\right]+\frac{a_1 a_2}{\kappa}. \end{equation}

Setting $a_1=a_2=1$ we get back the isotropic FP action as expected.

We are interested in couplings $\rho(q)$ on the lattice which are dimensionless quantities. In order to
achieve this, we rescale the above formula by dividing it by $a_1 a_2$, this absorbs the product in
the $\kappa$-term, also, the propagator $d(k)$ will become dimensionless:

\begin{equation}\label{aidk}
d(k)=\frac{1}{a_1a_2}\frac{1}{(q_1+\frac{2\pi}{a_1}l_1)^2+(q_2+\frac{2\pi}{a_2}l_2)^2}=\frac{1}{\xi^{-1}(p_1+2\pi
  l_1)^2+\xi(p_2+2\pi l_2)^2},\end{equation}
where $p_1$, $p_2$ are dimensionless momenta, $p_{\mu}\in (-\pi,\pi)$, and
$\xi=\frac{a_1}{a_2}$ is the anisotropy parameter as defined above.

\fancyhead[RE]{\nouppercase{\small\it Appendix \thechapter.\, The Quadratic Approximation for Gauge Fields}}
\chapter{The Quadratic Approximation for Gauge Fields} \label{app:quad}

In this Appendix we describe the calculation of the couplings $\rho_{\mu\nu}(r)$ in
the quadratic approximation to the classically perfect action, eq.~\ref{eq:quadapppos},
\begin{equation}
S^{\text{FP}}=\frac{1}{2N_c}\sum_{n,r}\rho_{\mu\nu}(r)\text{Tr}[A_\mu(n+r)A_\nu(n)]+\text{O}(A^3).
\end{equation}

\section{The Anisotropic Recursion Relation}

At the quadratic level, the recursion relation corresponding to eqs.~\ref{eq:FP_equation},
\ref{eq:blockkern}, \ref{eq:rgtnorm} reads in standard matrix notation
\begin{eqnarray}
\lefteqn{
\frac{1}{\xi V_B} \sum_{k_B} \text{Tr} \left\{ B^\dagger(k_B) \Xi \rho'(k_B)
  \Xi B(k_B)  \right\}   
 }
  \nonumber\\
 & = & \text{min}_{\{A\}} \left( \frac{1}{\xi V} \sum_{k} \text{Tr}
   \left\{ A^\dagger(k) \Xi \rho(k) \Xi A(k)  \right\} \right. \nonumber\\ 
 &  & \left. \hspace{0.5cm} + \frac{1}{\xi V_B} \sum_{k_B} \text{Tr} \left\{
   (\Gamma(k_B) - B(k_B))^\dagger \Xi \kappa \Xi (\Gamma(k_B) - B(k_B))
 \right\} \right),
\label{eq:fpquad}
\end{eqnarray}
where $V_B=\frac{1}{16}V$ is the volume of the blocked lattice, $B(k_B)$ denotes the blocked 
gauge configuration,  
$A_{\mu}$, $B_{\mu}$ are the gauge potentials of the configurations $U$, $V$,
\begin{equation}
V_\mu(n_B)=e^{iB_\mu(n_B)},\quad U_\mu(n)=e^{iA_\mu(n)}
\end{equation}
and 
\begin{equation}
\Xi  =  
\left(  
\begin{array}{cccc} 
    1  & & & \\
         & 1 & & \\
         & & 1 &  \\
         & & & \xi 
\end{array} 
\right).
\end{equation}

In the above formula  $\rho, \kappa$ are expressed in units of $a_s$ and the
blocking function $\omega(k)$ is a
dimensionless quantity, while
\begin{equation}
k_\mu = k_\mu^{\text{df}} a_\mu \quad \text{and} \quad X_\mu =
\frac{\prod_{\nu=0}^3
  a_\nu}{a_\mu} X^{\text{df}}_\mu,  
\end{equation}
where $X_\mu$ is a common symbol for the gauge potentials $A_\mu$, $B_\mu$ etc.

\section{The Calculation of the Couplings}

The blocking used in this work (see Section \ref{sec:spablock}) corresponds to 
\begin{equation}
\Xi \cdot  \Gamma(k_B) = \frac{1}{16} \sum_{l=0}^1 \omega(\frac{k_B}{2}+\pi l) \Xi
A(\frac{k_B}{2}+\pi l), \quad l=(l_1,l_2,l_3,l_4),
\label{eq:quadblock}
\end{equation}
with
\begin{equation}
\omega_{\mu\nu}=(1+e^{ik_\mu})\left[c_0\delta_{\mu\nu}+6c_1 \tau_{\mu\nu}^{(1)}(k)+12c_2 \tau_{\mu\nu}^{(2)}(k)
+8c_3 \tau_{\mu\nu}^{(3)}(k)\right].
\label{eq:rgt3kernel}
\end{equation}
The tensor $\omega_{\mu\nu}$ is fixed by the form of the blocking kernel, where up to this order the linear
contributions $\tau_{\mu\nu}^{(i)}$ to the fuzzy link operator have to be taken into account. They can be written
as
\begin{eqnarray}
\tau^{(1)}_{\mu \nu}(k) & \!\!\!=\!\!\! & \frac{1}{6}
\left[ \;
\widehat{k}_{\mu} \widehat{k}_{\nu}^* + 
\delta_{\mu \nu}
  (6-\gamma) 
\; \right], \nonumber \\
\tau^{(2)}_{\mu \nu}(k) & \!\!\!=\!\!\! & \frac{1}{24} 
\left[ \;
\widehat{k}_{\mu} \widehat{k}_{\nu}^* 
  (8 - \gamma + \widehat{k}_{\mu} \widehat{k}_{\mu}^*
           + \widehat{k}_{\nu} \widehat{k}_{\nu}^* ) + 
\delta_{\mu \nu} 
  (24 - 8\gamma+ \gamma^2 - \eta - 
   \gamma\widehat{k}_{\mu}\widehat{k}_{\mu}^*)
   \; \right], \nonumber\\
\tau^{(3)}_{\mu \nu}(k) & \!\!\!=\!\!\! & \frac{1}{48} 
\Big[ \;
\widehat{k}_{\mu} \widehat{k}_{\nu}^* 
  (24 - 6 \gamma + \gamma^2 - \eta 
    + 6 \widehat{k}_{\mu} \widehat{k}_{\mu}^*
    - 2 \gamma \widehat{k}_{\mu} \widehat{k}_{\mu}^*
    + 2 ( \widehat{k}_{\mu} \widehat{k}_{\mu}^* )^2
    + 6 \widehat{k}_{\nu} \widehat{k}_{\nu}^*  
 \nonumber\\ & & ~~~~~~~~~~~~~
    - 2 \gamma \widehat{k}_{\nu} \widehat{k}_{\nu}^* +
    2 ( \widehat{k}_{\nu} \widehat{k}_{\nu}^* )^2
    + 2 \widehat{k}_{\mu} \widehat{k}_{\mu}^*
        \widehat{k}_{\nu} \widehat{k}_{\nu}^* )  
                                                  \label{eq:tau}  \nonumber\\
 & & ~~~~
+ \delta_{\mu \nu} 
  ( 48 -24\gamma + 6 \gamma^2 -\gamma^3 
   -6\gamma \widehat{k}_{\mu} \widehat{k}_{\mu}^*
   +2\gamma^2\widehat{k}_{\mu} \widehat{k}_{\mu}^*
   -2\gamma( \widehat{k}_{\mu} \widehat{k}_{\mu}^* )^2 
 \nonumber\\ & & ~~~~~~~~~~~~~~
  + 3\gamma\eta - 2 \epsilon - 6 \eta - 
  2\eta \widehat{k}_{\mu} \widehat{k}_{\mu}^* )
\Big].
\end{eqnarray}

The lattice momentum $\widehat{k}_{\mu}$ is defined by
\begin{equation}
\widehat{k}_{\mu} = e^{ik_\mu}-1
\end{equation}
and the following definitions have been made for convenience:
\begin{equation} \label{eq:abbr}
\gamma \doteq |\hat{k}|^2 \doteq 
\sum_\mu \widehat{k}_{\mu} \widehat{k}_{\mu}^*,
\quad  
\eta  \doteq  \sum_\mu ( \widehat{k}_{\mu} \widehat{k}_{\mu}^* )^2, 
\quad 
\epsilon \doteq \sum_\mu ( \widehat{k}_{\mu} \widehat{k}_{\mu}^* )^3.
\nonumber
\end{equation}

The four parameters of the RGT in eq.~\ref{eq:rgt3kernel} have been optimised in \cite{Blatter:1996ti} 
and have the values:

\begin{equation}
c_1 = 0.07, \quad c_2 = 0.016, \quad c_3 = 0.008, \quad \kappa = 8.8.
\end{equation}

From eq.~\ref{eq:fpquad}, using the relation eq.~\ref{eq:quadblock}, we obtain a recursion relation for the inverse of the couplings 
$D_{\mu\nu}(k)=\rho^{-1}_{\mu\nu}(k)$:
\begin{equation}
D'_{\mu\nu}(k_B)=\frac{1}{16}\sum_{l=0}^{1}\left[\Xi\omega(\frac{k_B}{2}+\pi l) D(\frac{k_B}{2}+\pi l)\omega^\dagger
(\frac{k_B}{2}+\pi l)\Xi\right]_{\mu\nu}+\Xi\frac{1}{\kappa}\Xi.
\label{eq:quaditer}
\end{equation}

One way to find the FP solution for the propagator is to start from the anisotropic Wilson
propagator and to iterate the propagator to the FP. As 
we have to invert the coupling tensors $\rho_{\mu\nu}$ and
$\rho'_{\mu\nu}$ we have to introduce a temporary gauge fixing. The 
gauge--fixed anisotropic Wilson action reads
\begin{equation}
D^{(0)}_{\mu\nu}(k)=\frac{\delta_{\mu\nu}}{\sum_{i} |\widehat{k}_i|^2+\xi^2|\widehat{k}_4|^2}
+\alpha \frac{\widehat{k}_\mu\widehat{k}_{\nu}^*}{(\sum_{i} |\widehat{k}_i|^2+\xi^2|\widehat{k}_4|^2)^2}
\end{equation}

At the end the gauge--fixing will be removed, which corresponds to the limit $\alpha\rightarrow\infty$.

Using the gauge relation
\begin{equation}
\omega_{\mu\nu}(k)\widehat{k}_{\nu}=\widehat{2k_\mu}
\end{equation}
one can show that starting from the standard propagator, after an arbitrary number of iterations, the
propagator $D_{\mu\nu}(k)$ assumes the form
\begin{equation}
D_{\mu\nu}(k)=G_{\mu\nu}(k)+\alpha f(k)\widehat{k}_\mu \widehat{k}_\nu^*
\end{equation}
with $G_{\mu\nu}(k)$ and $f(k)$ independent of $\alpha$. It may now be shown \cite{Blatter:1996ti} that
in the limit $\alpha\rightarrow\infty$ the function $f(k)$ (which is the only part differing from the
isotropic case) drops out and the result for the couplings is
\begin{equation}
\rho_{\mu\nu}(k) = G^{-1}_{\mu\nu}(k) -
\frac{G^{-1}_{\mu\rho}(k)\widehat{k}_{\rho} \cdot
\widehat{k}_{\sigma}^*G^{-1}_{\sigma\nu}(k)}
{\widehat{k}_{\sigma}^*G^{-1}_{\sigma\rho}(k)\widehat{k}_{\rho}}.
\end{equation}

To construct the anisotropic propagator $\rho^{-1}(q)$ numerically we perform 3 recursions on a $32^4$
lattice using the $\kappa$ value obtained optimising the locality in the isotropic case, $\kappa=8.8$.
We study the dependence of the number of recursions and do not discover any notable changes if
we change it from 3 to 4 -- the FP quadratic propagator is thus reached already after 3 recursions.
The following steps, inverting $\rho^{-1}(q)$ to get the couplings in momentum space, removing
the gauge fixing (analytically) and the subsequent Fourier transformation do not change from the
isotropic case described in \cite{DeGrand:1995ji,Blatter:1996ti}. 

A priori there is no reason why the parameter $\kappa$ occurring in the recursion relation should be
chosen to have the same value for the spatial and the temporal part of the relation, i.e.~one may think
about changing $\kappa$ to the more general form
\[ \kappa \longrightarrow \tilde{\kappa}=\left(
\begin{array}{cccc}
    \kappa_s  & & & \\
         & \kappa_s & & \\
         & & \kappa_s &  \\
         & & & \kappa_t
\end{array}
\right).\]
There are two ``natural'' choices, either stick to the $\delta_{\mu\nu}$ form of $\kappa$ and set $\kappa_t=\kappa_s=8.8$
or change it to $\kappa_t=\xi^2\kappa_s=\xi^2\cdot 8.8$ to compensate for the insertion of the two $\Xi$-matrices
in eq.~\ref{eq:quaditer}. Performing numerical studies (in the full theory) it indeed turns out that the latter
choice is the one yielding actions that are well parametrisable.

\fancyhead[RE]{\nouppercase{\small\it Appendix \thechapter.\, The Norm of the Anisotropic APE Action}}
\chapter{The Norm of the Anisotropic APE Action} \label{app:aisym}

In this Appendix we derive the anisotropic normalisation condition which makes sure
that the action goes over to the continuum form for very smooth fields.

We first define for a general continuum gauge field a dimension 4 operator $R_0$:
\begin{equation}\label{eq:symr0}
R_0=-\frac{1}{2}\sum_{\mu\nu}\text{Tr}(\mathcal{F}_{\mu\nu}^2).
\end{equation}

To work out the form of the conditions for our parametrisation of the action we use constant
gauge non--Abelian gauge potentials, $\partial_\mu A_\nu=0$.
In the continuum we have
\begin{equation}
\mathcal{F}_{\mu\nu}=[A_\mu,A_\nu].
\end{equation}

The constant non--Abelian gauge--field is defined on the lattice as
\begin{equation}
\mathcal{A}_\mu=i\alpha_\mu\cdot\frac{1}{2}\sigma_\mu,
\end{equation}
where $\sigma_k$, $k$=1, 2, 3 are the Pauli matrices and $\sigma_4=(\sigma_1+\sigma_2+\sigma_3)/\sqrt{3}$.

To obtain the normalisation condition we expand our anisotropic gauge action, parametrised as described in
Section \ref{sec:par}, in powers of $\alpha_\mu$
keeping the spatial indices $k= 1, 2, 3$ and the temporal index 4 separate. This yields two operators
$R_0^{\text{s}}\sim O(a_s^4)$ and $R_0^{\text{t}}\sim O(a_s^2 a_t^2)$. Up to quartic order in $a_s$, $a_t$
the action can be expressed in $R_0^{\text{s}}$ and $R_0^{\text{t}}$:
\begin{eqnarray}
\mathcal{A}& = & \frac{R_0^{\text{t}}\xi}{2}\left[2 c_{12} p_{01}^{\text{tm}}+p_{10}^{\text{sp}} + p_{01}^{\text{sp}}\left(1+2 c_{11} + 2 \eta_1 c_{11}\right)\right] \nonumber\\
& & + \frac{R_0^{\text{s}}}{2\xi}\left[2\eta_3 c_{11} p_{01}^{\text{sp}}+
p_{10}^{\text{tm}}+p_{01}^{\text{tm}}\left(1+2c_{13}+\eta_2 c_{13}+\eta_4 c_{12}\right)\right],
\end{eqnarray}

where all parameters $\eta$, $c$ denote the constant ($0^{\text{th}}$ order) terms in the polynomials in $x_\mu$; 
the higher orders do not contribute due to $x_\mu\rightarrow 0$ for very smooth fields.

The coefficient of $R_0$ in eq.~\ref{eq:symr0} is required to be 1 and thus 
a simple comparison yields the two normalisation conditions for our parametrisation of the
anisotropic gauge action:

\begin{eqnarray}
p^{\text sp}_{01}+p^{\text sp}_{10}+2p^{\text sp}_{01}c_{11}+
2p^{\text tm}_{01}c_{12}+2p^{\text sp}_{01}c_{11}\eta_1 & = & \frac{1}{\xi},\\
p^{\text tm}_{01}+p^{\text tm}_{10}+2p^{\text tm}_{01}c_{13}+p^{\text tm}_{01}c_{13}\eta_2+
2p^{\text sp}_{01}c_{11}\eta_3+p^{\text tm}_{01}c_{12}\eta_4 & = & \xi.
\end{eqnarray}

These two conditions ensure that the parametrised action goes over to the continuum form with
correct anisotropy if the action is applied on smooth fields. They are required to be fulfilled exactly
for all our parameter sets.

\fancyhead[RE]{\nouppercase{\small\it Appendix \thechapter.\, APE Action Parameters}}
\chapter{Action parameters} \label{app:actions}

In this Appendix we collect the parameters of the FP and classically perfect actions that have
been used or constructed throughout this work. The isotropic FP action presented in
Section \ref{app:isoact} is the one that has been parametrised and examined closely in 
\cite{Wenger:2000aa,Niedermayer:2000yx,Niedermayer:2000ts}, results of these studies
are collected briefly in Chapter \ref{chap:iso}. The isotropic action $\mathcal{A}_5^{444}$
in Section \ref{app:isointact}
is an intermediate parametrisation valid on configurations typical for MC simulations that are minimised once,
parametrised during the cascade process for the isotropic action. It is used
in the spatial 
blocking (see Section \ref{sec:xi2constr}) to minimise the coarse $\xi=2$ configurations,
constructed using the ad-hoc anisotropic action presented
in Section \ref{app:adhoc}. The resulting $\xi=2$ perfect action for coarse
configurations is presented in Section \ref{app:xi2act}.

To repeat the blocking step, we need a $\xi=2$ action which is valid for $\xi=4$ configurations
minimised once in a purely spatial blocking step. Such an intermediate action has been constructed,
see Section \ref{app:xi2intact}. The resulting $\xi=4$ action is presented in Section \ref{app:xi4act}.

\section{The Isotropic FP Action} \label{app:isoact}

The parametrisation of the isotropic FP action is described in \cite{Wenger:2000aa,Niedermayer:2000yx,Niedermayer:2000ts} as well as in Chapter \ref{chap:iso}. It describes the FP action well 
in the range of lattice spacings $0.03\text{ fm} \lesssim a \lesssim 0.3\text{ fm}$.

The set of parameters consists of four non-linear parameters
$\eta^{(0)}$, $c_1^{(0)}$, $c_2^{(0)}$, $c_3^{(0)}$ describing the asymmetrically
smeared links $W_\mu^{(\nu)}$ and fourteen linear parameters $p_{kl}$ with $0
< k+l \leq 4$. This set approximates reasonably well the true FP action in the
range of $a$ given above.  (For smaller fluctuations --- occurring e.g.~in the intermediate
steps of the parametrisation --- we use polynomials for $\eta(x)$ and $c_i(x)$
up to forth order.)

The optimal non-linear parameters are found to be 
\begin{equation}
\eta^{(0)}=-0.038445 \,, \quad c_1^{(0)}=0.290643  \,, \quad
c_2^{(0)}=-0.201505 \,, \quad c_3^{(0)}=0.084679\,.
\nonumber
\end{equation}

The linear parameters are collected in Table \ref{tab:isoact}.

\begin{table}[htbp]
\renewcommand{\arraystretch}{1.3}
  \begin{center}
    \begin{tabular}{p{1.2cm}|rrrrr}
\hline\vspace{-0.05cm}
$p_{kl}$     & $l=0$ & $l=1$  & $l=2$ & $l=3$ & $l=4$ \\
\hline
 $k=0$  &            &  0.442827 & 0.628828 & -0.677790 &  0.176159 \\
 $k=1$  &  0.051944  & -0.918625 & 1.064711 & -0.275300 &           \\
 $k=2$  &  0.864881  & -0.614357 & 0.165320 &           &           \\
 $k=3$  & -0.094366  & -0.020693 &          &           &           \\
 $k=4$  &  0.022283  &           &          &           &           \\
\hline
  \end{tabular}
    \caption{{}The linear parameters $p_{kl}$ of the parametrised isotropic FP action.}
    \label{tab:isoact}
  \end{center}
\end{table}

\section{The Intermediate Isotropic Action $\mathbf{\mathcal{A}_5^{444}}$}\label{app:isointact}
The intermediate isotropic action has been parametrised during the construction of the isotropic
FP action. It is supposed to be valid on configurations that are obtained
by minimising configurations typical for MC simulations once.
It fulfills the $O(a^2)$ Symanzik conditions (see \cite{Wenger:2000aa,Niedermayer:2000yx})
and uses fluctuation polynomials $x_\mu(n)$ (see Section \ref{sec:isopar}), it thus smoothly approaches the
continuum (high $\beta$) limit and is expected to interpolate between the rather coarse configurations mentioned
above and the smooth limit. It is not intended to be used in MC simulations, its linear behaviour
(see Section \ref{sec:linfit}) is not checked, in addition the set of parameters is not as compact as might
be possible.

This action has been used for the (spatial) minimisation of coarse $\xi=2$ configurations to
describe the isotropic configurations on the r.h.s. of eq.~\ref{eq:FP_equation}.

The non-linear parameters describing order 3 polynomials are 

\begin{small}
\begin{tabular}{llll}
$\eta^{(0)}=0.082$,& $\eta^{(1)}=0.292353$,& $\eta^{(2)}=0.115237$,& $\eta^{(3)}=0.011456$,\\
$c_1^{(0)}=0.282$, & $c_1^{(1)}=-0.302295$, & $c_1^{(2)}=-0.302079$, & $c_1^{(3)}=-0.052309$,\\
$c_2^{(0)}=0.054$, & $c_2^{(1)}=0.298882$, & $c_2^{(2)}=-0.081365$, & $c_2^{(3)}=-0.023762$,\\
$c_3^{(0)}=-0.201671$, & $c_3^{(1)}=0.022406$, & $c_3^{(2)}=0.004090$, & $c_3^{(3)}=0.014886$,\\
$c_4^{(0)}=-0.008977$, & $c_4^{(1)}=0.245363$, & $c_4^{(2)}=0.140016$, & $c_4^{(3)}=0.028783$.\\
\end{tabular}
\end{small}

The linear parameters are collected in Table \ref{tab:isointact}.

\begin{table}[htbp]
\renewcommand{\arraystretch}{1.3}
  \begin{center}
    \begin{tabular}{p{1.2cm}|rrrrr}
\hline\vspace{-0.05cm}
$p_{kl}$     & $l=0$ & $l=1$  & $l=2$ & $l=3$ & $l=4$ \\
\hline
 $k=0$  &            &  0.629227 & -0.556304 &  0.186662 & -0.010110 \\
 $k=1$  & -0.368095  &  0.852428 & -0.199034 &  0.031614 &           \\
 $k=2$  &  0.389292  & -0.207378 & -0.010898 &           &           \\
 $k=3$  & -0.054912  &  0.039059 &           &           &           \\
 $k=4$  & -0.000424  &           &           &           &           \\
\hline
  \end{tabular}
    \caption{{}The linear parameters $p_{kl}$ of the parametrised intermediate isotropic FP action 
$\mathcal{A}_5^{444}$.}
    \label{tab:isointact}
  \end{center}
\end{table}

\section{``Ad-hoc'' Anisotropic Actions}\label{app:adhoc}
In the spatial blocking procedure described in Chapter \ref{ch:construction} one needs coarse
anisotropic ($\xi=$~2, 4, 6, $\ldots$) gauge configurations which are spatially minimised leading
to $\xi'=\xi/2$ configurations. As the goal of this step is to obtain a perfect anisotropic action
with anisotropy $\xi$ these coarse anisotropic configurations have to be produced using some other
action that is already present. This requirement might seem to endanger the whole ansatz,
however it is not crucial how the coarse configurations exactly look like as the perfectness of
the resulting coarse action comes from the perfect action on the fine configuration as well as from
the exactness of the RG transformation (see Section \ref{sec:renorm}). Still, we try to create coarse
configurations that might look similar to future ensembles produced using the perfect anisotropic
action on the coarse level and whose minimised configurations appear to have an anisotropy approximately
$\xi'=\xi/2$ (i.e.~which are isotropic in the case of the construction of the $\xi=2$ action).

In order to achieve this, we modify the isotropic FP action adding a term $(\xi^2-1)p_{10}^{st}$ (where
$p_{10}^{st}$ denotes the simple temporal plaquette). This modification turns the isotropic Wilson action
into the Wilson action with bare anisotropy $\xi$ and is expected to work approximately also for our FP action.
The main argument of using this coarse action and not e.g.~the anisotropic Wilson action is that due to
the spatial lattice spacing $a_s$ being larger than the temporal spacing $a_t$, the $O(a_s^2)$ artifacts are
also expected to be larger than the $O(a_t^2)$ effects. The modification described above should preserve
the FP properties corresponding to $a_s$ and is thus expected to be considerably better than the naive
anisotropic Wilson action.

Using this ad-hoc modification of the isotropic FP action (``ad-hoc'' action) in MC simulations shows (as expected) 
that (at least for small anisotropies $\xi\lesssim 5$) its properties are not comparable with the ones 
of the isotropic FP action, however the generated ensembles resemble to the ones generated with true
perfect anisotropic actions much more than ensembles generated by the Wilson action.

The anisotropies $\xi_{\text{ad-hoc}}$ that have to be used to generate coarse anisotropic configurations turning into
minimised configurations with anisotropy $\xi'$ are $\xi_{\text{ad-hoc}}\approx 3.2$ for $\xi'=1$ and
$\xi_{\text{ad-hoc}}\approx 6$ for $\xi'=2$. However, this value of $\xi_{\text{ad-hoc}}$ varies considerably
if $\beta$ is varied --- but as stated at the beginning of this section, the exact form of the coarse configurations
is not essential.

\section{The $\mathbf{\xi=2}$ Perfect Action} \label{app:xi2act}

The $\xi=2$ perfect action uses the parametrisation described in Sections \ref{sec:isopar},
\ref{sec:anisopar}.
The number of non-zero asymmetry values $\eta_i^{(0)}$ is 4, the parameters $c_i^{(0)}$ ($i=1,\ldots,3$) are splitted
into 3 parameters depending on the contribution to the smeared plaquette. The linear parameters
$p_{kl}$ are non-zero for $0<k+l\leq 4$ for spatial plaquettes and $0<k+l\leq 3$ for temporal plaquettes.

The non-linear parameters (constants in $x_\mu$) have the values 

\begin{small}
\begin{tabular}{llll}
$\eta_1=-0.866007$,& $\eta_2=-0.884110$,& $\eta_3=2.212499$,& $\eta_4=1.141177$;\\
$c_{11}=0.399669$, & $c_{12}=0.519037$, & $c_{13}=-0.071334$, &\\
$c_{21}=-0.076357$,& $c_{22}=-0.031051$,& $c_{23}=-0.282800$, &\\
$c_{31}= 0.032396$,& $c_{32}=-0.015844$,& $c_{33}= -0.046302$.&\\
\end{tabular}
\end{small}

The linear parameters are collected in Table \ref{tab:xi2act_linpar}

\begin{table}[htbp]
\renewcommand{\arraystretch}{1.3}
    \begin{tabular}{p{1.2cm}|rrrrr}
\hline\vspace{-0.05cm}
$p_{kl}^{ss}$     & $l=0$ & $l=1$  & $l=2$ & $l=3$ & $l=4$\\
\hline
 $k=0$  &            &  0.433417 & 0.098921 & -0.116251 & 0.023295\\
 $k=1$  &  0.217599  & -0.272668 & 0.248188 & -0.045278 &         \\
 $k=2$  &  0.316145  & -0.180982 & 0.028817          &  &         \\
 $k=3$  & -0.039521  & 0.003858  &        &          &          \\
 $k=4$  & 0.005443 & & & &\\
\hline
  \end{tabular}

    \begin{tabular}{p{1.2cm}|rrrr}
$p_{kl}^{st}$     & $l=0$ & $l=1$  & $l=2$ & $l=3$\\
\hline
 $k=0$  &            & -0.190195 & 0.554426 & -0.121766\\
 $k=1$  &  1.521212  & -0.328305 & 0.086655 &          \\
 $k=2$  &  0.011178  &  0.020932 &          &          \\
 $k=3$  &  0.022856  &           &          &          \\
\hline
  \end{tabular}

    \caption{The linear parameters of the $\xi=2$ parametrised classically perfect action.}
    \label{tab:xi2act_linpar}
\end{table}

\section{The $\mathbf{\xi=2}$ Intermediate Action} \label{app:xi2intact}

In order to be able to repeat the spatial blocking step constructing a $\xi\approx 4$ action
based on the $\xi=2$ perfect action we need a parametrisation of the $\xi=2$ action which is
valid on ($\xi\approx 2$) configurations that are obtained by spatially minimising coarse $\xi=4$ configurations
once. To construct such an action, we perform a non-linear fit to the derivatives of 5 sets of two configurations
each at $\beta=$ 6, 10, 20, 50, 100. The non-linear parameters are chosen to be linear in the fluctuation
parameter $x_\mu(n)$. Having four different parameters $\eta$ and splitting up $c_i$ into three
parameters (as it is done for all anisotropic parametrisations), this makes 20 non-linear parameters
to be fitted which is quite at the edge of what is still possible on our computers (see Section \ref{sec:nlfit}), that is
why we restrict the total number of configurations to 10. Rough checks performed on a larger number of configurations,
with an even larger number of parameters show however, that the resulting non-linear parameters are stable
and describe the data accurately.

The derivatives and action values of 5 sets of 10 configurations each are included in the linear fit
(where the relative weight of the action values is chosen to be $1.9\cdot 10^{-2}$ for the configurations
at $\beta=$ 50, 100 and $7.6\cdot 10^{-4}$ at $\beta=$ 6, 10, 20). A linear set where the parameters $p_{kl}$
are non-zero for $0<k+l\leq 3$ for spatial plaquettes and $0<k+l\leq 4$ for temporal plaquettes describes
the full action very well concerning this data. Again, this parametrisation is not intended to be used in
MC simulations, thus the linear behaviour of
the action is not checked.

The non-linear parameters (linear in $x_\mu$) have the values 

\begin{small}
\begin{tabular}{llll}
$\eta_1^{(0)}=-1.861267$,& $\eta_1^{(1)}=-0.327466$,& $\eta_2^{(0)}=-1.075610$,& $\eta_2^{(1)}=-0.550398$,\\
$\eta_3^{(0)}= 2.750293$,& $\eta_3^{(1)}= 0.089874$,& $\eta_4^{(0)}= 1.107017$,& $\eta_4^{(1)}= 0.265817$;\\
$c_{11}^{(0)}=0.520960$, & $c_{11}^{(1)}=0.006339$, & $c_{21}^{(0)}=-0.075219$, & $c_{21}^{(1)}=0.059506$\\
$c_{12}^{(0)}=0.266240$, & $c_{12}^{(1)}=0.121035$, & $c_{22}^{(0)}=-0.080771$, & $c_{22}^{(1)}=-0.021515$\\
$c_{13}^{(0)}=0.159372$, & $c_{13}^{(1)}=0.039564$, & $c_{23}^{(0)}=-0.043901$, & $c_{23}^{(1)}=0.009672$\\
\end{tabular}
\end{small}

The linear parameters are collected in Table \ref{tab:xi2intact_linpar}

\begin{table}[htbp]
\renewcommand{\arraystretch}{1.3}
    \begin{tabular}{p{1.2cm}|rrrr}
\hline\vspace{-0.05cm}
$p_{kl}^{ss}$     & $l=0$ & $l=1$  & $l=2$ & $l=3$\\
\hline
 $k=0$  &            &  0.088016 & 0.002225 & -0.000285 \\
 $k=1$  &  0.341850  & -0.015888 &-0.004087 &           \\
 $k=2$  & -0.053007  &  0.010121 &                   &  \\
 $k=3$  &  0.010500  &           &        &             \\
\hline
  \end{tabular}

    \begin{tabular}{p{1.2cm}|rrrrr}
$p_{kl}^{st}$     & $l=0$ & $l=1$  & $l=2$ & $l=3$ & $l=4$\\
\hline
 $k=0$  &            &  0.280043 & 5.077727 &  -13.714872 & 12.739964\\
 $k=1$  &  1.343946  & -6.934825 & 27.673937 & -32.288928 &\\
 $k=2$  &  2.069084  & -17.392027 & 28.248910 &           &\\
 $k=3$  &  3.691733  & -9.584760  &          &           &\\
 $k=4$  &  0.712244  & & & &\\
\hline
  \end{tabular}
    \caption{The linear parameters of the intermediate $\xi=2$ parametrised classically perfect action.}
    \label{tab:xi2intact_linpar}
\end{table}

\section{The $\mathbf{\xi=4}$ Perfect Action} \label{app:xi4act}

The $\xi=4$ perfect action uses the parametrisation described in Sections \ref{sec:isopar},
\ref{sec:anisopar}.
The number of non-zero asymmetry values $\eta_i^{(0)}$ is 4, the parameters $c_i^{(0)}$ ($i=1,\ldots,3$) are splitted
into 3 parameters depending on the contribution to the smeared plaquette. The linear parameters
$p_{kl}$ are non-zero for $0<k+l\leq 3$ for spatial plaquettes and $0<k+l\leq 2$ for temporal plaquettes.

The non-linear parameters (constants in $x_\mu$) have the values 

\begin{small}
\begin{tabular}{llll}
$\eta_1=-1.491457$,& $\eta_2=-1.115141$,& $\eta_3=1.510985$,& $\eta_4=7.721347$;\\
$c_{11}=2.014408$, & $c_{12}=0.128768$, & $c_{13}= 0.162296$, &\\
$c_{21}=-0.915620$,& $c_{22}= 0.134445$,& $c_{23}=-0.013383$, &\\
$c_{31}= 1.166289$,& $c_{32}= 0.061278$,& $c_{33}= 0.000759$, &\\
\end{tabular}
\end{small}

The linear parameters are collected in Table \ref{tab:xi4act_linpar}

\begin{table}[htbp]
\renewcommand{\arraystretch}{1.3}
    \begin{tabular}{p{1.2cm}|rrrr}
\hline\vspace{-0.05cm}
$p_{kl}^{ss}$     & $l=0$ & $l=1$  & $l=2$ & $l=3$\\
\hline
 $k=0$  &            &  0.027625 & 0.000052 &  0.000000\\
 $k=1$  &  0.072131  & -0.016852 & -0.000054 &\\
 $k=2$  &  0.036818  &  0.003558 & &\\
 $k=3$  & -0.007413  & & &\\
\hline
  \end{tabular}

    \begin{tabular}{p{1.2cm}|rrr}
$p_{kl}^{st}$     & $l=0$ & $l=1$  & $l=2$\\
\hline
 $k=0$  &            &  0.795779 &  0.621286 \\
 $k=1$  &  2.130563  & -0.286602 & \\
 $k=2$  &  0.076086  & & \\
\hline
  \end{tabular}

    \caption{The linear parameters of the $\xi=4$ parametrised classically perfect action.}
    \label{tab:xi4act_linpar}
\end{table}

\fancyhead[RE]{\nouppercase{\small\it Appendix \thechapter.\, Variational Techniques}}
\chapter{Variational Techniques} \label{app:vartech}

Normally, in a Monte Carlo simulation there are several operators
that have an overlap with a state one would like to measure. In the
case of glueballs, e.g., we have multiple shapes with combinations of
different orientations having the same quantum numbers. Another method
yielding several gauge operators with the same quantum numbers, is
smearing (see Section \ref{sec:smearing}). A priori it is unknown which
operators have the largest contribution to a given state, it is therefore
vital to have a method which takes into account the different operators
according to their overlap with the desired state. In this Appendix we
describe the variational techniques accomplishing this, used for all
mass / energy determinations (torelons, static $q\bar{q}$-potential,
glueballs).

\section{Obtaining Mass Estimates}

In a Monte Carlo simulation we measure the $N \times N$
correlation matrix of operators ${\cal O}_\alpha$
\begin{equation}
  \label{eq:full_correlation}
  C_{\alpha \beta}(t) =  \langle 0| {\cal O}_\alpha(t)
  {\cal O}^\dagger_\beta(0) |0 \rangle \,,
\end{equation}
where $N$ is the number of operators having the same quantum numbers
as the examined state.
To determine the coefficients $v_\alpha$ of the linear combination 
$\sum_{\alpha=1}^N v_\alpha {\cal O}_\alpha$ which has the largest 
overlap to the ground state relative to the excited states 
one has to minimise the effective mass given by
\begin{equation}
  \label{eq:minimizing_eff_mass}
  m(t_0,t_1) = 
-\ln\left[ \frac{(v, C(t_1)v)}{(v, C(t_0)v)}\right]/(t_1-t_0).
\end{equation}
The vector $v$ is obtained by solving the generalised eigenvalue equation
\cite{Michael:1985ne,Luscher:1990ck}
\begin{equation}
  \label{eq:gen_eigenvalue_problem}
  C(t_1) v = \lambda(t_0,t_1) C(t_0) v \,,
\end{equation}
where $0 \le t_0 < t_1$.

Assume first that only the lowest lying $N$ states contribute to
$C(t)$, i.e.,
\begin{equation}
\label{eq:C_N}
C_{\alpha \beta}(t)=\sum_{n=1}^{N} \text{e}^{-E_nt}\psi_{n\alpha}
\psi_{n\beta}^* \,,
\end{equation}
where $E_1 \le E_2 \le \ldots \le E_N$ are the energy levels in the
given symmetry channel and 
$\psi_{n\alpha}=\langle 0 | O_\alpha | n \rangle$ is the 
``wave function'' of the corresponding state.
The solution of eq.~\ref{eq:gen_eigenvalue_problem} is given
by the set of vectors $\{ v_n \}$ dual to the wave functions,
i.e.,~$(v_n,\psi_m)=\delta_{nm}$.
Multiplying eq.~\ref{eq:C_N} by $v_n$ one obtains
\begin{equation}
\label{eq:Cv}
C(t)v_n=\text{e}^{-E_nt}\psi_n = 
\text{e}^{-E_n(t-t_0)}\text{e}^{-E_nt_0}\psi_n=
\text{e}^{-E_n(t-t_0)}C(t_0)v_n \,.
\end{equation}
This gives $\lambda_n(t_0,t_1)=\exp( -E_n(t_1-t_0) )$ for the eigenvalues
in eq.~\ref{eq:gen_eigenvalue_problem}.
Of course, contributions from states with $n>N$ and statistical
fluctuations distort eq.~\ref{eq:C_N}, therefore the stability
of eq.~\ref{eq:gen_eigenvalue_problem} is an important issue.

Observe that eq.~\ref{eq:gen_eigenvalue_problem} is well defined only 
for positive definite $C(t_0)$. 
Because of statistical fluctuations, however, the measured correlation 
matrix $C(t_0)$ is not necessarily positive for $t_0>0$.
This is the reason why one usually considers only the $t_0=0$
case in applying the variational method, especially with a large
number of operators.
On the other hand, it is obvious that $C(0)$ is contaminated by highly 
excited states and contains only restricted information on the low lying 
part of the spectrum. Therefore it is desirable to take $t_0>0$.
This can be achieved in the following way \cite{Balog:1999ww}.

We first diagonalise $C(t_0)$,
\begin{equation}
  \label{eq:C0_diagonalization}
  C(t_0) \varphi_i  = \lambda_i \varphi_i, \quad \lambda_1 \geq \ldots \geq
  \lambda_N,
\end{equation}
and project the correlation matrices to the space of eigenvectors
corresponding to the $M$ highest eigenvalues,
\begin{equation}
  \label{eq:first_truncation}
  C^M_{ij}(t) = (\varphi_i, C(t) \varphi_j), \quad  i,j=1,\ldots, M.
\end{equation}
By choosing the operator space too large we introduce numerical instabilities
caused by very small (even negative) eigenvalues with large statistical errors
due to the fact that the chosen operator basis is not 
sufficiently independent on the given MC sample. 
By choosing $M$ appropriately we can get rid of those unstable
modes while still keeping all the physical information. In this way we 
render the generalised eigenvalue problem well defined.

Of course the final result should not depend on the choice of $M$ and one has
to take care in each case that this is really the case. Our observation is that for
any acceptable statistics one always finds a plateau in $M$ for which the
extracted masses are stable under variation of $M$.

In a next step we determine the vectors $v_n$, $n=1, \ldots, M$ through the
generalised eigenvalue equation in the truncated basis:
\begin{equation}
\label{genev_CM}
  C^M(t_1) v_n = e^{-E_n(t_1-t_0)} C^M(t_0) v_n \,.
\end{equation}
This equation yields the spectrum $E_n$. However, the procedure
--- although it is exact for a correlation matrix which
has {\em exactly} the form of eq.~\ref{eq:C_N} ---
is highly non-linear, and a small statistical fluctuation can be 
enhanced by it and cause a systematic shift in the energy values 
obtained, even when the instabilities are avoided by the truncation
to $M<N$.

In order to avoid this pitfall we use the (approximate) dual
vectors $v_n$ obtained from eq.~\ref{genev_CM} to restrict
the problem to an even smaller, therefore more stable subspace.

Define the new correlation matrix of size $K\times K$ (with $K\le M$) by
\begin{equation}
  \label{eq:second_truncation}
 C^K_{ij}(t) = (v_i, C^M(t) v_j), \quad i,j=1,\ldots, K \leq M \,.
\end{equation}
The steps performed until now can be thought of as a preparation for 
choosing the appropriate set of operators, i.e.~linear combinations
of original ${\cal O}_\alpha$ operators which effectively
eliminate the higher states. The correlation matrix
$C^K_{ij}(t)$ is then considered as a primary, unbiased object.

The next step is to fit $C^K_{ij}(t)$ in the range 
$t = t_{\text{min}} \ldots t_{\text{max}}$ using the ansatz
\begin{equation}\label{eq:ansaetze}
  \widetilde{C^K_{ij}}(t;\{\psi, E\}) = \sum_{n=1}^{K} e^{-E_n t} \psi_{ni} \psi_{nj}^* \,,
\end{equation}
where $\psi_{ni}$, $E_n$ are the free parameters to be fitted.

Usually we choose $K=1$  and $2$. For the $A_1^{++}$ glueball, however,
$K=2$ and $3$ are chosen since we do not subtract the vacuum contribution 
$\langle {\cal O}_\alpha \rangle \langle {\cal O}_\beta \rangle^*$ 
from the correlators but consider instead  the vacuum state
together with the glueball states in this channel
(cf.~remarks in Section \ref{sec:latgb}).

In the fitting step we use a correlated $\chi^2$ fit
which takes into account the correlation between $C^K_{ij}(t)$
and $C^K_{i'j'}(t')$, i.e.~using the inverse of the corresponding 
covariance matrix  $\text{Cov}(i,j,t; i',j',t')$ as a weight 
in the definition of $\chi^2$.
This has the advantage over the uncorrelated $\chi^2$
that the value of the latter can be artificially small
if the quantities to be fitted are strongly correlated.
Be warned however, that (as usually with sophisticated methods)
the correlated $\chi^2$ fit can have its own instabilities
if the number of data is not sufficiently large
\cite{Michael:1994yj,Michael:1995br}.

\begin{figure}[htbp]
\begin{center}
\epsfig{file=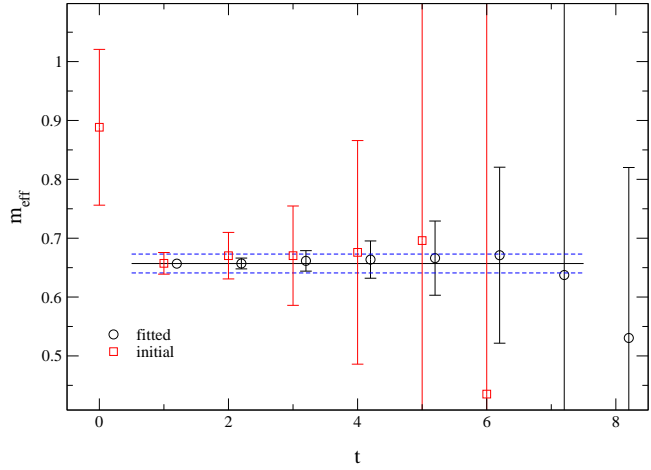,width=10cm}
\end{center}
\caption{The effective masses, as defined in eq.~\ref{eq:minimizing_eff_mass}, before (squares) and after
(circles) the fitting procedure for the ground state of the scalar glueball on the coarse lattice, 
at $\xi=2$, $\beta=3.15$. The solid and dashed horizontal lines mark the resulting energy and its error,
obtained performing a fit on $t=1,\ldots,7$.}
\label{fig:effmassplot}
\end{figure}

The result of this fitting process becomes apparent if one compares the initial effective mass,
eq.~\ref{eq:minimizing_eff_mass}, to the effective mass obtained using the fitted correlation
matrix elements $\widetilde{C^K_{ij}}(t)$ as displayed in Figure \ref{fig:effmassplot} for the
scalar glueball on the coarsest lattice ($\beta=3.15$) where the plateau is rather short.

\begin{figure}[htbp]
\begin{center}
\epsfig{file=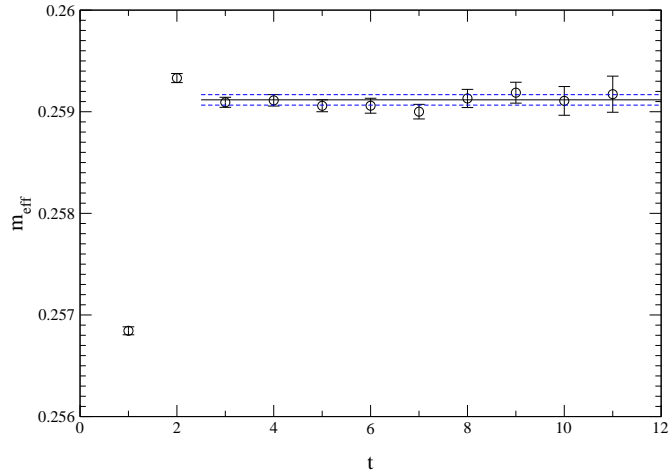,width=10cm}
\end{center}
\caption{The effective masses, as defined in eq.~\ref{eq:minimizing_eff_mass} after the fitting procedure
for the separation $R=1$ of the on-axis static quark-antiquark potential at $\beta=3.50$.The solid and dashed
horizontal lines mark the resulting energy and its error, obtained performing a fit on $t=3,\ldots,12$}
\label{fig:fineeffmassplot}
\end{figure}

Figure \ref{fig:fineeffmassplot} which displays only the final effective mass plots demonstrates that using anisotropic
lattices (even at $\xi=2$) may greatly improve the quality of the results.

\section{Choosing the Right Mass}

In the last Section, we described how one obtains estimates of a state's mass using sophisticated
variational methods, depending on input parameters such as the initial input operators, the number
$M$ of operators kept after the first step or the fitting range $t = t_{\text{min}} \ldots t_{\text{max}}$.
The number of degrees of freedom in the fit is very large, for typical mass determinations of glueballs
it may be of $O(10000)$ if one considers all combinations that might turn out to be sensible, i.e.~that
might result in a stable mass plateau and a good fit with small $\chi^2$.

\begin{figure}[htbp]
\begin{center}
\epsfig{file=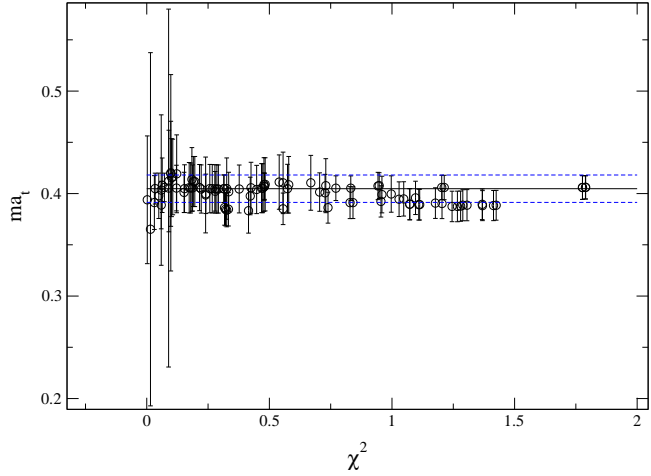,width=10cm}
\end{center}
\caption{The mass estimates of the scalar glueball mass on $\xi=2$, $\beta=3.50$ configurations. 16 different
fit ranges as well as 11 different truncations $M$ are included. The finally chosen value and its errors are
marked by the solid and the dashed lines, respectively.}
\label{fig:chi2m}
\end{figure}

The first choice one has to make, is the reduction of parameter sets examined to a feasible number, say
$O(100)$. This can normally be done quite easily, as the extracted mass should not depend too much on
whether, e.g., one chooses $M=13$ or $M=14$, whether one uses the fitting range $t = 1\ldots 5$ or
$t = 1\ldots 6$ etc. Figure \ref{fig:chi2m} displays the resulting estimates of the full variational method
with the condition $\chi^2/N_{\text{DF}}<2$ for the ground state of the $A_1^{++}$ glueball at 
$\beta=3.50$ (fixed initial number of operators $N=91$). We have included 16 different fit ranges as well 
as 11 different truncations $M$. In this case it turns out that it is unambiguous to pick a mass that is
consistent with all the determinations, marked by the solid line; the errors of the final result
are marked by the dashed lines.  

Of course, the picture does not look that nice in all the cases investigated. Especially for higher masses,
where the signal disappears in the noise after just a few time slices, one has to take care to study the behaviour
of the mass under the variation of all the parameters and to pick a region where the mass is stable for all sensible
choices of parameters. If this is not possible, one might specify an average mass estimate with an error that
comprises all the mass estimates coming from sensible choices of the parameters. However, in this case, the error
normally gets very large.

\fancyhead[RE]{\nouppercase{\small\it Appendix \thechapter.\, Representations of Two-Glueball States}}
\chapter{Representations of Two-Glueball States}\label{app:2gb}

In this Appendix we aim at calculating the representation of the cubic group $O_h$ to which a pair of
glueballs $(\Gamma_1^{PC},\Gamma_2^{P'C'})$ with opposite momenta $\vec{p}, -\vec{p}$ contributes. These
results are used to estimate the minimal energies of glueball pairs on the lattice appearing in each representation, see
Section \ref{sec:xi2gb}. The method has been described by Morningstar and Peardon in \cite{Morningstar:1999rf} and
is presented in more detail in the following.

\section{Point Groups Used}

\begin{figure}[h]
\begin{center}
  \includegraphics[width=5cm]{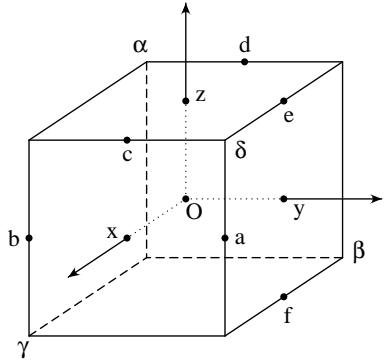}
\end{center}
\caption{The rotation axes $Oa, Ob, Oc ,Od, Oe, Of, Ox, Oy, Oz, O\alpha,
  O\beta, O\gamma$ and $O\delta$ of the cubic group $O_h$.}
\label{rotation_axes}
\end{figure}

Glueballs at rest transform according to the full cubic group $O_h$, see Section \ref{sec:latgb}. This group describes
the symmetries of a cube (see Figure \ref{rotation_axes}) and
contains 48 elements in 10 classes: the identity $E$, 8 three-fold ($120^\circ$) rotations $C_3$, 3 two-fold rotations ($180^\circ$) $C_2$
around
the $x,y,z$-axes, 6 two-fold rotations $C_2$ around the axes $a$--$f$ and 6 four-fold rotations ($90^\circ$) $C_4$; the other
five classes contain the same elements, each preceeded by the spatial inversion $I$.

Momentum breaks full $O_h$ symmetry. Depending on the orientation of the spatial momentum, different 
symmetries are preserved, the so-called \emph{little groups}, subgroups of $O_h$, describing these remnant symmetries are listed
 in Table \ref{tab:lgroups}. The group $C_{4v}$ has the symmetries of a regular 4-sided pyramid and contains
8 elements; accordingly the groups $C_{2v}$ and $C_{3v}$ have the symmetries of 2- and 3- sided pyramids respectively
and contain 4 and 6 elements. The little group $C_s$ contains only two operations, namely the identity $E$ and a
mirror reflection $\sigma$; the trivial group $C_1$ includes only the identity $E$.\footnote{These groups
describe important symmetries occurring in nature, namely the symmetries of molecules: e.g., $C_{2v}$ is the symmetry
group of molecules like $H_2 O$, $C_{3v}$ describes the symmetries of $NH_3$. They belong to a large set of groups
called molecular point groups. A subset of these groups are the 32 crystallographic point groups describing symmetries
of regular crystal lattices --- actually, all the point groups used in this work also belong to this subset.}

\begin{table}[htbp]
\renewcommand{\arraystretch}{1.5}
  \begin{center}
    \begin{tabular*}{10cm}[c]{c@{\extracolsep{\fill}}cc}
      \hline\vspace{-0.05cm} 
      $(n_x,n_y,n_z)$ & little group & international notation\\
      \hline
      (0,0,0) & $O_h$    & $\frac{4}{m}\bar{3}\frac{2}{m}$\\
      (0,0,n) & $C_{4v}$ & $4mm$\\
      (0,n,n) & $C_{2v}$ & $2mm$\\
      (n,n,n) & $C_{3v}$ & $3m$\\
      (0,m,n) & $C_{s}$  & $m$\\
      (m,m,n) & $C_{s}$  & $m$\\
      (l,m,n) & $C_1$    & $1$\\
      \hline
    \end{tabular*}
    \caption{The little groups for lattice glueball states having momentum $\vec{p}=2\pi(n_x,n_y,n_z)/S$ on a
      simple cubic spatial lattice with periodic boundary conditions. Note, that $l,m,n\ne 0$, $l\ne m, m\ne n, l\ne n$.
      The group $C_{s}$ is often denoted as $S_1$. The international notation for the groups is given together with
      the usual Schoenflies notation for convenience.}
    \label{tab:lgroups}
  \end{center}
\end{table}

\section{The Method}\label{sec:subsmeth}

In order to work out the representations of $O_h$ a pair of glueballs with net zero momentum is contributing to,
(1) we subduce the given representations of $O_h$, $\Gamma_1^{PC}$ and $\Gamma_2^{P'C'}$, into the little group
under consideration (depending on the momentum orientation). These representations which in general are reducible are
then (2) decomposed into direct sums of irreducible representations of the little group. 
Out of these sums, (3) we build pairs of irreducible representations each coming from one of the initial glueball
representations and build their direct product. The direct product is again decomposed into irreducible
representations of the little group, and finally, (4) the representation is induced back to $O_h$ and decomposed
into irreducible representations of this group. That this is possible follows from the net zero momentum of the
glueball pairs. In the following, we describe each of these steps in detail.

First, let us introduce the most important tool that will be used, namely \emph{characters}. The sum of all
eigenvalues of a matrix representation $D(G_a)$ of the group element $G_a$, called the character $\chi(G_a)$,
is invariant under similarity transformations $D'(G_a)=AD(G_a)A^{-1}$. Furthermore, all the elements
of a class $\mathcal{C}_p$ of a group have the same character $\chi_p$, because
\begin{eqnarray}
\chi(G_a) & \equiv & \sum_i D_{ii}(G_a) = \sum_i D_{ii}(G_m G_b G_m^{-1})\nonumber\\
          & =      & \sum_{ijk} D_{ij}(G_m) D_{jk}(G_b) D_{ki} (G_m^{-1})\nonumber\\
          & =      & \sum_{jk}  D_{jk}(G_b) D_{kj}(G_m^{-1}G_m)\nonumber\\
          & =      & \sum_j D_{jj}(G_b) \equiv \chi(G_b).
\end{eqnarray}
The characters of the classes of all the point groups used here are given in Tables \ref{tab:char_Oh}, \ref{tab:char_C4v},
\ref{tab:char_C2v}, \ref{tab:char_C3v}, \ref{tab:char_Cs}, \ref{tab:char_C1}.

These characters are very useful if one is reducing a representation, i.e.~decomposing a given (reducible)
representation into its irreducible contents
\begin{equation}
D=\bigoplus_{\alpha} m_{\alpha} T^{(\alpha)},
\end{equation}
where $\alpha$ runs only over the inequivalent irreducible representations (irreps) $T^{(\alpha)}$ and $m_{\alpha}$ gives
the number that this irrep occurs in the reduction. The character of the reducible representation $D$ is related
to the irreducible characters $\chi_p^{(\alpha)}$ by a similar equation:
\begin{equation}
\chi_p=\sum_{\alpha}m_{\alpha} \chi_p^{(\alpha)}.
\end{equation}
The vital task of deducing the coefficients $m_{\alpha}$ given the representation $D$ is facilitated a lot by the use
of characters. Using the orthogonality between different irreducible representations, one may show
\begin{equation}\label{eq:reduce}
\frac{1}{g}\sum_{p}c_p \chi_p^{(\beta)*} \chi_p = \frac{1}{g} \sum_{\alpha} m_{\alpha} \sum_p c_p \chi_p^{(\beta)*}
\chi_p^{(\alpha)}=\frac{1}{g} \sum_{\alpha} m_{\alpha} g \delta_{\alpha \beta}=m_{\beta},
\end{equation}
where $g$ is the order of the group, $c_p$ denotes the number of elements in a given class, $p$ runs over the number
of classes whereas $\alpha$ numbers the irreps. The coefficient $m_{\beta}$ is thus just the scalar product of the
character $\chi$ of the reducible representation with $\chi^{(\beta)}$, the character of the irrep\footnote{Note
that all the numbers $m_{\beta}$ must come out as positive integer numbers or zero --- this is a considerable
safeguard against errors!}.

Having prepared the most important tools, we may now start to apply the method step by step. As a random example, we
study the glueball pair $(T_1^{++}, A_1^{--})$ with momenta $p=\pm 2\pi/S (1,1,1)$.

\subsection{The Subduction into the Little Group}

First, we subduce the \emph{characters} of $O_h$ into the little group under consideration. This amounts to
simply picking the right columns out of Table \ref{tab:char_Oh} corresponding to the remnant symmetry operations
in the little group under consideration. These correspondences are listed in Table \ref{tab:class_corr}. For our
example, with the little group $C_{3v}$, we pick the columns denoted by $E$, $8C_3$ and $6IC_2$ corresponding
to the classes $E$, $2C_3$ and $3\sigma_v$ of the little group. 

\subsection{The Decomposition into Irreps}\label{sec:gdecomp}

The subduced representations, denoted by $O_h \downarrow G_l$, where $G_l$ denotes the little group, are in
general reducible in $G_l$. Applying eq.~\ref{eq:reduce} to the subduced characters and the respective characters
of the little group, we obtain the decomposition of all the initial representations into the irreps of the
little groups, listed in Tables \ref{tab:decomp_C4v}, \ref{tab:decomp_C2v}, \ref{tab:decomp_C3v}, \ref{tab:decomp_Cs},
\ref{tab:decomp_C1}. For our example groups, we build the scalar products of the subduced characters of $O_h$
prepared in the last section and the characters of $C_{3v}$ listed in Table \ref{tab:char_C3v} obtaining
Table \ref{tab:decomp_C3v}.

\subsection{Building Glueball Pairs}

Up to now, the steps performed referred to single glueballs having momentum $\vec{p}$. In order to obtain
physical objects transforming the same way as our single zero-momentum glueballs, we will now combine two
glueballs of opposite momentum. This can be obviously done in the little group corresponding to the momentum.
A pair of glueballs transforms according to the \emph{direct product} of two irreps. On the level of characters
this amounts to multiplying the characters of the irreps corresponding to the same group elements (or class).
Having in general reducible
representations  of the initial representations decomposed into irreps, there may be several direct products
to be taken into account. In our
example, $T_1^{++} \downarrow C_{3v} = A_2 \oplus E$ and $A_1^{--} \downarrow C_{3v} = A_2$. This leaves us with
the direct products $A_2 \otimes A_2$ and $E \otimes A_2$. Calculating the characters, one notices easily that
$A_2 \otimes A_2 = A_1$ and $E \otimes A_2 = E$. That the direct products are irreducible is true for most of the
cases; for the other cases, the reduction can be
performed exactly the same way as presented in Section \ref{sec:subsmeth} and applied in Section \ref{sec:gdecomp}.
For convenience, the multiplication tables (together with the decomposition into irreps) are given as
Tables \ref{tab:dprod_Oh}, \ref{tab:dprod_C4v}, \ref{tab:dprod_C2v}, \ref{tab:dprod_C3v}, \ref{tab:dprod_Cs}
(the table for the group $C_1$ is completely trivial and thus omitted).

In the case of both glueballs having momentum 0, i.e.~transforming according to $O_h$, the whole method
amounts to simply performing the direct products and reading off the resulting irreps of $O_h$ in Table \ref{tab:dprod_Oh}.

\subsection{The Induction back to the Cubic Group}

For most of the representations of the glueball pairs in the little groups, the induction (getting back onto
the level of the full cubic group $O_h$) is trivial, due to the \emph{Frobenius reciprocity theorem} that states
that the number of times a representation $D^{(\nu)}(G)$ occurs in $D^{(\mu)}\uparrow G$, the induced representation,
is equal to the number of occurrences of the irrep $D^{(\mu)}(G_s)$ in $D^{(\nu)}\downarrow G_s$, i.e.~if the representation
of the glueball pair in the little group is again an irrep of the little group, we can read off the contents of
the induced representation in terms of the irreps of $O_h$ again from Tables \ref{tab:decomp_C4v},
\ref{tab:decomp_C2v}, \ref{tab:decomp_C3v}, \ref{tab:decomp_Cs}, \ref{tab:decomp_C1}.

For the two cases ($E \otimes E$ in $C_{4v}$ and $E \otimes E$ in $C_{3v}$) where the direct product is \emph{not}
an irrep of the little group, we just give the results: $(E \otimes E) \uparrow O_h = E^+ \oplus E^-$ for $C_{4v}$,
whereas $E \otimes E$ contributes to all representations of $O_h$ for $C_{3v}$. The way of performing these inductions
is employing the left coset decomposition of $O_h$ in terms of the little group, see e.g.~\cite{Chen:1989,Elliott:1979}.

In our example, the irreps of the little group describing the glueball pair are $A_1$ and $E$. Using the Frobenius
theorem with Table \ref{tab:decomp_C3v} , we notice easily that the pair may contribute to the irreps 
$A_1^+, E^+, T_1^+, T_2^+, A_2^-, E^-, T_1^-, T_2^-$ of the cubic group $O_h$.

\subsection{A Note on Parity and Charge Conjugation}

In the previous sections, the parity quantum number $P=\pm$ we have always taken into account, whereas the
charge conjugation $C$ has been omitted. The reason for this is simple: once the momentum of a single glueball
is non-zero, it is no longer an eigenstate of parity, i.e.~that is why we have to consider the decomposition
into irreps of the little groups for $P=+$ and $P=-$ states separately. Charge conjugation is completely different:
the quantum number $C_2$ of the glueball pair is always just the product $C\cdot C'$ of the quantum numbers of the
initial states, therefore the results for $C=C'=+$ can be trivially generalised to the other three cases.

There is another important issue concerning parity, namely if the two glueballs are indistinguishable (having
the same $\Gamma^{PC}$), the final two-glueball state has to have positive parity due to the Bose-symmetry.
Hypothetic $P=-$ glueball pairs consisting of two states having the same quantum numbers thus have to be excluded.

For our example, all this means that because of the two representations, $T_1^{++}$ and $A_1^{--}$, being
different, Bose symmetry does not exclude any states; furthermore that the resulting states have charge
conjugation quantum number $C=-$. The two glueballs of our example, having momenta $\vec{p}$ parallel to the
space diagonal thus contribute to the representations $A_1^{+-}$, $E^{+-}$, $T_1^{+-}$, $T_2^{+-}$, $A_2^{--}$,
$E^{--}$, $T_1^{--}$, $T_2^{--}$.

\section{Results}

\begin{figure}[htbp]
\begin{center}
\includegraphics[width=11.5cm]{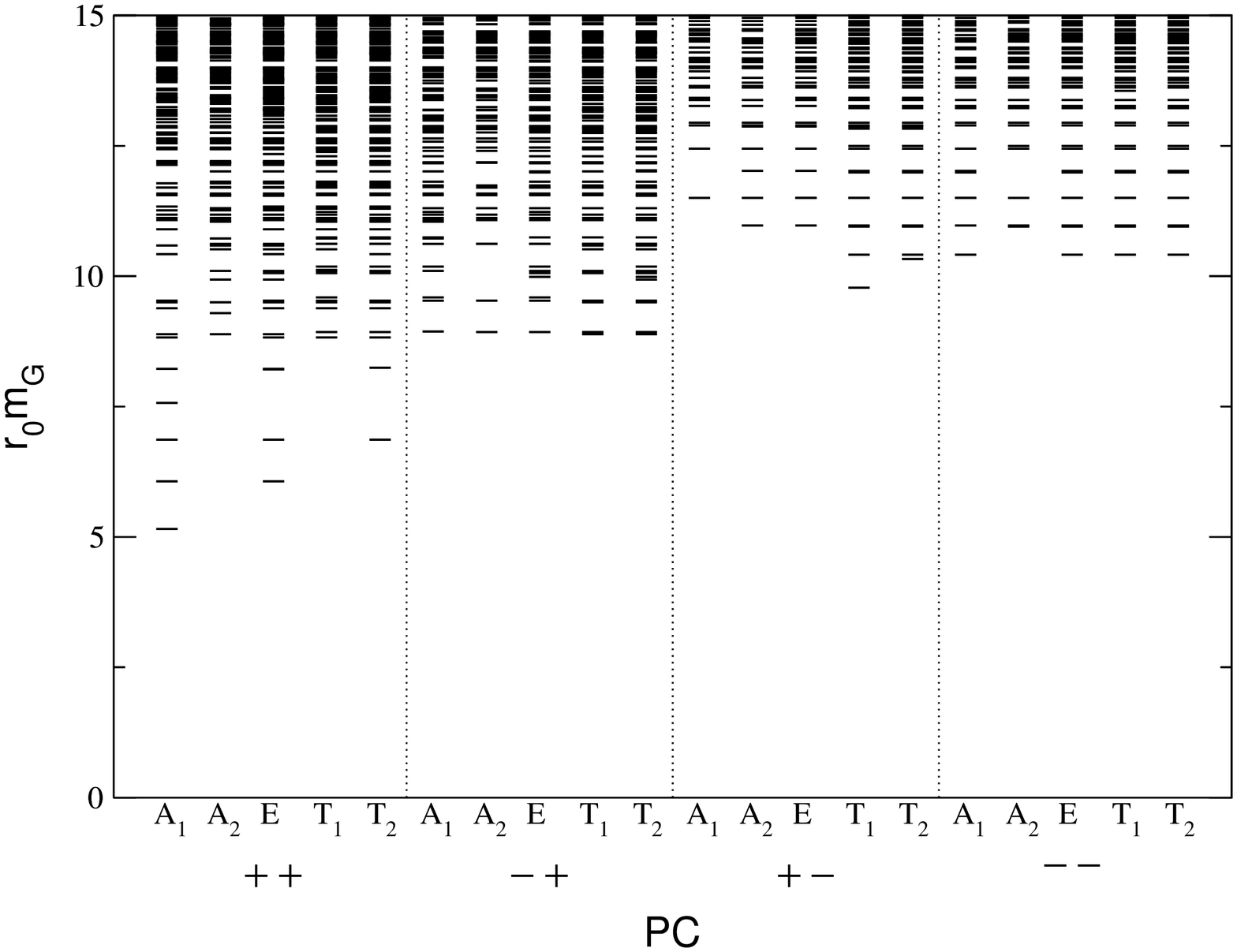}
\end{center}
\caption{The two-glueball spectrum on the $\beta=3.15$ lattice.}
\label{fig:2gb_b315}
\end{figure}

\begin{figure}[htbp]
\begin{center}
\includegraphics[width=11.5cm]{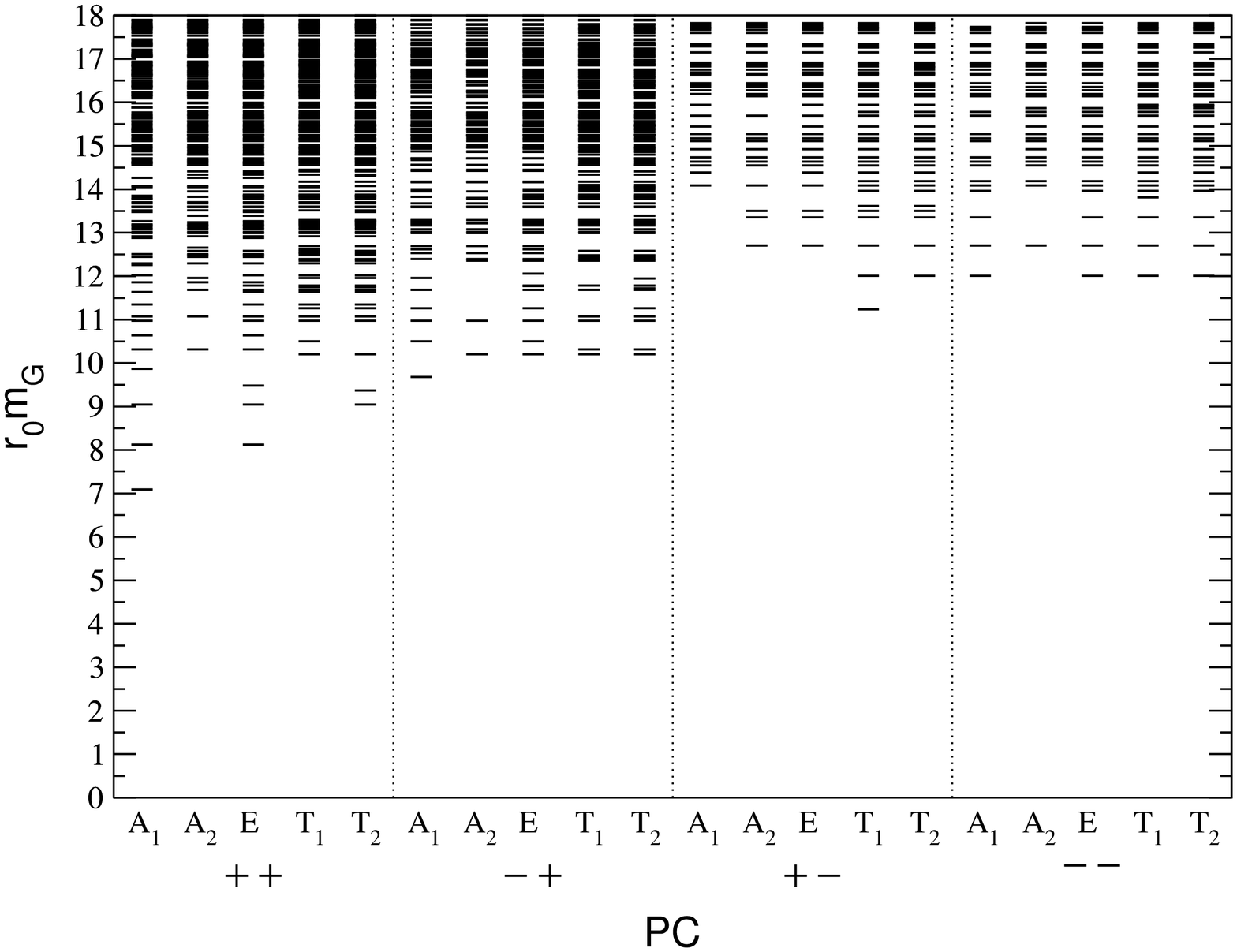}
\end{center}
\caption{The two-glueball spectrum on the $\beta=3.30$ lattice.}
\label{fig:2gb_b330}
\end{figure}

\begin{figure}[htbp]
\begin{center}
\includegraphics[width=11.5cm]{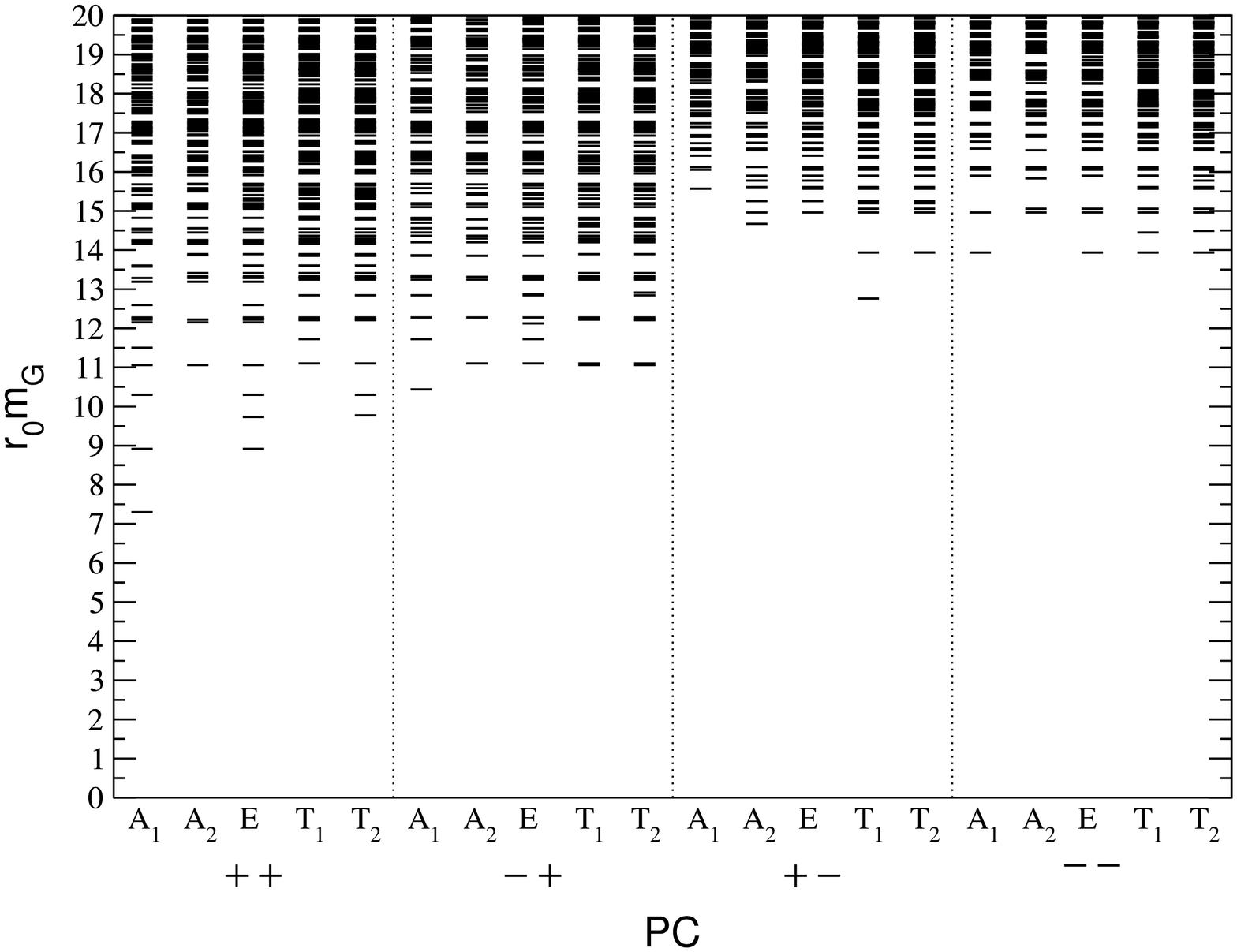}
\end{center}
\caption{The two-glueball spectrum on the $\beta=3.50$ lattice.}
\label{fig:2gb_b350}
\end{figure}

We calculate the energies of all possible glueball pairs on all our lattices for momenta up to $|\vec{p}|=3\cdot 2\pi/Sa_s$.
The resulting lowest-lying masses of glueball pairs, for all the representations where states can be measured
at a given $\beta$ are given in Tables \ref{tab:gbpair_b315}--\ref{tab:gbpair_b350}. The spectra are displayed
in Figures \ref{fig:2gb_b315}--\ref{fig:2gb_b350}. 

\begin{table}[htbp]
\renewcommand{\arraystretch}{1.5}
\centering
\begin{tabular*}{\textwidth}[c]{l@{\extracolsep{\fill}}|rrrrr|rrrrr}
$O_h$ & $E$ & $8C_3$ & $3C_2$ & $6C_2$ & $6C_4$ & \phantom{  }$I$ & $8IC_3$ & $3IC_2$ & $6IC_2$ & $6IC_4$\\
\hline
 $A_1^+$ & 1 &  1 &  1 &  1 &  1 &  1 &  1 &  1 &  1 &  1 \\
 $A_2^+$ & 1 &  1 &  1 & -1 & -1 &  1 &  1 &  1 & -1 & -1 \\
 $E^+  $ & 2 & -1 &  2 &  0 &  0 &  2 & -1 &  2 &  0 &  0 \\
 $T_1^+$ & 3 &  0 & -1 & -1 &  1 &  3 &  0 & -1 & -1 &  1  \\
 $T_2^+$ & 3 &  0 & -1 &  1 & -1 &  3 &  0 & -1 &  1 & -1 \\
\hline
 $A_1^-$ & 1 &  1 &  1 &  1 &  1 & -1 & -1 & -1 & -1 & -1 \\    
 $A_2^-$ & 1 &  1 &  1 & -1 & -1 & -1 & -1 & -1 &  1 &  1 \\
 $E^-  $ & 2 & -1 &  2 &  0 &  0 & -2 &  1 & -2 &  0 &  0 \\  
 $T_1^-$ & 3 &  0 & -1 & -1 &  1 & -3 &  0 &  1 &  1 & -1  \\
 $T_2^-$ & 3 &  0 & -1 &  1 & -1 & -3 &  0 &  1 & -1 &  1 \\
\hline
\end{tabular*}
\caption{The character table for the cubic point group $O_h$.}
\label{tab:char_Oh}
\end{table}

\begin{table}[htbp]
\renewcommand{\arraystretch}{1.5}
\centering
\begin{tabular}{l|ccccc}
$O$ & $A_1^+$ & $A_2^+$ & $E^+$ & $T_1^+$ & $T_2^+$\\
\hline
 $A_1^+$ & $A_1^+$ & $A_2^+$ & $E^+$ & $T_1^+$ & $T_2^+$\\
 $A_2^+$ &         & $A_1^+$ & $E^+$ & $T_2^+$ & $T_1^+$\\
 $E^+$   &         &         & $A_1^+\oplus A_2^+\oplus E$ & $T_1^+ \oplus T_2^+$ & $T_1^+ \oplus T_2^+$\\
 $T_1^+$ &         &         &       & $A_1^+\oplus E \oplus T_1^+ \oplus T_2^+$ & $A_2^+\oplus E \oplus T_1^+ \oplus T_2^+$\\
 $T_2^+$ &         &         &       &         & $A_1^+\oplus E \oplus T_1^+ \oplus T_2^+$\\
\hline
\end{tabular}
\caption{The direct products of the irreducible representations (irreps) of the cubic point group $O$,
decomposed back into direct sums of the irreps. For the (trivial) extension to the full cubic group $O_h$,
see text. Due to the obvious symmetry, only the upper diagonal part of the table is given.}
\label{tab:dprod_Oh}
\end{table}

\begin{table}[htbp]
\renewcommand{\arraystretch}{1.5}
\centering
\begin{tabular}{l|rrrrrrrrrr}
 & $E$ & $8C_3$ & $3C_2$ & $6C_2$ & $6C_4$ & \phantom{  }$I$ & $8IC_3$ & $3IC_2$ & $6IC_2$ & $6IC_4$\\
\hline
$C_{4v}$ & $E$ & & $C_2$ & & $2C_4$  & & & $2\sigma_v$ & $2\sigma_d$ & \\
$C_{2v}$ & $E$ & & & $C_2$ &         & & & $\sigma_x$ & $\sigma_y$ &   \\
$C_{3v}$ & $E$ & $2C_3$ &&        & & & & & $3\sigma_v$ & \\
$C_s$    & $E$ & &       &        & & & & & $\sigma$ &\\
$C_1$    & $E$ & & & & & & & & \\
\hline
\end{tabular}
\caption{The correspondence of the classes of the cubic point group $O_h$ to the classes of the little groups
$C_{4v}, C_{2v}, C_{3v}, C_s, C_1$.}
\label{tab:class_corr}
\end{table}

\begin{table}[htbp]
\renewcommand{\arraystretch}{1.5}
\centering
\begin{tabular}{l|rrrrr}
$C_{4v}$ & $E$ & $C_2$ & $2C_4$ & $2\sigma_v$ & $2\sigma_d$\\
\hline
 $A_1$ & 1 &  1 &  1 &  1 &  1 \\
 $A_2$ & 1 &  1 &  1 & -1 & -1 \\
 $B_1$ & 1 &  1 & -1 &  1 & -1 \\
 $B_2$ & 1 &  1 & -1 & -1 &  1 \\
 $E$   & 2 & -2 &  0 &  0 &  0 \\
\hline
\end{tabular}
\caption{The character table for the point group $C_{4v}$.}
\label{tab:char_C4v}
\end{table}

\begin{table}[htbp]
\renewcommand{\arraystretch}{1.5}
\centering
\begin{tabular}{l|ccccc}
$C_{4v}$ & $A_1$ & $A_2$ & $B_1$ & $B_2$ & $E$\\
\hline
 $A_1$ & $A_1$ & $A_2$ & $B_1$ & $B_2$ & $E$\\ 
 $A_2$ &       & $A_1$ & $B_2$ & $B_1$ & $E$\\
 $B_1$ &       &       & $A_1$ & $A_2$ & $E$\\
 $B_2$ &       &       &       & $A_1$ & $E$\\
 $E$   &       &       &       &       & $A_1 \oplus A_2 \oplus B_1 \oplus B_2$\\
\hline
\end{tabular}
\caption{The direct products of the irreducible representations (irreps) of the point group $C_{4v}$,
decomposed back into direct sums of the irreps. Due to the obvious symmetry, only the upper diagonal
part of the table is given.}
\label{tab:dprod_C4v}
\end{table}

\begin{table}[htbp]
\renewcommand{\arraystretch}{1.5}
\centering
\begin{tabular}{l|rrrrr}
$O_h$ irrep. & $A_1$ & $A_2$ & $B_1$ & $B_2$ & $E$\\
\hline
 $A_1^+$ & 1 & 0 & 0 & 0 & 0 \\
 $A_2^+$ & 0 & 0 & 1 & 0 & 0 \\
 $E^+  $ & 1 & 0 & 1 & 0 & 0 \\
 $T_1^+$ & 0 & 1 & 0 & 0 & 1 \\
 $T_2^+$ & 0 & 0 & 0 & 1 & 1 \\
 $A_1^-$ & 0 & 1 & 0 & 0 & 0 \\
 $A_2^-$ & 0 & 0 & 0 & 1 & 0 \\
 $E^-  $ & 0 & 1 & 0 & 1 & 0 \\
 $T_1^-$ & 1 & 0 & 0 & 0 & 1 \\
 $T_2^-$ & 0 & 0 & 1 & 0 & 1 \\
\hline
\end{tabular}
\caption{The decomposition of the subduced representations $O_h\downarrow C_{4v}$ in terms of the
irreducible representations of the little group.}
\label{tab:decomp_C4v}
\end{table}

\begin{table}[htbp]
\renewcommand{\arraystretch}{1.5}
\centering
\begin{tabular}{l|rrrr}
$C_{2v}$ & $E$ & $C_2$ & $\sigma_y$ & $\sigma_x$\\
\hline
 $A_1$ & 1 &  1 &  1 &  1 \\
 $B_1$ & 1 & -1 & -1 &  1 \\
 $A_2$ & 1 &  1 & -1 & -1 \\
 $B_2$ & 1 & -1 &  1 & -1 \\
\hline
\end{tabular}
\caption{The character table for the point group $C_{2v}$.}
\label{tab:char_C2v}
\end{table}

\begin{table}[htbp]
\renewcommand{\arraystretch}{1.5}
\centering
\begin{tabular}{l|cccc}
$C_{2v}$ & $A_1$ & $B_1$ & $A_2$ & $B_2$ \\
\hline
 $A_1$ & $A_1$ & $B_1$ & $A_2$ & $B_2$ \\
 $B_1$ &       & $A_1$ & $B_2$ & $A_2$ \\
 $A_2$ &       &       & $A_1$ & $B_1$ \\
 $B_2$ &       &       &       & $A_1$ \\
\hline
\end{tabular}
\caption{The direct products of the irreducible representations (irreps) of the point group $C_{2v}$,
decomposed back into direct sums of the irreps. Due to the obvious symmetry, only the upper diagonal
part of the table is given.}
\label{tab:dprod_C2v}
\end{table}

\begin{table}[htbp]
\renewcommand{\arraystretch}{1.5}
\centering
\begin{tabular}{l|rrrr}
$O_h$ irrep. & $A_1$ & $B_1$ & $A_2$ & $B_2$\\
\hline
 $A_1^+$ & 1 & 0 & 0 & 0 \\
 $A_2^+$ & 0 & 1 & 0 & 0 \\
 $E^+  $ & 1 & 1 & 0 & 0 \\
 $T_1^+$ & 0 & 1 & 1 & 1 \\
 $T_2^+$ & 1 & 0 & 1 & 1 \\
 $A_1^-$ & 0 & 0 & 1 & 0 \\
 $A_2^-$ & 0 & 0 & 0 & 1 \\
 $E^-  $ & 0 & 0 & 1 & 1 \\
 $T_1^-$ & 1 & 1 & 0 & 1 \\
 $T_2^-$ & 1 & 1 & 1 & 0 \\
\hline
\end{tabular}
\caption{The decomposition of the subduced representations $O_h\downarrow C_{2v}$ in terms of the
irreducible representations of the little group.}
\label{tab:decomp_C2v}
\end{table}

\begin{table}[htbp]
\renewcommand{\arraystretch}{1.5}
\centering
\begin{tabular}{l|rrr}
$C_{3v}$ & $E$ & $2C_3$ & $3\sigma_v$\\
\hline
$A_1$ & 1 &  1 &  1 \\
$A_2$ & 1 &  1 & -1 \\
$E$   & 2 & -1 &  0 \\
\hline
\end{tabular}
\caption{The character table for the point group $C_{3v}$.}
\label{tab:char_C3v}
\end{table}

\begin{table}[htbp]
\renewcommand{\arraystretch}{1.5}
\centering
\begin{tabular}{l|ccc}
$C_{3v}$ & $A_1$ & $A_2$ & $E$ \\
\hline
 $A_1$ & $A_1$ & $A_2$ & $E$ \\
 $A_2$ &       & $A_1$ & $E$ \\
 $E$   &       &       & $A_1 \oplus A_2 \oplus E$\\
\hline
\end{tabular}
\caption{The direct products of the irreducible representations (irreps) of the point group $C_{3v}$,
decomposed back into direct sums of the irreps. Due to the obvious symmetry, only the upper diagonal
part of the table is given.}
\label{tab:dprod_C3v}
\end{table}

\begin{table}[htbp]
\renewcommand{\arraystretch}{1.5}
\centering
\begin{tabular}{l|rrr}
$O_h$ irrep. & $A_1$ & $A_2$ & $E$ \\
\hline
 $A_1^+$ & 1 & 0 & 0 \\
 $A_2^+$ & 0 & 1 & 0 \\
 $E^+  $ & 0 & 0 & 1 \\
 $T_1^+$ & 0 & 1 & 1 \\
 $T_2^+$ & 0 & 1 & 1 \\
 $A_1^-$ & 0 & 1 & 0 \\
 $A_2^-$ & 1 & 0 & 0 \\
 $E^-  $ & 0 & 0 & 1 \\
 $T_1^-$ & 1 & 0 & 1 \\
 $T_2^-$ & 1 & 0 & 1 \\
\hline
\end{tabular}
\caption{The decomposition of the subduced representations $O_h\downarrow C_{3v}$ in terms of the
irreducible representations of the little group.}
\label{tab:decomp_C3v}
\end{table}

\begin{table}[htbp]
\renewcommand{\arraystretch}{1.5}
\centering
\begin{tabular}{l|rr}
$C_s$ & $E$ & $\sigma$\\
\hline
$A_1$ & 1 &  1 \\
$A_2$ & 1 & -1 \\
\hline
\end{tabular}
\caption{The character table for the point group $C_s$.}
\label{tab:char_Cs}
\end{table}

\begin{table}[htbp]
\renewcommand{\arraystretch}{1.5}
\centering
\begin{tabular}{l|cc}
$C_s$ & $A_1$ & $A_2$ \\
\hline
$A_1$ & $A_1$ & $A_2$ \\
$A_2$ & $A_2$ & $A_1$ \\
\hline
\end{tabular}
\caption{The direct products of the irreducible representations (irreps) of the point group $C_s$,
decomposed back into direct sums of the irreps.}
\label{tab:dprod_Cs}
\end{table}

\begin{table}[htbp]
\renewcommand{\arraystretch}{1.5}
\centering
\begin{tabular}{l|rr}
$O_h$ irrep. & $A_1$ & $A_2$ \\
\hline
 $A_1^+$ & 1 & 0 \\
 $A_2^+$ & 0 & 1 \\
 $E^+  $ & 1 & 1 \\
 $T_1^+$ & 1 & 2 \\
 $T_2^+$ & 2 & 1 \\
 $A_1^-$ & 0 & 1 \\
 $A_2^-$ & 1 & 0 \\
 $E^-  $ & 1 & 1 \\
 $T_1^-$ & 2 & 1 \\
 $T_2^-$ & 1 & 2 \\
\hline
\end{tabular}
\caption{The decomposition of the subduced representations $O_h\downarrow C_s$ in terms of the
irreducible representations of the little group.}
\label{tab:decomp_Cs}
\end{table}

\begin{table}[htbp]
\renewcommand{\arraystretch}{1.5}
\centering
\begin{tabular}{l|r}
$C_1$ & $E$ \\
\hline
$A_1$ & 1 \\
\hline
\end{tabular}
\caption{The character table for the point group $C_1$.}
\label{tab:char_C1}
\end{table}

\begin{table}[htbp]
\renewcommand{\arraystretch}{1.5}
\centering
\begin{tabular}{l|r}
$O_h$ irrep. & $A_1$ \\
\hline
 $A_1^+$ & 1 \\
 $A_2^+$ & 1 \\
 $E^+  $ & 2 \\
 $T_1^+$ & 3 \\
 $T_2^+$ & 3 \\
 $A_1^-$ & 1 \\
 $A_2^-$ & 1 \\
 $E^-  $ & 2 \\
 $T_1^-$ & 3 \\
 $T_2^-$ & 3 \\
\hline
\end{tabular}
\caption{The decomposition of the subduced representations $O_h\downarrow C_1$ in terms of the
irreducible representations of the little group.}
\label{tab:decomp_C1}
\end{table}

\clearpage

\fancyhead[RE]{\nouppercase{\small\it Appendix \thechapter.\, Collection of Results}}
\chapter{Collection of Results} \label{app:rescoll}

\begin{table}[h]
\renewcommand{\arraystretch}{1.4}
  \begin{center}
    \begin{tabular*}{\textwidth}[c]{c@{\extracolsep{\fill}}ccccc}
      \hline\vspace{-0.05cm} 
      $\beta$ & $p^2$ & $M$ & fit range & $a_t E(p^2)$ & $\chi^2/N_{\text{DF}}$\\
      \hline
      3.0  & 0 & 2 & 1--5 & 1.368(30) & 0.19\\
           & 1 & 3 & 1--6 & 1.372(17) & 0.49\\
           & 2 & 3 & 1--6 & 1.451(17) & 0.28\\
           & 4 & 3 & 1--8 & 1.570(21) & 0.46\\
           & 5 & 3 & 1--6 & 1.610(19) & 0.34\\
           & 8 & 2 & 1--3 & 1.707(29) & 0.05\\
      \hline
  3.15$^a$ & 0 & 3 & 2--5 & 0.688(16) & 0.11\\
           & 1 & 4 & 1--5 & 0.806(5)  & 0.06\\
           & 2 & 4 & 1--5 & 0.902(6)  & 0.59\\
           & 4 & 3 & 1--5 & 1.059(8)  & 2.35\\
           & 5 & 4 & 1--5 & 1.151(7)  & 0.16\\
           & 8 & 4 & 1--5 & 1.310(15) & 0.71\\
           & 9 & 4 & 1--5 & 1.356(23) & 1.58\\
           & 10 & 4 & 1--3 & 1.470(17) & 0.42\\
           & 13 & 3 & 1--3 & 1.621(23) & 0.01\\
           & 18 & 3 & 1--3 & 1.762(47) & 0.56\\
      \hline
  3.15$^b$ & 1 & 3 & 1--5 & 1.418(23) & 0.05\\
           & 2 & 3 & 1--5 & 1.541(37) & 0.53\\
           & 4 & 3 & 1--4 & 1.751(67) & 1.20\\
           & 5 & 3 & 1--5 & 1.922(104) & 1.08\\
      \hline
    \end{tabular*}
    \caption{Collection of results of the torelon measurements using the $\xi=2$ perfect action.
For each $\beta$-value and momentum $p^2=p_x^2+p_y^2$ we list the number of operators $M$ kept after
the first truncation in the variational method, the plateau region on which the fit of the correlators
to the form $Z(p^2)\exp(-t E(p^2))$ is performed (fit range), as well as the extracted energy together
with the $\chi^2$ per degree of freedom ($\chi^2/N_{\text{DF}}$). (\emph{Continued in Table \ref{tab:xi2torcoll2}})}
    \label{tab:xi2torcoll}
  \end{center}
\end{table}

\begin{table}[h]
\renewcommand{\arraystretch}{1.4}
  \begin{center}
    \begin{tabular*}{\textwidth}[c]{c@{\extracolsep{\fill}}ccccc}
      \hline\vspace{-0.05cm} 
      $\beta$ & $p^2$ & $M$ & fit range & $a_t E(p^2)$ & $\chi^2/N_{\text{DF}}$\\
      \hline
      3.3  & 0 & 3 & 1--10 & 0.318(6) & 1.39\\
           & 1 & 3 & 1--10 & 0.417(3) & 1.20\\
           & 2 & 3 & 1--5 & 0.493(3) & 0.13\\
           & 4 & 4 & 1--4 & 0.633(4) & 1.26\\
           & 5 & 5 & 1--10 & 0.688(3) & 0.84\\
           & 8 & 5 & 1--8 & 0.830(5) & 1.21\\
           & 9 & 5 & 1--6 & 0.877(6) & 0.42\\
           & 10 & 5 & 1--10 & 0.921(4) & 0.77\\
           & 13 & 5 & 1--10 & 1.036(6) & 0.30\\
           & 18 & 4 & 1--6 & 1.205(12) & 0.15\\
      \hline
   3.5$^a$ & 1 & 5 & 3--8 & 0.305(6) & 0.10\\
           & 2 & 5 & 1--10 & 0.407(4) & 1.65\\
           & 4 & 4 & 1--10 & 0.538(4) & 0.96\\
           & 5 & 5 & 1--10 & 0.586(3) & 1.16\\
           & 8 & 5 & 1--7 & 0.721(4) & 0.09\\
           & 9 & 5 & 2--6 & 0.733(10) & 0.11\\
           & 10 & 4 & 1--8 & 0.795(4) & 0.96\\
           & 13 & 5 & 1--6 & 0.901(4) & 1.26\\
           & 18 & 5 & 1--6 & 1.058(9) & 1.45\\
      \hline
      3.5$^b$ & 1 & 4 & 3--7 & 0.321(5) & 1.00\\
           & 2 & 5 & 1--8 & 0.432(3) & 0.81\\
           & 4 & 5 & 1--8 & 0.586(4) & 0.29\\
           & 5 & 5 & 1--7 & 0.656(3) & 1.24\\
           & 8 & 5 & 2--7 & 0.806(14) & 0.36\\
           & 9 & 5 & 1--7 & 0.859(7) & 0.74\\
           & 10 & 5 & 1--8 & 0.913(4) & 0.68\\
           & 13 & 5 & 2--5 & 1.010(16) & 0.32\\
           & 18 & 3 & 1--5 & 1.215(12) & 0.17\\
      \hline
    \end{tabular*}
    \caption{Collection of results of the torelon measurements using the $\xi=2$ perfect action.
For each $\beta$-value and momentum $p^2=p_x^2+p_y^2$ we list the number of operators $M$ kept after
the first truncation in the variational method, the plateau region on which the fit of the correlators
to the form $Z(p^2)\exp(-t E(p^2))$ is performed (fit range), as well as the extracted energy together
with the $\chi^2$ per degree of freedom ($\chi^2/N_{\text{DF}}$). (\emph{Continuation from Table \ref{tab:xi2torcoll}})}
    \label{tab:xi2torcoll2}
  \end{center}
\end{table}

\begin{table}[h]
\renewcommand{\arraystretch}{1.5}
  \begin{center}
    \begin{tabular*}{\textwidth}[c]{c@{\extracolsep{\fill}}ccccc}
      \hline\vspace{-0.05cm} 
      $\vec{r}$ & $|\vec{r}|$ & $M$ & fit range & $a_t V(\vec{r})$ & $\chi^2/N_{\text{DF}}$\\
      \hline
      (1,0,0) & 1 & 3 & 2--7 & 0.5053(3) & 0.70\\
      (1,1,0) & 1.414 & 3 & 3--5 & 0.7322(7) & 0.01\\
      (1,1,1) & 1.732 & 3 & 3--8 & 0.8902(14) & 0.60\\
      (2,0,0) & 2 & 3 & 2--8 & 0.9407(10) & 0.89\\
      (2,1,0) & 2.236 & 3 & 2--8 & 1.0642(11) & 0.36\\
      (2,1,1) & 2.449 & 2 & 2--5 & 1.1676(13) & 0.27\\
      (2,2,0) & 2.828 & 3 & 2--6 & 1.3082(18) & 0.17\\
      (3,0,0) & 3 & 3 & 2--6 & 1.3392(25) & 0.16\\
      (2,2,1) & 3 & 3 & 2--5 & 1.3867(20) & 0.26\\
      (2,2,2) & 3.464 & 3 & 2--6 & 1.5675(55) & 0.84\\
      (4,0,0) & 4 & 2 & 1--5 & 1.7429(24) & 0.14\\
      (4,2,0) & 4.472 & 3 & 2--6 & 1.9328(93) & 1.47\\
      (4,2,2) & 4.899 & 3 & 1--4 & 2.1197(34) & 0.15\\
      \hline
    \end{tabular*}
    \caption{Collection of results of the off-axis $q\bar{q}$ measurements using the $\xi=2$ perfect
action at $\beta=3.00$. For each separation vector we list the length of the vector in spatial lattice
units, the number of operators $M$ kept after
the first truncation in the variational method, the plateau region on which the fit of the correlators
to the form $Z(\vec{r})\exp(-t E(\vec{r}))$ is performed (fit range), as well as the extracted energy together
with the $\chi^2$ per degree of freedom ($\chi^2/N_{\text{DF}}$).}
    \label{tab:xi2offax_b300}
  \end{center}
\end{table}

\begin{table}[h]
\renewcommand{\arraystretch}{1.5}
  \begin{center}
    \begin{tabular*}{\textwidth}[c]{c@{\extracolsep{\fill}}ccccc}
      \hline\vspace{-0.05cm} 
      $\vec{r}$ & $|\vec{r}|$ & $M$ & fit range & $a_t V(\vec{r})$ & $\chi^2/N_{\text{DF}}$\\
      \hline
      (1,0,0) & 1 & 3 & 3--10 & 0.3006(1) & 0.20\\
      (1,1,0) & 1.414 & 3 & 3--10 & 0.3814(2) & 0.21\\
      (1,1,1) & 1.732 & 3 & 3--10 & 0.4250(3) & 0.65\\
      (2,0,0) & 2 & 3 & 6--10 & 0.4469(6) & 0.22\\
      (2,1,0) & 2.236 & 3 & 4--10 & 0.4752(4) & 0.41\\
      (2,1,1) & 2.449 & 3 & 4--10 & 0.4971(5) & 0.83\\
      (2,2,0) & 2.828 & 3 & 4--10 & 0.5309(7) & 0.59\\
      (3,0,0) & 3 & 3 & 4--10 & 0.5445(7) & 1.71\\
      (2,2,1) & 3 & 3 & 5--10 & 0.5456(8) & 0.62\\
      (2,2,2) & 3.464 & 3 & 4--10 & 0.5852(9) & 1.53\\
      (4,0,0) & 4 & 3 & 4--10 & 0.6269(1) & 0.51\\
      (3,3,0) & 4.243 & 3 & 5--10 & 0.6455(16) & 0.94\\
      (4,2,0) & 4.472 & 3 & 5--10 & 0.6630(14) & 0.86\\
      (4,2,2) & 4.899 & 3 & 5--10 & 0.6964(18) & 0.34\\
      (5,0,0) & 5 & 2 & 5--10 & 0.7044(22) & 1.24\\
      (3,3,3) & 5.196 & 3 & 4--10 & 0.7215(17) & 1.31\\
      (4,4,0) & 5.657 & 2 & 6--10 & 0.7509(41) & 1.23\\
      (4,4,2) & 6 & 3 & 5--10 & 0.7783(25) & 0.77\\
      \hline
    \end{tabular*}
    \caption{Collection of results of the off-axis $q\bar{q}$ measurements using the $\xi=2$ perfect
action at $\beta=3.30$. For each separation vector we list the length of the vector in spatial lattice
units, the number of operators $M$ kept after
the first truncation in the variational method, the plateau region on which the fit of the correlators
to the form $Z(\vec{r})\exp(-t E(\vec{r}))$ is performed (fit range), as well as the extracted energy together
with the $\chi^2$ per degree of freedom ($\chi^2/N_{\text{DF}}$).}
    \label{tab:xi2offax}
  \end{center}
\end{table}

\begin{table}[h]
\renewcommand{\arraystretch}{1.5}
  \begin{center}
    \begin{tabular*}{\textwidth}[c]{c@{\extracolsep{\fill}}ccccc}
      \hline\vspace{-0.05cm} 
      $\beta$ & $r$ & $M$ & fit range & $a_t V(r)$ & $\chi^2/N_{\text{DF}}$\\
      \hline
      3.15 & 1 & 5 & 3--10 & 0.3703(2) & 0.456\\
           & 2 & 5 & 3--7 &  0.6110(4) & 0.735\\
           & 3 & 5 & 3--10 & 0.8019(8) & 1.209\\
           & 4 & 5 & 3--7 & 0.9775(16) & 1.238\\
           & 5 & 5 & 4--9 & 1.129(8)   & 0.519\\
      \hline
      3.50 & 1 & 5 & 3--12 & 0.25910(5) & 1.163\\
           & 2 & 5 & 3--10 & 0.36490(12) & 0.769\\
           & 3 & 5 & 7--12 & 0.42168(38) & 0.664\\
           & 4 & 5 & 5--11 & 0.46587(51) & 0.205\\
           & 5 & 5 & 5--12 & 0.50388(76) & 1.139\\
           & 6 & 5 & 5--11 & 0.54051(96) & 0.853\\
      \hline
    \end{tabular*}
    \caption{Collection of results of the on-axis $q\bar{q}$ measurements using the $\xi=2$ perfect
action at $\beta=$ 3.15, 3.50. For each separation along the axes we list the number of operators $M$ kept after
the first truncation in the variational method, the plateau region on which the fit of the correlators
to the form $Z(r)\exp(-t E(r))$ is performed (fit range), as well as the extracted energy together
with the $\chi^2$ per degree of freedom ($\chi^2/N_{\text{DF}}$).}
    \label{tab:xi2onaxpot}
  \end{center}
\end{table}

\begin{table}[htbp]
  \begin{center}
    \renewcommand{\arraystretch}{1.5}
    \begin{tabular*}{\textwidth}{c@{\extracolsep{\fill}}cccccl}
      \hline
      Channel & $N$ & $t_0/t_1$ & $M$ & fit range &
      $\chi^2/N_{\text{DF}}$ & energies \\
      \hline
      $A_1^{++}$   & 87 & 1/2 & 12 & 1 -- 4 & 0.69 & \bf{0.645(10)} \\
    ${A_1^{++}}^*$ & 59 & 1/2 & 10 & 1 -- 3 & 1.23 & \bf{1.365(104)} \\
      $E^{++}$     & 47 & 1/2 & 7  & 1 -- 3 & 0.10 & \bf{1.405(29)} \\
      ${E^{++}}^*$ & 47 & 1/2 & 5  & 1 -- 3 & 0.17 & 1.985(140) \\
      $T_2^{++}$   & 47 & 1/2 & 10 & 1 -- 4 & 0.10 & \bf{1.416(32)} \\
    ${T_2^{++}}^*$ & 22 & 1/2 & 6  & 1 -- 3 & 1.20 & 1.990(131) \\
      $A_2^{++}$   & 21 & 1/2 & 2  & 1 -- 3 & 0.03 & \bf{1.675(147)} \\
      $T_1^{++}$   & 36 & 1/2 & 4  & 1 -- 3 & 0.34 & \bf{1.881(131)} \\
  ${T_1^{++}}^*$   & 36 & 1/2 & 6  & 1 -- 3 & 0.31 & 2.170(182)\\
      $A_1^{-+}$   & 25 & 1/2 & 3  & 1 -- 3 & 0.001 & \bf{1.588(114)} \\
      $E^{-+}$     & 25 & 1/2 & 2  & 1 -- 3 & 0.01 & 1.849(81) \\
      ${E^{-+}}^*$ & 25 & 1/2 & 2  & 1 -- 3 & 2.19 & 2.084(268) \\
      $T_2^{-+}$   & 64 & 1/2 & 4  & 1 -- 3 & 0.02 & \bf{1.850(83)} \\
      $T_1^{+-}$   & 61 & 1/2 & 4  & 1 -- 3 & 0.24 & \bf{1.798(59)} \\
      $T_2^{+-}$   & 18 & 1/2 & 4  & 1 -- 3 & 0.41 & \bf{1.935(177)} \\
    \hline
    \end{tabular*}
    \caption{{}Results from fits to the $\xi=2$, $\beta=3.15$ glueball correlators on
      the $8^3\times 16$ lattice in units of the temporal lattice spacing $a_t$:
      $t_0$/$t_1$ are used in the generalised eigenvalue
      problem, $N$ is the number of initial operators measured and $M$ denotes the number of operators kept after 
      the truncation in $C(t_0)$.}
    \label{tab:b315_fit_results}
  \end{center}
\end{table}

\begin{table}[htbp]
  \begin{center}
    \renewcommand{\arraystretch}{1.5}
    \begin{tabular*}{\textwidth}{c@{\extracolsep{\fill}}ccccl}
      \hline
      Channel & $t_0/t_1$ & $M$ & fit range &
      $\chi^2/N_{\text{DF}}$ & energies \\
      \hline
      $A_1^{++}$   & 1/2 & 9  & 2 -- 7 & 0.90 & \bf{0.590(17)} \\
    ${A_1^{++}}^*$ & 1/2 & 11 & 1 -- 3 & 1.57 & \bf{1.133(53)} \\
      $E^{++}$     & 1/2 & 11 & 1 -- 5 & 0.97 & \bf{0.983(17)} \\
      ${E^{++}}^*$ & 1/2 & 8  & 1 -- 3 & 0.90 & \bf{1.453(68)} \\
      $T_2^{++}$   & 1/2 & 12 & 1 -- 4 & 1.09 & \bf{0.965(15)} \\
    ${T_2^{++}}^*$ & 1/2 & 10 & 1 -- 4 & 1.68 & \bf{1.386(51)} \\
      $A_2^{++}$   & 1/2 & 4 & 1 -- 3 & 0.01 & 1.511(94) \\
      $T_1^{++}$   & 1/2 & 8 & 1 -- 5 & 0.84 & 1.492(62) \\
      $A_1^{-+}$   & 1/2 & 6 & 1 -- 4 & 4.38 & \bf{1.017(39)} \\
      $E^{-+}$     & 1/2 & 3 & 1 -- 4 & 0.74 & \bf{1.365(36)} \\
      $T_2^{-+}$   & 1/2 & 4 & 1 -- 5 & 1.16 & \bf{1.357(25)} \\
      $T_1^{+-}$   & 1/2 & 8 & 1 -- 4 & 0.95 & \bf{1.276(33)} \\
    \hline
    \end{tabular*}
    \caption{{}Results from fits to the $\xi=2$, $\beta=3.3$ glueball correlators on
      the $10^3\times 20$ lattice in units of the temporal lattice spacing $a_t$:
      $t_0$/$t_1$ are used in the generalised eigenvalue
      problem, $M$ denotes the number of operators kept after 
      the truncation in $C(t_0)$.}
    \label{tab:b330_fit_results}
  \end{center}
\end{table}

\begin{table}[htbp]
  \begin{center}
    \renewcommand{\arraystretch}{1.5}
    \begin{tabular*}{\textwidth}{c@{\extracolsep{\fill}}cccccl}
      \hline
      Channel & $N$ & $t_0/t_1$ & $M$ & fit range &
      $\chi^2/N_{\text{DF}}$ & energies \\
      \hline
      $A_1^{++}$   & 91  & 1/2 & 24  & 2 -- 7 & 0.31 & \bf{0.405(13)} \\
    ${A_1^{++}}^*$ & 110 & 0/1 & 110 & 4 -- 6 & 0.30 & \bf{0.720(158)} \\
      $E^{++}$     & 104 & 1/2 & 23  & 2 -- 5 & 0.16 & \bf{0.675(25)} \\
      ${E^{++}}^*$ & 104 & 1/2 & 14  & 1 -- 3 & 0.70 & \bf{1.183(55)} \\
      $T_2^{++}$   & 53  & 1/2 & 19  & 2 -- 5 & 0.31 & \bf{0.681(12)} \\
    ${T_2^{++}}^*$ & 53  & 1/2 & 22  & 1 -- 3 & 0.69 & \bf{1.128(23)} \\
      $A_2^{++}$   & 15  & 1/2 & 6   & 1 -- 3 & 0.05 & \bf{1.212(42)} \\
    ${A_2^{++}}^*$ & 15  & 1/2 & 6   & 1 -- 3 & 1.12 & \bf{1.568(130)} \\
      $T_1^{++}$   & 23  & 1/2 & 11  & 1 -- 4 & 0.05 & \bf{1.245(25)} \\
      $A_1^{-+}$   & 15  & 1/2 & 9   & 1 -- 3 & 0.17 & \bf{0.754(23)} \\
    ${A_1^{-+}}^*$ & 15  & 1/2 & 9   & 1 -- 3 & 0.61 & \bf{1.140(112)} \\
      $E^{-+}$     & 21  & 1/2 & 14  & 1 -- 3 & 0.27 & \bf{0.942(24)} \\
      ${E^{-+}}^*$ & 21  & 1/2 & 11  & 1 -- 3 & 1.20 & \bf{1.406(80)} \\
      $T_2^{-+}$   & 75  & 1/2 & 9   & 2 -- 5 & 0.32 & \bf{0.952(38)} \\
    ${T_2^{-+}}^*$ & 75  & 1/2 & 6   & 1 -- 3 & 1.70 & \bf{1.675(59)} \\
      $T_1^{+-}$   & 53  & 1/2 & 14  & 1 -- 3 & 0.69 & \bf{1.013(17)} \\
      $A_2^{+-}$   & 11  & 1/2 & 5   & 1 -- 3 & 0.07 & \bf{1.224(64)} \\
    ${A_2^{+-}}^*$ & 11  & 1/2 & 5   & 1 -- 3 & 0.79 & 1.726(147) \\
      $T_2^{+-}$   & 22  & 1/2 & 9   & 2 -- 4 & 0.33 & \bf{1.142(121)} \\
    ${T_2^{+-}}^*$ & 12  & 1/2 & 9   & 1 -- 3 & 1.78 & 1.663(139) \\
    $E^{+-}$       & 50  & 0/1 & 50  & 1 -- 3 & 1.01 & 1.379(34) \\
    $T_1^{--}$     & 80  & 0/1 & 80  & 2 -- 4 & 0.57 & 1.200(67) \\
    $T_2^{--}$     & 80  & 0/1 & 80  & 2 -- 5 & 0.20 & 1.205(81) \\
    $A_2^{--}$     & 35  & 0/1 & 35  & 1 -- 3 & 0.16 & 1.352(37) \\
    $A_1^{--}$     & 30  & 0/1 & 30  & 1 -- 4 & 0.14 & 1.703(99) \\
    \hline
    \end{tabular*}
    \caption{{}Results from fits to the $\xi=2$, $\beta=3.5$ glueball correlators on
      the $12^3\times 24$ lattice in units of the temporal lattice spacing $a_t$:
      $t_0$/$t_1$ are used in the generalised eigenvalue
      problem, $N$ is the number of initial operators measured and $M$ denotes the number of operators kept after 
      the truncation in $C(t_0)$.}
    \label{tab:b350_fit_results}
  \end{center}
\end{table}

\begin{table}[htbp]
  \begin{center}
    \renewcommand{\arraystretch}{1.5}
    \begin{tabular*}{\textwidth}{c@{\extracolsep{\fill}}ccc}
      \hline
      Channel & Glueballs & $p^2$ & $r_0\cdot m_{2G}$\\
      \hline
      $A_1^{++}$ & ($A_1^{++}$, $A_1^{++}$) & 0 & 5.16\\
      $E^{++}$   & ($A_1^{++}$, $A_1^{++}$) & 1 & 6.07\\
      $T_2^{++}$ & ($A_1^{++}$, $A_1^{++}$) & 2 & 6.86\\
      $A_2^{++}$ & ($A_1^{++}$, $E^{++}$)   & 1 & 8.89\\
      $T_1^{++}$ & ($A_1^{++}$, $A_1^{++}$) & 5 & 8.82\\
      $A_1^{-+}$ & ($A_1^{++}$, $A_1^{-+}$) & 0 & 8.94\\
      $E^{-+}$   & ($A_1^{++}$, $T_2^{++}$) & 1 & 8.93\\
      $T_2^{-+}$ & ($A_1^{++}$, $E^{++}$)   & 1 & 8.89\\
      $T_1^{+-}$ & ($A_1^{++}$, $T_1^{+-}$) & 0 & 9.78\\
      $T_2^{+-}$ & ($A_1^{++}$, $T_2^{+-}$) & 0 & 10.33\\
      \hline
    \end{tabular*}
    \caption{{}The energies of the lowest-lying two-glueball states with zero momentum for all the symmetry channels, given
in terms of $r_0$, using the masses determined in the $\beta=3.15$ simulation. The square of the momenta $p^2$ of 
the single glueballs building the pair is given in units of $(2\pi/8a_s)^2=(1.60/r_0)^2$.}
    \label{tab:gbpair_b315}
  \end{center}
\end{table}

\begin{table}[htbp]
  \begin{center}
    \renewcommand{\arraystretch}{1.5}
    \begin{tabular*}{\textwidth}{c@{\extracolsep{\fill}}ccc}
      \hline
      Channel & Glueballs & $p^2$ & $r_0\cdot m_{2G}$\\
      \hline
      $A_1^{++}$ & ($A_1^{++}$, $A_1^{++}$) & 0 & 7.10\\
      $E^{++}$   & ($A_1^{++}$, $A_1^{++}$) & 1 & 8.13\\
      $T_2^{++}$ & ($A_1^{++}$, $A_1^{++}$) & 2 & 9.04\\
      $A_2^{++}$ & ($A_1^{++}$, $E^{++}$)   & 1 & 10.32\\
      $T_1^{++}$ & ($A_1^{++}$, $T_2^{++}$) & 1 & 10.21\\
      $A_1^{-+}$ & ($A_1^{++}$, $A_1^{-+}$) & 0 & 9.68\\
      $E^{-+}$   & ($A_1^{++}$, $T_2^{++}$) & 1 & 10.21\\
      $T_2^{-+}$ & ($A_1^{++}$, $T_2^{++}$)   & 1 & 10.21\\
      $T_1^{+-}$ & ($A_1^{++}$, $T_1^{+-}$) & 0 & 11.24\\
      \hline
    \end{tabular*}
    \caption{{}The energies of the lowest-lying two-glueball states with zero momentum for all the symmetry channels, given
in terms of $r_0$, using the masses determined in the $\beta=3.30$ simulation. The square of the momenta $p^2$ of 
the single glueballs building the pair is given in units of $(2\pi/10a_s)^2=(1.98/r_0)^2$.}
    \label{tab:gbpair_b330}
  \end{center}
\end{table}

\begin{table}[htbp]
  \begin{center}
    \renewcommand{\arraystretch}{1.5}
    \begin{tabular*}{\textwidth}{c@{\extracolsep{\fill}}ccc}
      \hline
      Channel & Glueballs & $p^2$ & $r_0\cdot m_{2G}$\\
      \hline
      $A_1^{++}$ & ($A_1^{++}$, $A_1^{++}$) & 0 & 7.30\\
      $E^{++}$   & ($A_1^{++}$, $A_1^{++}$) & 1 & 8.93\\
      $T_2^{++}$ & ($A_1^{++}$, $T_2^{++}$) & 0 & 9.78\\
      $A_2^{++}$ & ($A_1^{++}$, $E^{++}$)   & 1 & 11.06\\
      $T_1^{++}$ & ($A_1^{++}$, $T_2^{++}$) & 1 & 11.11\\
      $A_1^{-+}$ & ($A_1^{++}$, $A_1^{-+}$) & 0 & 10.44\\
      $E^{-+}$   & ($A_1^{++}$, $T_2^{++}$) & 1 & 11.11\\
      $T_2^{-+}$ & ($A_1^{++}$, $E^{++}$)   & 1 & 11.06\\
      $T_1^{+-}$ & ($A_1^{++}$, $T_1^{+-}$) & 0 & 12.77\\
      $A_2^{+-}$ & ($A_1^{++}$, $A_2^{+-}$) & 0 & 14.67\\
      $T_2^{+-}$ & ($A_1^{++}$, $T_2^{+-}$) & 0 & 13.94\\
      $E^{+-}$   & ($A_1^{++}$, $T_1^{+-}$) & 2 & 14.97\\
      $T_1^{--}$ & ($A_1^{++}$, $T_1^{+-}$) & 1 & 13.94\\
      $T_2^{--}$ & ($A_1^{++}$, $T_1^{+-}$) & 1 & 13.94\\
      $A_2^{--}$ & ($A_1^{++}$, $T_1^{+-}$) & 2 & 14.97\\
      $A_1^{--}$ & ($A_1^{++}$, $T_1^{+-}$) & 1 & 13.94\\
      \hline
    \end{tabular*}
    \caption{{}The energies of the lowest-lying two-glueball states with zero momentum for all the symmetry channels, given
in terms of $r_0$, using the masses determined in the $\beta=3.50$ simulation. The square of the momenta $p^2$ of 
the single glueballs building the pair is given in units of $(2\pi/12a_s)^2=(2.57/r_0)^2$.}
    \label{tab:gbpair_b350}
  \end{center}
\end{table}

\begin{table}[htbp]
  \begin{center}
    \renewcommand{\arraystretch}{1.5}
    \begin{tabular*}{\textwidth}{c@{\extracolsep{\fill}}ccc}
      \hline
      Ratio & $\beta=3.15$ & $\beta=3.30$ & $\beta=3.50$\\
      \hline
      $m({A_1^{++}}^*)/m({A_1^{++}})$ & 2.12(19) & 1.92(15) & 1.78(45)\\ 
      $m({E^{++}})/m({A_1^{++}})$ &     2.18(8)  & 1.67(8)  & 1.67(12)\\
      $m({E^{++}}^*)/m({A_1^{++}})$ &   3.08(26) & 2.46(19) & 2.92(23)\\
      $m({T_2^{++}})/m({A_1^{++}})$ &   2.20(8)  & 1.64(7)  & 1.68(8)\\
      $m({T_2^{++}}^*)/m({A_1^{++}})$ & 3.09(25) & 2.35(15) & 2.79(15)\\
      $m({A_2^{++}})/m({A_1^{++}})$ &   2.60(27) & 2.56(23) & 2.99(20)\\
      $m({A_2^{++}}^*)/m({A_1^{++}})$ &          &          & 3.87(45)\\
      $m({T_1^{++}})/m({A_1^{++}})$ &   2.92(25) & 2.53(18) & 3.07(16)\\
      $m({A_1^{-+}})/m({A_1^{++}})$ &   2.46(21) & 1.72(12) & 1.86(12)\\
      $m({A_1^{-+}}^*)/m({A_1^{++}})$ &          &          & 2.81(37)\\
      $m({E^{-+}})/m({A_1^{++}})$ &     2.87(17) & 2.31(13) & 2.33(13)\\
      $m({E^{-+}}^*)/m({A_1^{++}})$ &   3.23(47) &          & 3.47(31)\\
      $m({T_2^{-+}})/m({A_1^{++}})$ &   2.87(17) & 2.30(11) & 2.35(17)\\
      $m({T_2^{-+}}^*)/m({A_1^{++}})$ &          &          & 4.14(28)\\
      $m({T_1^{+-}})/m({A_1^{++}})$ &   2.79(13) & 2.16(12) & 2.50(12)\\
      $m({A_2^{+-}})/m({A_1^{++}})$ &            &          & 3.02(26)\\
      $m({T_2^{+-}})/m({A_1^{++}})$ &            &          & 2.82(39)\\
      \hline
    \end{tabular*}
    \caption{{}Glueball mass ratios with the scalar glueball mass, $m_G/m({A_1^{++}})$.}
    \label{tab:a1pp_ratio}
  \end{center}
\end{table}

\begin{table}[htbp]
  \begin{center}
    \renewcommand{\arraystretch}{1.5}
    \begin{tabular*}{\textwidth}{c@{\extracolsep{\fill}}ccc}
      \hline
      Ratio & $\beta=3.15$ & $\beta=3.30$ & $\beta=3.50$\\
      \hline
      $m({A_1^{++}})/m({T_2^{++}})$ &   0.46(2)  & 0.61(3)  & 0.59(3) \\ 
      $m({A_1^{++}}^*)/m({T_2^{++}})$ & 0.96(10) & 1.17(7)  & 1.06(25)\\ 
      $m({E^{++}})/m({T_2^{++}})$ &     0.99(4)  & 1.02(3)  & 0.99(5) \\
      $m({E^{++}}^*)/m({T_2^{++}})$ &   1.40(13) & 1.51(9)  & 1.74(11)\\
      $m({T_2^{++}}^*)/m({T_2^{++}})$ & 1.41(19) & 1.44(8)  & 1.66(6) \\
      $m({A_2^{++}})/m({T_2^{++}})$ &   1.18(13) & 1.57(12) & 1.78(9) \\
      $m({A_2^{++}}^*)/m({T_2^{++}})$ &          &          & 2.30(23)\\
      $m({T_1^{++}})/m({T_2^{++}})$ &   1.33(12) & 1.55(9)  & 1.83(7) \\
      $m({A_1^{-+}})/m({T_2^{++}})$ &   1.12(11) & 1.05(6)  & 1.11(5) \\
      $m({A_1^{-+}}^*)/m({T_2^{++}})$ &          &          & 1.67(19)\\
      $m({E^{-+}})/m({T_2^{++}})$ &     1.31(9)  & 1.41(6)  & 1.38(6) \\
      $m({E^{-+}}^*)/m({T_2^{++}})$ &   1.47(22) &          & 2.06(15)\\
      $m({T_2^{-+}})/m({T_2^{++}})$ &   1.31(9)  & 1.41(5)  & 1.40(8) \\
      $m({T_2^{-+}}^*)/m({T_2^{++}})$ &          &          & 2.46(13)\\
      $m({T_1^{+-}})/m({T_2^{++}})$ &   1.27(7)  & 1.32(5)  & 1.49(5) \\
      $m({A_2^{+-}})/m({T_2^{++}})$ &            &          & 1.80(13)\\
      $m({T_2^{+-}})/m({T_2^{++}})$ &            &          & 1.68(21)\\
      \hline
    \end{tabular*}
    \caption{{}Glueball mass ratios with the tensor glueball mass, $m_G/m({T_2^{++}})$.}
    \label{tab:t2pp_ratio}
  \end{center}
\end{table}

\begin{table}[htbp]
  \begin{center}
    \renewcommand{\arraystretch}{1.5}
    \begin{tabular*}{\textwidth}{c@{\extracolsep{\fill}}cccc}
      \hline
      Channel & $J$ & terms in the fit & $\chi^2/N_{\text{DF}}$ & $r_0 m_G$\\
      \hline
      $A_1^{++}$   & 0 & $c$, $c_6$ & 0.99 & \bf{3.64(10)}\\
    ${A_1^{++}}^*$ & 0 & $c$, $c_6$ & 1.06 & \bf{6.90(38)}\\
      $E^{++}$     & 2 & $c$, $c_2$ & 1.06 & \bf{6.16(24)}\\
      ${E^{++}}^*$ & 2 & $c$, $c_2$ & 3.85 & \bf{10.59(63)}\\
      $T_2^{++}$   & 2 & $c$, $c_2$ & 1.98 & \bf{6.14(18)}\\
      $A_2^{++}$   & 3 & $c$, $c_2$ & 1.65 & \bf{11.65(57)}\\
      $A_1^{-+}$   & 0 & $c$        & 2.50 & \bf{6.46(18)}\\
      $E^{-+}$     & 2 & $c$, $c_2$ & 1.06 & \bf{8.73(34)}\\
      $T_2^{-+}$   & 2 & $c$, $c_2$ & 1.07 & \bf{8.77(39)}\\
    \hline
    \end{tabular*}
    \caption{{}Results of the continuum extrapolations of selected glueball representations in terms of the
hadronic scale $r_0$. The continuum spin assignment $J$, the terms in the fit, constant ($c$), $(a_s/r_0)^2$ ($c_2$) and $(a_s/r_0)^6$ ($c_6$) and
the goodness of the fit, $\chi^2/N_{\text{DF}}$, are also given.}
    \label{tab:cont_ex}
  \end{center}
\end{table}

\begin{table}[htbp]
  \begin{center}
    \renewcommand{\arraystretch}{1.5}
    \begin{tabular*}{\textwidth}{c@{\extracolsep{\fill}}ccccc}
      \hline
      $\Gamma^{PC}$ & $J$ & terms in the fit & $\chi^2/N_{\text{DF}}$ & $m_{\Gamma^{PC}}/m_{T_2^{++}}$ & $r_0 m_G$\\
      \hline
      $E^{++}$     & 2 & $c$ & 1.20 & 1.01(2) & 6.20(30)\\
      ${E^{++}}^*$ & 2 & $c$, $c_2$ & 2.20 & 1.74(12) & 10.68(105)\\
     ${T_2^{++}}^*$ & 2 & $c$, $c_2$ & 3.01 & 1.70(8) & \bf{10.44(80)}\\
      $A_2^{++}$   & 3 & $c$, $c_2$ & 1.11 & 1.90(11) & 11.66(102)\\
      $T_1^{++}$   & 3 & $c$, $c_2$ & 2.60 & 1.90(8) & \bf{11.66(83)}\\
      $A_1^{-+}$   & 0 & $c$        & 1.39 & 1.09(4) & 6.69(44)\\
      $T_2^{-+}$   & 2 & $c$, $c_6$ & 1.00 & 1.41(4) & 8.66(50)\\
      $T_1^{+-}$   & 1 & $c$, $c_2$ & 2.29 & 1.54(7) & \bf{9.45(71)}\\
    \hline
    \end{tabular*}
    \caption{{}Results of the continuum extrapolations of selected glueball ratios, $m_G/m_{T_2^{++}}$. The
 continuum spin assignment $J$, the terms in the fit, constant ($c$), $(a_s/r_0)^2$ ($c_2$) and
 $(a_s/r_0)^6$ ($c_6$) and the goodness of the fit, $\chi^2/N_{\text{DF}}$, are also given. The last
column lists the masses converted to units of $r_0^{-1}$ using the continuum result for the mass of the
tensor glueball $T_2^{++}$, values in bold face will be used further on.}
    \label{tab:cont_ex_ratio}
  \end{center}
\end{table}

\begin{table}[h]
\renewcommand{\arraystretch}{1.5}
  \begin{center}
    \begin{tabular*}{\textwidth}[c]{c@{\extracolsep{\fill}}ccccc}
      \hline\vspace{-0.05cm} 
      $\beta$ & $p^2$ & $M$ & fit range & $a_t E(p^2)$ & $\chi^2/N_{\text{DF}}$\\
      \hline
      3.0  & 1 & 5 & 3--10 & 0.317(5) & 1.34\\
           & 2 & 3 & 4--10 & 0.375(8) & 0.31\\
           & 4 & 3 & 3--11 & 0.487(7) & 0.65\\
           & 5 & 5 & 4--11 & 0.520(11) & 0.35\\
           & 8 & 5 & 3--7 &  0.663(17) & 0.09\\
           & 9 & 3 & 4--11 & 0.754(37) & 0.87\\
           & 10 & 5 & 4--11 & 0.678(32) & 1.56\\
      \hline
    \end{tabular*}
    \caption{Collection of results of the torelon measurement at $\beta=3.0$, using the $\xi=4$ perfect action.
For each momentum $p^2=p_x^2+p_y^2$ we list the number of operators $M$ kept after
the first truncation in the variational method, the plateau region on which the fit of the correlators
to the form $Z(p^2)\exp(-t E(p^2))$ is performed (fit range), as well as the extracted energy together
with the $\chi^2$ per degree of freedom ($\chi^2/N_{\text{DF}}$).}
    \label{tab:xi4tor}
  \end{center}
\end{table}

\end{appendix}

\fancyhead[RE]{\nouppercase{\small\it Bibliography}}

\end{document}